\documentclass[a4paper,10pt]{report}
\usepackage{graphicx}
\usepackage{bm}
\usepackage{amsmath}
\usepackage{amssymb}
\def\sub#1{_{\mathrm{#1}}}
\def\up#1{^{\mathrm{#1}}}

\title{Quantized vortices in superfluid helium and atomic Bose-Einstein condensates}

\author{Makoto Tsubota$^1$, Kenichi Kasamatsu$^2$, Michikazu Kobayashi$^3$ \\ \\
$^1$ Department of Physics, Osaka City University, Japan \\
$^2$ Department of Physics, Kinki University, Japan \\
$^3$ Department of Pure and Applied Sciences, University of Tokyo, Japan}

\begin{document}
\maketitle

\baselineskip 14pt
\setcounter{page}{1}

\chapter{Introduction}

Bose-Einstein condensation is often considered to be a macroscopic quantum phenomenon. 
This is because bosons occupy the same single-particle ground state below a critical temperature through Bose-Einstein condensation to form a macroscopic wave function (order parameter) extending over the entire system.
As a direct result of the formation of a macroscopic wave function, quantized vortices appear in the Bose-condensed system.  
A quantized vortex is a vortex of inviscid superflow, and any rotational motion of a superfluid is sustained by quantized vortices.
A quantized vortex is a stable and well-defined topological defect, very different from ordinary vortices in a conventional fluid. 
Hydrodynamics dominated by quantized vortices is called {\it quantum hydrodynamics}, and turbulence comprised of quantized vortices is known as {\it quantum turbulence} (QT). 
The studies of quantized vortices originally began in 1950s using superfluid $^4$He, and much theoretical, numerical, and experimental effort has been devoted to the field.
Superfluid $^3$He, discovered in 1972, presented a system with a variety of quantized vortices characteristic of p-wave superfluids.   
However, quantum hydrodynamics has become very important again in recent years, for two reasons.
The first reason was the appearance of new research activity on QT \cite{PLTP,TsubotaJPSJ}.
Before 1990, most studies of QT were limited to thermal counterflow in $^4$He in which the normal fluid and the superfluid flow oppositely through the injection of heat current.
Since counterflow turbulence has no classical analog, understanding the relationship between QT and traditional classical turbulence (CT) has been deserted. 
However, some experiments without counterflow appeared in the 1990s, demonstrating the Kolmogorov law of energy spectra which is one of the most important statistical laws in turbulence, thus opening a new stage of QT research.
The second reason is that Bose-Einstein condensation of cold atoms was realized in 1995.
This system enables the control and direct visualization of condensates, which was impossible in other quantum condensed systems, such as superfluid helium and superconductors.
The present article will review these recent studies on quantized vortices in both superfluid helium and cold atoms.  

 The contents of this article are as follows.
Section 2 deals with quantized vortices in superfluid helium.
A brief review of the research history and the dynamics of quantized vortices is followed by an important modern topic, QT, which includes the energy spectra issue, dissipation at very low temperatures, QT created by vibrating structures, and the visualization of quantized vortices and turbulence.
In Section 3, which begins with the basics on atomic Bose-Einstein condensates (BECs), we review the topics of vortices in single-component, two-component, spinor and dipolar BECs, and then discuss quantum turbulence.
Section 4 is devoted to a summary and conclusions.   

\chapter{Quantized vortices in superfluid helium}
Quantum hydrodynamics and quantized vortices have been long studied in superfluid $^4$He.
This section begins by describing the research history and some basic dynamics of quantized vortices, and then discusses the modern topics, QT.
\section{Research history}
Liquid $^4$He enters a superfluid state below the $\lambda$ point ($T_\lambda=$2.17 K) with
Bose--Einstein condensation of the $^4$He atoms. 
The characteristic phenomena of superfluidity were experimentally discovered  
in the 1930s by Kapitza \cite{Kapitza} and Allen {\it et al.}\cite{Allen}.
The hydrodynamics of superfluid helium are well described by the two-fluid model proposed by Landau \cite{Landau} and Tisza\cite{Tisza}. 
According to the two-fluid model, the system consists of an inviscid superfluid (density $\rho\sub{s}$) and 
a viscous normal fluid (density $\rho\sub{n}$) with two independent velocity fields $\mathbf{v}\sub{s}$ and $\mathbf{v}\sub{n}$. 
The mixing ratio of the two fluids depends on the temperature. 
As the temperature is reduced below the $\lambda$ point, the ratio of the superfluid component increases, and the fluid becomes entirely superfluid below about 1 K.
The two-fluid model successfully explained the phenomena of superfluidity, while it was known in 1940s that superfluidity breaks down when it flows fast \cite{GM} and this phenomenon was not explained through the two-fkuid model. 
This was later found to be caused by turbulence of the superfluid component due to random motion of quantized vortices.  

The $\lambda$ transition is closely related to the Bose-Einstein condensation of $^4$He atoms, as first proposed by London \cite{London}.
The Bose-condensed system exhibits the macroscopic wave function
$\Psi(\mathbf{x},t)=|\Psi(\mathbf{x},t)| e^{i \theta(\mathbf{x},t)}$ as an order parameter. 
The superfluid velocity field is given by $\mathbf{v}\sub{s}=(\hbar/m) \nabla \theta$, with boson mass $m$,
representing the potential flow. 
Since the macroscopic wave function should be single-valued for the space coordinate $\mathbf{x}$,
the circulation $\Gamma = \oint \mathbf{v} \cdot d\mathbf{\ell}$ for an arbitrary closed loop in the fluid
is quantized by the quantum $\kappa=h/m$.  A vortex with quantized circulation is 
called a quantized vortex. Any rotational motion of a superfluid is sustained only
by quantized vortices.

A quantized vortex is a topological defect characteristic of a Bose--Einstein condensate, and is different from a vortex in a classical viscous fluid. First, the circulation is quantized, which is contrary to a classical vortex that can have any circulation value. Second, a quantized vortex is a vortex of inviscid superflow. Thus, it cannot decay by the viscous diffusion of vorticity that occurs in a classical fluid.  Third, the core of a quantized vortex is very thin, of the order of the coherence length, which is only a few angstroms in superfluid $^4$He. Because the vortex core is very thin and does not decay by diffusion, it is always possible to identify the position of a quantized vortex in the fluid. These properties make a quantized vortex more stable and definite than a classical vortex.

The idea of quantized circulation was first proposed by Onsager, for a series of annular rings in a rotating superfluid \cite{Onsager}.
Feynman considered that a vortex in a superfluid can take the form of a vortex filament, with the quantized circulation $\kappa$ and a core of atomic dimension \cite{Feynman}.
Early experimental studies on superfluid hydrodynamics focused primarily on thermal counterflow. 
The flow is driven by an injected heat current, and the normal fluid and superfluid flow in opposite directions. The superflow was found to become dissipative when the relative velocity between the two fluids exceeds a critical value \cite{GM}. 
Gorter and Mellink attributed the dissipation to mutual friction between two fluids, and considered the possibility of superfluid turbulence. Feynman proposed a turbulent superfluid state consisting of a tangle of quantized vortices \cite{Feynman}. 
Hall and Vinen performed the experiments of second sound attenuation in rotating $^4$He, and found that the mutual friction arises from interaction between the normal fluid and quantized vortices \cite{HallVinen56a,HallVinen56b}; second sound refers to entropy wave in which superfluid and normal fluid oscillate oppositely, and its propagation and attenuation give the information of the vortex density in the fluid.
Vinen confirmed Feynman's findings experimentally, by showing that the dissipation arises from mutual friction between vortices and the normal flow \cite{Vinen57a,Vinen57b,Vinen57c,Vinen57d}. 
Vinen also succeeded in observing quantized circulation using vibrating wires in rotating superfluid $^4$He \cite{Vinen61}.
Subsequently, many experimental studies have examined superfluid turbulence (ST) in thermal counterflow systems, and have revealed a variety of physical phenomena\cite{Tough82}. Since the dynamics of quantized vortices are nonlinear and non-local, it has not been easy to quantitatively understand these observations on the basis of vortex dynamics. Schwarz clarified the picture of ST based on tangled vortices by numerical simulation of the quantized vortex filament model in the thermal counterflow \cite{Schwarz85,Schwarz88}. However, since the thermal counterflow has no analogy in conventional fluid dynamics, this study was not helpful in clarifying the relationship between ST and classical turbulence (CT). Superfluid turbulence is often called quantum turbulence (QT), which emphasizes the belief that it is comprised of quantized vortices.

Turbulence has long been one of the great mysteries in nature, with discussion dating back to the era of Leonardo da Vinci. Turbulence has been intensely studied in a number of fields, but it is still far from completely understood. This is primarily because turbulence is a complicated dynamical phenomenon with strong nonlinearity.  Comparing QT and CT demonstrates the importance of studying QT. Turbulence in a classical viscous fluid appears to be comprised of vortices, as pointed out by Da Vinci. However, these vortices are unstable, and appear and disappear repeatedly. Moreover, circulation is not conserved, nor is it identical for each vortex. QT consists of a tangle of quantized vortices that have the same conserved circulation. Thus, QT may present an easier system for study than CT because the elements are more definite and clear, which is confimred at relatively large scales by many works described later. However, the situation is not so simple. The severe constraint of quantum mechanics will be effective at small scales and lead to some definite difference from CT, and it is not so trivial what happens there.

Based on these considerations, QT research has headed in a new direction
since the mid 90s. One primary interest has been to understand the relationship between
QT and CT \cite{PLTP}.  
The energy spectrum of fully developed CT is known to obey the Kolmogorov law in an inertial range. The energy transfer in an inertial range is believed to be sustained by the Richardson cascade process, in which large eddies are broken up self-similarly into smaller eddies. Recent experimental and numerical studies have supported the Kolmogorov spectrum, even in QT. Another important problem is the dissipative mechanism in QT.  At a finite temperature, mutual friction works as a dissipative mechanism. However, it is not so easy to understand how the energy cascade is for the full range of the scales  and what mechanism causes dissipation at very low temperatures, at which the normal fluid component is negligible.

\section{Dynamics of quantized vortices} \label{sec-vortex-dynamics}
Quantum hydrodynamics, including QT, is reduced to the motion of quantized vortices.
Hence, understanding the dynamics of quantized vortices is a key issue in quantum hydrodynamics.
Two formulations are generally available for studying the dynamics of quantized vortices.  One is the vortex filament model, and the other is the Gross--Pitaevskii (GP) model. We will briefly describe these two formulations.
\subsection{Vortex filament model}
As described in Section 2. 1, a quantized vortex has quantized circulation. The vortex core is extremely thin, usually much smaller than other characteristic scales of vortex motion. These properties allow a quantized vortex to be represented as a vortex filament. In classical fluid dynamics \cite{Saffman}, the vortex filament model is just a convenient idealization; the vorticity in a realistic classical fluid flow rarely takes the form of clearly discrete vorticity filaments. However, the vortex filament model is accurate and realistic for a quantized vortex in superfluid helium.

The vortex filament formulation represents a quantized vortex as a filament passing through the fluid, having a definite direction corresponding to its vorticity. Except for the thin core region, the superflow velocity field has a classically well-defined meaning, and can be described by ideal fluid dynamics. The velocity at a point $\mathbf{r}$ due to a filament is given by the Biot--Savart expression:
\begin{equation}
\mathbf{v}\sub{s} (\mathbf{r}) = \frac{\kappa}{4\pi} \int_{\mathcal{L}} \frac{(\mathbf{s}_1 - \mathbf{r}) \times d\mathbf{s}_1}{|\mathbf{s}_1-\mathbf{r}|^3}, \label{eq-Biot-Savart}
\end{equation}
where $\kappa$ is the quantum of circulation. The filament is represented by the parametric form $\mathbf{s} = \mathbf{s}(\varsigma, t)$ with the one-dimensional coordinate $\varsigma$ along the filament. The vector $\mathbf{s}_1$ refers to a point on the filament, and the integration is taken along the filament. Helmholtz's theorem for a perfect fluid states that the vortex moves at the superfluid velocity. Calculating the velocity $\mathbf{v}\sub{s}$ at a point $\mathbf{r}=\mathbf{s}$ on the filament causes the integral to diverge as $\mathbf{s}_1 \rightarrow \mathbf{s}$. To avoid this divergence, we separate the velocity $\dot{\mathbf{s}}$ of the filament at the point $\mathbf{s}$ into two components \cite{Schwarz85}:
\begin{equation}
\dot{\mathbf{s}} = \frac{\kappa}{4\pi} \mathbf{s}^\prime \times \mathbf{s}^{\prime\prime} \ln \left(
\frac{2(\ell_+ \ell_-)^{1/2}}{e^{1/4} a_0} \right) + \frac{\kappa}{4\pi} \int_{\mathcal{L}}^\prime \frac{(\mathbf{s}_1 - \mathbf{r}) \times d\mathbf{s}_1}{|\mathbf{s}_1-\mathbf{r}|^3}. \label{eq-sdot}
\end{equation}
The first term is the localized induction field arising from a curved line element acting on itself, and $\ell_+$ and $\ell_-$ are the lengths of the two adjacent line elements after discretization, separated by the point $\mathbf{s}$. The prime denotes differentiation with respect to the arc length $\varsigma$. The mutually perpendicular vectors $\mathbf{s}'$, $\mathbf{s}''$, and $\mathbf{s}' \times \mathbf{s}''$ are directed along the tangent, the principal normal, and the binormal, respectively, at the point $\mathbf{s}$, and their respective magnitudes are 1, $R^{-1}$, and $R^{-1}$, where $R$ is the local radius of curvature. The parameter $a_0$ is the cutoff, corresponding to the core radius. Thus, the first term represents the tendency to move the local point $\mathbf{s}$ in the binormal direction with a velocity inversely proportional to $R$. The second term represents the non-local field obtained by integrating the integral of Eq. (\ref{eq-Biot-Savart}) along the rest of the filament, except in the neighborhood of $\mathbf{s}$. 

Neglecting the non-local terms and replacing Eq. (\ref{eq-sdot}) by $\dot{\mathbf{s}} = \beta \mathbf{s}' \times \mathbf{s}''$ is referred to as the localized induction approximation (LIA). Here, the coefficient $\beta$ is defined by $\beta = (\kappa/4\pi) \ln \left( c \langle R \rangle/a_0 \right)$, where $c$ is a constant of order 1 and $(\ell_+\ell_-)^{1/2}$ is replaced by the mean radius of curvature $\langle R \rangle$ along the length of the filament.  This approximation is believed to be effective for analyzing isotropic dense tangles due to cancellations between non-local contributions. However, the ILA lacks the interaction between vortices, not suitable for the description of a realistic vortex tangle.

A better understanding of vortices in a real system is obtained when boundaries are included in the analysis. For this purpose, a boundary-induced velocity field $\mathbf{v}\sub{s,b}$ is added to $\mathbf{v}\sub{s}$, so that the superflow can satisfy the boundary condition of an inviscid flow, that is, the normal component of the velocity should disappear at the boundaries. To allow for another, presently unspecified, applied field, we include $\mathbf{v}\sub{s,a}$. Hence, the total velocity $\dot{\mathbf{s}}_0$ of the vortex filament without dissipation is 
\begin{equation}
\dot{\mathbf{s}}_0 =\frac{\kappa}{4\pi} \mathbf{s}^\prime \times \mathbf{s}^{\prime\prime} \ln \left( \frac{2(\ell_+ \ell_-)^{1/2}}{e^{1/4} a_0} \right) + \frac{\kappa}{4\pi} \int_{\mathcal{L}}^\prime \frac{(\mathbf{s}_1 - \mathbf{r}) \times d\mathbf{s}_1}{|\mathbf{s}_1-\mathbf{r}|^3} + \mathbf{v}\sub{s,b}(\mathbf{s}) + \mathbf{v}\sub{s,a}(\mathbf{s}). \label{eq-s0dot}
\end{equation}
\noindent
At finite temperatures, it is necessary to take into account the mutual friction between the vortex core and the normal flow $\mathbf{v}\sub{n}$. 
Including this term, the velocity of $\mathbf{s}$ is given by 
\begin{equation}
\dot{\mathbf{s}} =\dot{\mathbf{s}}_0 + \alpha \mathbf{s}^\prime \times (\mathbf{v}\sub{n} - \dot{\mathbf{s}}_0) - \alpha^\prime \mathbf{s}^\prime \times [\mathbf{s}^\prime \times (\mathbf{v}\sub{n} - \dot{\mathbf{s}}_0)], \label{eq-sdotmf}
\end{equation}
where $\alpha$ and $\alpha'$ are temperature-dependent friction coefficients \cite{Schwarz85}, and $\dot{\mathbf{s}}_0$ is calculated from Eq. (\ref{eq-s0dot}). 

The numerical simulation method based on this model has been described in detail elsewhere \cite{Schwarz85,Schwarz88,Tsubota00,Adachi10}. A vortex filament is represented by a single string of points separated by a distance $\Delta\varsigma$. The vortex configuration at a given time determines the velocity field in the fluid, thus moving the vortex filaments according to Eqs. (\ref{eq-s0dot}) and (\ref{eq-sdotmf}). 
Vortex reconnection should be properly included when simulating vortex dynamics.  A numerical study of a classical fluid shows that the close interaction of two vortices leads to their reconnection, primarily because of viscous diffusion of the vorticity \cite{Boratav92}. Schwarz assumed that two vortex filaments reconnect when they come within a critical distance of one another, and showed that statistical quantities such as the vortex line density were not sensitive to how these reconnections occur \cite{Schwarz85,Schwarz88}.  Even after Schwarz's study, it remained unclear as to whether quantized vortices can actually reconnect. However, Koplik and Levine directly solved the GP equation to show that two closely quantized vortices reconnect, even in an inviscid superfluid \cite{Koplik93}. More recent simulations have shown that reconnections are accompanied by emissions of sound waves having wavelengths on the order of the healing length \cite{Leadbeater01,Ogawa02a}. 

Starting with several remnant vortices under thermal counterflow, Schwarz numerically studied how these vortices developed into a vortex tangle \cite{Schwarz88}.  The tangle was self-sustained by the competition between excitation due to the applied flow and dissipation through mutual friction. The numerical results were quantitatively consistent with typical experimental results.  Since Schwarz used the LIA, he could not obtain the statistical steady state without an artificial procedure. Performing the full nonlocal calculation, Adachi {\it et al.} succeeded in obtaining a steady state in counterflow turbulence that was consistent with typical observations. This was a significant accomplishment in numerical research \cite{Adachi10}.

Thus, the vortex filament model is very useful for QT, although it cannot describe phenomena directly related to vortex cores, such as reconnection, nucleation, and annihilation. These phenomena can be analyzed only by the GP model.

\subsection{The Gross-Pitaevskii model}

In a weakly interacting Bose system, the macroscopic wave function $\Psi(\mathbf{r},t)$ appears as the order parameter of Bose--Einstein condensation, obeying the Gross--Pitaevskii (GP) equation \cite{Gross1961,Pitaevskii1961}: 
\begin{equation}
i \hbar \frac{\partial \Psi(\mathbf{r},t)}{\partial t} = \left( - \frac{\hbar ^2}{2m}\nabla^2 + g |\Psi(\mathbf{r},t)|^{2}- \mu \right) \Psi(\mathbf{r},t). \label{eq-gp}
\end{equation}
Here $g=4\pi \hbar^2 m/a$ represents the strength of the interaction characterized by the s-wave scattering length $a $, $m$ is the mass of each particle, and $\mu$ is the chemical potential. In $\Psi = | \Psi | \exp (i \theta)$, the squared amplitude $|\Psi|^2$ is the condensate density, and the gradient of the phase $\theta$ gives the superfluid velocity  $\mathbf{v}\sub{s} = (\hbar/m) \nabla \theta$ is a frictionless flow of the condensate. This relation causes quantized vortices to appear with quantized circulation. The only characteristic scale of the GP model is the coherence length, defined by $\xi=\hbar/(\sqrt{2mg}| \Psi |)$, which gives the vortex core size. 

The GP model can explain not only the vortex dynamics but also vortex core phenomena such as reconnection and nucleation. However, strictly speaking, the GP equation is not applicable to superfluid $^4$He, which is not a weakly interacting Bose system. The GP model does not feature, for example, short-wavelength excitations such as rotons, which are present in superfluid $^4$He. The GP equation is applicable to Bose--Einstein condensation of a dilute atomic Bose gas \cite{PethickSmith}.

\section{Quantum turbulence}

In this section, we review recent developments in the study of QT.
``Quantum turbulence" is defined as the turbulent state of a quantum fluid, and this term is often used to emphasize the state is dominated by the behavior of quantized vortices at low temperatures, with little thermal effect \cite{PLTP}.
From a theoretical point of view, the properties of QT are often discussed in reference to statistical quantities such as the energy spectrum, which is important in turbulence.
Since quantized vortices in QT are definite and stable topological defects, and can be positively identified, the energy spectrum of QT is usually related to the dynamics of vortices, such as vortex cascades \cite{Vinen2007}.

The energy spectrum and the vortex cascade process are usually divided into two regions in wavenumber space.
The first region is called the classical region, and exists at wavenumbers below the inverse of the mean intervortex spacing.
The dynamics of vortices in the classical region are dominated by the Richardson cascade, in which large vortices are broken up self-similarly into smaller ones, or collective dynamics of aggregated quantized vortices at scales larger than the intervortex spacing.
Such behavior of vortices supports the analogy of QT to CT, namely, the Kolmogorov energy spectrum, which has been confirmed by several theoretical and experimental efforts \cite{Vinen2002}.
The second region is called the quantum region, in which vortex dynamics are dominated by the effects of the quantized circulation, specifically the Kelvin wave cascade of vortices \cite{Svistunov1995,Vinen2001}, which does no appear in CT.
The Kelvin wave cascade is also a very important concept in understanding the dissipation mechanism of QT at very low temperatures.

At finite temperatures, turbulent flow is carried by not only quantized vortices, but also by normal fluid.
On larger scales, superfluid and normal fluids are coupled together by the mutual friction between them, and behave as a classical fluid to show the Kolmogorov energy spectrum.
On small scales, turbulent flow is dissipated by the viscosity of the normal fluid, and Kelvin waves do not exist.

Another topic in QT is the decay mechanism of quantized vortices \cite{PLTP}.
The total vortex line length, or the vortex line density, can be experimentally observed, and the decay of vortices provides important information regarding QT.
Since the decay of vortices strongly depends on the initial QT state, we can indirectly determine the statistical properties, such as the energy spectrum, of vortices in the initial state.

This section is organized as follows.
First, we review the energy spectrum of QT, from the classical region to the quantum region, and the connection between these two regions, which remains an open question.
Next, we discuss the decay of quantized vortices in QT, recent experimental developments, QT created by vibrating structures and visualization of quantized vortices.

\subsection{Energy spectra in classical turbulence}

To discuss energy spectra of QT, we begin with a brief review of the energy spectrum in CT.
The dynamics of classical fluids are usually described by the Navier-Stokes equation \cite{Frisch1995}:
\begin{align}
\frac{\partial \mathbf{v}}{\partial t} + \mathbf{v} \cdot \nabla \mathbf{v} & = - \frac{1}{\rho} \nabla P + \nu \nabla^2 \mathbf{v} \label{eq-Navier-Stokes-velocity}
\end{align}
Here, $\mathbf{v} = \mathbf{v}(\mathbf{x},t)$ is the fluid velocity, $P = P(\mathbf{x},t)$ is the pressure, $\rho$ is the fluid density, and $\nu$ is the kinematic viscosity.
The flow of this fluid can be characterized by the ratio of the second term of the left-hand side of Eq. (\ref{eq-Navier-Stokes-velocity}), hereafter the inertial term, to the second term of the right-hand side, hereafter the viscous term.
This ratio is the Reynolds number: $R = \bar{v} D / \nu$.
Here $\bar{v}$ and $D$ are the characteristic velocity of flow and the characteristic scale, respectively.
When $\bar{v}$ increases to the point that the Reynolds number exceeds a critical value, the system enters a turbulent state in which the flow is highly complicated, with many eddies.

The energy spectrum can be obtained from the spatial Fourier transformation of the equal-time two-point velocity correlation:
\begin{equation}
F(\mathbf{k},t) = \frac{1}{2} \int d \mathbf{x} \: e^{-i \mathbf{k} \cdot \mathbf{x}} \int d \mathbf{y} \: \mathbf{v}(\mathbf{y},t) \cdot \mathbf{v}(\mathbf{y} + \mathbf{x},t).
\end{equation}
Here $k$ is the wave number from the Fourier transformation.
Kolmogorov proposed the concept of a globally steady state of fully developed turbulence \cite{Kolmogorov1941a,Kolmogorov1941b} in which energy is injected into the fluid at scales comparable to the system size in the energy-containing range.
In the inertial range, this energy is transferred to smaller scales without being dissipated.
In the inertial range, the system is assumed to be locally homogeneous and isotropic, and the angular dependence of $E(\mathbf{k},t)$ in wavenumber space becomes less important.
The integration over the angle is defined as the energy spectrum:
\begin{equation}
E(k,t) = \frac{1}{(2 \pi)^3} \int d \phi_{\mathbf{k}} d \theta_{\mathbf{k}} \: k^2 F(\mathbf{k},t). \label{eq-energy-spectrum}
\end{equation}
The energy spectrum holds the following relation with the spatial integration of the kinetic energy:
\begin{equation}
\int d k \: E(k,t) = \frac{1}{2 (2 \pi)^3} \int d \mathbf{k} \: |\mathbf{v}(\mathbf{k},t)|^2 = \frac{1}{2} \int d \mathbf{x} \: \mathbf{v}(\mathbf{x},t)^2 = E(t)
\end{equation}
Here, $E(t)$ is the total kinetic energy per unit mass, and $\mathbf{v}(\mathbf{k},t)$ is the Fourier transformation of the fluid velocity.
In the steady state, the averaged value of $E(k,t)$ over $t$ is regarded as the statistical value, together with a statistical law known as the Kolmogorov law:
\begin{equation}
E(k) = C \varepsilon^{2/3} k^{-5/3}. \label{eq-Kolmogorov-law}
\end{equation}
The energy transferred to smaller scales in the energy-dissipative range is dissipated at the Kolmogorov wave number $k\sub{K} = (\varepsilon/\nu)^{1/4}$ through the viscosity of the fluid, at the dissipation rate $\varepsilon$ in Eq. (\ref{eq-Kolmogorov-law}).
This spectrum of Eq. (\ref{eq-Kolmogorov-law}) is easily derived by assuming that $E(k)$ is locally determined by only $\varepsilon$ and $k$.
The Kolmogorov constant $C$ is a dimensionless parameter of order unity.

The inertial range is thought to be sustained by the self-similar Richardson cascade, in which large eddies are broken up into smaller ones with many reconnections \cite{Richardson2007}.
In CT, however, the Richardson cascade is not completely understood because it is impossible to definitively identify each individual eddy.

The Kolmogorov law is based on the assumption that the turbulence is locally homogeneous and isotropic.
However, actual turbulence is neither homogeneous nor isotropic and hence the energy spectrum deviates from eq. (\ref{eq-Kolmogorov-law}).
This phenomenon is called ``intermittency", which is closely related to the coherent dynamics of eddies \cite{Frisch1995}.

\subsection{Energy spectra in quantum turbulence: overall picture}

In QT, it is important to consider the energy spectrum for the following two reasons:
First, like CT, QT is a complicated dynamical phenomenon with many degrees of freedom, and it is very useful to consider the energy spectrum as the statistical law which extracts the important imformations for the dynamical properties of the turbulence.
Secondly, in QT, unlike in CT, vortices are definite and stable topological defects, which enables the study of actual vortex dynamics.
In particular, when there is no normal fluid component at near-zero temperatures, any rotational motion is carried by quantized vortices and their dynamics are not affected by thermal components, so the inherent dynamics of vortices in turbulence are revealed.
If QT has an analogy to CT, vortex dynamics in QT should show the real vortex dynamics in the Richardson cascade.
Therefore, QT is an ideal prototype for the study of the statistics of turbulence, such as the relationship between the Kolmogorov law in wavenumber space and the Richardson cascade in real space \cite{TsubotaJPSJ}.
Whether this scenario is true can be confirmed by investigating the energy spectrum and energy transfer in QT.

As in CT, the energy spectra in QT can also be obtained from the spatial Fourier transformation of the equal-time two-point superfluid velocity correlation: Eq. \eqref{eq-energy-spectrum}, where $\mathbf{v}(\mathbf{x},t)$ is replaced with the superfluid velocity $\mathbf{v}\sub{s}(\mathbf{x},t)$.

\begin{figure}[htb]
\centering
\includegraphics[width=0.75\linewidth]{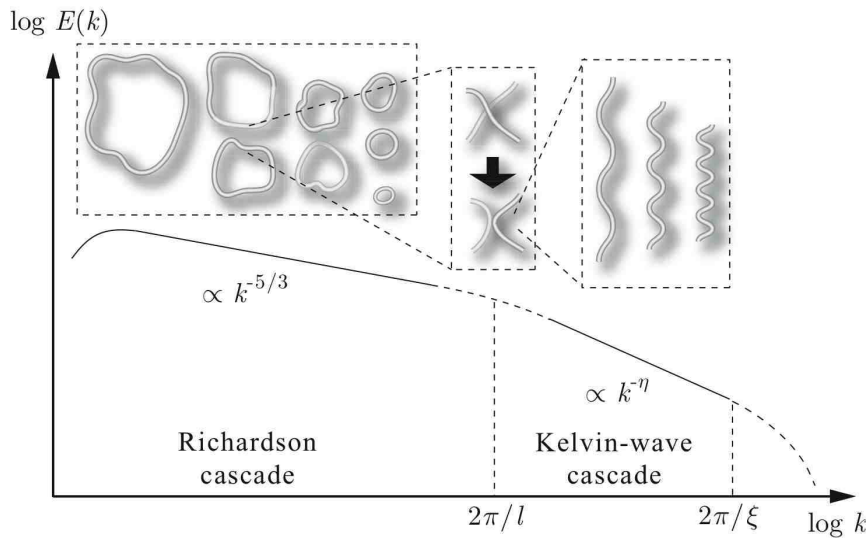}
\caption{\label{fig-two-cascade} Overall picture of the energy spectrum of QT at zero temperature.
The energy spectrum depends on the scale, and its properties change at about the scale of the mean intervortex spacing $l$.
When $k < k_l = 2 \pi / l$, a Richardson cascade of quantized vortices transfers energy from large to small scales, maintaining the Kolmogorov spectrum $E(k) = C \varepsilon^{2/3} k^{-5/3}$.
When $k > k_l$, energy is transferred by the Kelvin-wave cascade, which is a nonlinear interaction between Kelvin waves of different wavenumbers.
Eventually, energy is dissipated at scales of $\xi$ by the radiation of phonons.}
\end{figure}
Here, we summarize the overall picture of the energy spectrum of QT at zero temperature on the basis of theoretical and numerical studies (Fig. \ref{fig-two-cascade}) \cite{PLTP}.
In QT, vortices form a highly complicated tangled structure.
If the tangle is assumed to be homogeneous and isotropic, there are two characteristic length scales: one is the mean intervortex spacing $l = L^{-1/2}$ with a vortex line length density $L$ (the vortex line length per unit volume), and the other is the coherence length $\xi$ corresponding to the size of the vortex core.
Generally, $l$ is much larger than $\xi$; $l \gg \xi$.
Using the length scales $l$ and $\xi$, we can define the characteristic wave numbers $k_l = 2 \pi / l$ and $k_{\xi} = 2\pi / \xi$.

At length scales larger than $l$, the dynamics of QT are dominated by a tangled structure of many vortices.
Because vortex dynamics becomes collective at large scales, quantization of the circulation is not relevant, and the dynamics are similar to those of eddies in CT.
Therefore, this region can be referred to the classical region.
As a result, the energy spectrum $E(k)$ in the range of $k < k_l$ obeys the Kolmogorov law \eqref{eq-Kolmogorov-law}.
Vortices in QT sustain a Richardson cascade that transfers energy from smaller wavenumbers to larger ones without dissipation.
The Richardson cascade can be understood as large vortices breaking up into smaller ones in real space.
In the Richardson cascade process, the key vortex dynamic is reconnection; when two vortex lines approach each other, they become locally antiparallel and reconnect \cite{Koplik93,Leadbeater01,Ogawa02a}.
Reconnection causes topological changes of the vortex lines in the tangle, formations of distortion waves on the vortex lines (Kelvin waves), or fissions of vortex loops through self-reconnections, all of which are responsible for the cascade of energy towards smaller scales \cite{Tsubota00,Svistunov1995,Kivotides2001a}.
The Richardson cascade and the Kolmogorov energy spectrum have been confirmed by several numerical studies \cite{Nore1997a,Nore1997b,Araki2002,Kobayashi2005a,Kobayashi2005b,Yepez2009}.

At length scales comparable to $l$, reconnection is the dominant dynamic.
At these scales, two vortex lines separated by a distance $\sim l$ reconnect, forming small cusps on the vortex lines, which are regarded as the primary source of Kelvin waves in QT.
The wavelength of the created Kelvin waves is on the order of $l$.

At length scales smaller than $l$, which is referred to as the quantum region, the Richardson cascade is no longer dominant, and the quantized circulation and the motion of each vortex line becomes significant \cite{Svistunov1995,Samuels1990,Mitani2003,Kozik2004,Kozik2005,Nazarenko2006,Boffetta2009,Vinen2000}.
In this range, vortex dynamics are characterized by the cascade process of the Kelvin-waves which are formed by reconnection.
The nonlinear interaction of the Kelvin waves is the origin of the cascade from smaller to larger the wavenumbers.
The cascade dynamics of the Kelvin waves along a single vortex line has been confirmed, both numerically and theoretically.
The Kelvin-wave cascade in QT has also been studied.
The energy spectrum in the quantum region $k_l < k < k_{\xi}$ is theoretically predicted to obey a Kolmogorov-like power law: $E(k) \propto k^\eta$.
Several values have been proposed for the exponent $\eta$, which originates from both the configuration of the Kelvin waves and their cascade process.
The power-law behavior of the energy spectrum in this region has been confirmed numerically \cite{Yepez2009,Mitani2003,Kozik2004,Kozik2005,Nazarenko2006,Boffetta2009}.

In the region $k \sim k_\xi$, the Kelvin waves with wavelength of $\xi$ change to elementary excitations, such as phonons and rotons, in the primary decay process of QT near zero temperature \cite{Vinen2001}.

There is one open question regarding the energy spectrum in the region $k \sim k_l$, which is the transitional region between the Richardson cascade and the Kelvin-wave cascade.
A theoretical study proposed a bottleneck effect connecting the spectrum \eqref{eq-Kolmogorov-law} of the classical region to that of the quantum region \cite{L'vov2007}.
Another theoretical prediction was based on vortex reconnection dynamics on the scale $\sim l$, in which the transitional cascade process was predicted to occur by reconnection of the vortex bundle \cite{Kozik2008}.

\subsection{Energy spectra in quantum turbulence: classical region} \label{subsec-classical-region}

The energy spectra in the classical region $k < k_l$ has been a major problem in terms of the analogy of QT to CT.
In this region, the energy spectrum is determined by the collective behavior of many vortices, such as the vortex tangle and the aggregated bundle structure at scales larger than $l$.
Several numerical studies have calculated the energy spectrum in this region by simulating QT at zero temperature.

There are two formulations for studying the vortex dynamics in QT, as described in Sec. \ref{sec-vortex-dynamics}.
The first is the vortex filament model, in which the time evolution of each element $\mathbf{s}$ of a vortex filament follows Eq. \eqref{eq-sdotmf}.
At zero temperature there is no mutual friction, and we obtain $\alpha = \alpha^\prime = 0$ in \eqref{eq-sdotmf} and $\dot{\mathbf{s}} = \dot{\mathbf{s}}_0$.
In this model, the energy spectrum can be calculated directly from the configuration of the vortex filaments, but not through the superfluid velocity $\mathbf{v}\sub{s}(\mathbf{f})$ of Eq. \eqref{eq-Biot-Savart}.
The vorticity $\boldsymbol{\omega}(\mathbf{r}) = \nabla \times \mathbf{v}\sub{s}(\mathbf{r})$ is concentrated on the vortex filament:
\begin{align}
\boldsymbol{\omega}(\mathbf{r}) = \kappa \int d \varsigma \: \mathbf{s}^\prime(\varsigma) \delta(\mathbf{s}(\varsigma) - \mathbf{r}).
\end{align}
Here, $\varsigma$ is the arc length along the vortex filament.
The Fourier transformation of the vorticity $\boldsymbol{\omega}(\mathbf{k}) = i \mathbf{k} \times \mathbf{v}\sub{s}(\mathbf{k})$ becomes
\begin{align}
\boldsymbol{\omega}(\mathbf{k}) = \kappa \int d \varsigma \: \exp[-i \mathbf{k} \cdot \mathbf{s}(\varsigma)] \mathbf{s}^\prime(\varsigma).
\end{align}
Assuming fluid incompressibility $\nabla \cdot \mathbf{v}\sub{s}(\mathbf{r}) = 0$ and $\mathbf{v}\sub{s}(\mathbf{k}) = i \mathbf{k} \times \boldsymbol{\omega}(\mathbf{k}) / |\mathbf{k}|^2$, we obtain the energy spectrum as the integrated form over $\varsigma$:
\begin{align}
E(k) = \frac{\kappa^2}{2 (2\pi)^3} \int d \phi_{\mathbf{k}} d \theta_{\mathbf{k}} \: \int d \varsigma_1 d \varsigma_2 \: \exp[- i \mathbf{k} \cdot (\mathbf{s}(\varsigma_1) - \mathbf{s}(\varsigma_2))] \mathbf{s}^\prime(\varsigma_1) \mathbf{s}^\prime(\varsigma_2). \label{eq-filament-energy-spectrum}
\end{align}
This formula enables us to calculate the energy spectrum directly from the configuration of the vortex filament \cite{Araki2002}.

\begin{figure}[htb]
\centering
\includegraphics[width=0.65\linewidth]{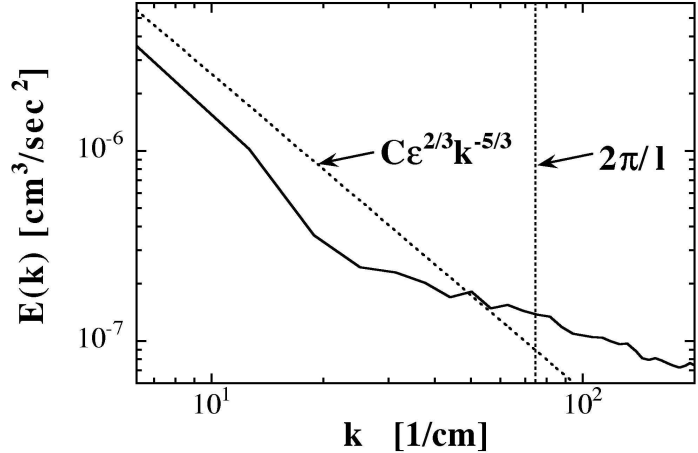}
\caption{\label{fig-Araki} Comparison of the energy spectrum (solid line) at $t = 70$ sec after the Taylor Green vortex with the Kolmogorov law \eqref{eq-Kolmogorov-law} (dotted line) with $C = 1$ and $\varepsilon = 1.287 \times 10^{-6}$ cm$^2$/sec$^3$ [Araki, Tsubota, and Nemirovskii: Phys. Rev. Lett {\bf 89} (2002) 145301, reproduced with permission. Copyright 2002 by the American Physical Society].}
\end{figure}
Using the vortex filament model, Araki {\it et al.} simulated QT with no mutual friction, starting from a Taylor Green vortex in a cube of size $D = 1.0$ cm \cite{Araki2002}.
A cutoff was introduced for the smallest vortices, the size of which was comparable to the space resolution $\Delta \varsigma = 1.83 \times 10^{-2}$, which was the only dissipation mechanism in the simulation.
The energy spectrum $E(k)$ below $k_l$ had a peak at the smallest $k$, and strongly depended on the initial configuration.
After about 70 sec, the system exhibited a nearly homogeneous and isotropic turbulent state, and lost its dependence on the initial configuration.
The energy dissipation rate $\varepsilon = - dE / dt$ became less dependent on time, and the system behaved as a quasi-steady turbulent state.
The energy spectrum obtained from Eq. \eqref{eq-filament-energy-spectrum} agreed with the Kolmogorov law \eqref{eq-Kolmogorov-law} at $k < k_l$ (Fig. \ref{fig-Araki}).
They also calculated the vortex length distribution $n(x)$, where $n(x) \Delta x$ is the number of vortex loops, the length of which ranged from $x$ to $x + \Delta x$.
In the range from $l$ to $D$, $n(x)$ obeyed the scaling property $n(x) \propto x^{-1.34 \pm 0.18}$, which supports the self-similar Richardson cascade from larger to smaller scales.
The energy transfer rate through the Richardson cascade is equal to the energy dissipation rate $\varepsilon = - dE / dt$ in Eq. \eqref{eq-Kolmogorov-law}.
Furthermore, the energy spectrum $E(k)$ in the region $k < k_l$ is determined by the energy dissipation rate, although the dissipation mechanism is effective only at wavenumbers corresponding to the cutoff $2 \pi / \Delta \xi$.
In the region of $k > k_l$, the energy spectrum deviates from the Kolmogorov law, showing a different power-law behavior.
Although this behavior may come from the velocity field near each vortex line, the simulation was not appropriate to examine the spectrum in this region because it did not have sufficient spatial resolution.

The other formulation is the GP model.
The GP equation \eqref{eq-gp} can be obtained from the energy functional:
\begin{align}
H = \int d\mathbf{x} \: \left[ \frac{\hbar^2}{2m} |\nabla \Psi(\mathbf{x},t)|^2 + \frac{g}{2} |\Psi(\mathbf{x},t)|^4 - \mu |\Psi(\mathbf{x},t)|^2 \right]. \label{eq-gp-Hamiltonian}
\end{align}
Because the vorticity $\boldsymbol{\omega}(\mathbf{x},t) = \nabla \times \mathbf{v}\sub{s}(\mathbf{x},t)$ vanishes everywhere in a singly-connected region of the fluid, all rotational flow is carried by quantized vortices.
In the core of each vortex, $\Psi(\mathbf{x},t)$ vanishes so that the circulation $\oint \mathbf{v}\sub{s}(\mathbf{x},t) \cdot d\mathbf{s}$ around the core is quantized by $h / m$.
The vortex core size is given by the healing length $\xi = \hbar / \sqrt{2 m g \rho\sub{s}}$, where the superfluid density $\rho\sub{s}$ is defined as the spatially averaged condensate density $|\Psi(\mathbf{x},t)|^2$.
When considering the energy spectrum, it should be noted that the hydrodynamics are compressible in the GP model.
The total energy of the GP equation per unit mass:
\begin{align}
E(t) = \frac{1}{m N} \int d\mathbf{x} \: \left[ \frac{\hbar^2}{2m} |\nabla \Psi(\mathbf{x},t)|^2 + \frac{g}{2} |\Psi(\mathbf{x},t)|^4 \right]
\end{align}
can be separated into the interaction energy $E\sub{int}(t)$, the quantum energy $E\sub{q}(t)$, and the kinetic energy $E\sub{kin}(t)$;
\begin{align}
& E\sub{int}(t) = \frac{g}{2 m N} \int d\mathbf{x} \: |\Psi(\mathbf{x},t)|^4 \\
& E\sub{q}(t) = \frac{(2 \pi)^2 \kappa^2}{2 N} \int d\mathbf{x} \: ( \nabla |\Psi(\mathbf{x},t)| )^2 \\
& E\sub{kin}(t) = \frac{(2 \pi)^2 \kappa^2}{2 N} \int d\mathbf{x} \: ( |\Psi(\mathbf{x},t)| \nabla \theta(\mathbf{x},t) )^2.
\end{align}
Here, $N \equiv \int d\mathbf{x} \: |\Psi(\mathbf{x},t)|^2$ is the total number of particles.
The kinetic energy can be further divided into a compressible part $E\sub{kin}\up{c}(t)$ due to compressible excitations, and an incompressible part $E\sub{kin}\up{i}(t)$ due to vortices \cite{Nore1997a,Nore1997b}:
\begin{align}
& E\sub{kin}\up{c,i}(t) = \frac{(2 \pi)^2 \kappa^2}{2 N} \int d\mathbf{x} \: [( |\Psi(\mathbf{x},t)| \nabla \theta(\mathbf{x},t) )\up{c,i} ]^2.
\end{align}
Here, $[\cdots]\up{c}$ denotes the compressible part and $\nabla \times [\cdots]\up{c} = 0$, and $[\cdots]\up{i}$ denotes the incompressible part $\nabla \cdot [\cdots]\up{i} = 0$.
Corresponding to each energy, there are several kinds of energy spectra.
The most important of these is the energy spectrum of the incompressible kinetic energy
\begin{align}
E\sub{kin}\up{i}(k,t) = \frac{\kappa^2}{2 (2 \pi) N} \int d \phi_{\mathbf{k}} d \theta_{\mathbf{k}} \: [\mathbf{p}\up{i}(\mathbf{k},t)]^2, \label{eq-incompressible-kinetic-energy-spectrum}
\end{align}
because it should obey the Kolmogorov law with the Richardson cascade of quantized vortices.
Here $\mathbf{p}\up{i}(\mathbf{k},t)$ is the Fourier transformation of the momentum density $\{|\Psi(\mathbf{x},t)| \nabla \theta(\mathbf{x},t)\}\up{i}$

Starting from a Taylor Green vortex, Nore {\it et al.} simulated QT by numerically solving the GP model \cite{Nore1997a,Nore1997b}.
After some time, the initial vortices became tangled, and the calculated spectrum obeyed the power-law behavior $E\sub{kin}\up{i}(k,t) \propto k^{-\eta(t)}$.
When vortices formed a tangle, the exponent $\eta(t)$ was about $5/3$, but this value did not hold for long because the turbulence was decaying with the conservation of total energy.
Because of energy conservation, the energy of the vortices $E\sub{kin}\up{i}(t)$ was transferred to compressible excitations $E\sub{kin}\up{i}(t)$ through repeated reconnections and disappearances of small vortex loops with sizes on the order of $\xi$.
This thermalization process hindered the vortex cascade process through interactions between vortices and compressible excitations, which caused $E\up{i}\sub{kin}(k,t)$ to deviate from the Kolmogorov law.

To avoid this thermalization, Kobayashi and Tsubota proposed a modified GP model, in which a dissipation term was introduced to remove the compressible excitations \cite{Kobayashi2005a,Kobayashi2005b}.
The characteristic wavelength of the compressible excitations is on the order of $\xi$, so the introduced dissipation was set to act only at scales smaller than $\xi$ and to not affect vortices.
The resulting GP equation in wavenumber space becomes:
\begin{align}\begin{split}
\hbar [i - \gamma(k)] \frac{\partial \Psi(\mathbf{k},t)}{\partial t} &= \left( \frac{\hbar ^2 k^2}{2m} - \mu(t) \right) \Psi(\mathbf{k},t) \\
& \quad + \frac{g}{(2 \pi)^6} \int d \mathbf{k}_1 d \mathbf{k}_2 \: \Psi(\mathbf{k}_1,t) \Psi^\ast (\mathbf{k}_2,t) \Psi(\mathbf{k} - \mathbf{k}_1 + \mathbf{k}_2,t). \label{eq-gp-dissipation}
\end{split}\end{align}
Here, $\Psi(\mathbf{k},t)$ is the Fourier transformation of $\Psi(\mathbf{x},t)$.
The dissipation term has the form $\gamma(k) = \gamma_0 \theta(k - 2 \pi / \xi)$ with the step function $\theta$, being effective only at $k > k_\xi$.
Introduction of $\gamma(k)$ conserves neither the energy $E$ nor the number of particles $N$.
However, when studying the hydrodynamics of turbulence, it is realistic to assume that the number of particles is conserved.
Hence, the time dependence of the chemical potential $\mu(t)$ was introduced to conserve the total number of particles $N$.
By numerically analyzing Eq. \eqref{eq-gp-dissipation}, they confirmed the Kolmogorov spectrum in two kinds of QT.

The first analysis was of decaying turbulence \cite{Kobayashi2005a}.
To obtain a turbulent state, an initial configuration was set to have a uniform density $|\Psi(\mathbf{x},t=0)|^2 = 1$ and the phase had a random spatial distribution.
Because the initial superfluid velocity $\mathbf{v}\sub{s}(\mathbf{x},t=0) = (\hbar / m) \nabla \theta(\mathbf{x},t=0)$ was also random, the initial wave function was dynamically unstable and soon produced fully developed turbulence with many quantized vortex loops.
They confirmed that only the compressible kinetic energy was decreased by the dissipation term, and that the incompressible kinetic energy $E\sub{kin}\up{i}(t)$ dominated the total kinetic energy $E\sub{kin}(t)$, demonstrating that the thermalization was effectively suppressed.
The obtained energy spectrum of the incompressible kinetic energy was consistent with the Kolmogorov law in the period during which the vortex tangle was formed, and this consistency lasted longer than it did without the dissipation term.
In this period, the energy dissipation rate $\varepsilon = - d E\sub{kin}\up{i}(t) / dt$ became nearly constant, and the system can be considered to occupy a quasi-steady turbulent state with a stationary Kolmogorov spectrum of $\varepsilon^{2/3} k^{-5/3}$.

\begin{figure}[htb]
\centering
\includegraphics[width=0.5\linewidth]{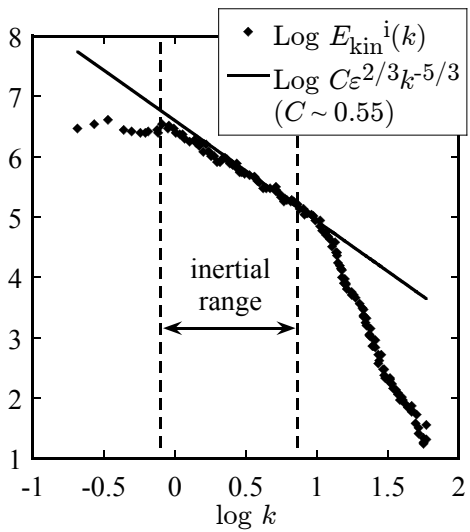}
\caption{\label{fig-Kobayashi} Incompressible energy spectrum $E\sub{kin}\up{i}(k)$ obtained by Kobayashi and Tsubota \cite{Kobayashi2005b}.
The plotted points are from an ensemble average of 50 randomly selected states.
The solid line is the Kolmogorov law [Kobayashi and Tsubota: J. Phys. Soc. Jpn. {\bf 74} (2005) 3248, reproduced with permission].}
\end{figure}
The second analysis was of steady turbulence, with not only energy dissipation but also large-scale energy injection \cite{Kobayashi2005b}.
The energy injection was accomplished by introducing the stochastic external potential $V(\mathbf{x},t)$ into the GP equation \eqref{eq-gp-dissipation}.
$V(\mathbf{x},t)$ has a Gaussian two-point correlation:
\begin{align}
\langle V(\mathbf{x},t) V(\mathbf{x}^\prime,t^\prime) \rangle = V_0^2 \exp \left[ - \frac{|\mathbf{x} - \mathbf{x}^\prime|^2}{2 X_0^2} - \frac{(t - t^\prime)^2}{2 T_0^2} \right], \label{eq-stochastic-potential}
\end{align}
where $X_0$ and $T_0$ are the characteristic spatial and time scales of the potential.
By introducing $V(\mathbf{x},t)$, energy was injected at the spatial scale of $X_0$, and quantized vortices of radius $X_0$ were effectively nucleated.
Starting from the uniform state $\Psi(\mathbf{x},t = 0) = 1$, they developed the GP equation \eqref{eq-gp-dissipation} with the potential \eqref{eq-stochastic-potential}, and obtained a steady QT in which $E(t)$, $E\sub{kin}(t)$, $E\sub{kin}\up{c}(t)$, and $E\sub{kin}\up{i}(t)$ were nearly constant.
In the simulation, the incompressible kinetic energy $E\sub{kin}\up{i}(t)$ was always dominant in the total kinetic energy $E\sub{kin}(t)$; the introduced potential contributed to the nucleation of vortices rather than that of compressible waves.
The obtained energy spectrum was consistent with the Kolmogorov law in the range $2 \pi / X_0 < k_\xi$ which is regarded as the inertial range of Richardson cascade of vortices. (Fig. \ref{fig-Kobayashi}).
They also calculated the flux $\Pi(k,t)$ of the incompressible kinetic energy from smaller to larger wavenumbers:
\begin{align}
\Pi(k,t) = \frac{(2 \pi)^2 \kappa^2}{N} \int d \mathbf{x} \: L_k [ \{ \mathbf{p}(\mathbf{x},t) \cdot \nabla \mathbf{v}\sub{s}(\mathbf{x},t) \}\up{i}] L_k[\{\mathbf{p}(\mathbf{x},t)\}\up{i}], \label{eq-incompressible-kinetic-energy-flux}
\end{align}
where $L_k$ is the operator for the low-pass filter:
\begin{align}
L_k[s(\mathbf{x})] = \frac{1}{(2 \pi)^3} \int_{|\mathbf{k}| < k} d\mathbf{k} \: \int d\mathbf{x}^\prime \: e^{i \mathbf{k} \cdot (\mathbf{x} - \mathbf{x}^\prime)} s(\mathbf{x}),
\end{align}
for an arbitrary function $s(\mathbf{x})$.
$\Pi(k,t)$ can be obtained from the scale-by-scale energy budget equation derived from the GP equation \eqref{eq-gp-dissipation}.
$\Pi(k,t)$ was nearly constant in the inertial range, and almost the same as the energy dissipation rate $\varepsilon = - d E\sub{kin}\up{i}(t) / dt$ after switching off the moving random potential.
This confirmed the picture of the inertial range and the Richardson cascade of quantized vortices in QT; the incompressible kinetic energy steadily flows in wavenumber space through the Richardson cascade at the constant rate$\Pi$, and finally dissipates at the rate $\varepsilon \simeq \Pi$.

In the above three numerical studies, the system size was not so large that the inertial range was less than one order in wavenumber space.
Furthermore, the mean intervortex spacing $l$ was close to the healing length $\xi$, and was too short to study the Kelvin wave cascade.
To obtain the energy spectrum of a wider range of wavenumber space, Yepez {\it et al.} performed a large-scale simulation of the GP model, the size of which was $(11.25)^3$ times larger than the above three simulations \cite{Yepez2009}.
They also used a highly accurate numerical code for the time development; the unitary quantum lattice gas algorithm, in which the time development of the order parameter $\Psi(\mathbf{x},t)$ in Eq. \eqref{eq-gp} could be performed as the unitary time evolution of qubits on a cubic lattice.
Starting from the initial state with 12 straight vortices, which consisted of three groups of four vortices aligned along $x$, $y$, and $z$ axes, they obtained QT with highly tangled vortices.
In this simulation, two length scales were defined; the inner radius of a vortex core $\xi$, and its outer radius $\pi \xi$.
On an $L^3$ grid, the wavenumber corresponding to the core's inner radius was $k\sub{inner} \simeq (\sqrt{3} / 2) L / \xi$, while the outer radius wavenumber was $k\sub{outer} \simeq k\sub{inner} / \pi$.
They found that the incompressible kinetic energy spectrum \eqref{eq-incompressible-kinetic-energy-spectrum} had three distinct power-law $k^{-\alpha}$ regions that ranged from the classical turbulent regime of Kolmogorov $\alpha = -5/3$ at large scales $k < k\sub{outer}$ to the quantum Kelvin-wave cascades $\alpha = 3$ at small scales $k > k\sub{inner}$.
There was a semiclassical region $6.34 \lesssim \alpha \lesssim 7.11$ connecting the Kolmogorov and Kelvin-wave spectra $k\sub{outer} < k < k\sub{inner}$.
Compared to previous simulations, this simulation supplied the Kolmogorov spectrum over a much wider inertial range, with about 2 orders in wavenumber space, and the Kelvin-wave spectrum in the quantum region.
Although they related the $k^{-3}$ spectrum at small scales $k > k\sub{inner}$ to the Kelvin-wave cascade, the length scale in this region is smaller than the vortex core size.
The one-dimensional picture of Kelvin wave is relevant in the scales much larger than the vortex core, and a nature of short waves propagating in three-dimensional space is considered to be dominant in the length scale smaller than the vortex core size.
They also did not discuss the mean intervortex spacing $l$, which is the key length scale to distinguish the classical and quantum resions.
Therefore, it is not so clear that the $k^{-3}$ spectrum reflects the Kelvin-wave cascade.

It should be noted that the Richardson cascade of quantized vortices is genuine in QT, and does not completely imitate that of CT.
The Kolmogorov constant $C$ in Eq. \eqref{eq-Kolmogorov-law} in QT is not necessarily equal to that of CT.
The obtained Kolmogorov constant was $C \simeq 0.7$ for the vortex filament model \cite{Araki2002}, and $C \simeq 0.32$ \cite{Kobayashi2005a} and $0.55$ \cite{Kobayashi2005b} for the decaying and steady QT in the GP model, which was smaller than that of CT, in which $C \simeq 1.5$.
These smaller Kolmogorov constants may be characteristic of QT.

\subsection{Energy spectra in quantum turbulence: quantum region}

In the region of $k > k_l$, the picture of aggregated vortices is no longer effective, and the motion of each vortex line becomes essential.
The most probable dominant dynamic of vortices is Kelvin waves, which originate from distortion waves on the vortex lines after their reconnection.
A Kelvin wave is a transverse, circularly-polarized wave motion, with the approximate dispersion relation for a rectilinear vortex:
\begin{align}
\omega_k = \frac{\kappa k^2}{4 \pi} \left[ \ln\Big(\frac{1}{k \xi}\Big) + c \right], \label{eq-Kelvin-dispersion}
\end{align}
with a dimensionless constant $c \sim 1$.
$k$ is the wavenumber of the Kelvin wave, and was different from that used for the energy spectrum.
Kelvin waves were theoretically proposed to exist in inviscid fluids \cite{Kelvin1880}, and were first observed by inducing torsional oscillations in a rotating superfluid $^4$He \cite{Hall1958,Hall1960}.

At finite temperatures where there is a significant fraction of normal fluid, Kelvin waves are damped by mutual friction.
On the other hand, at very low temperatures they can be damped only by the radiation of phonons.
Vinen estimated the rate of radiation, and found that it is extremely low unless the frequency is very high, typically on the order of 4 GHz for superfluid $^4$He \cite{Vinen2001}.
In these circumstances, low-frequency Kelvin waves can lose energy only by nonlinear coupling to waves of a different frequency, which is the Kelvin wave cascade from small to large wavenumbers.

An earlier numerical work showed the nonlinear interaction of Kelvin waves in the framework of the vortex filament model.
Samuels and Donnelly performed a simulation of a full Biot-Savart law \eqref{eq-sdot} and the local induction approximation (LIA), in which the second Biot-Savart term in \eqref{eq-sdot} was neglected, and found sideband instability in both cases above a critical amplitude of the Kelvin wave \cite{Samuels1990}: $A_0 > \lambda / (2 \pi n)$, where $A_0$ is the amplitude of the main helical wave, $\lambda$ is the wavelength, and $n$ is the number of half waves on the vortex.
The occurrence of sideband instability indicates the mixing of Kelvin waves of the main wavelength with those of smaller wavelengths via nonlinear interactions, and the transfer of energy from small to large wavenumbers.

Although the Kelvin-wave cascade seems to be a very important mechanism in QT at scales smaller than $l$, it is non-trivial problem about the actual cascade process.
Vinen {\it et al.} performed a numerical simulation of a Kelvin wave excited along a single vortex line using the vortex filament model \cite{Mitani2003}.
They considered a model system in which the helium was contained in a space between two parallel sheets separated by a distance $\ell\sub{B} = 1$ cm.
A single, initially rectilinear (along the $z$ axis), vortex was stretched between the two sheets.
The allowed wavenumbers of the Kelvin waves were $k = 2 \pi n / \ell\sub{B}$, where $n$ is a positive integer.
A Kelvin wave with a small wavenumber of $n_0$ was continuously driven, and all modes with $n$ exceeding a large critical value $n\sub{c}$ were strongly damped.
The simulation was based on the full Biot-Savart law, and the force driving one mode was of the form $V \rho \kappa \sin(k_0 z - \omega_0)$, where $k_0 = 2 \pi n_0 / \ell\sub{B}$, $\rho$ is the density of helium, and $\omega_0$ is related to the frequency $k_0$ by the dispersion relation \eqref{eq-Kelvin-dispersion}.
Damping was applied by a periodic smoothing process.
Starting from a straight vortex line, the total length of the vortex line evolved to reach a steady average value after application of the driving force, which suggests the existence of a steady state.
In the typical simulation, with $V = 2.5 \times 10^{-5}$ cm s$^{-1}$, $k_0 = 10 \pi $ cm$^{-1}$, and $k\sub{c} = 2 \pi n\sub{c} / \ell\sub{B} = 120 \pi$ cm$^{-1}$, they calculated the root mean square amplitudes $n_k(t) = \langle \zeta^\ast_k \zeta_k \rangle^{1/2}$ of the Fourier components of the displacement of the vortex.
Initially, only the mode that resonated with the drive was excited.
However, as time passed, nonlinear interactions led to the excitation of all modes.
The spectrum $n_k(t)$ also reached a steady state, and in this stage energy was injected at a certain rate at wavenumber $k_0$ and dissipated at the same rate at wavenumber $k\sub{c}$.
For large $k$, where the modes nearly formed a continuum, the steady state was observed to have, to a good approximation, a spectrum of the form
\begin{align}
n_k = A \ell\sub{B}^{-1} k^{-3}, \label{eq-single-Kelvin-spectrum}
\end{align}
with the dimensionless parameter $A$ of order unity.
They continued their simulations, changing the driving amplitude $V$ and the drive wavenumber $k_0$ to show that there was no effect on the steady state, which suggests that the scaling property of the spectrum \eqref{eq-single-Kelvin-spectrum} is universal.
The mean energy per unit length of a vortex in mode $k$ is related to $n_k$ by the equation \cite{Vinen2001}
\begin{align}
E\sub{K}(k) = \epsilon\sub{K} k^2 n_k, \label{eq-Kelvin-wave-energy}
\end{align}
where $\epsilon\sub{K}$ is an effective energy per unit length of vortex, given by
\begin{align}
\epsilon\sub{K} = \frac{\rho \kappa^2}{4 \pi} \left[ \ln\Big(\frac{1}{k a}\Big) + c_1 \right].
\end{align}
From Eqs. \eqref{eq-single-Kelvin-spectrum} and \eqref{eq-Kelvin-wave-energy}, the energy spectrum of the single Kelvin wave becomes $E\sub{K}(k) = A \epsilon\sub{K} (k \ell\sub{B})^{-1}$.

When two vortices reconnect, they twist to become locally antiparallel at the reconnection point and create small cusps or kinks after the reconnection, which were initially confirmed numerically by Schwarz \cite{Schwarz85,Schwarz88}, and later by Tsubota {\it et al.} in detail  \cite{Tsubota00} using the vortex-filament model.
Svistunov suggested that the relaxation process of these cusps or kinks causes the emission of Kelvin waves, and plays an important role in the decay of QT at low temperatures \cite{Svistunov1995}.
Following Svistunov, Vinen analyzed the energy spectrum of the Kelvin-wave cascade by introducing a ``smoothed" length of vortex line per unit volume after all the Kelvin waves were removed, and considered $E\sub{K}(k) dk$, the energy per unit length of the smoothed vortex lines associated with Kelvin waves in the range $k$ to $k + dk$ \cite{Vinen2001}.
By dimensional analysis, $E\sub{K}(k)$ was estimated as
\begin{align}
E\sub{K}(k) = A \rho \kappa^2 k^{-1} \label{eq-Vinen-Kelvin-spectrum}
\end{align}
with a constant $A$.

Kivotides {\it et al.} numerically confirmed the generation of Kelvin waves through reconnections using the vortex filament model \cite{Kivotides2001a}.
In their simulation, four vortex rings were placed symmetrically on opposite sides and oriented so that they all moved toward the center.
The four rings approached each other and underwent reconnections which induced cusp relaxation and the generation of large-amplitude Kelvin waves.
They calculated the energy spectrum $E(k)$ defined by Eq. \eqref{eq-energy-spectrum} for the superfluid velocity, and found that $E(k)$ developed approximately a $k^{-1}$ form after the reconnections occurred and the reconnected vortex lines began to move farther apart.
This supports the nonlinear transport of energy between different wavenumbers.
Considering that the fluctuations of the velocity field were induced by the Kelvin waves on the filament, this result is consistent with Vinen's analysis of Eq. \eqref{eq-Vinen-Kelvin-spectrum} with $E\sub{K}(k) \sim E(k)$.

Kozik and Svistunov analyzed the Kelvin-wave cascade using weak-turbulence theory \cite{Kozik2004}.
They employed the Hamiltonian representation of the vortex line motion; a Kelvin wave was excited along the $z$ direction and the position of the vortex line was specified in the parametric form $x = x(z)$, $y = y(z)$.
In terms of the complex canonical variable $w(z,t) = x(z,t) + i y(z,t)$, the Biot-Savart dynamics equation \eqref{eq-sdot}, acquires the Hamiltonian form $i \dot{w} = \delta H[w,w^\ast] / \delta w^\ast$ with
\begin{align}
H[w,w^\ast] = \frac{\kappa}{4 \pi} \int dz_1 dz_2 \: \frac{1 + \mathrm{Re}[w^{\prime \ast}(z_1) w^\prime(z_2)]}{\sqrt{(z_1 - z_2)^2 + |w(z_1) - w(z_2)|^2}}, \label{eq-Kelvin-wave-Hamiltonian}
\end{align}
where $w^\prime(z) = d w(z) / d z$.
Under the assumption that the amplitude of the Kelvin wave is small compared to its wavelength, the Hamiltonian \eqref{eq-Kelvin-wave-Hamiltonian} can be expanded in powers of $|w(z_1) - w(z_2)| / |z_1 - z_2|$: $H = E_0 + H_0 + H_1 + \cdots$ ($E_0$ is just a number). 
$H_0$ can be diagonalized by the Fourier transformation $w_k = \int dz \: w(z) e^{-ikz}$ to
\begin{align}
H_0 = \frac{\kappa}{4 \pi} \sum_k \omega_k w^\ast_k w_k,
\end{align}
where $\omega_k$ is the dispersion given in Eq. \eqref{eq-Kelvin-dispersion}.
By introducing kelvons, which are the quanta $\hat{w}_k$ for the amplitude of Kelvin waves $w_k$, the Hamiltonian can be canonically quantized with the kelvon annihilation operator $\hat{a}_k = \sqrt{\kappa \rho / 2 \hbar} \hat{w}_k$.
Because of the conservation of momentum and energy, the two-kelvon scattering channel is suppressed and three-kelvon scattering becomes the dominant process.
The effective vertex $V_{1,2,3}^{4,5,6}$ for the three-kelvon scattering process consists of two parts; a two-kelvon vertex in the second order of perturbation associated with $H_1$, and the bare three-kelvon vertex associated with $H_2$.
In the classical-field limit, the kinetic equation of the averaged kelvon occupation number $n_k = \langle a_k^\dagger a_k \rangle$ over the statistical ensemble is
\begin{align}
\dot{n}_1 = 216 \pi \sum_{2, \cdots, 6} |V_{1,2,3}^{4,5,6}|^2 \delta(\Delta \omega) \delta(\Delta k) (f_{4,5,6}^{1,2,3} - f_{1,2,3}^{4,5,6}). \label{eq-Svistunov-kinetic-equation}
\end{align}
with $\Delta k = k_1 + k_2 + k_3 - k_4 - k_5 - k_6$, $\Delta \omega = \omega_1 + \omega_2 + \omega_3 - \omega_4 - \omega_5 - \omega_6$, and $f_{1,2,3}^{4,5,6} = n_1 n_2 n_3 (n_4 n_5 + n_5 n_6 + n_6 n_4)$.
Here, $(1, \cdots, 6)$ denotes the set of the wave number $(k_1, \cdots, k_6)$.
The kinetic equation \eqref{eq-Svistunov-kinetic-equation} supports the energy cascade, if two conditions are satisfied.
The first is that kinetic time grows progressively smaller in the limit of large wavenumbers, which can be checked by a dimensional estimate.
The second is that the collision term is local in wavenumber space, which has been verified numerically.
By the dimensional analysis, Eq. \eqref{eq-Svistunov-kinetic-equation} yields
\begin{align}
\dot{n}_k \propto \omega_k^{-1} n_k^5 k^{16}, \label{eq-Svistunov-ndot}
\end{align}
with $k_1 \sim \cdots \sim k_6 \sim k$ and $|V| \sim k^6.$
The energy flux $\theta_k$ per unit vortex line length at momentum $k$ is defined as $\theta_k = L^{-1} \sum_{k^\prime < k} \omega_{k^\prime} \dot{n}_{k^\prime}$, where $L$ is the system size.
This implies $\theta_k \sim k \dot{n}_k \omega_k \sim n_k^5 k^{17}$ with Eq. \eqref{eq-Svistunov-ndot}.
The cascade requirement that $\theta_k$ is $k$ independent, without energy dissipation in the inertial range, leads to the spectrum: $n_k \propto k^{-17 / 5}$ and $H_0 \propto \sum_k \omega_k n_k \sim \sum_k k^{-7/5}$.
To investigate the above spectrum in more detail, Kozik and Svistunov performed a numerical simulation of the full Biot-Savart law of Equation \eqref{eq-sdot} over a wide range of scale \cite{Kozik2005}.
Starting from a single vortex line with the initial distribution $n_k \propto k^{-3}$, they found that the waves converted into a new power-law distribution $n_k \propto k^{-17 / 5}$.

For the spectrum of Kelvin waves, Nazarenko presented a nonlinear differential equation model, pointing out that turbulence displays a dual cascade behavior of both the direct energy cascade and an inverse cascade of wave action, which correspond to two constants of motion:
\begin{align}
E = \iint \frac{d\mathbf{s} d\mathbf{s}_0}{|\mathbf{s} - \mathbf{s}_0|} \label{eq-Biot-Savart-constant-1}
\end{align}
and
\begin{align}
\mathbf{P} = \int \mathbf{s} \times d\mathbf{s} \label{eq-Biot-Savart-constant-2}
\end{align}
for the full Biot-Savart equation \eqref{eq-sdot} \cite{Nazarenko2006}.
The nonlinear differential equation was constructed to preserve the main scaling of the original closure $n_k \propto k^{-17/5}$ as
\begin{align}
\dot{n} = \frac{C}{\kappa^{10}} \omega^{1/2} \frac{\partial^2}{\partial \omega^2} \left(n^6 \omega^{21/2} \frac{\partial^2}{\partial \omega^2} \frac{1}{n} \right), \label{eq-Nazarenko-DAM}
\end{align}
where $C$ is a dimensionless constant, $\omega = \omega(k) = \kappa k^2 / 4 \pi$ is the approximate dispersion of the Kelvin wave.
This equation preserves the energy $E = \int d \omega \: n \omega^{1/2}$, and the wave action $N = \int d \omega \: n \omega^{-1/2}$.
Equation \eqref{eq-Nazarenko-DAM} has both the direct cascade solution $n_k \propto k^{-17/5}$ and the inverse cascade solution $n_k \propto k^{-3}$.
It also has the same thermodynamic Rayleigh-Jeans solutions $n = T / (\omega + \mu)$ with constants representing temperature $T$ and chemical potential $\mu$.
This model was developed to include the effect of generation of Kelvin waves by reconnection, dissipation caused by mutual friction, and phonon radiation.

Boffetta {\it et al.} suggested a simpler model for the Kelvin-wave cascade than that of Svistunov {\it et al.} \cite{Kozik2004}, using a truncated expansion of the LIA \cite{Boffetta2009}.
As is the case in Eq. \eqref{eq-Kelvin-wave-Hamiltonian}, a Hamiltonian form for the LIA can be obtained as
\begin{align}
H[w] = 2 \frac{2 \kappa}{4 \pi} \ln\left(\frac{\ell}{\xi}\right)  \int dz \: \sqrt{1 + |w^\prime(z)|^2} = 2 \beta L[w], \label{eq-LIA-Hamiltonian}
\end{align}
where $\ell$ is a length on the order of the curvature radius, and $\beta = (\kappa / 4 \pi) \ln(\ell / \xi)$.
The Hamiltonian is proportional to the vortex length $L[w] = \int dz \: \sqrt{1 + |w^\prime(z)|^2}$.
The equation of motion is
\begin{align}
\dot{w} = \frac{i}{2} \left( \frac{w^\prime}{\sqrt{1 + |w^\prime|^2}} \right)^\prime
\end{align}
with $\beta = 1/2$.
As a consequence of the invariance under phase transformation, the total wave action $N[w] = \int dz \: |w|^2$ is also conserved.
In the framework of the LIA, the system becomes integrable with infinite numbers of conserved quantities, so that the energy and the wave action cannot cascade.
The integrability is broken by considering a truncated expansion of the Hamiltonian in power of wave amplitude $w^\prime(z)$ such as
\begin{align}
H\sub{exp}[w] = H_0 + H_1 + H_2 = \int dz \: \left(1 + \frac{|w^\prime|^2}{2} - \frac{|w^\prime|^4}{8}\right).
\end{align}
Taking into account only the six-wave resonant condition, and taking the Taylor expansion of $w(k,t)$ around $w(k,t) = c(k,0)$ ($w(k,t) = \int dz e^{- i k z} w(z,t)$ is the Fourier transformation of $w(z,t)$), one can obtain the resulting Hamiltonian expressed in $c_k$ and the equation of motion for $n_k = \langle |c_k|^2 \rangle$,
\begin{align}
& H_c = \int dk \: \omega_k |c_k|^2 + \int dk_{123456} \: C_{123456} \delta^{123}_{456} c_1^\ast c_2^\ast c_3^\ast c_4 c_5 c_6, \\
& \dot{n}_k = 18 \pi \int dk_{23456} \: |C_{k23456}|^2 \delta^{k23}_{456} \delta(\omega^{k234}_{456}) f_{k23456}, \label{eq-Boffetta-kinetics}
\end{align}
where $C_{123456}$ is the interaction coefficient, and
\begin{align}
& f_{k23456} = n_k n_2 n_3 n_4 n_5 n_6 \left[ \frac{1}{n_k} + \frac{1}{n_2} + \frac{1}{n_3} - \frac{1}{n_4} - \frac{1}{n_5} - \frac{1}{n_6} \right], \\
& \delta(\omega^{k23}_{456}) = \delta(\omega_k + \omega_2 + \omega_3 - \omega_4 - \omega_5 - \omega_6).
\end{align}
The kinetic equation \eqref{eq-Boffetta-kinetics} is the almost same as that of the Biot-Savart formulation \eqref{eq-Svistunov-kinetic-equation} by Svistunov, and a simple dimensional analysis of Eq. \eqref{eq-Boffetta-kinetics} gives $\dot{n}_k \propto k^{14} n_k^5$.
Besides the equilibrium Rayleigh-Jeans solution $n_k = T / (\omega_k + \mu)$, there are non-equilibrium steady state solutions of the kinetic equation \eqref{eq-Boffetta-kinetics} in wave turbulence theory, which rely on a constant flux in some inertial range and are known as Kolmogorov-Zakharov solutions \cite{Zakharov1992}.
Using a dimensional analysis, one can obtain the energy flux $\Pi_k\up{(H)} = \int dk^\prime \dot{n}_{k^\prime}\omega_{k^\prime} \sim k^{17} n_k^5$ and the wave action flux $\Pi_k\up{(N)} = \int dk^\prime \dot{n}_{k^\prime} \sim k^{15} n_k^5$.
Requiring the existences of two ranges of scales in which $\Pi_k\up{(H)}$ and $\Pi_k\up{(N)}$ are $k$-independent leads to the spectra:
\begin{subequations}\begin{align}
& n_k \sim k^{-17/5} \label{eq-Boffetta-cascade}, \\
& n_k \sim k^{-3} \label{eq-Boffetta-inverse-cascade},
\end{align}\label{eqs-Boffetta-spectra}\end{subequations}
respectively.
As in the analysis by Nazarenko \cite{Nazarenko2006}, the first spectrum is expected to be the direct cascade of energy flowing to large $k$, and the second is expected to be the inverse cascade of wave action flowing to small $k$.
Finally, we mention the energy spectra estimated from the spectra in \eqref{eqs-Boffetta-spectra}.
By using Eq. \eqref{eq-Kelvin-wave-energy} and the relation $E\sub{K}(k) \sim E(k)$ proposed by Kivotides {\it et al.}, we can obtain $E(k) \propto k^{-7/5}$ for the direct cascade and $E(k) \propto k^{-1}$ for the inverse cascade.

Yepez {\it et al.} numerically investigated the energy spectrum in the quantum region \cite{Yepez2009} using the Gross-Pitaevskii model.
The obtained spectrum $E(k) \propto k^{-3}$ is inconsistent with those of the above analytical works, and the energy spectra in the quantum region remains as an open question.
As described in the previous subsection, however, the relationship between the Kelvin-wave spectrum and the mean intervortex spacing $l$ was not mentioned, and it is not clear whether the obtained spectrum can appropriately be regarded as the spectrum of the Kelvin-wave cascade.

\subsection{Energy spectra in classical-quantum crossover}

As discussed in the previous sections, there are two different types of energy spectra in the classical ($k < k_l$) and quantum ($k_l < k< k_\xi$) regions.
An important question arises: How do these two energy spectra connect to each other at the length scale $l$?
Although there are several theoretical and numerical reports on this region, consistency among this work has not yet been obtained, and the problem remains controversial.
In analysis of classical-quantum crossover, $\Lambda = \ln (l/\xi)$ appears to be an important parameter.
In typical $^4$He experiments, $\Lambda$ is about 15.

\begin{figure}[htb]
\centering
\includegraphics[width=0.55\linewidth]{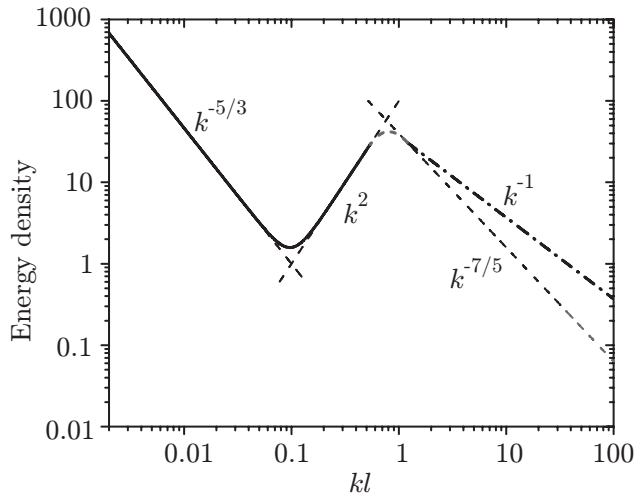}
\caption{\label{fig-L'vov-2007} The energy spectra $E(k)$ in the classical, $k < 1/l$, and quantum $k > 1/l$ ranges, proposed by L'vov  {\it et al.} \cite{L'vov2007}.
The two straight solid lines in the classical range indicate pure Kolmogorov scaling $E(k) \propto k^{-5/3}$ and pure thermodynamic scaling $E(k) \propto k^2$.
In the quantum range, the dashed line indicates the Kelvin wave cascade spectrum (slope $-7/5$, whereas the dash-dotted line marks the spectrum corresponding to the noncascading part of the vortex tangle energy (slope $-1$) [L'vov, Nazarenko, and Rudenko: Phys. Rev. B {\bf 76} (2007) 024520, reproduced with permission. Copyright 2007 by the American Physical Society].}
\end{figure}

L'vov {\it et al.} suggested a bottleneck crossover between the two regions \cite{L'vov2007}.
For a given mean intervortex spacing $l$, the mean vorticity in the system $\langle |\omega| \rangle$ is given by $\langle |\omega| \rangle \simeq \kappa l^{-2}$.
The energy spectrum in the classical region is of the Kolmogorov form (Eq. \eqref{eq-Kolmogorov-law}): $E\sub{cl}(k) \simeq \varepsilon^{2/3} k^{-5/3}$, where the Kolmogorov constant $C$ is approximately unity.
Assuming that $\langle |\omega| \rangle$ is dominated by the classical-quantum crossover scale, we obtain
\begin{align}
\langle |\omega| \rangle^2 \simeq \int^{1/l} k^2 E\sub{cl}(k) \simeq \varepsilon^{2/3} l^{-4/3}.
\end{align}
In the quantum region, the reformulated kelvon occupation number $n_k$ from the analysis by Kozik and Svistunov is $n_k \simeq (l \varepsilon)^{1/5} \kappa^{2/5} k^{-17/5}$ from dimensional analysis, and the corresponding energy spectrum becomes
\begin{align}
E\sub{qu}(k) \simeq \Lambda \left(\frac{\kappa^7 \varepsilon}{l^8}\right)^{1/5} k^{-7/5},
\end{align}
which is consistent with Eq. \eqref{eq-Boffetta-cascade}.
This equation also gives the mean vorticity as
\begin{align}
\langle |\omega| \rangle^2 \simeq \frac{E\sub{qu}(k=1/l)}{l^3} \simeq \Lambda \left(\frac{\kappa^7 \varepsilon}{l^{16}} \right)^{1/5}.
\end{align}
If the energy flux $\varepsilon$ is the same in both the classical and the quantum regions, the ratio between $E\sub{cl}(1/l)$ and $E\sub{qu}(1/l)$ at the crossover of $k \simeq 1/l$ is $E\sub{qu}(1/l) / E\sub{cl}(1/l) \simeq \Lambda^{10/3} \gg 1$.
This large mismatch indicates that the energy flux carried by classical hydrodynamic turbulence cannot fully propagate through the crossover region, and that larger scale hydrodynamic motion will increase in energy up to the level $E\sub{qu}(1/l)$; this is called the bottleneck effect (Fig. \ref{fig-L'vov-2007}).
To get a qualitative picture of the bottleneck, L'vov {\it et al.} used the warm cascade solutions following from the model:
\begin{align}
\varepsilon = - \frac{1}{8} \sqrt{k^{13} F_k} \frac{d F_k}{d k}, \quad F_k = \frac{E(k)}{k^2}, \label{eq-Leith-67}
\end{align}
for the energy flux of hydrodynamic turbulence \cite{Connaughton2004}.
Here $F_k$ is the three-dimensional spectrum of turbulence.
For a constant energy flux, the solution of Eq. \eqref{eq-Leith-67} is
\begin{align}
F_k = \left[\frac{24 \varepsilon}{11 k^{11/2}} + \Bigg(\frac{T}{\pi \rho}\Bigg)^{3/2} \right]^{2/3}, \label{eq-Leith-67-solution}
\end{align}
where $\rho$ is the fluid density.
While \eqref{eq-Leith-67-solution} coincides with the Kolmogorov energy spectrum at low $k$, the large $k$ range comprises a thermalized portion of the spectrum with equipartition of energy, characterized by an effective temperature $T$.

\begin{figure}[htb]
\centering
\includegraphics[width=0.5\linewidth]{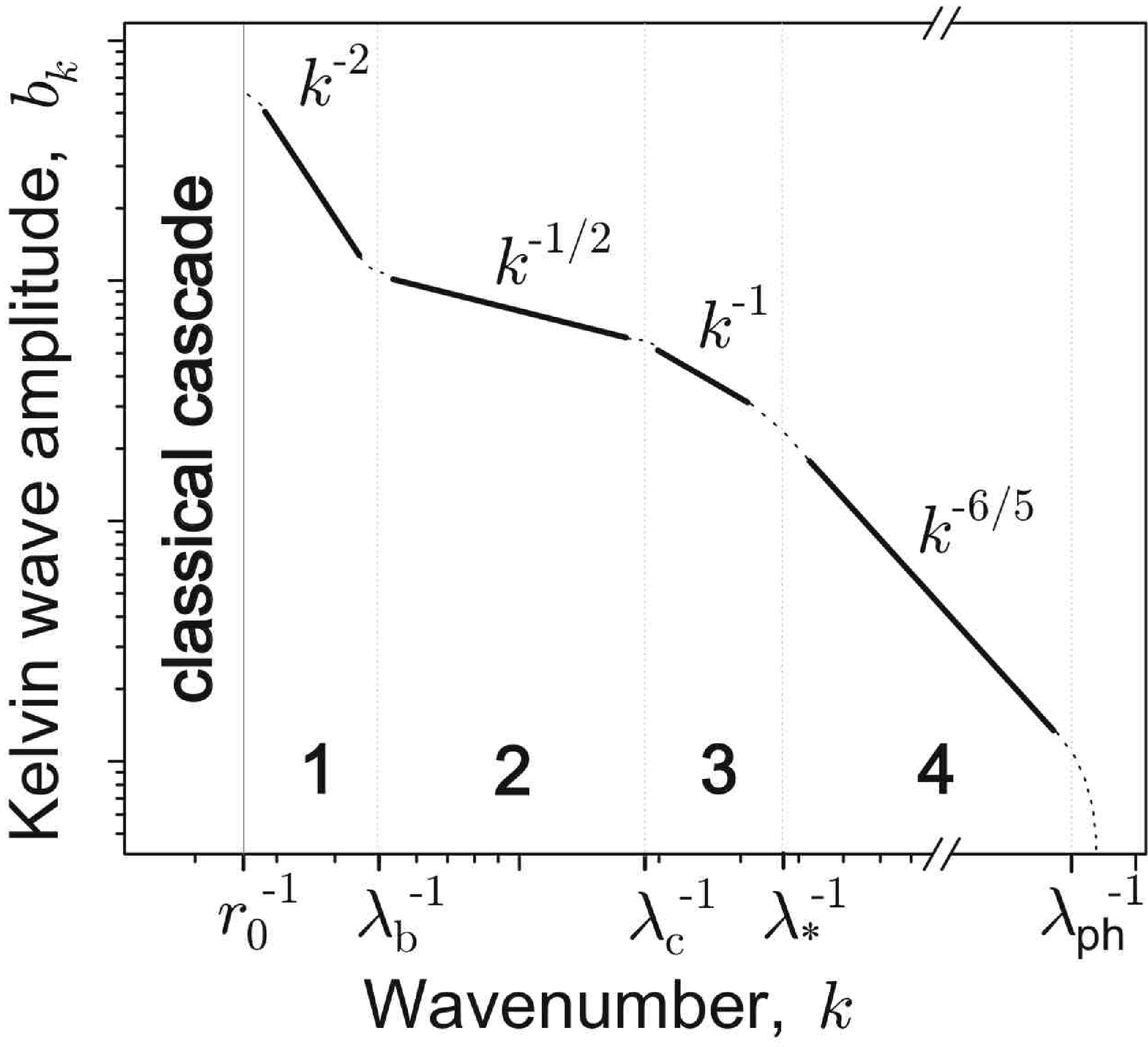}
\caption{\label{fig-Kozik-2008} Spectrum of Kelvin waves in the quantized regime proposed by Kozik  {\it et al.} \cite{Kozik2008}.
The inertial range consists of a chain of cascades driven by different mechanisms: (1) reconnections of vortex-line bundles, (2) reconnections between nearest-neighbor vortex lines in a bundle, (3) self-reconnections on single vortex lines, and (4) nonlinear dynamics of single vortex lines without reconnections [Kozik and Svistunov: Phys. Rev. B {\bf 77} (2008) 060502(R), reproduced with permission. Copyright 2008 by the American Physical Society].}
\end{figure}
The analysis by L'vov {\it et al.} was based on the assumption that the coarse-grained macroscopic description of quantized vorticity is effective down to the scale of $l$.
Based on the LIA, Kozik and Svistunov suggested a different picture for the crossover region, in which the locally-induced motion of the vortex lines emerges at the scale of $r_0 \sim \Lambda^{1/2} l$, and the crossover range is divided into three subranges, $r_0^{-1} < k < \lambda\sub{b}^{-1}$, $\lambda\sub{b}^{-1} < k < \lambda\sub{c}^{-1}$, and $\lambda\sub{c}^{-1} < k < \lambda_\ast^{-1}$, where $\lambda\sub{b} \sim \Lambda^{1/4} l$, $\lambda\sub{c} \sim l / \Lambda^{1/4}$, and $\lambda_\ast \sim l / \Lambda^{1/2}$ (Fig. \ref{fig-Kozik-2008}) \cite{Kozik2008}.
In the first region, $r_0^{-1} < k < \lambda\sub{b}$, polarized vortex lines are organized in bundles and reconnect with other bundles to form Kelvin waves with amplitude $b_k \sim r_0^{-1} k^{-2}$, where $b_k$ is defined as $b_k^2 \sim L^{-1} \sum_{q \sim k} \langle \hat{a}^\dagger_q \hat{a}_q \rangle = L^{-1} \sum_{q \sim k} n_q \sim k n_k$.
Here, $\hat{a}_q$ is the kelvon annihilation operator and $n_q$ is the kelvon occupation number.
In the second region, $\lambda\sub{b}^{-1} < k < \lambda\sub{c}^{-1}$, the cascade is supported by nearest-neighbor reconnections in a bundle, and $b_k \sim l (\lambda\sub{b} k)^{-1/2}$.
In the third range, $\lambda\sub{c}^{-1} < k < \lambda_\ast^{-1}$, the cascade is driven by self-reconnection of vortex lines, giving $b_k \sim k^{-1}$.
The quantum region for the Kelvin-wave cascade of a single vortex line begins from $\lambda_\ast^{-1}$, giving $b_k \propto k^{-6/5}$.
The Kelvin-wave spectrum $b_k$ smoothly connects these ranges, and there is no bottleneck effect in the model.
Although they emphasized that the energy spectrum $E(k)$ is practically meaningful only in the classical region, we can estimate $E(k)$ from their model: $E(k) \sim k^{-3}$ in $r_0^{-1} < k < \lambda\sub{b}^{-1}$, $E(k) \sim k^{0}$ in $\lambda\sub{b}^{-1} < k < \lambda\sub{c}^{-1}$, and $E(k) \sim k^{-1}$ in $\lambda\sub{c}^{-1} < k < \lambda_\ast^{-1}$.

The theoretical works of both L'vov {\it et al.} and Kozik {\it et al.} were based on the idea that quantized vortices are locally polarized in a tangle, forming bundles.
However, the nature of their polarization is not clear.
Furthermore, the energy spectra estimated from their model were inconsistent with those given from the numerical work of the GP model by Yepez et al \cite{Yepez2009} which also remained a problem in this wavenumber region as we commented in the subsection \ref{subsec-classical-region}.
More quantitative analysis and detailed numerical work is needed to clarify the nature of the crossover region.

\subsection{Energy spectra in quantum turbulence at finite temperatures}
Experimental studies measured the energy spectrum of QT at finite temperatures, and supported the Kolmogorov spectrum directly or indirectly \cite{Smith1993,Maurer1998,Stalp1999,Skrbek2000a,Skrbek2000b,Stalp2002,Skrbek2003}.
These experiments were also consistent, in the sense that they showed similarities between QT and CT.
Vinen theoretically considered this similarity, and proposed that the superfluid and the normal fluid are likely to be coupled together by mutual friction at scales larger than the intervortex spacing $l$, and would thus behave like a classical fluid and show a Kolmogorov energy spectrum where the mutual friction does not cause dissipation \cite{Vinen2000}.
This idea was confirmed by Kivotides {\it et al.} through numerical simulation of coupled dynamics of a vortex filament and a normal fluid \cite{Kivotides2007}, and by L'vov {\it et al.} through theoretical analysis of the two fluid equations \cite{L'vov2006}.
Although they also showed a decoupling of the two fluids at small scales, this remains a controversial topic.

\subsection{Experimental study of quantum turbulence}
\begin{figure}[htb]
\centering
\includegraphics[width=0.5\linewidth]{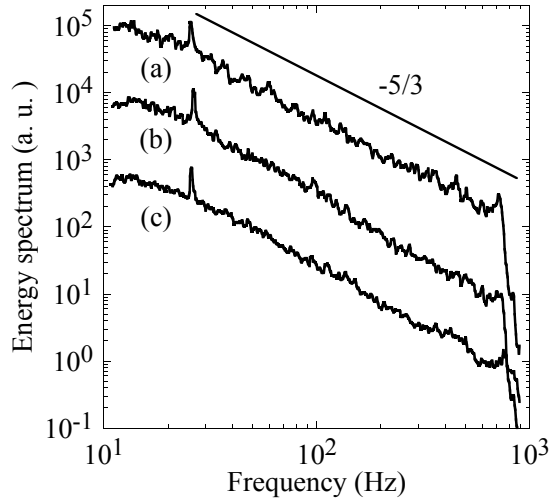}
\caption{\label{fig-Maurer} Energy spectra obtained using counter-rotating disks at (a) 2.3 K, (b) 2.08 K, and (c) 1.4 K \cite{Maurer1998}.}
\end{figure}

Early experimental studies of QT focused on thermal counterflow, in which the normal fluid and superfluid flow in opposite directions \cite{Donnelly1991}.
However, as thermal counterflow has no analogy with conventional fluid dynamics, it has not enabled an understanding of the relationship between QT and CT.
In the mid-90s, QT experiments were performed that did not involve thermal counterflow.
Maurer and Tabeling studied QT of superfluid $^4$He that was produced in a cylinder 8 cm in diameter and 20 cm high and driven by two counter-rotating disks \cite{Maurer1998}.
They observed local pressure fluctuations, and converted these into an energy spectrum.
The experiments were done at three temperatures: 2.3 K ($\rho\sub{s} / \rho = 0$), 2.08 K ($\rho\sub{s} / \rho = 0.05$), and 1.4 K ($\rho\sub{s} / \rho = 0.9$), which are both above and below the $\lambda$ point.
In all cases, a Kolmogorov energy spectrum was observed (Fig. \ref{fig-Maurer}).
Above the $\lambda$ point, the obtained spectrum is reasonable because the system is a classical viscous fluid.
Below the $\lambda$ point, however, it is debatable whether the obtained spectra are consistent with the Kolmogorov law and independent of the ratio between $\rho\sub{s}$ and $\rho\sub{n}$, namely, the dissipative mechanism.
Their observations were understood on the basis of the idea that the two fluids were probably coupled together by mutual friction, and behaved as a one-component fluid \cite{Vinen2000}.

Afterwards, a series of experiments with superfluid $^4$He above 1 K were performed by the Oregon group \cite{Smith1993,Stalp1999,Skrbek2000a,Skrbek2000b,Stalp2002,Skrbek2003}.
The helium was contained in a channel with a square (1 cm $\times$ 1 cm) cross section, along which a grid was pulled at a constant velocity.
This method of creating QT is similar to a method of generating homogeneous, isotropic turbulence in a classical fluid \cite{Frisch1995}.
A pair of second-sound transducers placed in the walls of the channel detected the appearance of a vortex tangle by second-sound attenuation \cite{Vinen2002}.
The decay of the vorticity of the tangle behind the grid was observed.
To obtain the energy spectrum from this result, the authors made some assumptions.
In a fully developed CT, the energy dissipation rate becomes $\varepsilon = \nu \langle \omega^2 \rangle$, where $\langle \omega^2 \rangle$ is the mean square vorticity \cite{Hinze1975}.
A similar formulation could be satisfied in fully developed QT above 1 K, in which $\kappa^2 L^2$ is a measure of $\langle \omega^2 \rangle$.
They assumed that the dissipation rate in the experiment became
\begin{align}
\varepsilon = \nu^\prime \kappa^2 L^2, \label{eq-quasi-classical-dissipation-rate}
\end{align}
with an effective kinematic viscosity $\nu^\prime$.
They also assumed that the flows of the superfluid and normal fluid were coupled together to behave as a one-component fluid by mutual friction at scales larger than $l$, and Eq. \eqref{eq-quasi-classical-dissipation-rate} could be directly connected with the observation of second-sound attenuation by choosing a suitable value of $\nu^\prime$ as a function of temperature \cite{Stalp2002}.
The result suggests that $\kappa L$ decays as $\kappa L \propto t^{-3/2}$, which is indirectly consistent with the Kolmogorov energy spectrum in QT.

The relationship between $\kappa L \propto t^{-3/2}$ and the Kolmogorov energy spectrum was reviewed by Vinen and Niemela \cite{Vinen2002}.
When we assume that the flow motion in the inertial range $D^{-1} < k < k\sub{d}$ with $D^{-1} \ll k\sub{d}$ dominates the overall fluid dynamics and gives a Kolmogorov energy spectrum \eqref{eq-Kolmogorov-law}, the total energy can be approximated by
\begin{align}
E = \int_{D^{-1}}^{k\sub{d}} C \varepsilon^{2/3} k^{-5/3} \sim \frac{3}{2} C \varepsilon^{2/3} D^{2/3}.
\end{align}
If the turbulence decays slowly, the dissipation rate $\varepsilon$ can be written as
\begin{align}
\varepsilon = - \frac{d E}{dt} = -C \varepsilon^{-1/3} D^{2/3} \frac{d \varepsilon}{dt}.
\end{align}
The solution of this differential equation is $\varepsilon = 27 C^3 D^2 / (t + t_0)^3$, with a constant $t_0$.
This result and Eq. \eqref{eq-quasi-classical-dissipation-rate} give
\begin{align}
L = \frac{(3 C)^{3/2} D}{\kappa \nu^{\prime 1/2}} (t + t_0)^{-3/2}. \label{eq-decay-dense-vortices}
\end{align}
The behavior $L \sim t^{-3/2}$ for large $t$ suggests that the Kolmogorov energy spectrum in QT is independent of the value of $\nu^\prime$, namely the temperature.

This type of decay $L \sim t^{-3/2}$ was also observed in turbulence by an impulsive spin down for superfluid $^4$He \cite{Walmsley2007}, in turbulence by injecting negative ions into superfluid $^4$He \cite{Walmsley2008}, and for grid turbulence in superfluid $^3$He-B \cite{Bradley06}, with little normal fluid at low temperatures, which supports the classical analogue of QT with no normal fluid component.

The Helsinki group experimentally studied vortex dynamics of propagation into a region of vortex-free flow in a rotating superfluid $^3$He-B \cite{Eltsov2007}.
They measured the velocity of the vortex front toward the metastable region, and determined the rate of dissipation as a function of temperature.
The results showed a transition from laminar through quasiclassical turbulent to quantum turbulent flow with decreasing temperature.
Below $0.25 T\sub{c}$ with a superfluid critical temperature $T\sub{c}$, there was a peculiar decrease of dissipation.
As one possibility, the authors suggested that the energy flux toward small scales propagates to the scale $l$, and vortex discreteness and quantization effects become important.
The dominant part of the energy loss was the Kelvin-wave cascade below the scale $l$.
As proposed by L'vov {\it et al.}, Kelvin waves are much less efficient in downscale energy transfer than classical turbulence, which leads to a bottleneck accumulation of kinetic energy \cite{L'vov2007}.

\subsection{Decay of vortices in quantum turbulence at low temperatures} \label{subsec-decay-of-vortices}

In this section, we consider the decay process of QT at low temperatures, at which the normal fluid component is negligible and mutual friction does not occur.
In this case, there is no dissipation in vortex lines at large scales and dissipation occurs only at scales comparable to the core radius $\xi$.
Energy at large scales cannot dissipate, but flows to smaller scales via the Richardson cascade and the Kelvin-wave cascade \cite{Vinen2002}.

Feynman first proposed a dissipation mechanism of QT at zero temperature \cite{Feynman}.
In this mechanism, large vortex loops are broken up into smaller loops via the Richardson cascade, and the smallest vortex rings, with radii comparable to the atomic scale, decay into excitations, such as rotons.
This mechanism is not currently accepted.
Later, Vinen considered the decay of superfluid turbulence at finite temperatures in an examination of his experimental results from thermal counterflow, and proposed Vinen's equation \cite{Vinen57c}.
At finite temperatures, vortices in turbulence are randomly spaced with no polarization, and there is only one length scale $l$.
The energy and the decay of vortices spreads over a wide range of scales.
The decay of the total energy is expressed in terms of a characteristic velocity $v\sub{s} = \kappa / 2 \pi l$ and a characteristic time constant $\tau = l / v\sub{s}$ as
\begin{align}
\frac{d v\sub{s}^2}{d t} = - \chi \frac{v\sub{s}^2}{\tau} = - \chi \frac{v\sub{s}^3}{l},
\end{align}
where $\chi$ is a temperature dependent dimensionless parameter \cite{Vinen57c}.
Incorporating the vortex line density $L = l^{-2}$, this equation becomes
\begin{align}
\frac{d L}{d t} = - \chi \frac{\kappa}{2 \pi} L^2.
\end{align}
This is a type of Vinen's equation without a generation term proportional to $(v\sub{s} - v\sub{n}) L^{3/2}$ due to thermal counterflow.
The solution of this equation is
\begin{align}
\frac{1}{L} = \frac{1}{L_0} + \chi \frac{\kappa}{2 \pi} t, \label{eq-decay-dilute-vortices-Vinen}
\end{align}
where $L_0 = L(t=0)$.
At large $t$, $L$ behaves as $L \propto t^{-1}$.
The experimental observations of thermal counterflow can be explained by this solution, and $\chi$ can be obtained as a function of temperature.
Although the decay of QT at finite temperatures is due to mutual friction, this solution can describe QT at zero temperature without mutual friction.
Using a vortex filament model, Tsubota {\it et al.} performed numerical simulations of QT without mutual friction, and estimated the value of $\chi$ \cite{Tsubota00}.
Equation \eqref{eq-decay-dilute-vortices-Vinen} can also be formulated by assuming that the motion of the quantized vortices dominates the fluid dynamics and their energy $\Lambda/(4\pi) \kappa^2 L$ is their primary contribution to the total energy \cite{Walmsley2008}.
By using Eq. \eqref{eq-quasi-classical-dissipation-rate}, we obtain
\begin{align}
\frac{d E}{dt} = \frac{\Lambda}{4 \pi} \kappa^2 \frac{d L}{dt} = - \nu^\prime \kappa^2 L^2.
\end{align}
The solution of this differential equation is
\begin{align}
\frac{1}{L} = \frac{1}{L_0} + \frac{\Lambda}{4 \pi \nu^\prime} t^{-1}, \label{eq-decay-dilute-vortices}
\end{align}
which is the same result as Eq. \eqref{eq-decay-dilute-vortices-Vinen}.
The effective kinematic viscosity $\nu^\prime$ in Eq. \eqref{eq-decay-dense-vortices} and that in Eq. \eqref{eq-decay-dilute-vortices} are naturally different.
We denote them as $\nu^\prime\sub{K}$ for Eq. \eqref{eq-decay-dense-vortices} and $\nu^\prime\sub{V}$ for Eq. \eqref{eq-decay-dilute-vortices} \cite{Walmsley2008}.

Two important questions arise: What causes the difference in decay between Eqs. \eqref{eq-decay-dilute-vortices} ($L \propto t^{-1}$) and \eqref{eq-decay-dense-vortices} ($L \propto t^{-3/2}$), and what is the origin of the decay mechanism at zero temperature?
The answer to the first question comes from the structural differences of vortex tangles in QT.
When a vortex tangle supports the Kolmogorov energy spectrum, vortices form an inertial range in which the tangle is self-similar, creating polarized vortex bundle structures that are different from a completely random distribution of vortex lines in QT at finite temperatures.
In this case, QT decays as $L \propto t^{-3/2}$.
However, when a vortex tangle is dilute and random, with no correlation, there is only one length scale $l$ in the tangle and QT decays as $L \propto t^{-1}$, even at zero temperature.
This different behavior of $L$ enables us to understand the structure of a vortex tangle in QT.
The behavior of $L \propto t^{-1}$ has been observed in turbulence by injecting negative ions into $^4$He \cite{Walmsley2008} and by a vibrating grid in $^3$He-B \cite{Bradley06}.
As discussed in the previous subsection, the behavior of $L \propto t^{-3/2}$ has been observed in turbulence by an impulsive spin down of superfluid $^4$He, by injecting negative ions into superfluid $^4$He \cite{Walmsley2007}, and by a vibrating grid in superfluid $^3$He-B \cite{Bradley06}.
Bradley {\it et al.} also observed the crossover from $L \propto t^{-3/2}$ to $L \propto t^{-1}$ behavior of grid turbulence in $^3$He \cite{Bradley06}.
The structure of a generated vortex tangle depends on the velocity of the grid, and the decay of the turbulence becomes $L \propto t^{-3/2}$ for a dense vortex tangle with a rapidly oscillating grid, and $L \propto t^{-1}$ for a dilute vortex tangle with a slowly oscillating grid.
By using Eqs. \eqref{eq-decay-dense-vortices} and \eqref{eq-decay-dilute-vortices}, Walmsley and Golov estimated the effective kinematic viscosity $\nu^\prime\sub{K}$ and $\nu^\prime\sub{V}$ for turbulence in $^4$He and showed that they are almost same at high temperatures $T > 1.0$ K for a normal fluid, and that $\nu^\prime\sub{K}$ becomes much smaller than $\nu^\prime\sub{V}$ at low temperatures with no normal fluid \cite{Walmsley2008}.

For the second question, there are several possible answers.
The first is acoustic emissions at vortex reconnections, similar to that of eddies in a classical fluid.
Numerical simulations of the GP model support acoustic emissions at every reconnection \cite{Leadbeater01,Ogawa02a}.
The energy dissipation for each reconnection is about $3 \xi$ times the vortex line energy per unit length: $\sim \Lambda \kappa^2 / 4 \pi$.
In the case of superfluid $^4$He, however, $\xi$ is quite small and the dissipation due to reconnections can be negligible.
This dissipation mechanism may emerge for the turbulent state of atomic Bose-Einstein condensates \cite{PethickSmith}.
By using the GP model, Kobayashi and Tsubota investigated the decay from fully developed turbulence and reported both $L \propto t^{-3/2}$ to $L \propto t^{-1}$ behaviors \cite{Kobayashi2006}.
In their simulation, $\xi$ is close to the system size, and acoustic emission upon reconnection is the main dissipation mechanism.
Another possible answer is the radiation of phonons from high-frequency Kelvin waves \cite{Vinen2001}.
Vortex reconnections excite Kelvin waves whose wavelength is on the order of $l$ \cite{Schwarz85,Svistunov1995,Tsubota00,Kivotides2001a}.
Although Kelvin waves with wavelength $\sim l$ cannot cause effective radiation, the Kelvin-wave cascade creates waves with much shorter wavelengths \cite{Vinen2001,Mitani2003,Kozik2004,Kozik2005,Nazarenko2006,Boffetta2009}.

In actual experiments, one must also consider vortex diffusion as an origin of the decrease of $L$.
When a vortex tangle is inhomogeneous or local and the probe observing the turbulence is also local, a vortex tangle may escape from the observable region.
Using the vortex filament model, Tsubota {\it et al.} numerically studied the diffusion of an inhomogeneous vortex tangle \cite{Tsubota2003}.
The effects of diffusion can be quantitatively evaluated by the equation:
\begin{align}
\frac{\partial L(\mathbf{x},t)}{\partial t} = - \chi \frac{\kappa}{2 \pi} L(\mathbf{x},t)^2 + D \nabla^2 L(\mathbf{x},t).
\end{align}
Here, $L(\mathbf{x},t)$ is the space-dependent line length density, and $D$ is the diffusion constant.
The numerical simulation indicated that $D \sim 0.1 \kappa$

\subsection{Quantum turbulence created by vibrating structures}
Recently, vibrating structures, such as discs, spheres, grids, wires, and tuning forks, have been widely used for research into QT \cite{VinenPLTP}. Despite detailed differences between the structures, the experiments have shown some surprisingly common phenomena. This trend started with the pioneering observation of QT on an oscillating microsphere by J\"ager {\it et al.}\cite{Jager95}. 
Subsequently, many groups have experimentally investigated the transition to turbulence in superfluid $^4$He and $^3$He-B by using grids \cite{Nichol04a,Nichol04b,Charalambous06,Bradley05a,Bradley05b,Bradley06}, wires \cite{Fisher01,Bradley04,Yano07,Hashimoto07,Goto08}, and tuning forks \cite{Blazkova07b,Blazkova09}.  The details of these observations were described in a review article \cite{VinenPLTP}.

\begin{figure}[htb] \centering \begin{minipage}[t]{0.8\linewidth} \begin{center} \includegraphics[width=.8\linewidth]{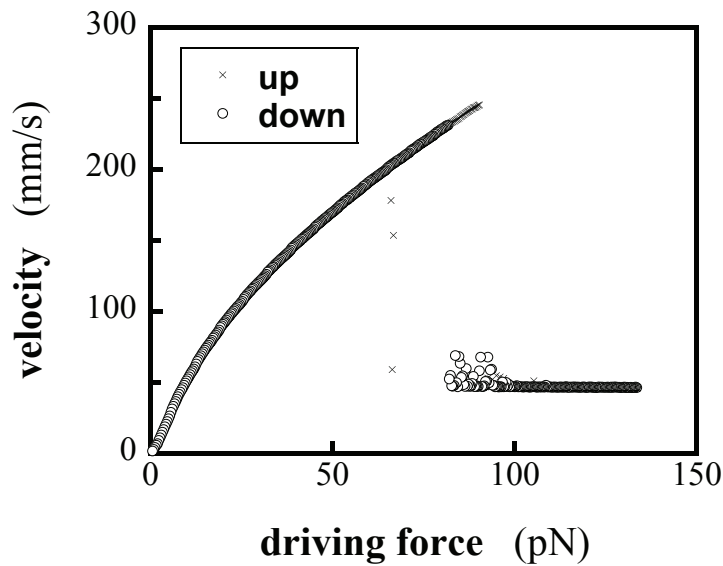} \end{center} \end{minipage} 
\caption{Typical response of a vibrating wire in superfluid $^4$He at 30 mK. Courtesy of H. Yano.} \label{Yano}\end{figure}

Here, we will describe briefly the essence of the observations by referring to a typical result from Yano {\it et al.}\cite{Yano09}. A thin superconducting (typically NbTi) wire with a micron-size radius is formed into a semicircle, and its two edges are attached to the wall of a vessel. The wire vibrates in resonance with a Lorentz force due to an alternating electric current under a static magnetic field.  Figure \ref{Yano} shows the response velocity of the wire as a function of the driving force.
When the driving force is relatively low, the wire moves smoothly.
However, if the driving force exceeds some value, the velocity suddenly drops, indicating that the wire does not vibrate so much despite an increase of the driving force. 
The energy injected from the drive must go somewhere, and the only possible escape is the excitation of quantum turbulence.  The response of the wire clearly shows a hysteresis between the upward and downward sweeps of the driving force.

Experimental studies conducted by many groups reported common behaviors, independent of the structural details, such as type, shape, and surface roughness. However, there were also some differences. The key points are the critical velocity for the onset of turbulence, the shift of the resonance frequency, and the hysteresis and the drag coefficient. 

The observed critical velocities were in the range from 1 mm/s to approximately 200 mm/s, much lower than the velocity necessary for intrinsic nucleation of quantized vortices, which is on the order of 10 m/s. It is known that there are initially some remnant vortices in superfluid $^4$He\cite{Awschalom84}, and the transition to turbulence should come from their extension or amplification.  This scenario was confirmed by numerical simulations using the vortex filament model \cite{Hanninen07} and an experiment using a wire free from remnant vortices \cite{Goto08}, which will be described later. However, it should be noted that this scenario is applicable only to $^4$He. The transition to turbulence observed in the B phase of superfluid $^3$He \cite{Bradley05b,Bradley06} can be related with the intrinsic nucleation of vortices, which is known to occur near a solid surface at very small velocities, on the order 4 mm/s, depending on the surface roughness \cite{Ruutu97}. This issue was discussed in detail in a review article \cite{VinenPLTP}.    
The dependence of the critical velocity $v\sub{c}$ on the oscillation $\omega$ is very important. Most experiments have provided evidence for a universal scaling property $v\sub{c} \sim \sqrt{\kappa \omega}$, independent of the geometry of the vibrating structure. This relation is trivial from dimensional analysis, but it is necessary to investigate its physical origin. H\"anninen and Schoepe discussed the scaling property \cite{Hanninen10}. This relation was obtained from the "superfluid Reynolds number" $Re=v\ell/\kappa$ with some characteristic length scale $\ell$ \cite{Volovik03}. It is also possible to understand the scaling by extending the scenario of the dynamical behavior of a vortex tangle in counterflow turbulence at constant velocity \cite{Kopnin04} to the case of oscillating flows. However, more detailed investigation of the issue is required.

The transition to turbulence is accompanied with a characteristic shift of the resonance frequency. In the wire experiments Yano {\it et al.} observed that the resonance frequency decreased with increasing drive force in the laminar regime, but increased in the turbulent regime \cite{Yano07}. Bradley {\it et al.} also observed an upward shift of the resonance frequency at the onset of turbulence \cite{Bradley05a}. Such an increase in the resonance frequency is surprising, because it indicates either an increased stiffness or a decreased effective mass of the wire. If vortices were amplified and a tangle surrounds the wire, the effective mass should increase. The reasons for the frequency shift are not yet resolved, and more study is required.

Many groups have reported some hysteresis in the transition to turbulence, which arises from the stability of the laminar and turbulent states. By using an oscillating microsphere,  J\"ager {\it et al.} observed that the transition from laminar to turbulent response was accompanied by significant hysteresis at low temperatures \cite{Jager95}. Below 0.5 K, turbulence was observed to be unstable and the flow switched intermittently between turbulent and laminar flow. Schoepe analyzed the lifetime of turbulence in connection with the statistical fluctuations of the vortex line length density \cite{Schoepe04}.   Yano {\it et al.} \cite{Yano09} and Bradley {\it et al.} \cite{Bradley05a} also observed hysteresis when using vibrating wires.

The net drag force is another important quantity when we consider the transition to QT. The drag force $F$ is often expressed in terms of a drag coefficient $C\sub{D}$, defined by the equation \cite{Batchelor}
\begin{equation}  F=\frac12 C\sub{D} \rho A U^2, \label{drag} \end{equation}
where $\rho$ is the density of the fluid, $A$ is the projected area of the object normal to the flow, and $U$ is the flow velocity. The behavior of $C\sub{D}$ is well known in a steady classical flow past an obstacle. In the case of laminar potential flow of an ideal fluid past a cylinder, the flow is symmetric about the plane through the center of the cylinder and normal to the flow. As a result, the net force vanishes with $C\sub{D}=0$ (the d'Alembert paradox).  For laminar viscous flow, the drag force is approximately proportional to $U$ so that $C\sub{D} \sim U^{-1}$ (the Stokes' formula). In the case of flow at a high Reynolds number, the flow behind the obstacle accompanies a wake, which destroys the symmetry and leads to the d'Alembert paradox, and $C\sub{D}$ becomes of order unity. The value of $C\sub{D}$ depends on the geometry of the obstacle \cite{Batchelor}; $C\sub{D}$ is approximately unity for a disc or cylinder, and about 0.3 for a sphere. Also, in quantum turbulence created by vibrating structures, it is possible to obtain the drag coefficient from the dependence of velocity on the driving force, as in Fig. \ref{Yano}. Much of the data on microspheres \cite{Jager95}, mm-scale spheres \cite{Hemmati09},  grids \cite{Nichol04a,Nichol04b}, and quartz forks \cite{Blazkova09} show the classical analogue behavior in which the drag coefficient is about $U^{-1}$ in the laminar regime and of order unity in the turbulent regime. This is another classical analogue of QT. Although the experimental results are trivial, we need a clearer understanding of the phenomena.  The drag coefficient characteristic of the turbulent regime was numerically confirmed by simulation of the vortex filament model \cite{Fujiyama09}.    

The remnant vortices play several important roles in these phenomena, which are discussed in the remainder of this subsection.

\begin{figure}[htb] \centering \includegraphics[width=.95\linewidth]{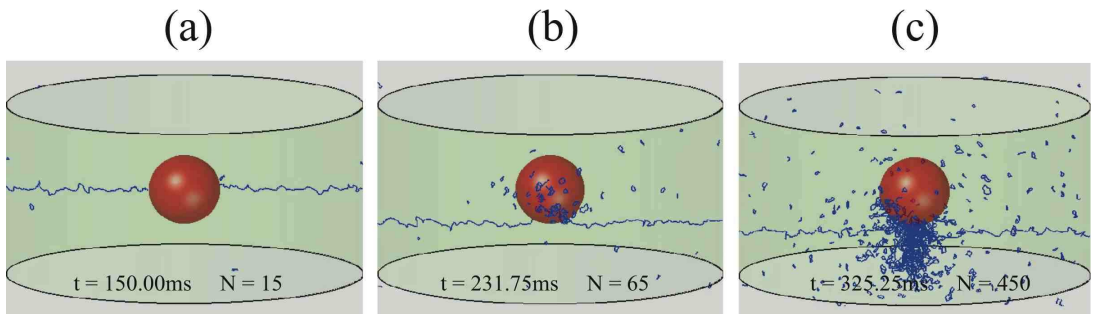}
\caption{Evolution of the vortex line near a sphere of radius
100~$\mu$m in an oscillating superflow of 150 mm$^{-1}$ at 200~Hz
[H\"anninen, Tsubota and Vinen: Phys. Rev. B \textbf{75} (2007) 064502, reproduced with permission. Copyright 2007 by the American Physical Society].} \label{Hanninen}\end{figure}

H\"anninen {\it et al.} performed a numerical simulation using the vortex filament model, and described how remnant vortices develop into a tangle under an AC superflow, as shown in Fig. \ref{Hanninen} \cite{Hanninen07}.  A smooth solid sphere of radius 100~$\mu$m was placed in a cylindrical vessel filled with superfluid $^4$He.  Generally, we do not know how remnant vortices remain in a given geometry, but here they were assumed to initially extend between the sphere and the vessel wall. When an oscillating superflow of 150 mm$^{-1}$ at 200~Hz was applied, Kelvin-waves resonant with the flow were gradually excited along the remnant vortices. The amplitude of the Kelvin-waves grew large enough to lead to self-reconnection and the emission of small vortex rings.  These vortices gathered around the stagnation points, repeatedly reconnected, and were amplified by the flow, eventually developing into a vortex tangle. This simulation likely captured the essence of what occurred in the experiments. However, the simulation was not necessarily satisfactory, and some problems remain. The first is the critical velocity. The situation of Fig. \ref{Hanninen} is quite similar to the experiment using a microsphere \cite{Jager95} and gives a critical velocity of about 120 mm$^{-1}$, which is much larger than the observed value of about 40 mm$^{-1}$. The second problem is hysteresis. The simulation does not produce any hysteresis between upward and downward sweeps. If the velocity is taken above the critical value for the formation of a tangle, and then reduced below this value, the vortices immediately decay, eliminating the turbulent state. These two difficulties may arise from the artificially smooth surface of the sphere. A solid surface is generally rough, and even tiny bumps can act as pinning sites for quantized vortices with such a thin core \cite{Schwarz85,Tsubota93}.  If such pinning effects on the surface are taken into account, they should induce additional disturbances of vortices and help the transition to turbulence, thereby reducing the critical velocity. Regarding the hysteresis, some vortices may be trapped by pinning cites through the downward sweep, which can make the physics different from the upward sweep. However, we do not know how to consider the effect of surface roughness, namely multi-pinning cites for numerical vortex dynamics, and the development of applicable methods is urgently needed.  

The roles of remnant vortices were revealed by several interesting experiments by the Osaka City University group. First, Hashimoto {\it et al.} succeeded in setting a vibrating wire free from remnant vortices using a unique idea\cite{Hashimoto07}. They added a small chamber with a pinhole in a sample cell, and filled the chamber extremely slowly with superfluid $^4$He. The vortices were likely filtered out by the pinhole, providing a vibrating wire free from remnant vortices. The wire never caused a transition to turbulence, even when the velocity exceeded 1 m/s. Next, Goto {\it et al.} set two wires in the small chamber. By using the same procedure, they obtained one wire (labeled A) free from remnant vortices and another (labeled B) with remnant vortices \cite{Goto08}.  Wire B can create turbulence by itself from remnant vortices and the resulting tangle emits vortex rings. Although wire A alone never caused turbulence, it could create turbulence if it received seed vortices from wire B. This observation clearly demonstrated that the turbulence arises from remnant vortices. This behavior was confirmed by numerical simulation of the vortex filament model \cite{Goto08,Fujiyama09}.   

\subsection{Visualization of quantized vortices and turbulence}
There has been little direct experimental information regarding the flow in superfluid $^4$He. This is mainly because usual flow visualization techniques are not applicable to cryogenic superfluids. However, this situation is rapidly changing \cite{SciverPLTP}. For QT, one can seed the fluid with tracer particles in order to visualize the flow field and possibly quantized vortices, which are observable by appropriate optical techniques.  

A significant contribution was made by Zhang and Van Sciver \cite{Zhang05a}. Using a particle image velocimetry (PIV) technique with 1.7-$\mu$m-diameter polymer particles, they visualized a large-scale turbulent flow both in front of and behind a cylinder in a counterflow in superfluid $^4$He at finite temperatures. In classical fluids, such large-scale turbulent structures are seen downstream of objects such as cylinders, with the structures periodically detaching to form a vortex street. In the present case of $^4$He counterflow, the locations of the large-scale turbulent structures were relatively stable, and they did not detach and move downstream, although local fluctuations in the turbulence were evident. 

\begin{figure}[htb] \centering \begin{minipage}[t]{0.8\linewidth} \begin{center} \includegraphics[width=.8\linewidth]{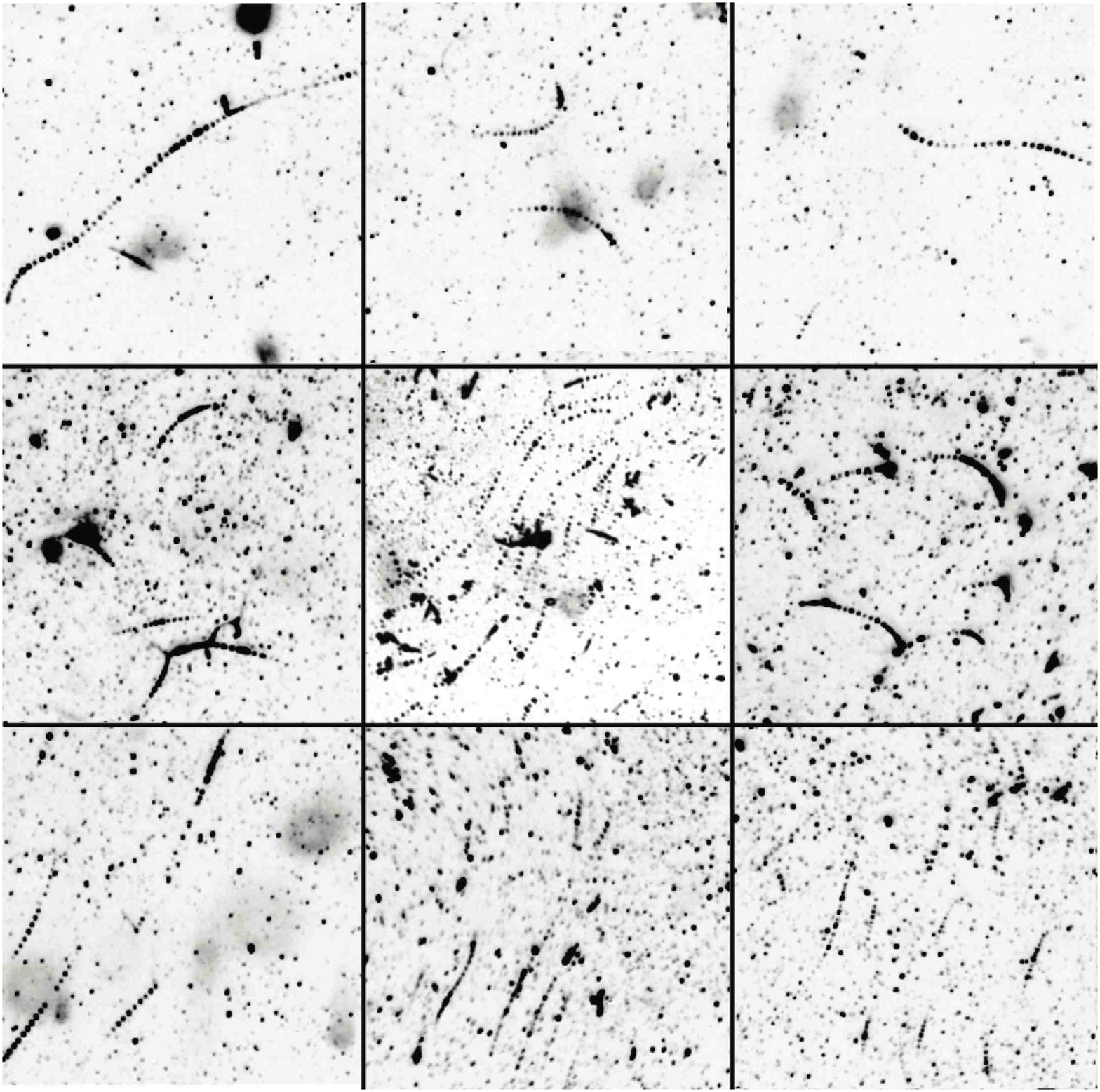} \end{center} \end{minipage} 
\caption{Visualization of quantized vortices by intensity-inverted images. Each image shows the full field of view of 6.78 $\times$ 6.78 mm$^2$. Multiple particles are often trapped in each visible vortex, and trapped particles tend to be uniformly spaced along the vortex core \cite{Paoletti08}. Courtesy of E. Fonda, K. Gaff, M. S. Paoletti, and D. P. Lathrop.} \label{Lathrop}\end{figure}

Another significant contribution was the visualization of quantized vortices by Bewley {\it et al.} \cite{Bewley06}.  In their experiments, the liquid helium was seeded with solid hydrogen particles smaller than 2.7 $\mu$m at a temperature slightly above $T_\lambda$, after which the fluid was cooled to below $T_\lambda$.  When the temperature was above $T_\lambda$, the particles were seen to form a homogeneous cloud that dispersed throughout the fluid. However, on passing through $T_\lambda$, the particles coalesced into web-like structures, as shown in Fig. \ref{Lathrop}. Bewley {\it et al.} suggested that these structures represent decorated quantized vortex lines. They reported that the vortex lines appear to form connections rather than remaining separated, and were homogeneously distributed throughout the fluid. The fork-like structures may indicate that several vortices were attached to the same particle, as indicated by numerical simulations of vortex pinning \cite{Tsubota94}. 
Using the same technique, Paoletti {\it et al.} obtained the trajectories of tracer particles visualized thermal counterflow \cite{Paoletti08}. The observed trajectories showed two distinct types of behavior. One group consisted of trajectories that moved in the direction of the heat flux, while the other consisted of those that opposed this motion. The former trajectories were smooth and uniform, but the latter could be quite erratic. Particles of the former trajectories were probably dragged by the normal fluid, while those of the latter were trapped in vortex tangles.

Here, it is necessary to understand whether such tracer particles follow the normal flow or the superflow, or an even more complex flow. Poole {\it et al.} studied this problem theoretically and numerically, and showed that the situation changes depending on the size and mass of the tracer particles\cite{Poole05}. However, the situation is so complicated that many issues remain unanswered \cite{Sergeev06,Kivotides07,Kivotides08,Barenghi09}.

\chapter{Quantized vortices in atomic Bose-Einstein condensates}

\section{Introduction}
The achievement of Bose--Einstein condensation in trapped atomic gases at ultra-low temperatures 
\cite{Anderson95,Bradley,Davis} has stimulated intense experimental and theoretical activity in 
modern physics. In this section, 
we review the physics of quantized vortices in atomic-gas BECs.
While quantized vortices have been thoroughly studied in superfluid helium \cite{Doneley}, 
there has been a resurgence of interest in vortices in atomic BECs because of the peculiar features 
of this system: 
\begin{enumerate}
\item {\it Small gas parameter} \\
Because the gas is dilute, the GP equation (\ref{eq-gp}) gives a quantitatively accurate 
description of the static and dynamic properties of the 
atomic condensate \cite{PethickSmith,PitaevskiiStringari}. 
Therefore, the vortex structure and dynamics can be 
discussed by a more fundamental approach than with superfluid helium, in which 
the interaction cannot be described in such a simple local form. 
Furthermore, the diluteness of a gas leads to a relatively large healing length that characterizes the vortex 
core size, thus enabling the direct experimental visualization of vortex cores 
\cite{Matthews,Madison,Abo,Madison01,Hodby} 
\item {\it Trapping potential }\\
The finite size effect of the trapping potential and the associated density inhomogeneity 
yield new characteristics of vortex physics. This also provides a new spectroscopy 
to study characteristic vortex dynamics through collective motion of the inhomogeneous condensate
\cite{Chevy,Haljan01,Bretin03,Hodby2}. 
\item {\it Laser manipulation and rotation}\\
The manipulation of a condensate wave function via external 
(magnetic or optical) fields provides a versatile scheme to control the vortex states. 
Laser beams can yield effective repulsive or attractive forces on atoms, 
creating confining potentials, periodic lattice potentials, and 
impurity/obstacle potentials for atomic condensates. 
Vortex formation can be achieved by the mechanical rotation of a 
laser-created optical spoon \cite{Madison,Abo,Madison01} or the transfer of orbital angular 
momentum from a Laguerre-Gaussian laser beam \cite{Anderson3}. 
The tunability of the rotation over a wide range is useful to study various vortex phases 
and their dynamics in trapped BECs. This provides an opportunity to study 
a rapidly rotating condensate with a dense vortex lattice \cite{Abo,Haljan,Bretin04}, 
which is not easy to accomplish in superfluid helium systems. 
Such rapidly rotating BECs present a so-called mean-field quantum-Hall regime described 
by the orbitals of the lowest Landau levels (LLL), and also a regime dominated by 
strong many-body correlation \cite{Cooperrev,Viefersrev,Bloch}. 
Moreover, laser-induced coherent transition of atoms between different hyperfine levels 
can control the phase profile in such a way that the condensate has a circulation \cite{Matthews}. 
Recently, the laser field has played the role of an artificial vector potential \cite{Lin,Lin2}, which could be 
an alternative means of manipulating the rotational properties of condensates. 
\item {\it  Detection} \\
The vortex cores can be directly visualized through the observed density profile 
by a time-of-flight (TOF) technique \cite{Matthews,Madison,Abo,Madison01}. 
The TOF technique involves switching off the trapping 
field (magnetic or optical) at time $t=0$ and taking an image of the BEC several 
(typically 5 to 25) milliseconds later. Switching off the trap allows the atomic gas to expand 
until a laser beam probe becomes available to observe the density profile. 
Images are taken by shining a resonant laser beam into the atomic 
gas and using a CCD camera to observe the shadow cast by the absorption of photons, 
from which the integrated atomic density can be determined. 

The matter-wave interference technique is also applicable to vortex 
detection \cite{Inouye2,Chevy2}. The presence of vortices can be revealed by 
finding the dislocations in the interference fringes between expanding condensates \cite{Bolda}. 
In addition, the Bragg scattering of matter waves from laser beams can be used to detect the vortex state, 
because the behavior of the scattered condensate is sensitive to the spatial phase 
distribution of the initial state \cite{Blakie,Muniz}. 
\item {\it Multicomponent condensates} \\
Since it is possible to load and cool atoms in more than one hyperfine spin state 
or more than one atomic element in the same trap, multicomponent BECs with internal 
degrees of freedom can be created experimentally 
\cite{Myatt,Hall1,Stenger,Barrett,Modugno2,Schmaljohann,Chang,Kuwamoto,Thalhammer,Papp2}.
Multicomponent condensates allow the formation of various unconventional topological defects 
with complex properties that arise from interactions between different components, 
providing a new platform for the study of unconventional vortices. This content relates closely 
to other fields of physics, such as superfluids $^{3}$He \cite{Vollhardt}, 
unconventional superconductors \cite{supercond}, 
quantum Hall systems \cite{Girvin}, as well as high-energy physics 
and cosmology \cite{Volovik}. 
Atomic BEC is an extremely flexible system for the study of such topological defects, since optical
techniques allow the control of the condensate wave functions. 
An external field can couple the internal sublevels of the condensate 
and cause a coherent transition of the population. 
This coherent transition can be used to control the spatial variation 
of the condensate wave functions, resulting in an ``imprinting'' of 
a phase pattern onto the condensate. In most schemes, the spatial 
configuration of the field, the intensity and detuning of the laser fields, 
and the phase relationship between the different fields must 
be carefully controlled to create the correct phase pattern of topological excitations 
through the complex internal dynamics.
\item {\it  Feshbach resonance} \\
The most salient feature of the cold atom system is that a field-induced Feshbach resonance can tune 
the s-wave scattering length between atoms \cite{Inouye}, which determines the strength of atom-atom interaction.
A Feshbach resonance occurs when a quasi-bound molecular state in a closed channel has an energy 
equal to that of two colliding atoms in an open channel. This technique can change the 
scattering length over a wide range, from negative values to positive ones, thus 
creating condensates with strongly attractive interactions \cite{Donley}, 
strongly repulsive interactions \cite{Papp}, or dominant long-range dipole-dipole 
interactions \cite{Lahaye}. 
It also provides new physics associated with the formation of quasi-bound 
molecules, provided by the coherent transition between an atomic condensate 
and a molecular one \cite{Donley2}. 
In the case of fermions, this has been used to induce pairing of two fermions, 
which results in the formation of fermion condensates by controlling the character 
of the two-body interactions from strongly-bound 
bosonic molecules to weakly-interacting fermion (Cooper) pairs \cite{Inguscio}. 
Vortices in these exotic condensates are expected to show quite intriguing properties. 
A review of Feshbach resonance was recently given by Chin {\it et al}., \cite{Chin}.
\end{enumerate}

There are several excellent reviews of quantized vortices in ultracold atomic BECs. 
The basics of the theoretical formulation and the early research on the physics of 
quantized vortices were reviewed by Fetter and Svidzinsky \cite{Fetterrev}. 
Recent progress of the extensive study of quantized vortices 
is described in Ref. \cite{Kasamatsurev,Fetterrev2}. 
These reviews were mainly concerned with the properties of a conventional single-component BEC, 
whose order parameter is scalar and interaction is 
characterized only by an s-wave scattering length.  
In this chapter, therefore, we will also discuss vortex physics in unconventional BEC's, 
which was not covered by these reviews. 
We begin with the basic theory and experiments on quantized vortices 
in conventional BECs. 
Then, we describe vortices in some unconventional condensates realized in ultra-cold 
atomic systems, such as spinor condensates and dipolar condensates. 

\section{Vortices in single-component Bose-Einstein condensates}
First, we review vortices in a single-component BEC in a trapping potential 
from both theoretical and experimental points of view. 
A more detailed review and references can be found in Ref. \cite{Kasamatsurev,Fetterrev2}. 
The contents of this section will give the basic information necessary for subsequent sections. 
After reviewing the theoretical formulation for trapped 
BECs, we describe the properties of a single vortex and a vortex lattice in trapped BECs, 
together with various interesting current topics.  

\subsection{Theoretical description of ultra-cold atomic-gas BECs }
\subsubsection{Gross-Pitaevskii equation} 
The many-body Hamiltonian for bosons in a trapping 
potential $V_{\rm ex}({\bf r})$ is given by 
\begin{equation}
\hat{H} = \int d{\bf r} \hat{\psi}^{\dagger}({\bf r}) \left[ 
- \frac{\hbar^{2} \nabla^{2}}{2m} +V_{\rm ex} \right] \hat{\psi}({\bf r})
+ \frac{1}{2} \int d{\bf r} \int d{\bf r}'  
\hat{\psi}^{\dagger}({\bf r}) \hat{\psi}^{\dagger}({\bf r}')
V({\bf r} -{\bf r}') \hat{\psi}({\bf r}') \hat{\psi}({\bf r}) \label{GPfunctional}
\end{equation}
with the bosonic field operator $\hat{\psi}({\bf r})$ 
and the two-body interaction $V({\bf r} -{\bf r}')$. The dynamics of $\hat{\psi}$ are governed 
by the Heisenberg equation $i \hbar \partial \hat{\psi}/\partial t = [\hat{\psi}, \hat{H}]$. 
In the low temperature limit, the field operator can be approximated by a 
classical field known as ``condensate wave function" $\Psi = \langle \psi \rangle$, 
and the trapped BECs can thus be described by the GP equations:
\begin{equation}
i \hbar \frac{\partial \Psi({\bf r},t)}{\partial t} = \left[ - \frac{\hbar^{2} 
\nabla^{2}}{2m} +V_{\rm ex} + g |\Psi({\bf r},t)|^{2} \right] \Psi({\bf r},t). \label{GPeq}
\end{equation} 
The trapping potential typically has the form of a harmonic oscillator 
$V_{\rm ex} = (m/2) ( \omega_{x}^2 x^2 + \omega_{y}^2 y^2 + \omega_{z}^2 z^2)$. 
Comparison of the kinetic and trap energies introduces a characteristic 
length scale $a_{\rm ho} = \sqrt{\hbar / m \omega}$ 
with  $\omega = (\omega_{x} \omega_{y} \omega_{z})^{1/3}$. 
The time-dependent GP equation (\ref{GPeq}) can be used to 
explore the dynamic behavior of the condensate, which is characterized by variations of the order parameter 
over distances larger than the mean distance between atoms. 
This equation is valid when the s-wave scattering length is much smaller than the average 
distance between atoms, and the number of atoms in the condensate is much larger than unity. 
The success of the quantitative description of the trapped BECs 
is described in the books by Pethick and Smith \cite{PethickSmith}, 
and Pitaevskii and Stringari \cite{PitaevskiiStringari}.

To determine the ground state of a trapped BEC, 
we can write the condensate wave function as $\Psi({\bf r},t)=\Phi({\bf r}) e^{-i \mu t/\hbar}$, 
where $\Phi({\bf r})$ obeys the time-independent GP equation
\begin{equation}
\left[ - \frac{\hbar^{2} \nabla^{2}}{2m} +V_{\rm ex} + g |\Phi({\bf r})|^{2} \right] \Phi({\bf r}) 
= \mu \Phi({\bf r}).  \label{staGPeq}
\end{equation}
Here, $\Phi$ is normalized to the number of condensed particles $\int d {\bf r} |\Phi({\bf r})|^{2} = N_{0}$, 
which determines the chemical potential $\mu$. In the dilute limit (the gas parameter $\bar{n} |a|^3$ 
is typically less than 10$^{-3}$, where $\bar{n}$ is the average density of the gas) 
the condensed particles $N_{0}$ can be approximated as the total number $N$,
because depletion of the condensate is small as $N^{\prime}=N-N_0\propto \sqrt{\bar n|a|^3}N\ll N$.  
The time-independent GP equation (\ref{staGPeq}) 
is also derived by minimizing the GP energy functional
\begin{equation}
E[\Phi, \Phi^{\ast}] = \int d {\bf r} \Phi^{\ast} \left( -\frac{\hbar^{2} \nabla^{2}}{2m}  
+ V_{\rm ex} + \frac{g}{2} |\Phi|^{2} \right) \Phi \equiv E_{\rm kin} + E_{\rm tr} +E_{\rm int}, \label{Hamilfunctional}
\end{equation}
subject to the constraint of a fixed particle number $N$. This constraint is taken into account 
by the Lagrange multiplier $\mu$ as $\delta (E - \mu N)/ \delta \Phi^{\ast} = 0$, 
where $\mu$ is the chemical potential that ensures a fixed $N$.

Although the exact solutions of the ground state can be generally obtained 
only by solving Eq. (\ref{staGPeq}) numerically, an approximate analytic solution can be found 
when the interaction energy $E_{\rm int}$ is much larger than $E_{\rm kin}$ \cite{Baym}. 
To see this, let us neglect the anisotropy of the harmonic potential and assume that the cloud 
occupies a region of radius $\sim R$, so that $n \sim N/R^{3}$. Thus, the scale of the harmonic 
oscillator energy per particle is $\sim m \omega^{2} R^{2} /2 $, while each particle experiences an 
interaction with the other particles of energy $\sim gN/R^{3}$. By equating these energies, the radius 
is found to be $R \sim a_{\rm ho} (8 \pi N a/a_{\rm ho})^{1/5}$. 
Since the kinetic energy is of order $\hbar^{2}/2mR^{2}$, 
the ratio of the kinetic to interaction (or trap) energies is $\sim (N a/a_{\rm ho})^{-4/5}$. 
In the limit $Na/a_{\rm ho} \gg 1$, which is relevant to most of the experiments, 
the repulsive interactions significantly expand the condensate, so that the kinetic energy associated 
with the density variation becomes negligible compared to the trap and interaction energies. 
As a result, the kinetic-energy operator can be omitted in Eq. (\ref{staGPeq}), which gives the 
Thomas--Fermi (TF) parabolic profile for the ground-state density
\begin{equation}
n({\bf r}) \simeq |\Phi_{\rm TF}({\bf r})|^2 = \frac{\mu - V_{\rm ex}({\bf r})}{g} 
\Theta\left[ \mu - V_{\rm ex}({\bf r}) \right], \label{TFparabola}
\end{equation}
where $\Theta(x)$ denotes the step function. The resulting ellipsoidal density in three-dimensional (3D) space 
is characterized by two types of parameters: the central density $n_{0} = \mu/g$ and 
the three condensate radii $R_{j}^2 = 2\mu/m\omega_{j}^{2}$ ($j=x,y,z$). 
The chemical potential $\mu$ is determined by the normalization $\int d {\bf r} n({\bf r}) = N$ 
as $\mu = (\hbar \omega/2) (15 N a/a_{\rm ho})^{2/5}$. 
In the TF regime, the time-dependent GP equation (\ref{GPeq}) reduces to the incompressible 
hydrodynamic equation for superfluids \cite{PitaevskiiStringari}. 

\subsubsection{The Bogoliubov-de Genne equation}
The spectrum of elementary excitations of the condensate is an essential ingredient  in calculations 
of thermodynamic properties. To study the low-lying collective-excitation spectrum of trapped  BECs, 
the Bogoliubov-de Gennes (BdG) equation coupled with the GP equation is a general formalism. 
Let us consider the equation of motion for a small perturbation around the stationary state $\Phi$, 
which is a solution of Eq. (\ref{staGPeq}). The wave function may take the form 
$\Psi({\bf r},t) = [ \Phi({\bf r}) + u_{j} ({\bf r}) e^{-i \omega_{j} t} 
- v_{j}^{\ast} ({\bf r}) e^{i \omega_{j} t}]  e^{-i \mu t}$. 
Inserting this ansatz into Eq. (\ref{GPeq}), and retaining terms up to 
the first order in $u$ and $v$, we obtain the BdG equation:
\begin{eqnarray}
\left(
\begin{array}{cc}
L({\bf r}) & - g \Phi({\bf r})^{2}  \\
g \Phi^{\ast}({\bf r})^{2} & - L ({\bf r})
\end{array}
\right)  \left(
\begin{array}{c}
u_{j}({\bf r}) \\
v_{j}({\bf r})
\end{array}
\right) = \hbar \omega_{j} \left(
\begin{array}{c}
u_{j}({\bf r}) \\
v_{j}({\bf r})
\end{array} \right), \label{BdG}
\end{eqnarray}
where $L ({\bf r})= - \hbar^{2} \nabla^{2}/2m + V_{\rm ex} - \mu + 2g |\Phi({\bf r})|^{2}$, 
and $\omega_{j}$ are the eigenfrequencies related to the quasiparticle normal-mode functions 
$u_{j}({\bf r})$ and $v_{j}({\bf r})$. The mode functions are subject to the orthogonality and symmetry 
relations $\int d {\bf r} [u_{i} u_{j}^{\ast} - v_{i} v_{j}^{\ast}] = \delta_{ij}$ and 
$\int d {\bf r} [u_{i} v_{j}^{\ast} - v_{i} u_{j}^{\ast}] = 0$.

Since the energy $\hbar \omega_{j}$ of this quasiparticle is defined with respect to the condensate energy 
(Eq. (\ref{Hamilfunctional}) with the stationary solution $\Phi$), the presence of quasiparticles with negative 
frequencies implies an ``thermodynamic" instability for the solution $\Phi$; if there is energy 
dissipation in the system, the excitation of negative-energy modes lowers the total energy, 
and the modes grow spontaneously to relax $\Phi$ into a more stable state. 
In other words, there is a path in configuration space 
along which the energy decreases. This argument is closely related to the Landau 
criterion for superfluidity (Landau instability). On the other hand,   
one can note that, since the matrix element of Eq. (\ref{BdG}) 
is non-Hermitian, the eigenfrequencies can assume complex values. 
Therefore, the small-amplitude fluctuations of the corresponding eigenmodes 
grow exponentially, even for the energy-conserving time development. 
This is known as ``dynamical instability". These instabilities play a key role 
in vortex dynamics. 

\subsection{Vortex experiments in rotating BECs}
The first experimental detection of a vortex in an atomic BEC was made by Matthews {\it et al.} 
in 1999 \cite{Matthews}. Their study involved condensates of $^{87}$Rb atoms residing 
in two hyperfine states. We describe this experiment in the next section, and summarize here 
the experiments on vortices in single-component BECs under rotation. 

\begin{figure}[htbp] 
\begin{center} 
\includegraphics[angle=0,width=13cm]{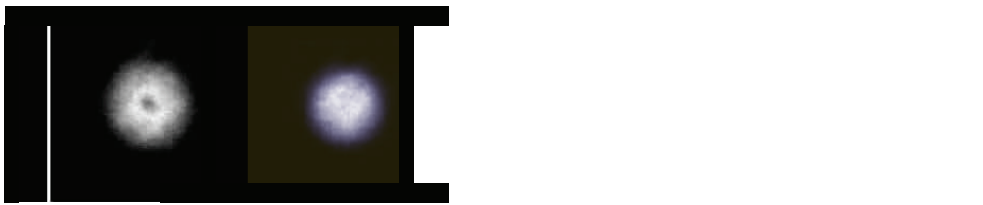} 
\end{center} 
\caption{Typical density profiles of a rotating condensate in the plane 
perpendicular to the rotation axis, taken by TOF
measurement. From left to right, the rotation frequency incrases. 
Courtesy of J. Dalibard.} 
\label{vorarray} 
\end{figure}
An intuitive method of creating vortices is to stir a condensate mechanically with a rotating ``bucket". 
Using this scheme, Madison {\it et al.} observed the formation of vortices 
in a single-component elongated BEC \cite{Madison} [see Fig. \ref{vorarray}]. 
The condensate was trapped in a static axisymmetric magnetic trap 
which was deformed by a nonaxisymmetric attractive dipole potential created with stirring laser beams. 
This combined potential produced a cigar-shaped harmonic trap with a slightly anisotropic transverse profile. 
By rotating the orientation of the transverse anisotropy at a frequency $\Omega$, 
they observed the dynamic formation of a vortex above a certain critical value of $\Omega$. 
When $\Omega$ was increased further, multiple vortices appeared, forming a triangular lattice. 
The quantized vortices could be directly visualized as ``dips'' in the transverse density profile 
of the TOF image. Following the experiments of the ENS group, other groups have 
also observed quantized vortices using slightly different methods based on the concept 
of the rotating bucket. Abo-Shaeer {\it et al.} observed a vortex lattice consisting of up 
to 100 vortices in a $^{23}$Na BEC \cite{Abo}. 
Hodby {\it et al.} created a vortex lattice by rotating the anisotropic magnetic trap directly 
without using a laser beam, which is similar to the rotating bucket experiment \cite{Hodby}. 
Rotating an optical spoon made by multiple-spot laser beams or narrow focusing 
beams has been also used to nucleate vortices \cite{Raman2}.

Haljan {\it et al.} created a vortex state 
by cooling an initially rotating thermal gas in a static confining potential \cite{Haljan}. 
Thermal gas above the transition temperature was rotated by a slightly anisotropic trap. 
After recovering the anisotropy of the potential, the rotating thermal gas was evaporatively 
cooled until most of the atoms were condensed. Although the number of atoms decreased 
through the evaporative cooling, the condensate continued to rotate because the angular 
momentum per atom did not change, and thus the vortex lattice was created. 
Since atoms can be selectively removed during evaporation, 
spinning up of the condensate can be efficiently achieved by removing atoms 
extending in the axial direction, as opposed to the radial direction, 
allowing a BEC with a high rotation rate to be obtained. 
Various properties of rapidly rotating BECs were studied, 
e.g., equilibrium properties \cite{Coddington2}, 
collective dynamics of a vortex lattice \cite{Engels,Coddington}, 
vortex aggregation \cite{Engels2}, and the lowest Landau level regime \cite{Schweikhard}.

\subsection{Structure of a single vortex}
As a simple example, let us consider the structure of a single vortex in a condensate trapped by an axisymmetric 
harmonic potential $V_{\rm ex}(r,z)=m \omega_{\perp}^{2} (r^{2}+\lambda^{2}z^{2})/2$ with the transverse trapping 
frequency $\omega_{\perp}$ and the aspect ratio 
$\lambda=\omega_{z}/\omega_{\perp}$. The condensate wave function with a straight vortex line located 
along the $z$-axis takes the form $\Phi({\bf r}) = \phi(r,z) e^{iq\theta}$ with a winding number $q$ 
and cylindrical coordinate $(r,\theta,z)$. The velocity field around the vortex line 
is ${\bf v}_{\rm s} = (q \hbar/m r) \hat{\theta}$. 
Equation (\ref{staGPeq}) for a real function $\phi$ becomes
\begin{equation}
\left[ -\frac{\hbar^{2}}{2m} \left( \frac{\partial^{2}}{\partial r^{2}} + \frac{1}{r} \frac{\partial}{\partial r} 
+ \frac{\partial^{2}}{\partial z^{2}} \right) + \frac{q^{2} \hbar^{2}}{2mr^{2}} + V_{\rm ex}(r,z) +g n \right] \phi
= \mu \phi. \label{singletimeindGPrz}
\end{equation}
Here, the centrifugal term $q^{2} \hbar^{2}/2mr^{2}$ arises from the azimuthal motion of the condensate. 
Equation (\ref{singletimeindGPrz}), which must be solved numerically, gives the structure of the vortex state. 
Asymptotically, the solution takes $n (r,z) \simeq n_{0} ( r/\xi )^{2q^{2}}$ for $ r \ll \xi $, and 
$n(r,z) = n_{0} ( 1 - r^{2}/R_{\perp}^{2} - z^{2}/R_{z}^{2} - q^{2} \xi^{2}/r^{2} )$ 
for $\xi \ll r < R_{\perp}$. 
The latter can be obtained in the TF limit $N a/a_{\rm ho} \gg 1$ for Eq. (\ref{singletimeindGPrz}), 
where $n_{0}=\mu/g$ is the density at the center of the vortex-free TF profile. 
We have defined the TF radii $R_{\perp}^{2} = 2 \mu / m \omega_{\perp}^{2}$,  
$R_{z}^{2} = 2 \mu / m \omega_{z}^{2}$ and the healing length 
$\xi = (\hbar^{2}/2 m g n_{0})^{1/2}$. 
The condensate density vanishes at the center, 
out to a distance of order $\xi$, 
whereas the density in the outer region has the form of an upward-oriented parabola. 
Hence, the healing length $\xi$ characterizes the vortex core size; for typical BEC parameters, 
$\xi \sim 0.2$ $\mu$m. In the TF limit, the core size is very small because 
$\xi/R_{\perp} = \hbar \omega_{\perp} /2 \mu = (15 Na/a_{\rm ho})^{-2/5} \ll 1$. 
Increasing the winding number $q$ widens the core radius due to centrifugal effects.

This axisymmetric solutions holds for a BEC in a pancake trap with $\lambda \gg 1$, 
but for a cigar-shaped trap with $\lambda \ll 1$ the axisymmetry is spontaneously broken 
as a result of {\it vortex bending}. Numerical simulation of the 
3D GP equation revealed that the vortex bending is held stationary for the ground state 
of a rotating cigar-shaped condensate \cite{Garcia-Ripoll01,Modugno03,Aftalion03}.
This bending is a symmetry breaking effect, which happens even in a completely axisymmetric setup. 
Evidence of vortex bending in the ground state was observed by the ENS group 
\cite{Rosenbusch02}. 

\subsection{Vortex stability}\label{singlecompstability}
Stability of the vortex state should be ensured to demonstrate that 
a BEC exhibits the superfluidity characterized by a ``persistent current". 
At zero temperature, the BdG analysis shows that the excitation spectrum 
of a condensate with a centered vortex in a cylindrical system has 
at least one negative energy mode with a positive norm $\int d {\bf r} (|u|^{2}-|v|^{2}) > 0$, 
localized at the core (so-called ``anomalous modes") \cite{Dodd97}. This implies that the single-vortex 
state is thermodynamically unstable \cite{Rokhsar97}. In fact, in the absence of rotation, 
the centered vortex would decay by spiraling outward due to dissipation by the 
thermal atoms.  

As demonstrated in the experiments, imposing rotation on the system is a direct way 
to achieve vortex stabilization. 
If the system is under rotation, it is convenient to consider the corresponding rotating frame; 
for a rotation frequency ${\bf \Omega}=\Omega \hat{\bf z}$, the integrand of the GP energy 
functional (\ref{Hamilfunctional}) 
acquires an additional term:
\begin{equation}
E' = \int d {\bf r} \Psi^{\ast} \left( -\frac{\hbar^{2} \nabla^{2}}{2m} + V_{\rm ex} 
+ \frac{g}{2} |\Psi|^{2} - \Omega L_{z} \right) \Psi , \label{GPenergyrot}
\end{equation}
where $L_{z}=- i \hbar (x \partial_{y} - y \partial_{x})$ is the angular momentum operator 
along the $z$-axis. The corresponding GP equation becomes
\begin{equation}
i \hbar \frac{\partial \Psi({\bf r},t)}{\partial t} = \left[ - \frac{\hbar^{2} \nabla^{2}}{2m} 
+V_{\rm ex} + g |\Psi({\bf r},t)|^{2}  - \Omega L_{z} \right] \Psi({\bf r},t) . \label{GPeqrot}
\end{equation}

The energy associated with a single vortex line is predominantly contributed 
by the kinetic energy of the superfluid flow in the vortex, which can be estimated as
\begin{equation}
E_{1} = \int \frac{1}{2} m n v_{\rm s}^{2} d {\bf r} \simeq \frac{m \bar{n}}{2} 2R_{z} 
\int^{R_{\perp}}_{\xi} v_{\rm s}^{2} 2 \pi r d r = q^{2} R_{z} \frac{2 \pi \hbar^{2} \bar{n}}{m} 
\ln \left( \frac{R_{\perp}}{\xi} \right), \label{onevorenrgy}
\end{equation}
where $\bar{n}$ is the mean uniform density. Since $E_{1} \propto q^{2}$, the energy 
cost to create one $q=2$ vortex is higher 
than that to create two $q=1$ vortices, thus vortices with $q>1$ are energetically unfavorable. 
Therefore, a stable quantized vortex usually has $q=1$, and we will mainly concentrate 
on the $q=1$ vortex in the following discussion. 

If there is a quantized vortex along the trap axis, $\langle L_{z} \rangle = N \hbar$, 
the corresponding energy of the system in the rotating frame is $E_{1}'=E_{1} - N \hbar \Omega$. 
The difference between $E_{1}'$ and the vortex-free energy $E_{0}'$ determines 
the energetic favorability of a vortex entering the condensate. 
Since $E_{0}'$ is equal to the energy $E_{0}$ in the laboratory frame, the difference 
is given by $\Delta E' = E_{1}'-E_{0}' = E_{1}-E_{0} - N \hbar \Omega$. 
Thus, the critical rotation frequency $\Omega_{c}$ for the energetic stability of
a vortex line is given by $\Omega_{c} = (E_{1}-E_{0})/N \hbar$. 
When $\Omega$ exceeds $\Omega_{c}$, 
the single vortex state is {\it thermodynamically} stable.

To calculate $E_{1}$ more precisely, it is necessary to take into account the inhomogeneous effects 
of condensate density \cite{Lundh}. In the TF limit, for a condensate in a cylindrical trap 
$\omega_{z}=0$ (an effective 2D condensate), the critical frequency is given by 
$\Omega_{c} = (2 \hbar/ m R_{\perp}^{2}) \ln (0.888 R_{\perp}/\xi)$. 
For an axisymmetric trap $V_{\rm ex}(r,z)$, the critical frequency is 
$\Omega_{c} = (5\hbar / 2 m R_{\perp}^{2}) \ln (0.671 R_{\perp}/\xi)$. 
For a nonaxisymmetric trap, $\Omega_{c}$ is slightly modified by a small numerical factor \cite{Svidzinsky,Feder}.

\subsection{Vortex nucleation}
The critical rotation frequency $\Omega_{c}$ only indicates the energetic stability of the central vortex state. 
Vortex nucleation in a non-rotating condensate occurs 
when the trap is rotated at a higher frequency than $\Omega_{c}$ to overcome the energy barrier that stops 
the transition from the nonvortex state to the vortex state \cite{Isoshima}. 
The threshold of the rotation frequency for instability, leading to vortex nucleation, is related to the excitation 
of surface modes of the condensate \cite{Feder3,Dalfovo2,Anglin,Williams2,Simula2}. 
According to the Landau criterion for rotationally invariant systems, 
the critical rotation frequency is given by $\Omega_{v} = {\rm min}(\omega_{l} / l)$, 
where $\omega_l$ is the frequency 
of a surface mode with the profile $\sim e^{i l \theta}$ and the 
quantum number $l$ of the azimuthal ($\theta$-) direction. 
Above $\Omega_{v}$, some surface modes with negative energy appear 
in the spectrum of a non-rotating condensate \cite{Isoshima,Dalfovo2}, 
and their growth may trigger vortex nucleation. 
The negative-energy modes can grow exponentially in the presence of dissipation, 
caused, for example, by interactions with thermal atoms \cite{Williams2}. 

The frequency $\Omega_{v}$ can explain vortex nucleation by a rotating 
thermal gas \cite{Haljan,Anglin}, but cannot yet explain the results 
of external stirring potentials \cite{Madison,Abo,Madison01,Hodby}. 
For example, in the case of the ENS group \cite{Chevy}, 
the number of nucleated vortices had a peak near $\Omega = 0.7 \omega_{\perp}$. 
Their experiments demonstrated that instability occurs when a particular surface 
mode is resonantly excited by a deformed rotating potential. The rotating potential of the ENS group mainly 
excited the surface mode with $l =2$ (quadrupole mode). In a rotating frame with frequency $\Omega$, 
the frequency of the surface mode is increased by $- l \Omega$. The resonance thereby occurs close to the rotation 
frequency $\Omega = \omega_l / l$. In the TF limit, the dispersion relation for the surface mode is given 
by $\omega_{l}=\sqrt{l}\omega_{\perp}$ with the trapping frequency $\omega_{\perp}$ \cite{Stringari}. 
Hence, it is expected that the quadrupole 
mode with $l =2$ is resonantly excited at $\Omega = \omega_{\perp}/ \sqrt{2} \simeq 0.707 \omega_{\perp}$. 
A theoretical study revealed that when the quadrupole mode is resonantly excited, an imaginary component 
in the excitation frequency appears in surface modes with high multipolarity \cite{Sinha}. 
This indicates that dynamic instability 
can trigger vortex nucleation even at zero temperature. This picture was further supported by an experiment of 
the MIT group \cite{Raman2}, in which surface modes with higher multipolarity ($l = 3, 4$) were resonantly 
excited using multiple laser-beam spots. The largest number of vortices were generated at frequencies 
close to the expected values $\Omega = \omega_{\perp}/\sqrt{l}$. 
However, a recent theory proposed an alternative mechanism, noting that 
the single vortex state can be regarded as the first Zeeman-like excited state, and 
its resonance frequency is quantitatively consistent with all of the above experiments \cite{Reinisch}.

\begin{figure}[htbp] 
\begin{center} 
\includegraphics[angle=0,width=13cm]{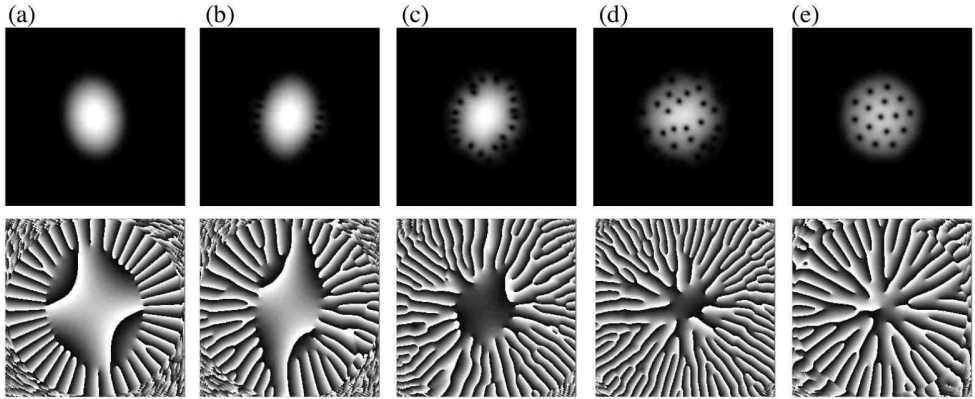} 
\end{center} 
\caption{Dynamics of vortex lattice formation in a rotating BEC. 
The graphs show both the condensate density (upper panels) and phase (lower panels), 
being at (a) $t$ = 300 ms, (b) 370 ms, (c) 385 ms, (d) 410 ms, and 
(e) 550 ms after the start of rotation. 
The phase varies continuously from 0 (white) to 2 $\pi$ (black). 
The simulations were performed using the 2D GP equation with the experimentally 
accessible parameters; see Ref. \cite{Tsubota} for details.} 
\label{vorlatdyn} 
\end{figure}
The observation of vortex nucleation and lattice formation \cite{Madison01} has been well reproduced by 
numerical simulations of the time-dependent GP equation (\ref{GPeqrot}) 
\cite{Tsubota,Kasamatsu,Penckwitt,Lobo,Kasamatsu2,Parker,Wright}. 
The simulation results clarified the following dynamics. 
When a sufficiently high rotation is applied to an initially non-rotating condensate, 
it becomes elliptical and undergoes a quadrupole oscillation [Fig. \ref{vorlatdyn}(a)]. 
Then, the boundary surface of the condensate becomes unstable, and 
generates ripples that propagate along the surface [Fig. \ref{vorlatdyn}(b)]. 
It is also possible to identify quantized vortices in the phase profile. 
As soon as the rotation starts, many vortices appear in the low-density region 
outside of the condensate [Fig. \ref{vorlatdyn}(a)]. Since quantized vortices are excitations, 
their nucleation increases the energy of the system. 
Because of the low density in the outskirts of the condensate, however, 
their nucleation contributes little to the energy or angular momentum. 
Since these vortices outside of the condensate are not observed in the 
density profile, they are called ``ghost vortices". 
Their movement toward the Thomas-Fermi surface excites ripples [Fig. \ref{vorlatdyn}(c)]. 
It is not easy for these ghost vortices to enter the condensate, 
because that would increase both the energy and angular momentum. 
Only some vortices enter the condensate cloud to become ``real vortices" with 
the usual density profile of quantized vortices [Fig. \ref{vorlatdyn}(d)], eventually forming 
a vortex lattice [Fig. \ref{vorlatdyn}(e) and (f)].

\subsection{Dynamics of a single vortex}\label{Dynamicssinglevortex}
\subsubsection{Precession of an off-centered vortex}
Precession of a vortex core upon the off-center of the condensate is a 
simple example of vortex motion. Core precession can be described in terms 
of a Magnus force effect. A net force on a quantized vortex core creates a pressure imbalance, 
resulting in core motion perpendicular to both the force and the vortex quantization axis. 
In the case of trapped BECs, these net forces can be caused by either condensate 
density gradients \cite{Svidzinsky,Jackson,McGee} 
or drag due to thermal atoms \cite{Fedichev}. The former may effectively act as 
a buoyancy of the vortex. Typically, the total buoyant force is towards the 
condensate surface, and the net effect is a precession of the core around the condensate 
center via the Magnus effect. The latter causes radial drag and spiraling of the core 
towards the condensate surface due to energy dissipation.

Core precession was experimentally investigated by Anderson {\it et al.} \cite{Anderson00}. 
Initially, a vortex was prepared to be off-centered, and 
its precession frequency was determined from the vortex position subtracted directly 
from the snapshots of the density profile. The vortex core precessed 
in the same direction as the vortex fluid flow around the core. 
These features can be understood as a result of the anomalous mode (with negative energy 
and positive norm); when the energy is negative, the precession has the same direction 
as the vortex flow. The observed excitation frequency also agrees with the theory \cite{Svidzinsky} 
and the numerical simulations \cite{Jackson,McGee,Feder01}. 

\subsubsection{Vortex wave}
\begin{figure}[htbp] 
\begin{center} 
\includegraphics[angle=0,width=11cm]{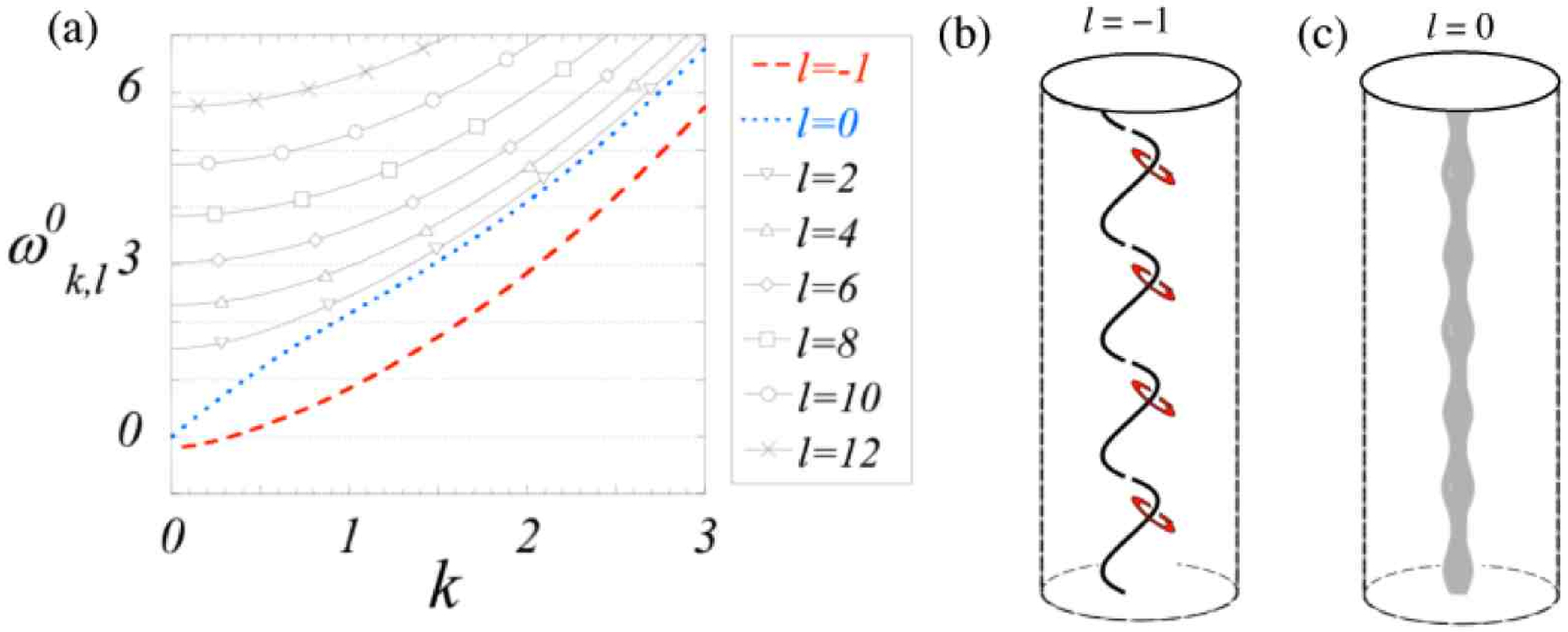} 
\end{center} 
\caption{(a) Dispersion relation of the axisymmetric single-vortex state of Kelvin waves 
with $l=-1$ (dashed line), varicose waves with $l=0$ (dotted line), and several surface waves 
with $l>0$ (solid lines), calculated for a cylindrically symmetric condensate with a vortex (see Ref. \cite{Takeuchi09}). 
(b) and (c) show schematic illustrations of the Kelvin and varicose waves of the vortex line, respectively.} 
\label{Kelvindis} 
\end{figure}

As in the studies of superfluid helium, vortex waves are also an interesting subject 
in trapped BECs. The dispersion relation of the vortex wave can be obtained by BdG 
analysis of a single vortex state, as shown in Fig. \ref{Kelvindis}(a). 
In this calculation, the perturbation can be taken as $\delta \Psi = e^{i \theta} 
[u_{k_{z},l} e^{i(k_{z}z+l\theta-\omega t)}-v^{\ast}_{k_{z},l} e^{-i(k_{z} z+l\theta-\omega t)}]$ 
in the cylindrical system, where $k_{z}$ and $l$ refer to the wave number and the 
angular quantum number along the $z$-axis. The lowest modes along the radial 
direction can be classified into three groups: the Kelvin wave ($l=-1$) [Fig. \ref{Kelvindis}(b)], 
the varicose wave ($l=0$) [Fig. \ref{Kelvindis}(c)], and the surface waves ($l \neq 0, -1$). 
Because of the finite-size of the system, the dispersion of the Kelvin wave is well-described by \cite{Simula08}
\begin{equation}
\omega(k_{0}+k_{z}) = \omega_{0} + \frac{\hbar k_{z}^{2}}{2m} \ln \left( \frac{1}{r_{c} k_{z}} \right) 
\hspace{4mm} (k r_{c} \ll 1), 
\end{equation}
where $r_{c}$ is on the order of the core size, $k_{0} \sim 2\pi/R_{z}$ 
the smallest wave number, and $\omega_{0} < 0$ the frequency of 
the anomalous mode described above. The varicose wave with $l=0$ is known in classical fluids 
as the axisymmetrically propagating mode, 
in which the core diameter of a vortex oscillates along the vortex line.

The collective modes of the condensate have 
a strong influence on the excitation of vortex waves due to the finite size effect. Bretin {\it et al.} observed 
that when superimposed $l=\pm2$ quadrupole modes were excited with equal 
amplitudes in the single vortex state, the oscillation of the $l=-2$ mode decayed faster than that of 
the $l=+2$ mode \cite{Bretin03}. 
This indicates that the $l=-2$ mode decays to Kelvin modes through a non-linear Beliaev 
process, which is supported by theoretical analysis based on the BdG equation \cite{Mizushima03} and 
by numerical simulations \cite{Simula08}. 

When an additional velocity $V$ is applied along 
the $z$-axis, the dispersion of the vortex waves behaves as $\omega - V k_{z}$, and the frequency 
can become negative above the critical velocity. This is known as the Donnely-Glaberson 
instability in superfluid helium \cite{Tsubota03}, 
and it results in the amplification of Kelvin waves \cite{Takeuchi09}. 
This could induce reconnections of adjoining vortex lines, and eventually a turbulent state. 
  
\subsubsection{Splitting of a multiply-charged vortex}
As seen in Eq. (\ref{onevorenrgy}), the energy cost to create a $q>1$ multiply-charged vortex is unfavorable 
compared to $q$ singly-charged vortices. This raises an interesting question: What happens when such 
an unstable vortex is created? A multiply-charged vortex can be created 
experimentally in atomic BECs by using topological phase imprinting \cite{Leanhardt}, and causes 
interesting disintegration dynamics \cite{Shin} (The physics of topological phase imprinting 
will be described in Sec. \ref{Spin1}).

The stability of a multiply-charged vortex in a trapped BEC shows an 
interesting interaction dependence, which originates from the finite size effect \cite{Pu99}. 
BdG analysis for the cylindrical system revealed the appearance 
of complex eigenvalue modes, which implies that the multiply-charged vortex 
is dynamically unstable. For a vortex with a winding number $q$, the conservation of angular momentum 
leads to constraint of the normal mode functions 
$u_{j} ({\bf r}) = u_{j} (r) e^{i (q+2) \theta}$ and $v_{j} ({\bf r}) = v_{j} (r) e^{i (q-2) \theta}$. 
For $q=2$ there are alternating stable and unstable regions with respect to 
the interaction parameter $a n_{z} = a \int |\Psi|^{2} dx dy$; the first (second) 
region appears for $0 < an_{z} < 3$ ($11.4 < an_{z} < 16$). 
Numerical simulations demonstrated that, when the system is in the unstable region, 
a doubly-charged vortex decays into two singly-charged vortices \cite{Mottonen03}.

The experiment by Shin {\it et al.} investigated the splitting process of a doubly-charged vortex 
and its characteristic time scale as a function of $an_{z=0}$ \cite{Shin}. The results showed that 
the doubly-charged vortex decayed, but that its lifetime increased monotonically 
with $an_{z=0}$, showing no periodic behavior. Numerical simulation of the 3D GP equations 
also displayed this mysterious phenomenon, demonstrating that the detailed dynamical behavior 
of a vortex along the entire $z$-axis is relevant in the characterization of the splitting 
process in an axially-elongated condensate \cite{Huhtamaki06,Mateo06}. The vortex begins to split 
near both condensate edges in the $z$-direction, and then the splitting propagates to the center.  

Isoshima {\it et al.} studied the splitting dynamics of quadruply-charged ($q = 4$) vortices, both 
theoretically and experimentally \cite{Isoshima07}. 
In this case, a rich variety of splitting patterns with $l$-fold symmetry was expected \cite{Kawaguchi04}. 
They initially prepared an off-centered $q=4$ vortex, and observed that the vortex split into 
an array of four linearly-aligned $q=1$ vortices. This is because the perturbation with 
two-fold rotational symmetry ($l=2$) is the most unstable mode under the experimentally 
relevant parameters. These observations were well-reproduced by numerical simulation 
of the GP equation, where the initial displacement of the vortex position was crucial 
to causing splitting instability due to the conservation of angular momentum.  

\subsection{Vortex lattices}\label{fastrotatinLatice}
At very high frequencies $\Omega$, a rotating superfluid mimics rigid body rotation with 
$\nabla \times {\bf v}_{\rm s} = 2 {\bf \Omega}$, by forming a triangular vortex lattice 
which arranges itself parallel to the rotation axis \cite{Yarmchuk79}. 
These properties closely resemble the magnetic flux-line lattice in type-II superconductors 
predicted by Abrikosov \cite{Abrikosov}. Here, we summarize the physics of vortex lattices 
in trapped BECs. Since various features of the rapidly rotating BECs were reviewed 
thoroughly by Cooper \cite{Cooperrev} and Fetter \cite{Fetterrev2}, in this section 
we describe only the theoretical basics that will be useful in the following 
sections. 

\subsubsection{Basics}
For vortex lines parallel to the $z$-axis and located at ${\bf r}_{i} = (x_{i}, y_{i})$, 
the vorticity is given by the form 
$\nabla \times {\bf v}_{s} = \sum \kappa \delta^{(2)} ({\bf r}_{i}) \hat{\bf z}$. Therefore, 
the average vorticity per unit area is given by $\nabla \times {\bf v}_{\rm s} = \kappa n_{v} \hat{\bf z}$, 
where $n_{v}$ is the number of vortices per unit area. 
Hence, the density of the vortices is related to the rotational frequency 
$\Omega$ as $n_{v} = 2 \Omega/\kappa$ (``Feynman's rule") \cite{Feynman}. 
This relation can be used to estimate the maximum possible number of vortices in a given area 
as a function of $\Omega$.
The vortex lattice can be characterized 
by the nearest-neighbor lattice spacing $\sim b=(\hbar/m\Omega)^{1/2}$ defined by the area 
per vortex $n_{v}^{-1} = \pi b^{2}$, and by the radius of each vortex core $r_{c} \sim \xi$.

\subsubsection{Thomas-Fermi regime}
The GP energy functional of Eq. (\ref{GPenergyrot}) in a rotating frame can be rewritten as
\begin{equation}
E' = \int d {\bf r} \left( \frac{\hbar^{2}}{2m} \left| \left(- i \nabla 
- \frac{m}{\hbar} {\bf \Omega} \times  {\bf r} \right) \Psi \right|^{2} 
+ V_{\rm eff} |\Psi|^{2} + \frac{g}{2} |\Psi|^{4} \right), \label{GPenergyrot2}
\end{equation}
where $V_{\rm eff} = m (\omega_{\perp}^{2} - \Omega^{2}) r^{2} /2 + m \omega_{z}^{2} z^{2} / 2$ 
is the effective trapping potential combined with the centrifugal potential; 
the rotation effectively releases the radial potential, and causes it to vanish at $\Omega = \omega_{\perp}$. 
Because the first term in Eq. (\ref{GPenergyrot2}) reads 
$\hbar^{2} (\nabla |\Psi|)^{2} / 2m + m ({\bf v}_{s} - {\bf \Omega} \times {\bf r})^{2} |\Psi|^{2} /2 $, 
it can be neglected when both the TF limit and the rigid-body rotation limit 
${\bf v}_{s} = {\bf \Omega} \times {\bf r}$ are satisfied. 
This TF approximation is valid when the lattice spacing $b$ is much larger than the 
vortex core size $r_{c}$, giving an effective coarse-grained description of the 
rotating BECs. The TF radius is given by 
$R_{\perp}(\Omega) = R_{\perp} / [1-(\Omega/\omega_{\perp})^{2}]^{3/10}$ 
with $R_{\perp}$ for a non-rotating condensate, providing an aspect ratio 
$\lambda_{\rm rb} = R_{\perp}(\Omega)/R_{z} = \lambda / [1-(\Omega/\omega_{\perp})^{2}]^{1/2}$. 
Thus, measurement of $\lambda_{\rm rb}$ can give the rotation rate of the condensate \cite{Haljan,Raman2}. 
Also, in the high rotation limit $\Omega \rightarrow \omega_{\perp}$, 
the condensate flattens out and reaches an interesting quasi-2D regime; 
current experiments have reached $\Omega/\omega_\perp\approx 0.995$ \cite{Coddington2}.

\subsubsection{Mean-field quantum Hall regime}
Note that the first term in Eq. (\ref{GPenergyrot2}) can be identified as the Hamiltonian 
$H_{L} = (-i \hbar \nabla -e {\bf A}/c)^{2}/2m$ of a charge $-e$ particle moving in the $xy$ plane under 
a magnetic field $B \hat{\bf z}$ with a vector potential ${\bf A} = (mc/e) {\bf \Omega} \times {\bf r}$. 
If interaction is neglected ($g=0$), the eigenvalues of the Hamiltonian of Eq. (\ref{GPenergyrot2}) forms 
Landau levels as $\epsilon_{n, m, n_{z}} / \hbar = \omega_{\perp} +  n (\omega_{\perp} + \Omega) 
+ m (\omega_{\perp} - \Omega) + (n_{z}+1/2)\omega_{z}$, 
where $n$ is the Landau level index, $m$ indexes the substates within the $n$-th Landau level, and 
$n_{z}$ is the index of the states along the $z$-axis. 
The lowest energy states of two adjacent Landau levels are separated by $\hbar(\omega_\perp + \Omega)$, 
whereas the distance between two adjacent substates in a given Landau level is $\hbar(\omega_{\perp} - \Omega)$; 
when $\Omega = \omega_\perp$, all states in a given Landau level are degenerate. 
This corresponds to a situation where the centrifugal force exactly balances the trapping force 
in the $x$-$y$ plane, and only the Coriolis force remains. The system is then invariant under translation, 
and therefore exhibits macroscopic degeneracy. This formal analogy has led to the prediction that 
quantum Hall-like properties would emerge in rapidly rotating BECs; see \cite{Cooperrev,Viefersrev} 
and references therein for the details of this strongly correlated phase in rapidly rotating bosons. 

The quantum-Hall formalism provides a useful mean-field description of rotating 
BECs with vortex lattices \cite{Ho}. Interaction will lead to the mixing of different ($n$, $m$, $n_{z}$) states. 
However, because the averaged density $\bar{n}$ of the system drops 
as $\Omega \to \omega_\perp$, the interaction energy $\sim g \bar{n}$ can become low compared 
to $2 \hbar \omega_{\perp}$ and $\hbar \omega_{z}$. In this limit, particles should condense into 
the lowest Landau levels (LLL) with $n=0$. The system then enters the ``mean field'' quantum Hall regime, 
where the wave function can be described by only the LLL orbitals with the form 
\begin{equation}
\Psi_{\rm LLL} =\sum_{m \geq 0} a_{m} \psi_{m}({\bf r}) 
= A \prod_{j}( z - z_{j}) e^{-r^2/2 a_{\rm ho}^2}, \label{LLLansatz}
\end{equation}
where $z=x+iy$, $z_{j}$ are the positions 
of vortices (zeros), and $A$ is a normalization constant. By minimizing Eq. (\ref{GPenergyrot2}) using 
the ansatz $\Psi_{\rm LLL}$ with respect to $z_{j}$, one can analytically describe 
the vortex structure \cite{Watanabe,Cooper1,Aftalion4,Sonin2,Aftalion5,Cozzini06}, although 
this method is effective only near the limit $\Omega \simeq \omega_{\perp}$. 
However, this would be impractical 
for numerical calculations because of the time and accuracy required. 
A similar approach was applied to 
investigate the ground state of (not rapidly) rotating BECs with very weak interaction 
\cite{Butts,Kavoulakis,Vorov}, where the coefficients $a_{m}$ of the harmonic oscillator basis 
$\psi_{m}$ were minimized. 

Schwaikhard {\it et al}. created rapidly rotating BECs by spinning up the 
condensates to $\Omega/\omega_{\perp} > 0.99$ \cite{Schweikhard}, and found 
some evidence that the condensate entered the LLL regime: 
(1) The 2D signature of a rapidly rotating BEC was confirmed by the excitation of an 
axial breathing mode ($m_{z}=0$). For a BEC in the axial TF regime, an axial breathing 
frequency $\omega_{\rm B} = \sqrt{3} \omega_{z}$ has been predicted in the limit 
$\Omega/\omega_{\perp} \rightarrow 1$ \cite{Cozzini03}, whereas 
$\omega_{\rm B} =2 \omega_{z}$ is expected for a noninteracting gas 
with $\mu < \hbar \omega_{z}$. Schweikhard {\it et al}. observed a crossover 
of $\omega_{\rm B}$ from $\sqrt{3} \omega_{z}$ to $2 \omega_{z}$ with increasing 
$\Omega$ ($\mu \sim 3 \hbar \omega_{z}$). 
(2) A signature of the LLL regime is that the vortex core size $r_{c}$ is similar to 
the vortex separation distance $b=(\hbar/m\Omega)^{1/2}$. The theory 
\cite{Cozzini06,Fischer,Baym3,Watanabe06} predicted that the vortex cores 
begin to shrink as the intervortex spacing becomes comparable to the healing length $\xi$, 
and eventually the core radius becomes proportional to the intervortex spacing. 
The saturation of the fractional area 
$A = r_{c}^{2}/b^{2}$ with increasing $\Omega$ was actually observed, indicating 
that the system was in the LLL regime.   
(3) The observed global density profile remained a parabolic TF profile even 
for $\Omega \to \omega_\perp$, although it was expected to be a Gaussian \cite{Ho} 
(see Eq. (\ref{LLLansatz})) in the LLL regime. However, very small distortions of the vortex 
lattice from a perfect triangle could result in large changes in the global density distribution, 
such as from a Gaussian to a TF form \cite{Watanabe,Cooper1,Aftalion4}. 

\subsubsection{Tkachenko oscillation of a vortex lattice}
It should be possible to propagate collective waves transverse to the 
vortex lattice in the superfluid, in so-called Tkachenko (TK) modes \cite{Tkachenko2}. 
For incompressible superfluids, the dispersion law is given 
by $\omega_{\rm TK} (k) = \sqrt{\hbar \Omega/4m} \ k$. The TK mode of 
a vortex lattice in a trapped BEC was observed experimentally \cite{Coddington}, 
and was identified by the sinusoidal displacement of vortex cores with an origin 
at the center of the condensate. 

To explain the observed frequency of the TK mode $\omega_{(n,m)}$, where the quantum 
number $(n,m)$ refers to the radial and angular nodes, the effects of compressibility should 
be taken into account. According to the elastohydrodynamic approach developed by 
Baym \cite{Baym03}, the TK frequency is described by the compressional modulus 
$C_{1}$ and the shear modulus $C_{2}$ of the vortex lattice included in the elastic energy 
\begin{equation}
E_{\rm el} = \int d {\bf r} \left\{ 2 C_{1} (\nabla \cdot {\bf \epsilon})^{2} 
+ C_{2} \left[ \left( \frac{\partial \epsilon_{x}}{\partial x} 
- \frac{\partial \epsilon_{y}}{\partial y} \right)^{2} 
+ \left( \frac{\partial \epsilon_{x}}{\partial y} + \frac{\partial \epsilon_{y}}{\partial x} \right)^{2} \right] \right\},
\end{equation}
where ${\bf \epsilon}({\bf r},t)$ is the continuum displacement field of the vortices 
from their home positions. In the incompressible TF regime, $C_{2}=-C_{1}=n \hbar \Omega/8$. 
Compressibility yields two branches in the energy spectrum: The upper branch follows the dispersion 
law $\omega_{+}^{2} = 4 \Omega^{2} + c^{2} k^{2}$, where $c=\sqrt{gn/m}$ is the velocity of sound, which is the standard inertial mode of a rotating fluid with a gap at $k=0$. 
The lower frequency branch corresponds to the TK mode with the dispersion 
$\omega_{-}^{2} = (\hbar \Omega/4m) [c^{2} k^{4}/(4\Omega^{2} + c^{2} k^{2})]$. 
For large $k$, this reproduces the original TK frequency $\omega_{\rm TK} $, 
but for small $k$ it exhibits the quadratic behavior $\omega_{-} \simeq \sqrt{\hbar/16 m \Omega} c k^{2}$. 
The transition between $k^{2}$ and $k$ dependence takes place 
at $k \sim \Omega/c > R_{\perp}^{-1}$. This suggests that the effects of compressibility, 
which characterize the $k^{2}$ dependence, play a crucial role in the TK mode. 
Thus, this regime is distinguished from the usual incompressible TF regime, 
and is referred to as the ``soft" TF regime. Including finite compressibility, this can explain the 
observed values of $\omega_{(1,0)}$ \cite{Baym03,Cozzini04,Sonin05}. 
First-principles simulations based on the GP and BdG formalism also agreed excellently 
with the experimental data \cite{Mizushima04,Baksmaty04}.

\subsection{Various topics on vortices in single-component BECs}
\subsubsection{Vortices in an anharmonic potential}
\begin{figure}[htbp] 
\begin{center} 
\includegraphics[angle=0,width=13cm]{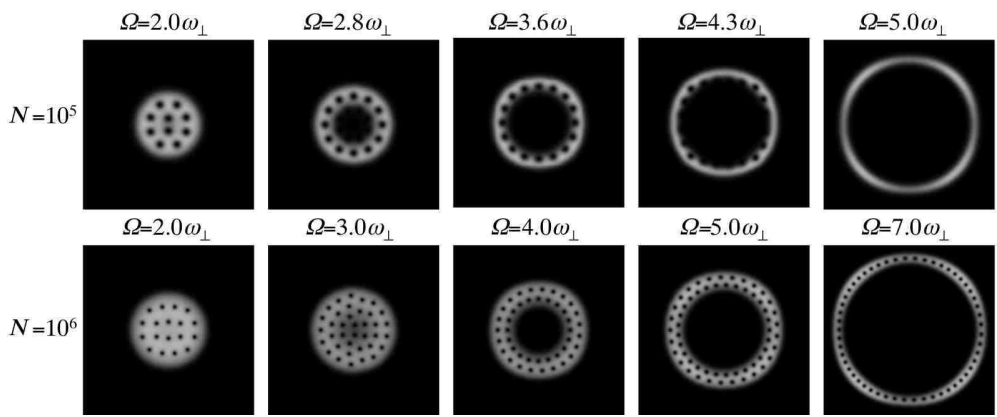} 
\end{center} 
\caption{Vortex structure of a rapidly rotating BEC in a combined harmonic-quadratic potential. 
This calculation was done by 2D simulations of the GP equation with the parameter values 
$m=m_{^{87}{\rm Rb}}$, $\omega_{\perp} =10 \times 2 \pi$ Hz, $a=5.61$ nm 
and $\lambda = 0.5$. The upper and lower panels correspond to $N =10^{5}$ and  
$N =10^{6}$, respectively. } 
\label{giantvor} 
\end{figure}

For a rotating condensate with a frequency $\Omega$ in a harmonic potential 
$(1/2)m\omega_{\perp}^{2}r^{2}$, the centrifugal potential effectively reduces the confinement, 
preventing a BEC from rotating at $\Omega$ beyond $\omega_{\perp}$. 
This restriction can be avoided by introducing 
an additional quartic potential \cite{Fetter,Lundh2,Kasamatsu1}, 
so that the combined trapping potential in the $xy$ plane becomes 
$V_{\rm ex}(r) = (1/2) m \omega_{\perp}^{2} ( r^{2}  + \lambda r^{4}/a_{\rm ho}^{2})$, 
where the dimensionless parameter $\lambda$ represents the relative strength of the quartic potential. 
The properties of a rotating condensate in an anharmonic potential have attracted 
considerable theoretical attention \cite{Fetter05,Danaila05,Fu06,Blanc08}. 
The vortex phases in an anharmonic trap are quite different from those in a harmonic trap, 
as shown in Fig. \ref{giantvor}, because it is possible to rotate the system arbitrarily fast. 
For small $\Omega$, the equilibrium 
state is the usual vortex lattice state. As $\Omega$ increases, the vortices begin to 
merge in the central region, and the centrifugal force pushes the particles 
towards the edge of the trap. This results in a new vortex state consisting of 
a uniform lattice (multiple circular arrays of vortices) with a central density hole. 
The central hole becomes larger with increasing $\Omega$, and the condensate 
forms an annular structure with a single circular array of vortices. 
A further increase of $\Omega$ stabilizes a {\it giant vortex}, where all vortices are 
concentrated in the single hole. 

A combined harmonic-quartic potential was constructed by Bretin {\it et al}., 
who superimposed a blue detuned laser with the Gaussian profile \cite{Bretin04}. 
Since the waist $w$ of the beam propagating along the $z$-axis was larger than the 
condensate radius, the potential created by the laser $U_{0} \exp(-2 r^{2}/w^{2})$ 
can be written as $U(r) \simeq U_{0} (1 - 2 r^{2} / w^{2} + 2  r^{4} / w^{4})$. 
The second term leads to a reduction of the transverse trapping frequency 
$\omega_\perp$, and the third term provides the desired quartic confinement, 
giving $\omega_{\perp}/2\pi=65$ Hz and $\lambda \simeq 10^{-3}$ for $V_{\rm ex}(r)$. 
For $\Omega <\omega_{\perp}$, a vortex lattice was clearly observed. 
When $\Omega >\omega_{\perp}$, however, the vortices became gradually more difficult 
to observe and the images became less clear for $\Omega=1.05 \omega_{\perp}$. 
The most plausible explanation of this observation is that the vortices were not in equilibrium, 
and were strongly bent \cite{Danaila05}.  

\subsubsection{Vortex pinning in an optical lattice}
Rotating BECs combined with a co-rotating optical lattice are an interesting system, 
which has two competing length scales: vortex separation and the periodicity 
of the optical lattice. The structure of the vortex lattice is strongly dependent on 
the externally applied optical lattice. Various vortex phases appear, depending 
on the number of vortices per pinning center, i.e., the filling factor. 
Tung {\it et al.} created a rotating optical lattice using a rotating mask \cite{Tung06}, 
which provided a periodic pinning potential that was stationary 
in the corresponding rotating frame of vortices. The authors observed a structural crossover 
from a triangular to a square lattice with increasing potential amplitude of the optical lattice, 
which was consistent with theoretical predictions \cite{Reijnders04,Pu05}. 
In a deep optical lattice, the condensates are well localized at each potential minimum, so 
that the system can be regarded as a Josephson-junction array (JJA). A rotating 
Josephson-junction array can have characteristic vortex patterns with the unit-cell 
structure \cite{09Kasamatsu}, and was realized experimentally by Williams {\it et al}. \cite{10Williams}.

\subsubsection{Thermally-activated vortex generation}
In 2D systems with continuous symmetry, true long-range order is destroyed 
by thermal fluctuations at any finite temperature. For 2D Bose systems, 
a quasi-condensate can be formed with a correlation decaying algebraically 
in space, where superfluidity is still expected below a certain critical temperature. 
This 2D phase transition is closely connected with the emergence of thermally 
activated vortex--antivortex pairs, known as the Berezinskii--Kosterlitz--Thouless (BKT) 
phase transition occurring at $T=T_{\rm BKT}$ \cite{Berezinskii,Kosterlitz}. 
For $T<T_{\rm BKT}$, there are no isolated free vortices; vortices only exist as bound pairs, 
consisting of two vortices with opposite circulations. 
The contribution of these vortex pairs to the decay of the correlation is negligible, 
and the algebraic decay is dominated by phonons. For $T>T_{\rm BKT}$, the vortex pairs 
are separated, and the free vortices form a disordered gas of phase defects and give rise to 
an exponential decay of the correlation.

Recently, a BKT transition was experimentally observed in ultracold atomic 
gases \cite{Stock05,Hadzibabic06}. 
In the experiment, a 1D optical lattice was applied to an elongated condensate, 
splitting the 3D condensate into an array of independent quasi-2D BECs. 
The interference technique revealed the temperature dependence of an exponent 
of the first-order correlation function of the fluctuating 2D bosonic field \cite{Polkovnikov}. 
A universal jump in the superfluid density characteristic of the BKT transition was confirmed 
by observing a sudden change in the exponent, in which a finite size effect caused 
a finite-width crossover, rather than a sharp transition. The microscopic origin 
of this transition was clarified 
from the image of the interference of the two 2D condensates. 
If isolated free vortices were present in either of the two condensates, the interference fringes 
would exhibit dislocations. Such a dislocation was observed in the high-$T$ region of the 
crossover, as predicted theoretically by classical field simulations \cite{Simula3}. 

\subsubsection{Vortex formation during a rapid quench} \label{subsec-rapid-quench}
One universal element of many continuous phase transitions 
is the spontaneous formation of topological defects during a quench through 
the critical point, known as the Kibble-Zurek mechanism \cite{Kibble76,Zurek85}. 
The rapid symmetry-breaking phase transition in the early universe is expected to have produced 
causally disconnected domains of a true vacuum state, and their phase mismatch 
would leave behind topological defects like cosmic strings or monopoles. Since the 
cosmological phase transition may be second-order, it is especially significant to perform 
experiments in a superfluid system whose transition 
is also second-order \cite{Bauerle96,Ruutu96}. 

Ultracold atomic BECs offer an ideal opportunity 
for studying the Kibble-Zurek mechanism because the temperature, 
strength of interaction, and external parameters such as the magnetic field 
and trapping potentials can be changed in a time shorter than the characteristic 
time scale of topological defect formation.  
A quench through the BEC transition temperature was realized by using a rapid ramping of the rf-field 
for evaporative cooling. Weiler {\it et al.} studied 
the spontaneous formation of vortices in BEC both experimentally and theoretically \cite{Weiler08}. Vortices appeared spontaneously 
during condensate formation, without the application of any external rotation.
This experiment was compared to a simulation of BEC growth dynamics, 
calculated using the stochastic projected GP equation \cite{Gardiner02}, 
\begin{equation}
d \psi({\bf r},t) = {\mathcal P} \left\{ -\frac{i}{\hbar} L_{\rm GP} \psi({\bf r},t) dt 
+ \frac{\gamma({\bf r})}{k_{\rm B} T} (\mu - L_{\rm GP}) \psi({\bf r},t) dt + dW_{\rm \gamma}({\bf r},t) \right\}
\end{equation}
which was derived from first principles using the Wigner phase-space representation \cite{Gardiner03}. 
Here, the states of the trapped system are divided into the ``coherent" region 
and the ``incoherent" region by the high-energy cut-off of the harmonic oscillator mode. 
The first term on the right describes unitary evolution of the classical field 
in the coherent region according to the GP equation. 
The second term represents growth processes, that is, collisions that 
transfer atoms from the thermal bath to the classical field and vice versa, and the form of 
$\gamma({\bf r})$ may be determined from kinetic theory \cite{Bradley08}. 
The third term is the complex-valued noise associated with condensate growth, 
which is consistent with the fluctuation-dissipation theorem. 
The projection operator ${\mathcal P}$ restricts the dynamics to the coherent region. 
The simulation results show that, after the temperature quench, the condensate domains begin to grow in 
various regions of the fluctuating classical field, and several vortices and anti-vortices 
(like a quantum turbulence) are formed. Eventually, a single vortex remains in the 
well-developed condensate. 

\subsubsection{Vortex formation during the interference of BECs}
In a related work, vortices were formed by merging three uncorrelated BECs that 
were initially separated by a triple-well potential \cite{Scherer}. This may 
verify one side of the KZ mechanism: Can the causally disconnected 
domains generate a net circular flow, i.e., vortices, through a merging process? 
When the potential barrier created by a repulsive laser beam was removed, 
the three BECs merged. Depending on the relative phases between the condensates 
and the rate of barrier removal, vortices formed stochastically even without an applied rotation. 
If the condensates have random initial phases, a vortex should appear with a 
probability of 0.25 \cite{Scherer,Kasamatsu02}, which was observed with a slow removal rate 
of the barrier. With quick barrier removal, the resulting interference fringes could generate 
pairs of vortices and anti-vortices via ``snake instability" \cite{Feder00,Carretero08}, but these soon 
disappear due to pair annihilation. These properties were studied by 
extensive numerical simulations of the GP equation \cite{Carretero,Ruben}

\section{Vortices in two-component condensates}
We now address the properties of vortices in multicomponent BECs. 
Although there have been many theoretical predictions of exotic vortices in multicomponent BECs,
much experimental work remains to be performed. Here, we primarily describe 
theoretical studies of the topic, along with a few related experimental results 
\cite{Matthews,Leanhardt2,Schweikhard2,Sadler,Leslie09}. 
In this section, we focus on the properties of vortices in two-component BECs. 

\subsection{Multicomponent Bose-Einstein condensates}
Before discussing vortices, we summarize the types of multicomponent 
BECs and describe how they are obtained.
In this system, one can load and cool atoms in more than one hyperfine spin state 
or more than one atomic element in the same trap, thereby
experimentally realizing multicomponent condensates. 
The simplest and most intuitive example 
is a mixture of different kinds of atoms. However, due to the 
effect of strong inelastic collisions between different kinds of atoms, 
this example is generally difficult to realize. One successful example is the 
$^{87}$Rb-$^{41}$K mixture \cite{Modugno2}. Recently, a BEC mixture of isotopes of Rb atoms 
($^{87}$Rb-$^{85}$Rb) was realized by two groups \cite{Thalhammer,Papp2}. 
These experiments demonstrated that the interacting properties of two atomic 
species were tuned via a Feshbach resonance, 
opening a new avenue to study superfluid mixtures in a well-controlled manner.  

A common method used to realize a condensate mixture of ultra-cold atoms is to utilize 
the hyperfine spin states of atoms. The resulting BECs can have internal degrees 
of freedom, due to these hyperfine states. A hyperfine-Zeeman sublevel 
of an atom with total electronic 
angular momentum ${\bf J}$ and nuclear spin ${\bf I}$ may be labeled by the 
projection $m_{F}$ of total atomic spin ${\bf F} = {\bf I} + {\bf J}$ on the axis of the 
field ${\bf B}$ and by the total $F$, which can take a value from $|I-J|$ to $|I+J|$. 
This is because the hyperfine coupling, which is proportional to ${\bf I} \cdot {\bf J}$, is 
much larger than the typical temperature of an ultra-cold atomic system. 
The hyperfine state is denoted by $| F, m_{F} \rangle$ with $m_{F} = -F, -F+1, \cdots , F-1, F$. 
The simultaneous trapping of atoms with different hyperfine sublevels makes it 
possible to create multicomponent (often called ``spinor'') BECs with internal degrees 
of freedom, characterized by multiple order parameters 
\cite{Hall1,Stenger,Barrett,Schmaljohann,Chang,Kuwamoto}.
The order parameter of a BEC with hyperfine spin $F$ has $2F+1$ component 
$\Psi_{m_{F}}$ with respect to the basis vectors 
$| m_{F} \rangle$ defined by $\hat{F}_{z} | m_{F} \rangle = m_{F} | m_{F} \rangle$, 
which can be expanded as $| \Psi \rangle 
= \sum_{m_{F}=0,\pm 1, \cdots \pm F} \Psi_{m_{F}} | m_{F} \rangle$. 

Two-component mixtures of BECs with different hyperfine spin states have been created 
in the laboratory for the systems of $^{87}$Rb \cite{Hall1,Maddaloni00,Mertes07}
and $^{23}$Na \cite{Miesner99}. In Ref. \cite{Hall1}, the
two-component BECs of $^{87}$Rb atoms consisted of the hyperfine spin states 
$|F=1,m_{F}=-1 \rangle \equiv |1 \rangle$ and $|F=2,m_{F}=1 \rangle \equiv |2 \rangle$.  
The MIT group used an optical trap, all spin manifold $F=1$ spin states of $^{23}$Na 
($m_{F} = 0, \pm1$) were simultaneously trapped in a single trap. 
In this case, the spin exchange interactions (see Sec. \ref{Spin1}) gave rise to spin mixing dynamics 
among the different components. 
They made use of a quadratic Zeeman effect for atoms by applying a magnetic field to 
prevent the $|m_{F}=-1 \rangle$ component from appearing. 
As a result, the system can be regarded as a two-component BEC. 

When the trapping potential $V_{\rm ex}^{i}({\bf r})$ ($i=1,2$) is produced by a magnetic field, 
the atoms of each hyperfine state have their own potential energy 
because the potential energy depends on the magnetic moment of the atom. 
For example, the trapping potential for atoms in the $| F_{i},m_{F_i} \rangle$ state is 
expressed as a function of the magnetic field $|B({\bf r})|$ as \cite{Ho2}
\begin{eqnarray}
V_{\rm ex}^{i}({\bf r}) &=&  g_{F_i} \mu_{B} m_{F_i} |B({\bf r})| 
\simeq g_{F_i} \mu_{B} m_{F_{i}} \biggl[ B_{0} + \frac{1}{2} (K_{x}x^{2}+K_{y}y^{2}+K_{z}z^{2}) \biggr] \nonumber \\
 & \equiv & V_{0}^{i} + \frac{1}{2} m_{i} ( \omega_{ix}^{2}  x^{2} + \omega_{iy}^{2}  y^{2} + \omega_{iz}^{2}  z^{2} ), 
\label{trap2compda}
\end{eqnarray}
where $g_{F_i}$ is the $g$-factor of the $i$-th atom, $\mu_{B}$ is the Bohr magneton,
$V_{0}^{i} = g_{F_i} \mu_{B} m_{F_i} B_{0}$, and $\omega_{ik}$ ($k=x,y,z$) 
is the trapping frequency 
satisfying the relation $m_{i}\omega_{ik}^{2} = g_{F_i} \mu_{B} m_{F_i} K_{k}$.
The potential minima $V_{0}^{i}$ for BECs with different values 
of $g_{F_i} m_{F_i}$ do not coincide. 
This positional shift is also influenced by a gravitational effect, the difference in nuclear magnetic 
moments, and nonlinearity of the Zeeman shift \cite{Myatt,Hall1}. If the potential is created by 
a polarized optical laser \cite{Stenger,Barrett,Schmaljohann,Chang}, 
atoms in all hyperfine states share the same confining potential. 

\subsection{Coupled Gross-Pitaevskii equations}
First, we describe the mean-field theory for two-component BECs, which 
is a natural extension from that of single-component BECs. 
The condensates are confined by trapping potentials 
$V_{\rm ex}^{i}({\bf r})$ ($i=1,2$), which are assumed to rotate 
at a rotation frequency $\Omega$ about the $z$ axis as ${\bf \Omega}=\Omega \bf{\hat{z}}$. 
The GP energy functional of the two-component BEC in the rotating frame is 
\begin{equation}
E = \int d {\bf r} \biggl[ \sum_{i=1,2} \Psi_{i}^{\ast} 
\biggl( - \frac{\hbar^{2} \nabla^{2}}{2m_{i}} + V_{\rm ex}^{i}({\bf r}) 
-\Omega L_{z} + \frac{g_{i}}{2} |\Psi_{i}|^{2} \biggr) \Psi_{i} 
+ g_{12} |\Psi_{1}|^{2} |\Psi_{2}|^{2} \biggr].  
\label{energyfunctio2}
\end{equation}
Here, $m_{i}$ is the mass of the $i$-th atom. The coefficients $g_{i}$ and $g_{12}$ 
represent the atom-atom interactions between atoms in the same component, 
and atoms in different components, respectively. 
These are expressed in terms of the corresponding $s$-wave scattering lengths 
$a_{1}$, $a_{2}$, and $a_{12}$ as $g_{i} = 4 \pi \hbar^{2} a_{i}/m_{i}$, and 
$g_{12} = 2 \pi \hbar^{2} a_{12}/m_{12}$, where $m_{12}^{-1} = m_{1}^{-1} + m_{2}^{-1}$ 
is the reduced mass. The time-dependent coupled GP equations for two-component BECs can be obtained 
using a variational procedure $i \hbar \partial \Psi_{i} = \delta E / \delta \Psi_{i}^{\ast}$ as 
\begin{equation}
i \hbar \frac{\partial \Psi_{i}}{\partial t} = \left( -\frac{\hbar^{2} \nabla^{2} }{2m_{i}} 
+ V_{\rm ex}^{i}({\bf r}) + g_{i}|\Psi_{i}|^{2} + g_{12}|\Psi_{3-i}|^{2} 
-\Omega L_{z} \right) \Psi_{i}, \label{bingp1td}
\end{equation}
Substituting $\Psi_{i} ({\bf r},t) = \Phi_{i}({\bf r}) e^{-i \mu_{i} t / \hbar}$ yields the time-independent 
coupled GP equations 
\begin{eqnarray}
\left( -\frac{\hbar^{2} \nabla^{2}}{2m_{i}} + V_{\rm ex}^{i}({\bf r}) + g_{i}|\Phi_{i}|^{2} 
+ g_{12}|\Phi_{3-i}|^{2} -\Omega L_{z} \right) \Phi_{i} =\mu_{i} \Phi_{i}, \label{bingp1}
\end{eqnarray}
where we have introduced the Lagrange multiplier $\mu_{i}$, which represents the chemical potential 
and was determined so as to satisfy the conservation of particle number 
$N_{i}= \int d {\bf r} |\Phi_{i}|^{2}$ 
for each component. 

It is instructive to examine the ground state without rotation ($\Omega=0$). 
Compared to single-component BECs, two-component BECs exhibit a rich 
variety of ground state structures, depending on the various parameters of the system \cite{Ho2}. 
The intercomponent interaction $g_{12}$ plays an especially important role; 
phase separation of homogeneous two-component BECs occurs 
when $g_{12}^{2}>\sqrt{g_{1}g_{2}}$ \cite{Tim98,Ao98}, where strong intercomponent repulsion 
overcomes the intracomponent repulsion. 
In $^{87}$Rb, the scattering lengths are nearly equal 
($a_{1} = 5.67$ nm, $a_{2} = 5.34$ nm, 
and $a_{12} = 5.50$ nm), thus $g_{12}^{2} \simeq \sqrt{g_{1}g_{2}}$, but the small relative 
displacement of the trapping minima leads to a spatial separation between the two components \cite{Hall1}. 
Even without the trap displacement, the two components phase separate into 
oscillating ring structures \cite{Mertes07}.
In contrast, for the $^{23}$Na-mixture, $a_{1}=2.75$ nm, 
$a_{2}=2.65$ nm, and $a_{12}=2.75$ nm, thus 
$g_{12}^{2} > \sqrt{g_{1}g_{2}}$, causing a well-separated spin domain 
to form \cite{Stenger,Miesner99}. 
A recent experiment demonstrated the control of the miscibility 
of a $^{85}$Rb-$^{87}$Rb mixture via Feshbach resonance \cite{Papp2}. 

\subsection{A single vortex in a two-component BEC} \label{SingletwoBEC}
\subsubsection{Experimental creation} 
The first observation of a quantized vortex in an atomic BEC was achieved in 
a two-component BEC consisting of $^{87}$Rb atoms with hyperfine spin states
$|F=1,m_{F}=-1 \rangle \equiv |1 \rangle$ and $|F=2,m_{F}=1 \rangle \equiv |2 \rangle$ 
\cite{Matthews}, which were confined simultaneously in almost identical 
magnetic potentials. Initially, condensed atoms were trapped in one state, such as 
the $|1\rangle$ state. Then, a two-photon microwave field was applied, inducing 
coherent Rabi transitions of atomic populations between the $|1\rangle $ state 
and the $|2\rangle$ state. For a homogenous system in which both components 
have uniform phases, the interconversion takes place at the same rate everywhere. 
However, time variation of the spatially inhomogeneous potential changes 
the nature of the interconversion. Let us consider a co-rotating frame 
of an off-centered perturbation potential at the rotation frequency $\Omega'$. 
In this frame, the energy of a vortex with one unit of angular momentum is shifted 
by $\hbar \Omega'$ relative to its value in the laboratory frame. When this energy 
shift is compensated for by the sum of the detuning energy of an applied microwave 
field and the small chemical potential difference between the vortex and non-vortex states, 
a resonant transfer of population can occur between the non-vortex $| 1 \rangle$ state 
and the vortex $| 2 \rangle$ state. 
The rotating perturbation created by a spatially inhomogeneous laser beam 
was rotated around the initial non-rotating component $| 1\rangle$. 
By adjusting $\Omega'$ and the detuning, the $|2\rangle$ component was resonantly 
transferred to a state with unit angular momentum by precisely controlling the time 
at which the coupling drive was turned off \cite{Williams99}. 
This procedure resulted in a ``composite'' vortex, where the $|2\rangle$ component 
had a vortex at the center, while the non-rotating $|1\rangle$ component occupied 
the center and acted as a pinning potential to stabilize the vortex core. 

It is possible to prepare the initial condensate into either the $| 1 \rangle $ or $| 2 \rangle $ state, 
and to then form a vortex in the $| 2 \rangle $ or $| 1 \rangle $ state.
It was observed that the stability of the vortex states was strongly dependent 
on the atom-atom interactions \cite{Matthews}.
The vortex was stable only when the vortex was in the $|1\rangle $ state and the core was in the $|2\rangle $ state. 
The other state, with the vortex in the $|2\rangle $ state, exhibited an instability, 
with the $|2 \rangle$ vortex sinking inward toward the trap center and breaking up. 
The details are discussed below. 

\subsubsection{Equilibrium structure}
Here, we present theoretical studies on the structures of a single vortex state 
in two-component BECs. 
The analysis is based on the coupled GP equations (\ref{bingp1}). 
As observed in Ref. \cite{Matthews}, the vortex 
structure consists of one circulating component that surrounds the other 
non-rotating component, where both wave functions have axisymmetric profiles. 
We assume that the condensate wave functions have one quantized vortex at the center 
as $\Phi_{i}({\bf r})= \phi_{i}(r,z) e^{i q_{i} \theta}$, where $n_{i}=\phi_{i}^{2}$ is the condensate 
density and the trapping potential is given by 
$V_{\rm ex}^{i}({\bf r}) = m_{i} (\omega_{i \perp}^{2} r^{2} + \omega_{i z}^{2} z^{2}) / 2$. 
Thus, the axisymmetric (singly-charged) vortex states are characterized 
by $(q_{1}, q_{2}) = (1,0)$, $(0,1)$, $(1,1)$. 

\begin{figure}[htbp] 
\begin{center} 
\includegraphics[angle=0,width=13cm]{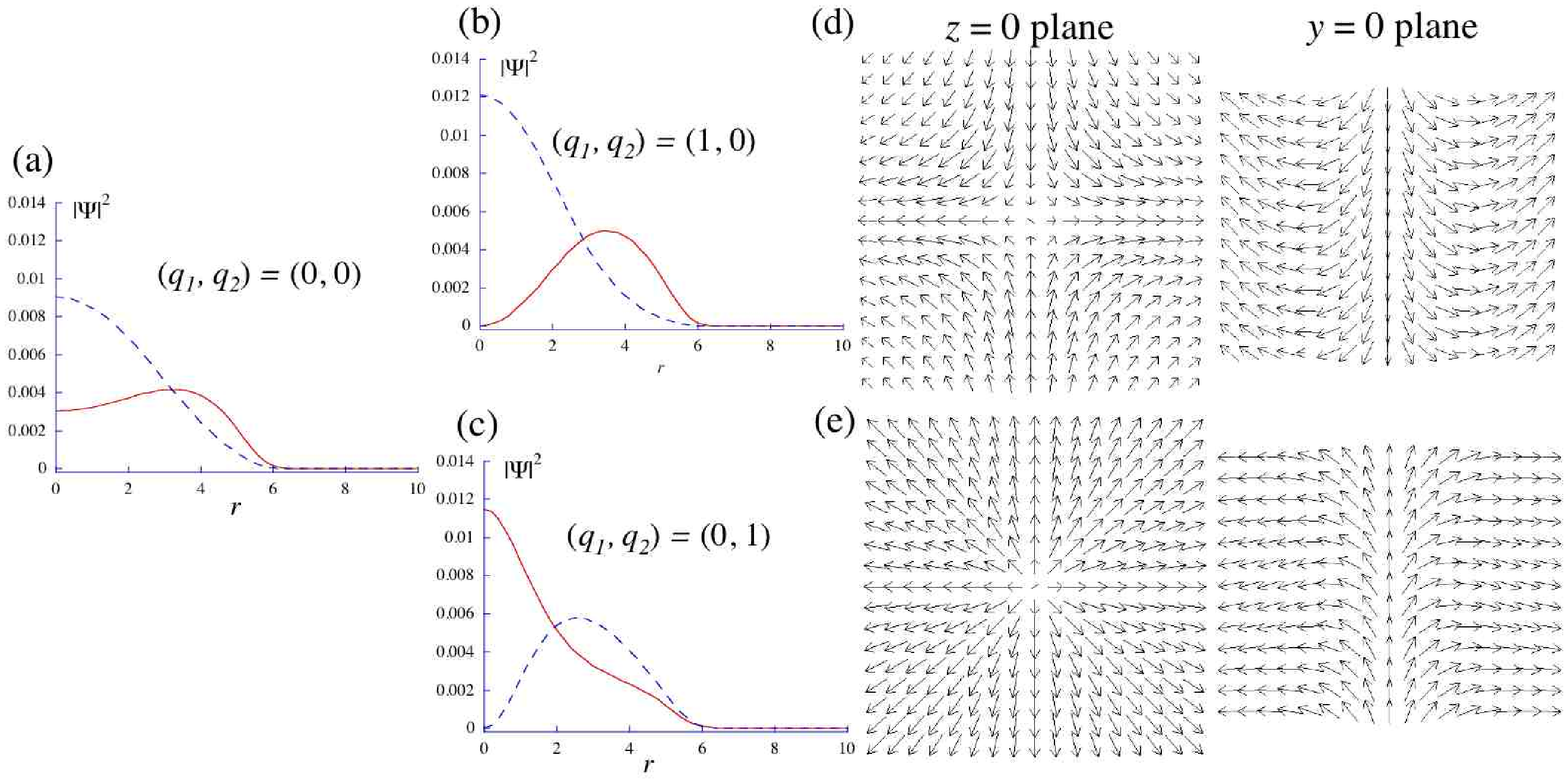} 
\end{center} 
\caption{
Stationary solutions of two-component BECs with 
winding number $(q_{1},q_{2}) = (0,0)$, $(1,0)$, and $(0,1)$. The parameters 
of the $^{87}$Rb atoms in $| 1, -1 \rangle$ and $| 2, 1 \rangle$ states were used. 
The trapping potential was oblate as $\omega_{\perp} = 5 \times 2 \pi$ Hz 
and $\omega_{z} = 10 \times 2 \pi$ and the particle number $N_1=N_2=10^5$. 
In (a)-(c), the radial density profile is shown. The corresponding (pseudo)spin textures 
in the $z=0$ and $y=0$ planes are shown for (d) $(1,0)$ and (e) $(0,1)$ states.} 
\label{twocomposingle}
\end{figure}
For $^{87}$Rb atoms in non-rotating traps, in order to decrease the intracomponent 
mean field energy, the $\Psi_{1}$ component with the larger intracomponent scattering 
length $a_{1}$ forms a shell outside the $\Psi_{2}$ component occupying the central part, 
as shown in Fig. \ref{twocomposingle}(a). 
For $(q_{1}, q_{2}) = (1,0)$ [Fig. \ref{twocomposingle}(b)], a density depletion associated with the 
vortex core appears in the density of $\Psi_{1}$ along the $z$-axis. 
The $\Psi_{2}$ component then fills the vortex core, 
because of intercomponent repulsion. As a result, the characteristic size of the core 
becomes larger than the healing length $\xi=\hbar/\sqrt{2 m_1 g_{1} n_{1}}$ of 
the single component. 
In this case, the centrifugal force associated 
with the vortex causes the $\Psi_{1}$ component to expand radially, which allows a 
decrease in the intracomponent mean-field energy of the $\Psi_{1}$ component, 
rather than that of the nonvortex state. 

On the other hand, the structure of $(q_{1}, q_{2}) = (0,1)$ is different 
from the $(q_{1}, q_{2})=(1,0)$ solution [see Fig. \ref{twocomposingle}(c)]. 
The core size becomes smaller than that of the $(1,0)$ solution. 
While a fraction of the $\Psi_{1}$ component fills the vortex core, the
excessive $\Psi_{1}$ component extends outward, resulting in an 
increase in size of the region of coexistence with the rotating $\Psi_{2}$ component. 
As discussed below, 
this configuration is {\it dynamically} unstable \cite{Ripoll00,Skrybin00}. 
Also, the $(q_{1}, q_{2}) = (1,1)$ state is unstable because the 
two components fully overlap, which is energetically unfavorable. 
In the presence of a slow rotation, the 
thermodynamically stable configuration of $(1,1)$ is that in which the two vortices in each component 
are displaced from the center, which decreases the size of the overlapping region. 

It is instructive to represent the spinor order parameter of two-component BECs 
using the ``pseudospin" \cite{Leonhardt00,Mueller04,Kasamatsu05}, which 
allows us to analyze this system as a spin-1/2 BEC. 
We introduce a normalized spinor $\bm{\zeta}$ to represent the two-component 
wave functions as 
$[\Psi _1,\Psi _2]^T= \sqrt[]{n_{\rm T}} e^{i \Theta /2} \bm{\zeta}$, where
\begin{eqnarray}
\bm{\zeta} = \left( 
\begin{array}{c}
\zeta_{1} \\
\zeta_{2}
\end{array} 
\right)
= \left[ 
\begin{array}{c}
\cos(\theta /2)e^{-i\varphi /2} \\
\sin(\theta /2)e^{i\varphi /2} \\
\end{array} 
\right]
\end{eqnarray}
and $|\zeta_{1}|^{2}+|\zeta_{2}|^{2}=1$. 
Here, we assume that $\Psi_{1}$ and $\Psi_{2}$ represent the up and down components of the 
spin-1/2 spinor, respectively. Hence, the four degrees of freedom of the original 
wave functions $\Psi_{j} = \sqrt{n_{j}}e^{i\theta_{j}}$ 
(their amplitudes $n_{j}$ and phases $\theta_{j}$) are expressed 
in terms of the total density $n_{\rm T} = n_{1} + n_{2}$, the total 
phase $\Theta=\theta_{1}+\theta_{2}$, 
and the polar and azimuthal angles ($\theta$, $\varphi$) of the local
pseudospin ${\bf S}=(S_{x},S_{y},S_{z})$, defined as 
\begin{eqnarray}
{\bf S}={\bm \zeta}^{\dagger}  \bm{\sigma} {\bm \zeta}
= \left[ 
\begin{array}{c}
\zeta_{1}^{\ast} \zeta_{2} + \zeta_{2}^{\ast} \zeta_{1} \\
-i(\zeta_{1}^{\ast} \zeta_{2} - \zeta_{2}^{\ast} \zeta_{1})\\
|\zeta_{1}|^{2} - |\zeta_{2}|^{2}
\end{array} 
\right]
= \left[ 
\begin{array}{c}
\sin \theta \cos \varphi \\
\sin \theta \sin \varphi \\
\cos \theta
\end{array} 
\right] , \label{spindef}
\end{eqnarray}
where ${\bm \sigma}$ is the Pauli matrix, $\cos \theta = (n_{1}-n_{2})/n_{\rm T}$, 
$\varphi = \theta_{2}-\theta_{1}$, and $|{\bf S}|^2=1$. 

In the pseudospin picture, the axisymmetric vortex is analogous to a spin texture, 
extensively studied in superfluid $^{3}$He. 
When the wave function for the $(1,0)$ case is parameterized as 
$( \Psi_{1}, \Psi_{2} ) = \sqrt{n_{\rm T}} ( e^{i \theta} \cos (\beta(r)/2) , \sin (\beta(r)/2) )$,
the configuration satisfying the boundary condition 
$\beta(0)=\pi$ and $\beta(\infty)=0$ is referred to as 
an Anderson-Toulouse vortex \cite{Anderson77}, 
and $\beta(\infty)=\pi/2$ is a Mermin-Ho vortex \cite{Mermin76}. 
The spin texture for the $(1,0)$ solution is shown 
in Fig. \ref{twocomposingle} (d). Here, the $\Psi_{1}$ component vanishes 
at the center, so that the pseudospin points down at $r=0$ ($\beta(0)=\pi$)
according to the definition of Eq. (\ref{spindef}). 
The spin aligns with a hyperbolic distribution as $(S_{x},S_{y}) \propto (-x,y)$ 
around the center in the $x$-$y$ planes. 
At the edge of the atomic cloud, the $\Psi_{2}$ component 
vanishes, and the pseudospin points up. In between, 
the bending angle $\beta(r)=\cos^{-1} S_{z}$ decreases smoothly from 
$\pi$ at $r=0$ as $r$ increases, as seen in the profile in the $y=0$ plane. 
For the $(0,1)$ state, the texture exhibits a radial-disgyration 
as shown in Fig. \ref{twocomposingle} (e), 
where the spin at $x=y=0$ points up and aligns as $(S_{x},S_{y}) \propto (x,y)$ 
in the $x$-$y$ plane. Note that the spin rotates $\pi/2$ 
as going radially outward from the center. 
This configuration corresponds to a Mermin-Ho vortex. 
In the case of superfluid $^{3}$He, an MH vortex is stabilized by the boundary 
condition of the $\hat{\bf l}$-vector imposed by a cylindrical vessel. 
However, in an atomic-BEC system there is 
no constraint at the boundary; the value $\beta(r)$ at the boundary $r=R$ should 
be determined self-consistently. 
The value $\beta(R)$ can change arbitrarily with $\Omega$ 
by varying the ratio $N_{1}/N_{2}$ or $g_{1}/g_{2}$ \cite{Kasamatsu05}, 
which implies that an intermediate configuration between a Mermin-Ho vortex 
($\beta(R)=\pi/2$) and an Anderson-Toulouse vortex ($\beta(R)=0$) 
can be thermodynamically stable. 

In terms of the spin-1/2 BEC, since the density $n_{\rm T}$ does not vanish at the vortex 
core, these vortices can be called {\it coreless} vortices. We can define 
an effective velocity field for the spinor BEC as 
\begin{equation}
{\bf v}_{\rm eff} = 
\frac{\hbar}{2im} \sum_{j=1,2} \left( \zeta_{j}^{\ast} \nabla \zeta_{j} 
- \zeta_{j} \nabla \zeta_{j}^{\ast} \right)  
= \frac{\hbar}{2m} \left( \nabla \Theta - \cos \theta \nabla \varphi \right) ,  \label{effvelo} 
\end{equation}
which depends on the gradient of the total phase $\Theta=\theta_{1}+\theta_{2}$ 
and that of the angle of the pseudospin. By neglecting the singular contribution of the 
velocity due to the total phase $\Theta$, we can obtain the vorticity 
\begin{eqnarray}
{\bm \omega}_{\rm eff} = \nabla \times {\bf v}_{\rm eff} 
= \frac{\hbar}{2m} (\nabla \theta) \times ({\sin} \theta \nabla \varphi) 
=  \frac{\hbar}{4m} \epsilon_{\alpha \beta \gamma} 
S_\alpha \nabla S_\beta \times \nabla S_\gamma 
\label{vorticityeff}
\end{eqnarray}
with the Levi-Civita symbol $\epsilon_{\alpha \beta \gamma}$. 
The last equality is known as the Mermin-Ho relation \cite{Mermin76}. 
The vortex was characterized by the topological charge (or the Pontryagian index) \cite{Rejanbook}
\begin{equation}
Q \equiv \int d^{2} r q({\bf r}) = \frac{m}{h} \int d^{2} r \omega_{{\rm eff} z} 
= \frac{1}{8 \pi} \int d^{2} r \left( \epsilon_{\alpha \beta \gamma} 
S_\alpha \nabla S_\beta \times \nabla S_\gamma \right)_{z},
\label{topologicalnumber}
\end{equation}
where $q({\bf r})$ is the topological charge density. 
The topological charge density $q({\bf r})$ characterizes the spatial distribution 
of the vorticity of the coreless vortex, which is distributed around the center, 
and $|{\bf v}_{\rm eff}|$ vanishes 
at the center, contrary to the case of a conventional vortex in a single-component condensate. 

\begin{figure}[htbp] 
\begin{center} 
\includegraphics[angle=0,width=13cm]{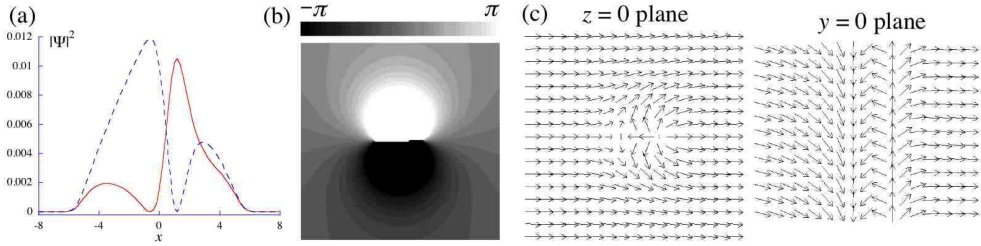} 
\end{center} 
\caption{
A simulated $(1,1)$ state, referred to as a vortex molecule \cite{Kasamatsu04}. 
The parameters of the $^{87}$Rb atoms in $| 1, -1 \rangle$ and $| 2, 1 \rangle$ states were used. 
The trapping potential is oblate $\omega_{\perp} = 5 \times 2 \pi$ Hz 
and $\omega_{z} = 20 \times 2 \pi$ and the total particle number $N=2 \times 10^5$. 
To stabilize the vortex molecules, the rotation $\Omega = 0.2\omega_{\perp}$ and the 
Rabi coupling with $\omega_{R}=0.1\omega_{\perp}$ is applied. Then, the balance of 
the population ($N_{1} \simeq N_{2}$) is controlled by the detuning $\bar{\Delta}=0.08$, which denotes the 
difference between the frequency of the rf-field and the energy difference between 
the two internal states. (a) Cross section of the condensate density along the $x$-axis. 
(b) Grayscale plot of the relative phase around the vortex molecule in the $z=0$ plane. 
(c) The spin textures in the $z=0$ and $y=0$ planes.} 
\label{vortexmolecule}
\end{figure}
It is interesting to consider the nonaxisymmetric $(1,1)$ state in terms of the pseudospin, 
where the pair of coreless vortices can be identified as a pair of Mermin-Ho vortices 
(meron-pair) \cite{Mermin76}, because the spins are oriented 
along the $x$-$y$ plane, far from the two vortices. An example is shown in Fig. \ref{vortexmolecule}. 
Note that the boundary condition for the pseudospin is not well-defined in our system. 
If we introduce the {\it internal Josephson coupling} between the two components, 
the spin should be oriented along the $x$-axis, far from the defect. 
This can be achieved by applying an external driving field that couples the two hyperfine 
levels of the atoms. When the atoms are condensed, coherent Rabi-oscillation 
between the two components occurs \cite{Matthews99}. 
If the field strength is increased gradually from zero and its frequency is gradually ramped 
to resonance, a stationary nearly-weight superposition of the two components can be obtained.  
This Josephson coupling energy, 
given by $E_{\rm J} = -\hbar \omega_{\rm R} (\Psi_{1}^{\ast} \Psi_{2} + {\rm c.c}) 
= - 2 \hbar \omega_{\rm R} \sqrt{n_{1} n_{2}} \cos \varphi = - 2 \hbar \omega_{\rm R} n_{\rm T} S_{x}$ 
with the Rabi frequency $\omega_{\rm R}$, acts as an external field that aligns the 
spin along the $x$-axis. Therefore, the 
meron-pair is stabilized \cite{Kasamatsu04}. In the presence of the Josephson coupling, 
the two vortices are bounded by the domain wall of the {\it relative phase} \cite{Son02}, 
shown in Fig. \ref{vortexmolecule}(b), whose binding energy can be controlled 
by the Josephson coupling strength.  

\subsubsection{Vortex stability and dynamics}
Unstable vortex dynamics were experimentally observed to 
depend on which component had a vortex \cite{Matthews}. 
As seen in the ground state configuration of Fig. \ref{twocomposingle}(a), 
if a component with smaller $g_{i}$ 
is displaced from the trap center, it will tend to return to the trap center. This effect is sometimes 
called the ``buoyancy" effect in trapped two-component BECs \cite{McGee}, and plays an 
important role in the vortex dynamics. 

Quantitatively, the instability can be understood from the BdG analysis of the vortex state 
\cite{Ripoll00,Skrybin00}. Numerical simulations showed that for equal populations 
$N_{1}=N_{2}=N$, and with the coupling constants $g_{1}>g_{12}>g_{2}$, 
the axisymmetric vortex 
$(1,0)$ state is stable, while the $(0,1)$ state is {\it dynamically} unstable. 
Among the normal modes of the $(1,0)$ case, 
there is a negative eigenvalue. This mode, which is a dipole-type mode, 
belongs to a perturbation that causes an initial displacement of the vortex from the trap center. 
However, the drift instability is affected by dissipation; the configuration is dynamically 
stable without dissipation, and furthermore, it can be energetically stabilized by 
rotation \cite{Kasamatsu05}. 
In the $(0,1)$ case, on the other hand, there are normal modes with complex frequencies. 
The unstable modes have a similar shape as the negative-eigenvalue modes of the $(1,0)$ case, 
which means that the perturbations also displace the vortex from the center. 
The imaginary part of the eigenvalues implies that vortices with unit charge in $|2\rangle $ 
are dynamically unstable under a generic perturbation of the initial configuration 
and this instability grows without dissipation. 
This result is consistent with the JILA experiments \cite{Matthews}. 

To study the dynamics beyond the linear stability, one begins with the time-dependent coupled GP equations 
(\ref{bingp1td}). Numerical simulations of the nonlinear vortex dynamics caused by large perturbations 
were performed by Garc\'{i}a-Ripoll and P\'{e}rez-Garc\'{i}a \cite{Ripoll00}. The 
linearly stable state $(1,0)$ is robust and survives even under a wide range of perturbations; 
the vortex only shows a precession around the center. In contrast, the unstable 
configuration $(0,1)$ gives rise to recurrent dynamics: the density of the $|1\rangle $ component and the 
vortex in $|2\rangle $ oscillate synchronously. These oscillations grow in amplitude 
until the vortex spirals outward. Since Eq. (\ref{bingp1td}) conserves the total angular momentum 
$\langle L_{z} \rangle= \int d {\bf r} ( {\Psi}_{1}^{*} L_{z} \Psi_{1} + {\Psi}_{2}^{*} L_{z} \Psi_{2})$, 
the vortex is transferred from $|2\rangle $ to $|1\rangle $. 
Although the dynamics are not completely periodic, this mechanism exhibits some 
recurrence, and the vortex eventually returns to $|2\rangle $. The unstable dynamics 
of $| 2\rangle$ with a multiply-quantized vortex such as $(q_1,q_2)=(0,2), (0,3)$ 
were investigated by Skryabin \cite{Skrybin00}. 

The effect of the intercomponent interaction was also demonstrated by the precession 
of an off-centered vortex \cite{Anderson00}. The experiment observed that the 
precession frequency of the {\it filled}-core (coreless) vortex was slower than that 
of the empty core. The slower precession of filled cores can be understood 
in terms of the buoyancy effect. Because of its slightly smaller scattering length, 
the $| 2 \rangle$ component has a negative buoyancy with respect to the $| 1 \rangle$ component, 
and consequently tends to sink inward toward the center of the condensate. 
With increasing amounts of the $|2\rangle$ component in the core, the inward force 
on the core begins to counteract the effective outward buoyancy of the vortex 
(described in Sec. \ref{Dynamicssinglevortex}), 
resulting in a reduced precession velocity. 
It was predicted that with a filling component of sufficiently negative buoyancy in the core, 
the core precession may stop, or even precess in a direction opposite to that 
of the fluid flow\cite{McGee}. 

\subsection{Vortex lattices in two-component BECs}
Vortex lattices in rotating two-component BECs are more complicated than 
those in single-component condensates because the system has two-component order parameters 
and are coupled by intercomponent interactions. 
To simplify the problem, each component has an equal number of bosons, and the trapping potentials 
of the two components are identical. Then, the two components will have the same size 
and the same density of vortices. In this case, one would expect that each component will 
contain identical but {\it staggered} vortex lattices, with one lattice displaced relative 
to the other because of the intercomponent repulsive interaction. 
We will focus on the case of $g_{1} = g_{2} \neq g_{12}$, and 
vary the parameter $\delta = g_{12}/\sqrt{g_{1}g_{2}}$.  
As described above, the condition of phase separation for non-rotating 
condensates is given as $\delta >1$. 
The presence of a vortex lattice naturally modulates the density of each component, 
with the high density regions of one component overlapping with the low density regions of the other
(the vortex core is filled by the other component). 
Thus, the system is effectively phase-separated whenever staggered vortex lattices are present, 
even for $\delta<1$. 

The first theoretical study of the lattice structure was done by Mueller and Ho within the 
mean-field quantum Hall regime \cite{Mueller02}, extending the analysis 
in Sec. \ref{fastrotatinLatice} to a two-component system. 
In this analysis, (i) the rotation frequency was assumed to be very close 
to the radial trapping frequency ($\Omega \simeq \omega_{\perp}$), 
(ii) the vortex lattice was assumed to be perfectly regular, (iii) the range $\delta > 1$ 
could not be described, because a non-periodic structure appears as shown below. 
For $\delta \leq 0$, the two components prefer to overlap and form a single 
triangular lattice. As $\delta$ increases, the lattices undergo several structural transitions. 
There are five distinct phases of the lattice: triangular, displaced triangular, 
rhombic, square, and rectangular with increasing $\delta$. 
Ke\c{c}eli and Oktal calculated the dispersion relation of the TK mode for each lattice type, 
and discussed its relation to the structural phase transition \cite{Keceli06}. 

\begin{figure}[htbp] 
\begin{center} 
\includegraphics[angle=0,width=13cm]{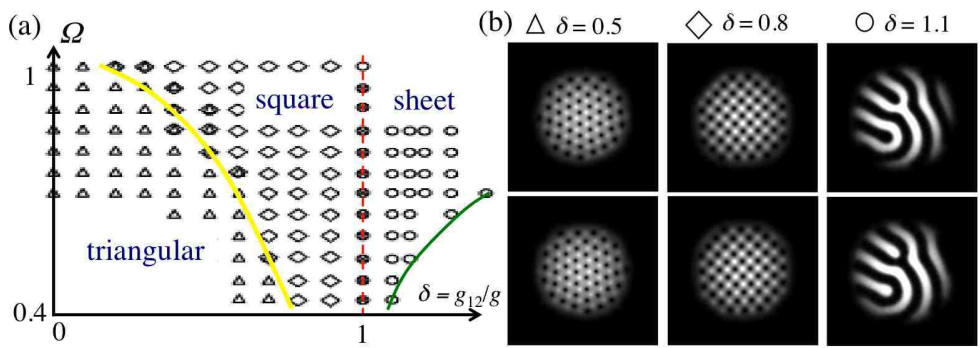} 
\end{center} 
\caption{(a) $\Omega$-$\delta$ phase diagram for vortex states in rotating two-component BECs, 
obtained by 2D numerical simulations of Eq. (\ref{bingp1}): 
$\triangle$: triangular lattice, $\diamond$: square lattice, $\circ$: vortex sheet; 
see Ref. \cite{Kasamatsu03} for details.
Because of the continuous change between triangular and square lattices, their boundary 
is shown by a combination of $\triangle$ and $\diamond$. 
The plots at $\Omega=1$ show the results obtained by Ref. \cite{Mueller02}
based on the LLL approximation. (b) Typical vortex states in each phase for 
$\Omega = 0.8\omega_{\perp}$ and $\delta=0.5$, 0.8, 1.1.} 
\label{twocomplattice} 
\end{figure}
Beyond the above restrictions (i)-(iii), one needs full numerical calculations 
of the coupled GP equations (\ref{bingp1}) \cite{Kasamatsu03}. 
Figure \ref{twocomplattice} shows the numerically obtained phase diagram of vortex states 
in the $(\delta, \Omega)$ plane. 
In the overlapping region $\delta<1$, two types of regular vortex lattices were obtained. 
For $\delta=0$, the two components did not interact. Therefore, triangular vortex lattices 
formed, as in a single component BEC. As $\delta$ increased, the positions 
of the vortex cores in one component gradually shifted away from those of the other component 
and the triangular lattices became distorted. Eventually, the vortices in each component 
formed a square lattice rather than a triangular one. The two vortex lattices were interlaced 
in such a manner that a peak in the density of one component was located at the density 
minimum of the other. The stable region of the square lattice 
depended not only on $\delta$, but also on $\Omega$. While an increase in $\delta$ did 
cause the deformation of the lattices from triangular to square, the transition occurred 
at a significantly higher value of $\delta$ than that of the LLL result ($\delta=0.373$ \cite{Mueller02}); 
for example, the transition occurred at $\delta \simeq 0.65$ for $\Omega=0.7$. 
This implies that an increase in rotation frequency causes a transition from 
triangular to square lattices. The excitation spectrum of these equilibrium states were 
calculated numerically by Woo {\it et al.} \cite{Woo07}, showing complicated behavior 
consisting of the TK modes, the hydrodynamic modes, and the surface modes. 

When $\delta$ exceeds unity, the condensates undergo phase separation to spontaneously 
form domains having the same spin component. 
Concurrently, vortices begin to overlap. 
In the strongly phase-separated region $\delta>1$ ($c_{2}<0$), 
the domains of the same spin component, at which the other-component vortices are located, 
merge further, resulting in the formation of ``serpentine" vortex sheets \cite{Kasamatsu03}. 
A typical example is shown in Fig. \ref{twocomplattice}. 
Singly-quantized vortices line up in sheets, and the sheets of components 1 and 2 are 
interwoven. The sheet distance can be estimated from the competition of the kinetic 
energy of the superflow and the surface tension of the domain wall. 
The detailed discussion of the vortex sheet was reported in Ref. \cite{Kasamatsu09}.

Schweikhard {\it et al.} observed an interlaced vortex lattice in a two-component BEC 
consisting of two hyperfine levels 
of the $^{87}$Rb atom: $ |1\rangle$ and $|2\rangle$ \cite{Schweikhard2}. 
They initially created a regular triangular vortex lattice 
in a BEC in the $|1\rangle$ state. Then, a short pulse of the coupling drive 
transferred a $80-85\%$ fraction of the population into the $|2\rangle$ state. 
After a variable wait time, they took images of the condensates, revealing interesting 
dynamics of the structural transition of the lattices. 
For the first period $0.1-0.25\,\rm{s}$, 
there was little dynamical behavior in either component, and certainly 
no structural transition in the vortex lattice. 
From $0.25 - 2\,\rm{s}$, turbulent behavior appeared in both components 
in which vortex visibility degraded significantly, exhibiting a transition from overlapping 
hexagonal vortex lattices to interlaced square lattices. From $2 - 3\,\rm{s}$, 
square lattices emerged from the turbulent state. 
From $3 - 5.5\,\rm{s}$ stable square lattices were observed in both components. 
At this stage, despite the large ($80-85\,\%$) initial population transfer to the state $|2\rangle$, 
the number of $|2\rangle$ atoms decreased rapidly because of a larger inelastic scattering 
loss of the $|2\rangle$ state than 
of the $|1\rangle$ state.  As the $|2\rangle$ state population continued to decay, 
the vortex lattice reverted to a hexagonal lattice in the $|1\rangle$ state. 

\section{Vortices in spinor condensates} \label{Spin1}
An optical trap removes the restriction of the confinable hyperfine spin states 
of atoms, thus realizing new multicomponent BECs with internal degrees 
of freedom, referred to as ``spinor BECs" \cite{Stenger,Barrett,Schmaljohann,Chang,Kuwamoto}. 
In contrast to the two-component BECs, interatomic interactions allow 
for a coherent transfer of population 
between different hyperfine spin states (spin-exchange collisions), which yields a fascinating 
physics of both ground-state properties and spin dynamics. 
Recently, an excellent pedagogical review appeared, addressing 
various problems on spinor BECs \cite{Ueda10}. In this section, we focus on the studies 
of quantized vortices in such spinor BECs with $F=1$ and $F=2$ hyperfine spin states. 

\subsection{General formulation of spin-$F$ Bose systems}
To describe the zero-temperature properties of spinor BECs, 
a generalized GP mean-field model was introduced by Ohmi and Machida \cite{Ohmi98} 
and Ho \cite{Ho98} for $F = 1$ spinor BECs. Here, we consider the most general case: 
spin-$F$ bosons under a magnetic field ${\bf B}$. 
The second quantized Hamiltonian of the system is $\hat{H} 
= \hat{H}_{0} + \hat{H}_{\rm int}$. The single-particle part $\hat{H}_{0}$ 
can be written as  
\begin{equation}
\hat{H}_{0} = \int d{\bf r} \sum_{\alpha,\beta=-F}^{F} \hat{\psi}_{\alpha}^{\dag}({\bf r}) 
\left( \hat{h} + g_{F} \mu_{\rm B} {\bf B} \cdot {\bf F}_{\alpha \beta} 
\right) \hat{\psi}_{\beta} ({\bf r}),
\label{spin1singlehamilton}
\end{equation}
where $\hat{\psi}_{\alpha}({\bf r})$ is the field annihilation operator for an atom 
in the hyperfine state $| F, \alpha \rangle$ ($\alpha = -F, -F+1, \cdots , F-1, F $), 
satisfying the commutation relation 
$[\hat{\psi}_{\alpha}({\bf r}),\hat{\psi}_{\beta}^{\dagger}({\bf r}')] 
= \delta_{\alpha \beta} \delta ({\bf r}-{\bf r}')$ 
and $[\hat{\psi}_{\alpha}({\bf r}),\hat{\psi}_{\beta}({\bf r}')] = 0$, and 
$\hat{h} =  - \hbar^2 \nabla^2 / 2m + V_{\rm ex}({\bf r})$ with an optical potential 
$V_{\rm ex}({\bf r})$ that can confine the atoms independently of their hyperfine 
spin states. The spin matrix ${\bf F}_{\alpha \beta}=[(F_{x})_{\alpha \beta}, 
(F_{y})_{\alpha \beta}, (F_{z})_{\alpha \beta}]$ is the spin-$F$ 
representation of the $SU(2)$ Lie algebra. 

The interaction Hamiltonian $\hat{H}_{\rm int} = \int d {\bf r}_{1} d {\bf r}_{2} 
V_{F} ( {\bf r}_{1} - {\bf r}_{2})$ is governed by a two-body 
interaction $V_{F} ( {\bf r}_{1} - {\bf r}_{2})$ that is invariant under spin rotation 
and preserves the hyperfine spin of the individual atoms. In a low-energy limit, 
this is of the form 
\begin{equation}
V_{F} ( {\bf r}_{1} - {\bf r}_{2}) = \delta ( {\bf r}_{1} - {\bf r}_{2}) \sum^{2F}_{f=0} g_{f} {\mathcal P}_f, 
\end{equation}
where $g_{f} = 4 \pi \hbar^{2} a_{f}/m$ for atoms with an s-wave scattering length $a_{f}$ 
corresponding to a channel with a total hyperfine spin $f$ of two colliding atoms. 
The operator ${\mathcal P}_{f}$ projects 
the pair of atoms to a state with total hyperfine spin $f$, written in the 
second quantized form as ${\mathcal P}_{f} = \sum_{\alpha=-f}^{f} 
\hat{O}^{\dagger}_{f,\alpha} \hat{O}_{f,\alpha}$, 
where $\hat{O}_{f,\alpha} = \sum_{\alpha_{1},\alpha_{2}} 
\langle f \alpha | F, \alpha_{1}; F, \alpha_{2} \rangle 
\hat{\psi}_{\alpha_{1}}^{\dagger} \hat{\psi}_{\alpha_{2}}$ with the Clebsch-Gordon coefficient 
$\langle f \alpha | F, \alpha_{1}; F, \alpha_{2} \rangle$, forming a total spin $f$ state from 
two spin-$F$ particles. The sum is taken for even numbers of $f$, because 
only symmetric spin channels are allowed for bosons. 

\subsection{Mean-field theory for spin-1 BECs}
\subsubsection{Generalized GP equation}
For the spin-1 case, the field operator has three components: 
$(\hat{\psi}_{1}, \hat{\psi}_{0}, \hat{\psi}_{-1})$. The interaction can be expressed as 
\begin{equation}
V_{F=1} = \frac{g_{n}}{2} \sum_{\alpha,\beta=-1}^{1} \hat{\psi}_{\alpha}^{\dagger} 
\hat{\psi}_{\beta}^{\dagger}  \hat{\psi}_{\beta} \hat{\psi}_{\alpha} 
+ \frac{g_{s}}{2} \sum_{\alpha,\alpha ', \beta, \beta '= -1}^{1} \hat{\psi}_{\alpha}^{\dagger} 
\hat{\psi}_{\beta}^{\dagger} {\bf F}_{\alpha \alpha '} \cdot {\bf F}_{\beta \beta '}
\hat{\psi}_{\beta '} \hat{\psi}_{\alpha '}, 
\label{spin1interaction}
\end{equation}
where we have omitted $\delta ({\bf r}_{1} - {\bf r}_{2})$. 
The collision coefficients are written as $g_{n} = (g_{0}+ 2 g_{2})/3$ 
and $g_{s} = (g_{2} - g_{0})/3$. The term proportional to $g_n$ is
symmetric in the spin indices and represents the spin-independent
contact interaction. The term proportional to $g_s$, on the other
hand, is spin-dependent and represents the short-range
spin-exchange interaction. The sign of $g_s$ determines the nature
of the spin-exchange coupling: a negative (positive) $g_s$ represents
ferromagnetic (antiferromagnetic) coupling. 
The spinor BEC of $F=1$ $^{23}$Na atoms 
was demonstrated to have an antiferromagnetic interaction \cite{Stenger}, 
while the $F=1$ $^{87}$Rb condensate has a ferromagnetic interaction \cite{Barrett}. 
Although $g_{n} \gg g_{s}$ in most systems, the $g_{s}$-term plays a crucial role 
in determining the properties of spinor BECs. 

In the zero-temperature limit, the total Hamiltonian reduces 
to an energy functional by taking the expectation value $E = \langle \hat{H} \rangle$ 
and introducing the condensate wave function $\Psi_{\alpha} = \langle \hat{\psi}_{\alpha} \rangle$, 
which amounts to neglecting the effects of the non-condensed gas. 
For the condensates trapped by an optical potential and in a uniform magnetic field 
${\bf B} = B \hat{\bf z}$, we can obtain
\begin{eqnarray}
E[{\bf \Psi}] = \int d {\bf r} 
\left[ \sum_{\alpha=-1}^1 \Psi_{\alpha}^\ast \hat{h} \Psi_{\alpha} - p \langle F_{z} \rangle 
+ q  \langle F_{z}^{2} \rangle
+\frac{g_n}{2} n^2 + \frac{g_s}{2} | {\bf S} |^2 \right],
\label{spin1energyfunc}
\end{eqnarray}
where $n_{\rm T}({\bf r}) = \sum_{\alpha=-1}^{1} | \Psi_\alpha({\bf r}) |^2$ is the total particle density, and 
${\bf S} = (\langle F_{x} \rangle, \langle F_{y} \rangle, \langle F_{z} \rangle)$ is the spin density 
vector defined by $\langle F_{i} ({\bf r}) \rangle = \sum_{\alpha, \beta=-1}^{1} \Psi_\alpha^{\ast}({\bf r}) 
(F_{i})_{\alpha \beta} \Psi_\beta({\bf r})$ ($i = x, y, z$). 
In addition to the linear Zeeman term with $p=-g_{F} \mu_{B} B$, we have introduced an additional quadratic 
Zeeman term characterized by $q = (g_{F} \mu_{B} B)^{2}/E_{\rm hf}$ 
with the hyperfine splitting $E_{\rm hf}$; 
these coefficients can be varied arbitrarily in experiments \cite{Stenger}, 
and play an important role in determining the properties of spinor BECs. 

The time evolution of the wave function is governed by 
$i \hbar \partial \Psi_\alpha ({\bf r}) / \partial t = \delta E / \delta \Psi_\alpha^{\ast}$; 
substituting Eq. (\ref{spin1energyfunc}) into the right-hand side gives
\begin{equation}
i \hbar \frac{\partial \Psi_\alpha}{\partial t}
= \hat{h} \Psi_\alpha + g_n n_{\rm T} \Psi_\alpha 
+\sum_{\beta = -1}^{1} \left[- p  ( F_{z} )_{\alpha \beta} 
+ q ( F_{z}^{2} )_{\alpha \beta} 
+ g_s  {\bf S} \cdot {\bf F}_{\alpha \beta} 
\right] \Psi_{\beta},
\label{spin-1GPE}
\end{equation}
which are the multicomponent GP equations that describe the mean-field properties 
of spin-1 BECs \cite{Ohmi98,Ho98}.
In a stationary state, we substitute $\Psi_\alpha ({\bf r}, t) = \Phi_\alpha ({\bf r}) 
e^{-i \mu t / \hbar} $
into Eq. (\ref{spin-1GPE}) to obtain
\begin{eqnarray}
 \hat{h} \Phi_\alpha + g_n n_{\rm T} \Phi_\alpha 
+\sum_{\beta = -1}^{1} \left[- p  ( F_{z} )_{\alpha \beta} 
+ q ( F_{z}^{2} )_{\alpha \beta} 
+ g_s  {\bf S} \cdot {\bf F}_{\alpha \beta} 
\right]  \Phi_{\beta} = \mu \Phi_\alpha.
\label{stationaryGPE(f=1)}
\end{eqnarray}
The chemical potential $\mu$ is determined so as to conserve the total particle number 
$N= \int d {\bf r} n_{\rm T} ({\bf r}) $. In addition, the spin dynamics governed 
by Eq. (\ref{spin-1GPE}) conserves 
the total magnetization $M = \int d {\bf r} \langle F_{z} ({\bf r}) \rangle$, because the interactions conserve 
the total spin of two colliding atoms. 
Therefore, when we search for the ground state at a given magnetization $M$, 
we must replace $E$ with $E-\lambda M$, where $\lambda$ is a Lagrange multiplier.
$p$ is then replaced with $\tilde{p}=p+\lambda$, which is determined as a function of $M$.

\subsubsection{Ground state properties}
It is easy to identify the ground state of a uniform system when the Zeeman terms 
are negligible ($p=q=0$), because we can set the kinetic energy and $V_{\rm ex}({\bf r})$ to zero.
From the energy functional (\ref{spin1energyfunc}), the ground state is 
ferromagnetic ($|{\bf S}|=n_{\rm T}$) for $g_s < 0$, and polar or antiferromagnetic 
($|{\bf S}|=0$) for $g_s > 0$. Introducing the normalized spinor ${\bm \zeta}$ as 
${\bf \Psi}({\bf r}) = \sqrt{n_{\rm T}({\bf r})} {\bm \zeta} ({\bf r}) = \sqrt{n_{\rm T}({\bf r})} 
( \zeta_{+} ({\bf r}), \zeta_{0} ({\bf r}), \zeta_{-} ({\bf r}))^{T}$, where $|{\bm \zeta}|$ = 1, one can 
write the order parameter of the ferromagnetic phase as $\sqrt{n_{\rm T}}(1,0,0)^T$, 
while that of the polar (antiferromagnetic) phase is given by $\sqrt{n}(0,1,0)^T$ 
($\sqrt{n_{\rm T}/2}(1,0,1)^T$) \cite{Ohmi98,Ho98}. Here, the polar and antiferromagnetic state are 
related to the $\pi/2$-rotation of the spin vector about the $y$-axis, 
which is degenerate in the absence of a magnetic field. Because of the symmetry of the
Hamiltonian, all states obtained by the gauge transformation $e^{i \chi}$ and the spin rotation 
$U(\alpha,\beta,\gamma) = e^{-i F_{z} \alpha} e^{-i F_{y} \beta} e^{-i F_{z} \gamma}$ 
of these ground states are degenerate, where $(\alpha, \beta, \gamma)$ are the Euler angles. 
Thus, the general forms of the order parameter can be written as 
\begin{eqnarray}
{\bf \Psi}^{\rm ferro}=\sqrt{n_{\rm T}} e^{i (\chi- \gamma)} \left(
\begin{array}{c}
  e^{-i \alpha} \cos^{2} \frac{\beta}{2}  \\
 \sqrt{2} \cos \frac{\beta}{2} \sin \frac{\beta}{2}  \\
   e^{i \alpha} \sin^{2} \frac{\beta}{2}
\end{array}
\right) ,  \hspace{3mm}
{\bf \Psi}^{\rm polar}=\sqrt{n_{\rm T}} e^{i \chi} \left(
\begin{array}{c}
- \frac{e^{-i \alpha} \sin \beta}{\sqrt{2}}     \\
  \cos \beta \\
  \frac{e^{i \alpha} \sin \beta }{\sqrt{2}} 
\end{array}
\right) \label{generalop}
\end{eqnarray}
for the ferromagnetic and polar states, respectively. The symmetry group of the ferromagnetic 
state is $SO(3)$, since the distinct configurations of ${\bm \zeta}$ are given by the full 
range of Euler angles. Note that the phase change by $\chi$ and the spin rotation 
by $-\gamma$ play an equivalent role, which represents a ``spin-gauge" symmetry \cite{Ho96}. 
That of the polar state is $U(1) \times S^{2} / Z_{2}$, where $U(1)$ 
denotes the phase angle $\chi$ and $S^{2}$ is a surface of a unit sphere denoting 
all orientations $(\alpha,\beta)$ of the spin quantization axis. Moreover, 
a discrete symmetry $Z_{2}$ indicates that ${\bf \Psi}^{\rm polar}$ 
is invariant under $\pi$ gauge transformation combined with $\pi$ spin rotation around 
an axis perpendicular to the quantization axis 
$\hat{\bf n} = (\cos \alpha \sin \beta, \sin \alpha \sin \beta, \cos \beta)$, 
i.e., $\chi \rightarrow \chi + \pi$ and $\hat{\bf n} \rightarrow -\hat{\bf n}$ \cite{Zhou01}.

\begin{figure}[htbp] 
\begin{center} 
\includegraphics[angle=0,width=13cm]{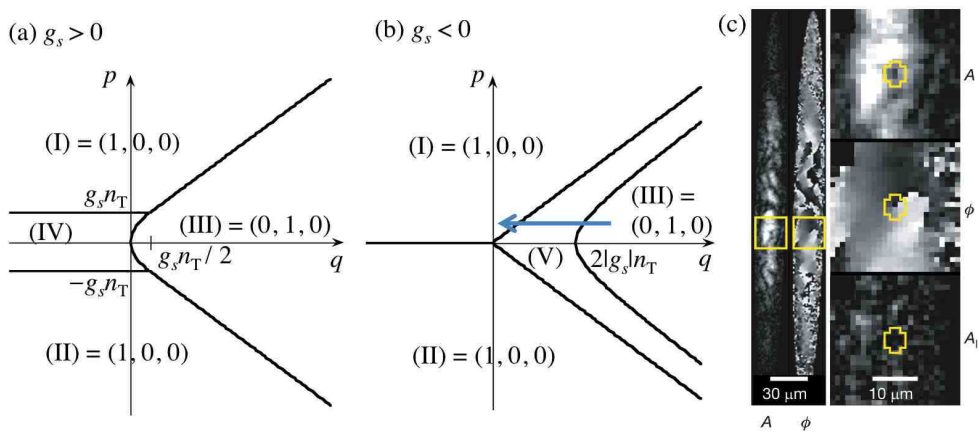} 
\end{center} 
\caption{$p$-$q$ phase diagram of a uniform spin-1 BEC with (a) $g_{s}>0$ 
and (b) $g_{s}<0$. For (a) $g_{s}>0$, the ferromagnetic phase and the polar phase are separated by 
a clear boundary $|p| = q + g_{s}$, which indicates that the $\Psi_{0}$ and $\Psi_{\pm 1}$ components are 
immiscible. For small bias fields, with $q<g_{s}n_{\rm T}/2$ and $|p|<g_{s} n_{\rm T}$, 
there is a region (IV), where the $\Psi_{\pm 1}$ components are mixed and the ratio of the populations 
is given by $|\Psi_{+}|^{2}/|\Psi_{-}^{2}|=(2g_{s} + p)/(2g_{s} - p)$. 
The boundary between (III) and (IV) lies at $|p| = 2 \sqrt{cq}$. 
For (b) $g_{s}<0$, The polar state occurs for $|p| \leq \sqrt{q (q-4 |g_{s}|)}$, 
and the ferromagnetic state occurs for $|p| > q$. 
Here, all three components are generally miscible, because the polar state 
is skirted by a region of a broken-axisymmetry phase (V) 
where all components have non-zero population. 
In (c), the in-situ observation of spontaneously generated spin vortices 
in a ferromagnetic spinor BEC \cite{Sadler} is shown, 
where a magnetic field was quenched along the arrow in (b). 
The picture represents the profile of the magnitude of the 
transverse magnetization $A$, its orientation $\phi$, and
the magnitude of the longitudinal magnetization $A_{l}$ 
[Sadler {\it et al}.: Nature (London) \textbf{443} (2006) 312, 
reproduced with permission. Copyright 2006 by the Nature Publishing Group].} 
\label{spin1phasediagram} 
\end{figure}
In the presence of an external magnetic field, the ground state 
becomes more complicated due linear and quadratic Zeeman effects.
The ground-state phase diagram in a parameter space of $(p,q)$ was studied in Ref. \cite{Stenger}.
Since $g_n n_{\rm T}$ can be absorbed into $\mu$ by defining $\tilde{\mu}\equiv\mu-g_n n_{\rm T}$, 
the ground state corresponds to the spinor, which minimizes 
the spin-dependent portion of the energy density $E_{s} / V = g_{s} |{\bf S}|^{2} /2 - p \langle  F_{z} \rangle 
+ q \langle F_{z}^{2} \rangle$. 
A phase diagram of the ground state is shown in Fig. \ref{spin1phasediagram} (a) and (b); 
the details of which are described in Ref. \cite{Ueda10,Murata07}.
Briefly, there are five distinct states (I) - (V) depending on the values of $(g_{s},p,q)$: 
(I) and (II) are ferromagnetic phases, whereas (III) is polar. In (IV), the longitudinal 
magnetization depends on $p$ as $S_z=p/g_s$. In (V), the magnetization is 
tilted from the $z$-axis and its azimuthal angle is spontaneously chosen. This is 
referred to as the broken-axisymmetry phase \cite{Murata07}. 
The rotational symmetry of the order parameter about the magnetic-field axis 
is broken in states (III) and (V). 

\subsection{Topological vortex formation in spinor BECs}
The internal degrees of freedom in BECs provide a peculiar way to generate 
vortices, which is called ``topological phase imprinting" \cite{Nakahara00,Isoshima00,Nakahara08}. 
The MIT group \cite{Shin,Leanhardt,Leanhardt2} and the Kyoto university group 
\cite{Isoshima07,Kumakura06} used this method to create a vortex in a trapped BEC. 
In a magnetic field, atoms can be trapped by the Zeeman interaction 
of the electron spin with an inhomogeneous magnetic field. Atoms with 
electron spins parallel to the magnetic field are attracted 
to the minimum of the magnetic field (weak-field seeking state), 
while those with an electron spin antiparallel are repelled (strong-field seeking state). 
In the experiment at MIT, $^{23}$Na BECs were prepared in weak-field seeking states, 
a $|F,m_F\rangle = |1,-1\rangle$ or $|2,+2\rangle$ hyperfine state, 
and confined in a Ioffe--Pritchard magnetic trap, described by 
${\bf B} = B_{\perp}' (x \hat{\bf x} - y \hat{\bf y}) + B_{z}(t) \hat{\bf z}$. 
An essential idea in topological phase imprinting is azimuthally dependent 
adiabatic inversion of the bias field $B_{z}(t)$. 

To interpret the mechanism, let us consider a BEC with $F=1$. 
The basis vectors in the representation $(\Psi_{1}, \Psi_{0}, \Psi_{-1})$ are $\{|\pm \rangle, |0 \rangle\}$. 
We introduce another set of basis vectors $|x \rangle, |y \rangle$ and $|z \rangle$, 
which are defined by $F_x |x \rangle= F_y |y \rangle=F_z |z \rangle=0$; 
these vectors are related to the previous vectors 
as $|\pm1\rangle = \mp (1/\sqrt{2})\left( |x \rangle \pm i|y \rangle\right)$ 
and $|0 \rangle=|z \rangle$. 
When the $z$-axis is taken to be parallel to the uniform magnetic field, 
the order parameter of the weak-field seeking state takes the form 
$\Psi_{-1} = \psi$ and $\Psi_{0}=\Psi_{1}=0$, or in vectorial form 
as ${\bf \Psi} =(\psi/\sqrt{2}) \left(\hat{\bf x} -i \hat{\bf y} \right)$. 
When the magnetic field points in the direction 
${\bf B} = B (\cos \alpha \sin \beta, \sin \alpha \sin \beta, \cos \beta)$, 
a rotational transformation with respect to the Euler angle $(\alpha, \beta, \gamma)$ 
gives ${\bf \Psi} = (\psi/\sqrt{2}) e^{-i \gamma} (\hat{\bf m} + i \hat{\bf n})$, 
where $\hat{\bf m} = (\cos \alpha \cos \beta, \sin \alpha \cos \beta, -\sin \beta)$ 
and $\hat{\bf n}= (\sin \alpha, - \cos \alpha, 0)$. 
The unit vector $\hat{\bf l} = \hat{\bf m} \times \hat{\bf n} 
= (-\cos \alpha \sin \beta, - \sin \alpha \sin \beta, - \cos \beta)$ corresponds to the direction 
of the spin ${\bf S}$, which is antiparallel to ${\bf B}$. 
The three real vectors $\{\hat{\bf l}, \hat{\bf m}, \hat{\bf n}\}$ form a triad, 
analogous to the order parameter of the orbital part of superfluid $^{3}$He. 
The same amplitudes in the basis $\{|0 \rangle, |\pm \rangle\}$ are 
$\Psi_{1} = (\psi/2) (1 - \cos \beta) e^{-i \alpha - i \gamma} $, 
$\Psi_0 = - (\psi/\sqrt{2}) \sin \beta e^{- i \gamma}$, 
and $\Psi_{-1} = (\psi/2) (1 + \cos \beta) e^{i \alpha - i \gamma}$.

When the field $B_z$ is strong compared to the transverse quadrupole field 
$B_{\perp}' r$, the trapped condensate initially has an order parameter 
${\bf \Psi} = (\psi/\sqrt{2})(\hat{\bf x} - i \hat{\bf y})$ without vorticity, 
i.e., $\Psi_{-1} = \psi$ and $\Psi_{1} = \Psi_{0} = 0$. 
This configuration corresponds to $\beta=0$ and $\gamma = \alpha = \theta$, 
where $\theta$ is the azimuthal angle.  The adiabatic condition with respect 
to the change of $B_z$ is required for atoms to remain in the weak-field seeking 
state, so that $\hat{\bf l}$ is always antiparallel to ${\bf B}$. 
Thus, when $B_{z}$ is gradually changed in the opposite $-\hat{\bf z}$ direction, 
the $\hat{\bf l}$-vector points $+\hat{\bf z}$ so that $\beta= \pi$, 
but $\gamma = \alpha = \theta$ is maintained by the effect of $B_{\perp}'$. 
Substituting these angles into $\Psi_{\pm 1}$ and $\Psi_{0}$, we obtain $\Psi_{-1} = \Psi_0 = 0$ 
and $\Psi_{1}  =  \psi e^{-2i \theta}$, which corresponds to a vortex with a winding number $q=2$. 
This result can be reinterpreted in terms of Berry's phase \cite{Leanhardt,Ogawa02b}. 
In general, when the hyperfine spin is $F$, we obtain a vortex with a winding number 
$2F$ since $\Psi_{-F}$ and $\Psi_{F}$ acquire phases $F(\alpha-\gamma)$ 
and $F(-\alpha-\gamma)$, respectively \cite{Shin,Kumakura06}.

When the bias field is turned off at $B_{z} = 0$, a spin texture 
known as cross disgyration appears in the Ioffe--Pritchard trap. 
Here, the angle $\beta$ increases from $0$ to $\pi/2$, whereas $\gamma = \alpha = \theta$. 
The $\hat{\bf l}$-vector aligns with a 
hyperbolic distribution around the singularity at the center by following the direction of 
a local magnetic field of the Ioffe--Pritchard trap. 
This spin texture has been observed as a coreless vortex composed of 
three-component order parameters $\Psi_{\pm 1, 0}$ \cite{Leanhardt2}, 
and its theoretical account will be given below.

If one inverts the bias field back to the initial state after the vortex creation, 
the induced vortex unwinds itself, and 
the resulting state is non-rotating. On the other hand, if higher order 
magnetic fields are employed, it is possible to gain more 
than $2 F \hbar$ of angular momentum per particle through a single inversion of the bias field. 
Hence, by using, e.g., a hexapole field in the radial direction and 
inverting the bias field $B_{z}$, a 4$F$-quantum vortex is produced. 
Switching to a quadrupole field and inverting the bias field back to the original value, 
one loses only $2 F \hbar$ of angular momentum and the final state has a total 
angular momentum of $2 F \hbar$. This process can be repeated, and each cycle 
increases the angular momentum of the system by $2 F \hbar$ per particle. 
The idea of cyclicly pumping vortices into a condensate was theoretically proposed 
in Ref. \cite{Mottonen07,Xu08}. 

\subsection{A single vortex (a few vortices) in spin-1 BECs} 
Next, we consider the vortex states in slowly rotating spin-1 BECs.  
The characteristics of the vortex states are strongly dependent 
on the sign of $g_{s}$, i.e., ferromagnetic ($g_{s}>0$) or antiferromagnetic ($g_{s}<0$). 
This can be seen by their superfluid velocity ${\bf v} = (\hbar/m) \zeta^{\dagger} \nabla \zeta$; 
from Eq. (\ref{generalop}), they are 
\begin{equation}
{\bf v}^{\rm ferro} = \frac{\hbar}{m} \left[ \nabla(\chi - \gamma) - \cos \beta \nabla \alpha \right], \hspace{5mm}
{\bf v}^{\rm polar} = \frac{\hbar}{m} \nabla \chi, 
\label{spin1velocity}
\end{equation}
where the angles are assumed to be spatially varying functions. 
For the ferromagnetic case, the supercurrent can be induced by not only 
a spatial variation of the phase $\chi$, but also by that of the
spin, similar to the pseudospin interpretation of two-component BECs 
described in Sec. \ref{SingletwoBEC}. 
On the other hand, the polar case resembles that of scalar BECs. However, it allows 
half of the unit of quantization due to order parameter symmetry, as shown below. 

The structure and stability of the axisymmetric vortex states for various $\Omega$ and 
total magnetizations $M$ were thoroughly investigated in Ref. \cite{Isoshima01,Isoshima02}. 
The equilibrium state can be obtained by minimizing the energy 
in the rotating frame: $E - \lambda M- \Omega \langle L_{z} \rangle$.
The axisymmetric vortex configuration is classified by the combination 
of the winding number $q_{j}$ of the condensate wave function 
$\Psi_{j}=\sqrt{n_{j}(r)} e^{i (\chi_{j} + q_{j} \theta)}$ with the 
cylindrical coordinate ($r,\theta,z$), 
where the homogeneity of the wave functions 
along the $z$-axis is assumed, and $\chi_{j}$ is the overall phase of the 
$j$-th component. The phases are constrained to minimize the spin-exchange 
interaction energy in Eq. (\ref{spin1energyfunc}):
\begin{eqnarray}
E_{s} = \frac{g_{s}}{2} |{\bf S}|^{2} 
= \frac{g_{s}}{2} \left\{ (n_{1} - n_{-1})^{2} +
2 n_{0} \left[ n_{1} + n_{-1} 
+ 2 \sqrt{n_{1} n_{-1}} \cos (\bar{\chi} + \bar{q} \theta) \right] \right\}, 
\label{spinexint}
\end{eqnarray}
where $\bar{\chi} = \chi_{1} + \chi_{-1} - 2 \chi_{0}$ and 
$\bar{q} = q_{1} + q_{-1} - 2 q_{0}$. 
To minimize $E_{s}$, $\chi_{j}$ and $q_{j}$ should satisfy 
$2 \chi_{0} = \chi_{1} + \chi_{-1} + n' \pi$ and 
$2 q_{0} = q_{1} + q_{-1}$, 
where $n'$ is an odd (even) integer for antiferromagnetic 
(ferromagnetic) interaction. The global phase $\bar{\chi}$ has no effect on the vortex 
structure, and was therefore set to zero. 
The possible combination of $q_{j}$ satisfying the above relation gives this system 
a characteristic vortex structure. If the values of $q_{j}$ are restricted 
to $|q_{j}| \leq 1$, we have $ (q_{1},q_{0},q_{-1})=\pm (1,1,1)$, 
$ \pm (1,0,-1)$ and $\pm (1,1/2,0)$, where cases with negative sign 
are omitted in the following. For $(1,1/2,0)$, the value $1/2$ is not allowed in 
an axisymmetric system, so that the $\Psi_{0}$-component vanishes; 
we denote this state as $(1, \times ,0)$. 

Unfortunately, the assumption of axisymmetry restricts the possible ground states, 
limiting the stable vortex configurations. The vortex states without assuming axisymmetry were considered by some 
authors \cite{Yip99,Mizushima02a,Mizushima02b,Martikainen02,Bulgakov03,Mizushima04b}, 
who constructed phase diagrams of vortex states for various parameter spaces in various situations.
The details for each vortex state are summarized below.

\subsubsection{$(1, 1, 1)$ vortex}
In this case, all components have singly-quantized vortices at the center, and the 
densities of the three components are fully overlapped. 
As a result, the total density is equivalent that of a conventional BEC 
with an empty vortex core. 

With a fixed $M$, this axisymmetric vortex state cannot be stabilized in any parameter region for 
both $g_{s} < 0$ and $g_{s} > 0$ \cite{Isoshima02}. However, as in the (1,1) 
vortex state in two-component BECs shown in Fig. \ref{twocomposingle}, by displacing the vortex cores of each 
component from the center of the trap, the (1,1,1) vortex can be stabilized in a 
nonaxisymmetric coreless configuration \cite{Mizushima02b}. 
For $g_{s} > 0$, the vortex cores are displaced such 
that the condensates have two singularities through an overlap 
of the $\Psi_{1}$ and $\Psi_{-1}$ components, or they have three singularities 
that form a triangular configuration. For $g_{s} < 0$, two singularities of 
$\Psi_{1}$ and $\Psi_{-1}$ are displaced from the center, while $\Psi_{0}$ 
with a singularity at the center prevents phase separation. 
For all of these cases, the breaking of axisymmetry leads to a 
smooth variation of the total density by decreasing the overlap 
area of $\Psi_{\pm 1}$ and $\Psi_{0}$ for $g_{s} > 0$, and 
$\Psi_{1}$ and $\Psi_{-1}$ for $g_{s} < 0$. 

\subsubsection{$(1, 0, -1)$ vortex}
In this configuration, the non-rotating $\Psi_{0}$ component occupies 
the central region of the vortex core, which is made up of the $\Psi_{1}$ and $\Psi_{-1}$ 
components with opposite circulations.  
This vortex state is referred to as a ``polar-core" vortex. 
Since the condensate at the center of the trap consists only of the polar state, 
the spin texture has a singularity, although it can vary continuously around the vortex core. 
This vortex state is thermodynamically stable in the high magnetization region, which 
is independent of the strength and sign of the spin exchange interaction \cite{Mizushima02b}. 
Also, it becomes stable in a non-rotating harmonic trap with an applied Ioffe-Pritchard 
magnetic field \cite{Bulgakov03}. 

Interestingly, this vortex state can be generated by the intrinsic dynamical instability 
of a ferromagnetic spinor BEC. Saito {\it et al.} demonstrated that 
an initially prepared $\Psi_{0}$ without vorticity can generate transverse 
magnetization due to ferromagnetic interaction \cite{Saito06}. Since spin conservation prohibits 
the appearance of longitudinal magnetization, the dynamical instability results in a polar-core vortex by 
spontaneously breaking the chiral symmetry of the initial state, where the
$\Psi_{\pm 1}$ component can take either $q=\pm 1$ circulation. 
Since oppositely rotating $\Psi_{\pm 1}$ components have the same amplitude due to $M=0$, the vortex carries 
no net mass current but a spin current with one quantum of circulation.
This spontaneous formation of such polar-core ``spin" vortices was demonstrated experimentally 
by the Berkeley group, using an instantaneous quench from the polar phase 
to the ferromagnetic phase \cite{Sadler} [See Fig. \ref{spin1phasediagram}] . 

\subsubsection{$(1, \times, 0)$ vortex}
This vortex state consists of a non-rotating $\Psi_{-1}$ component at the center filling the vortex core of the 
$\Psi_{1}$ component. This is similar to the (1,0) vortex of two-component BECs, but 
its interpretation is quite different. 
In the (1,$\times$,0) vortex, the spin component is suddenly reversed near the
vortex core of the $\Psi_{1}$ component because the absence of the $\Psi_{0}$ component 
means that $S_{x} = S_{y} = 0$, which cannot yield a continuous variation of the spin texture. 
This vortex state is thermodynamically stable only for the polar case 
($g_{s}>0$) \cite{Isoshima02,Mizushima02b}.

This vortex state is called the half-quantum vortex state (``Alice state") \cite{Isoshima02,Leonhardt00}. 
To see this, let us consider ${\bf \Psi}^{\rm polar}$ and a loop that encircles a 
vortex with a fixed radius. Each point on the loop can be specified by its azimuthal 
angle $\theta$. The single-valuedness of the order parameter ${\bf \Psi}^{\rm polar}$ in Eq. (\ref{generalop}) 
is met if we take, for example, $-\alpha=\chi=q \theta /2$ and $\beta=\pi/2$.
The order parameter at $r \to \infty$ is given by
${\bf \Psi}^{\rm polar} = \sqrt{n_{\rm T}/2} (-e^{i q \theta}, 0,  1)$. Thus, from Eq. (\ref{spin1velocity}), 
the circulation becomes $\oint {\bf v}^{\rm polar} \cdot d{\bm \ell} = (h/2m) q$, 
quantized in units of $h/(2m)$ rather than the usual $h/m$. 
The underlying physics for this half quantum number
are based on the discrete $Z_2$ symmetry of ${\bf \Psi}^{\rm polar}$, which is 
invariant under $\pi$ gauge transformation ($\chi \to \chi+\pi$) 
combined with a $\pi$ spin rotation around an axis perpendicular to 
$\hat{\bf n}=(\cos \alpha \sin \beta, \sin \alpha \sin \beta, \cos \beta)$: 
$\hat{\bf n} \to -\hat{\bf n}$. 
Thus, the polar phase of a spin-1 BEC can host a half-quantum vortex \cite{Zhou01,Leonhardt00}.
The dynamical creation of half-quantum vortices under an external rotating 
potential was demonstrated in numerical simulations \cite{Ji08,Chiba08}. 

The half-quantum vortex state is unstable for ferromagnetic spinor BECs. 
The axisymmetry of the transverse magnetization is spontaneously broken 
to form three-fold domains \cite{Hoshi08}. 
This originates from the topological structure of the half-quantum vortex and 
spin conservation. 

\subsubsection{(0, 1, 2) vortex} 
In ferromagnetic interactions, a more stable vortex configuration occurs 
when the winding number $q_{j}$ exceeds unity. Mizushima {\it et al.} 
found that ferromagnetic interaction supports the thermodynamic stability of
the $(0,1,2)$ axisymmetric vortex state over wide parameter regions \cite{Mizushima02a,Mizushima02b}. 
The central region of the harmonic trap is occupied by a non-rotating $\Psi_{1}$ component. 
The $\Psi_{0}$ component with $q_{0}=1$ is pushed outward, 
while the $\Psi_{-1}$ component with $q_{-1}=2$ occupies the 
outermost region. This configuration is favorable for the ferromagnetic case, because 
it is more favorable than spatial phase separation of $\Psi_{1}$ and $\Psi_{-1}$; the presence of vortices with 
different $q_{j}$ effectively causes phase separation in the radial direction. 
Only a (0,1,2) vortex can have a nonsingular continuous spin texture under slow rotation. 
Mizushima {\it et al}., performed extensive numerical studies to determine the $\Omega$-$M$ phase 
diagram of the vortex states, including the $(2,1,0)$ state referred to 
as the mixed-twist texture \cite{Mizushima04b}.  

When the wavefunction ${\bf \Psi}^{\rm ferro}$ is parameterized by the spatially-dependent 
bending angle $\beta(r)$ and $\chi=0$ and $\alpha = -\gamma = \theta$ in Eq. (\ref{generalop}), 
we have ${\bf \Psi}^{\rm ferro} = \sqrt{n_{\rm T} (r)}$ ($\cos^{2} \frac{\beta(r)}{2}$,
$e^{i \theta} \sqrt{2} \sin \frac{\beta(r)}{2} \cos \frac{\beta(r)}{2}$,
$e^{2 i \theta} \sin^{2} \frac{\beta(r)}{2}$ ),
where $\beta(r)$ is an increasing function of $r$ from $\beta(0) = 0$ 
and $0<\beta(r)<\pi$. 
The spin direction is related to the $\hat{\bf l}$-vector, and is given 
as $\hat{\bf l}(r)=\hat{\bf z} \cos \beta(r) 
+ \sin \beta(r) (\cos \theta \hat{\bf x} + \sin \theta \hat{\bf y})$, 
where $\beta$ varies from $\beta (0) =0$ to $\beta (R)=\pi/2$ ($=\pi$) for 
a Mermin-Ho (Anderson-Toulouse) vortex.
As in two-component BECs [see Sec. \ref{SingletwoBEC}], the $(0,1,2)$ vortex 
can be interpreted as a vortex with an intermediate 
boundary condition ($\pi/2 < \beta(R) < \pi$); 
we can control the value of $\beta(R)$ from the Mermin-Ho condition 
to the Anderson-Toulouse condition by merely changing the total magnetization $M$ \cite{Mizushima02a}. 
In another numerical study, Martikainen {\it et al.} analyzed the coreless vortex state 
as a function of rotational frequency without fixing the total magnetization \cite{Martikainen02}, 
finding that $\beta(R)$ increases with increasing $\Omega$ and that the upper value of $\beta(R)$ is $3\pi/4$, 
above which additional vortices nucleate. This implies that an Anderson-Toulouse vortex 
can never be the ground state of the system. 

Because this coreless vortex contains a doubly-charged vortex in one of the components, 
it is interesting to study whether the coreless vortex state inherits the 
dynamical instability of doubly-charged vortices in scalar condensates, as 
shown in Sec. \ref{Dynamicssinglevortex}. 
Pietil\"{a} {\it et al.} studied the low-lying excitations of the $(0,1,2)$ state for $g_{s}<0$ 
in the presence of an Ioffe-Pritchard trap based on the BdG approach. The phase diagram of 
the dynamical stability was explored 
for the bias field $B_z$ and the transverse quadrupole field $B_{\perp}'$ \cite{Pietila08}. 
Takahashi {\it et al.} considered the same problem for both $g_{s}<0$ and $g_{s}>0$ 
without the Ioffe-Pritchard trap, finding that the dynamical instability is suppressed (enhanced) 
for $g_{s}<0$ ($g_{s}>0$) \cite{Takahashi09}. The unstable dynamics are associated with 
the vortex-splitting of the $q=2$ vortex, and phase-separation in the $g_{s} > 0$ case. 

\subsection{Vortex lattices in F=1 spinor BECs}
Vortex lattices in rapidly rotating spin-1 BECs can possess richer vortex phases 
than those in two-component BECs, but their classification is very complicated. 
Kita {\it et al.} studied vortex-lattice structures of antiferromagnetic spin-1 BECs 
using the phenomenological Ginzburg-Landau equation \cite{Kita02}, 
which is formally similar to the mean-field quantum Hall approach. 
This study revealed that the conventional Abrikosov lattice with hard vortex cores 
is unstable, and the vortex cores shift their locations. 
The system has many metastable configurations of vortices, depending sensitively 
on the ratio $g_{s}/g_{n}$. The vortices are characterized by the distribution of 
magnetization and the difference in the number of circulation of quanta per unit cell, 
all of which makes the characterization of the vortex lattice structure complicated. 

Further discussion on the phase diagram of the ground state in rotating spin-1 
bosons was reported in Reijnders {\it et al.} \cite{Reijnders04b}. 
This paper also describes a study of vortex-lattice structures with the 
mean-field quantum Hall approach over wide parameter regions 
and the exact diagonalization study of the quantum liquid phase. 
A similar analysis was done by Mueller for slowly rotating spin-1 BECs \cite{Mueller04}.
In addition, Mizushima {\it et al.} performed numerical simulations of the 
GP equations (\ref{stationaryGPE(f=1)}) for a ferromagnetic spinor BEC 
in a fast-rotating regime and also in an external magnetic field \cite{Mizushima04b}. 
For $M=0$, the equilibrium state is a square lattice constructed from two sublattices of the 
Mermin-Ho vortex and the mixed-twisted vortex, and the local spin on the two 
sublattice sites are locked in alternate directions $\hat{\bf z}$ and $-\hat{\bf z}$. 
For $M \neq 0$, the composite lattice of a coreless vortex and a polar-core vortex was 
found. 

\subsection{Quenched spinor BEC: Kibble-Zurek mechanism}
Topological defect formation via the Kibble-Zurek mechanism can occur 
in a ferromagnetic phase of a spinor BEC. 
In an experiment by the Berkeley group \cite{Sadler}, $^{87}$Rb atoms 
(with $g_{s}<0$) in the $m=0$ state were prepared 
by applying a magnetic field along the $z$-axis. This initial state corresponds to the point 
in region (III) of the phase diagram in Fig \ref{spin1phasediagram}(b). 
The strength of the magnetic field was then suddenly decreased toward region (I) 
along a constant-$p$ line of the phase diagram. 
The $m=0$ state was no longer the ground state, and a spontaneous transverse 
magnetization emerged due to the conservation of longitudinal magnetization. 
Since this phase transition was triggered by a sudden change in a magnetic field, 
this can be regarded as a {\it quantum} quench at $T=0$. 
The magnetization dynamics were observed by a spin-sensitive {\it in situ} measurement to 
form complicated ferromagnetic domains. Remarkably, polar-core spin vortices 
were identified in some snapshots of the spin distribution, as shown in 
Fig. \ref{spin1phasediagram}(c), consistent with the prediction \cite{Saito06}. 

Theoretical analysis of this experiment was done by several 
authors \cite{Saito07a,Lamacraft07,Uhlmann07,Damski07,Saito07b}, in relation to 
the Kibble-Zurek mechanism. The numerical simulations revealed that the creation 
of the spin vortices was caused by instability of the initially generated solitons, 
whose pattern and characteristic length scales were strongly dependent on not only the 
quench time but also the properties of the initial noise. 

\subsection{Vortices in F=2 spinor condensates}
\begin{figure}[htbp] 
\begin{center} 
\includegraphics[angle=0,width=13cm]{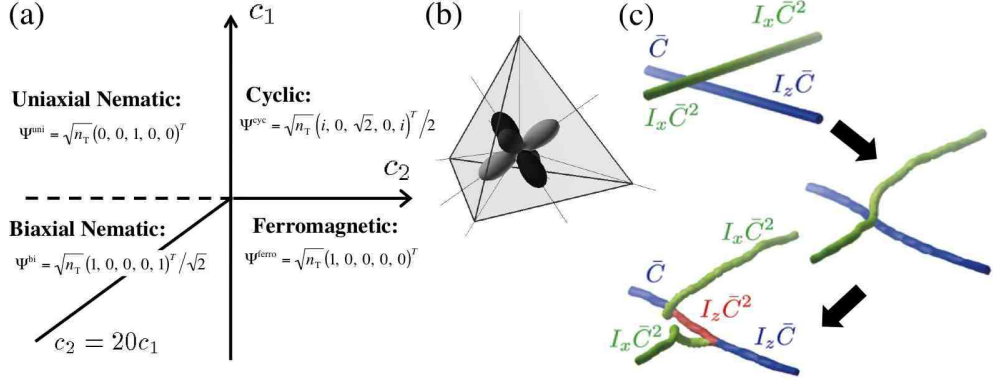} 
\end{center} 
\caption{(a) Mean-field phase diagram of spin-2 BECs. (b) Schematic illustration 
showing that the symmetry axes of the cyclic order parameter constitute a tetrahedron. 
(c) Numerical simulation 
of the collision dynamics of non-Abelian vortices. After the collision, two vortices 
get connected and a rung appears between them. 
The topological charges for each vortex line are denoted 
[Kobayashi {\it et al}.: Phys. Rev. Lett. \textbf{103} (2009) 115301, 
reproduced with permission. Copyright 2009 by the American Physical Society].} 
\label{spin2phase} 
\end{figure}

For the $F=2$ case, the field operator has five components 
$(\hat{\psi}_{2},\hat{\psi}_{1},\hat{\psi}_{0},\hat{\psi}_{-1},\hat{\psi}_{-2})$. 
The interaction becomes
\begin{eqnarray}
V_{F=2} = \frac{c_0}{2} \sum_{\alpha,\beta=-2}^{2} \hat{\psi}_{\alpha}^{\dagger} 
\hat{\psi}_{\beta}^{\dagger}  \hat{\psi}_{\beta} \hat{\psi}_{\alpha} 
+ \frac{1}{2} \sum_{\alpha,\alpha ', \beta, \beta '= -2}^{2} \biggl( c_{1} \hat{\psi}_{\alpha}^{\dagger} 
\hat{\psi}_{\beta}^{\dagger} {\bf F}_{\alpha \alpha '} \cdot {\bf F}_{\beta \beta '}
\hat{\psi}_{\beta '} \hat{\psi}_{\alpha '} \nonumber \\ 
+5 c_{2} \hat{\psi}_{\alpha}^{\dagger} \hat{\psi}_{\beta}^{\dagger} 
\langle 2, \alpha; 2, \alpha ' | 0, 0 \rangle \langle 0, 0 | 2, \beta; 2, \beta ' \rangle
\hat{\psi}_{\beta '} \hat{\psi}_{\alpha '} \biggr). 
\end{eqnarray}
The collision coefficients are given by $c_{0} = (4 g_{2}+ 3 g_{4})/7$, $c_{1} = - (g_{2} - g_{4})/7$, 
and $c_{2} = (g_{0} - g_{4})/5-2(g_{2} - g_{4})/7$. 
The energy functional $E = \langle \hat{H} \rangle$ for spin-2 BECs can be written as
\begin{eqnarray}
E[{\bf \Psi}] = \int d{\bf r} 
\left( \sum_{\alpha=-2}^2 \Psi_{\alpha}^\ast \hat{h} \Psi_{\alpha} 
+ \frac{c_{0}}{2} n_{\rm T}^{2} + \frac{c_{1}}{2} |{\bf S}|^2 + \frac{c_2}{2} | A_{20} |^2 \right),
\label{spin2energyfunc}
\end{eqnarray}
where $n_{\rm T}({\bf r}) = \sum_{\alpha=-2}^{2} | \Psi_\alpha({\bf r}) |^2$ is the total particle density and 
${\bf S} = (\langle F_{x} \rangle, \langle F_{y} \rangle, \langle F_{z} \rangle)$ is the spin density 
vector defined by $\langle F_{i} ({\bf r}) \rangle = \sum_{\alpha, \beta=-2}^{2} \Psi_\alpha^{\ast}({\bf r}) 
(F_{i})_{\alpha \beta} \Psi_\beta({\bf r})$ ($i = x, y, z$). 
The new feature of spin-2 BECs, not found in spin-1 BECs, is the presence of the spin-singlet 
pair amplitude $A_{20} = (2 \Psi_{2} \Psi_{-2} - 2 \Psi_{1} \Psi_{-1} + \Psi_{0}^{2})/\sqrt{5}$. 
Figure \ref{spin2phase}(a) shows a mean-field phase diagram in the absence of a magnetic field.
There are four distinct phases: ferromagnetic, uniaxial nematic, biaxial nematic, and cyclic, where 
the order parameters of the last three phases are characterized by a discrete symmetry. 
According to mean-field theory, the uniaxial nematic phase and biaxial phase are degenerate in the 
absence of a magnetic field, however, quantum fluctuations induce a quantum phase 
transition between the two phases, lifting the degeneracy \cite{Turner2007,Song2007}.
For $c_{1}>0$ and $c_{2}>0$, both ${\bf S}$ and $A_{20}$ vanish, 
which yields the cyclic phase concentration characterized by the spin-singlet ``trios" peculiar to spin-2 BECs 
\cite{Ueda02}; one representation for the cyclic phase is given by 
${\bf \Psi}^{\rm cyc} = \sqrt{n_{\rm T}} (i/2,0,1/\sqrt{2},0,i/2)^{T}$.
Although several experiments demonstrated that spinor BECs with an $F=2$ manifold 
of $^{87}$Rb atoms were characterized by biaxial-nematic spinor dynamics, \cite{Schmaljohann,Kuwamoto}, the possibility of the cyclic phase has not yet been excluded due to complications arising from quadratic Zeeman effects and hyperfine-spin-exchange relaxations \cite{Tojo2009}.

The vortex states in spin-2 BECs are strongly affected by the symmetry of the 
order parameters in each phase. For ferromagnetic phases, the properties of the 
vortex states are not so different from those of spin-1 BECs. However, the symmetry 
of the biaxial nematic phase and the cyclic phase can yield the exotic vortices 
governed by the non-Abelian group 
\cite{Semenoff07,Barnett08,Turner09,Kobayashi09,Huhtamaki09}. 
Here, we focus on the vortex 
states in the cyclic phase, and briefly summarize the salient results. 

\subsubsection{Vortices in cyclic condensates}
As in the spin-1 BEC, through the $U(1)$ gauge transformation and the $SO(3)$ spin rotation, 
it is possible to transform from one to another cyclic state, 
$\Psi^{\rm cyc \prime} = e^{i \chi} e^{-i {\bf F} \cdot \hat{\bf n} \theta} \Psi^{\rm cyc}$, 
where $\hat{\bf n}$ and $\theta$ are the unit vector of the rotational axis 
and angle of the spin rotation, respectively.
The order parameter $\Psi^{\rm cyc}$ is invariant under the following 12 
transformations: $\bf{1}$, $I_x = e^{i F_x \pi}$, $I_y = e^{i F_y \pi}$, $I_z = e^{i F_z \pi}$, 
$\bar{C} = e^{2 \pi i / 3} e^{- 2 \pi i (F_x + F_y + F_z) / 3 \sqrt{3}}$, $\bar{C}^2$, 
$I_x \bar{C}$, $I_y \bar{C}$, $I_z \bar{C}$, $I_x \bar{C}^2$, $I_y \bar{C}^2$, 
and $I_z \bar{C}^2$ \cite{Semenoff07}.
The overbar was added to emphasize that the operation includes not only a rotation 
in spin space but also a gauge transformation.
These 12 transformations form the non-Abelian tetrahedral group $T$, 
shown in Fig. \ref{spin2phase} (b). 

Topological charges of vortices can be classified by 12 generators 
$\bf{1}$, $I_x$, $I_y$, $\cdots$ of the non-Abelian group $T$. 
Vortices are also classified into four conjugacy classes: 
(I): integer vortex; $\bf{1}$, (II): 1/2 - spin vortex; $I_x$, $I_y$, and $I_z$, 
(III): 1/3 vortex; $\bar{C}$, $I_x \bar{C}$, $I_y \bar{C}$, and $I_z \bar{C}$, 
and (IV): 2/3 vortex; $\bar{C}^2$, $I_x \bar{C}^2$, $I_y \bar{C}^2$, and $I_z \bar{C}^2$.
Topological charges in the same conjugacy class transform into one 
another under the arbitrary global gauge transformation 
and the spin rotation, denoted by $\hat{S}$.
The order parameters for straight vortices along the $z$-axis 
in each conjugacy class can be written in cylindrical coordinates $(r, \theta, z)$ as
\begin{eqnarray}
\Psi  = \frac{1}{2} \sqrt{n_{\rm T}(r)} e^{i n_1 \theta} \hat{S} 
 \times \left\{ 
 \begin{array}{l}
\displaystyle \left( i f(r) e^{2 i (n_2 + 1) \theta}, 0, \sqrt{2} h(r), 0, i f(r) e^{- 2 i (n_2 + 1) \theta} \right)^{T} \\
\displaystyle \left( i f(r) e^{i (2 n_2 + 1) \theta}, 0, \sqrt{2} h(r), 0, i f(r) e^{- i (2 n_2 + 1) \theta} \right)^{T} \\
\displaystyle \frac{2}{\sqrt{3}} \left( f(r) e^{i (2 n_2 + 1) \theta}, 0, 0, \sqrt{2} g(r) e^{- i n_2 \theta}, 0 \right)^{T} \\
\displaystyle \frac{2}{\sqrt{3}} \left( f(r) e^{i (2 n_2 - 1) \theta}, 0, 0, \sqrt{2} g(r) e^{- i n_2 \theta}, 0 \right)^{T}
\end{array} \right. 
\label{eq-vortex-wave-function}
\end{eqnarray}
for (I), (II), (III), and (IV), respectively, where the vortex is located at $r = 0$.
Here, $n_1$ and $n_2$ are the integer winding numbers, and $f$, $g$, 
and $h$ are real functions that satisfy $[f^2 + h^2] / 2 = [f^2 + 2 g^2] / 3 = 1$ 
and $f(r \to \infty) = g(r \to \infty) = h(r \to \infty) = 1$.
At the vortex core, the cyclic order parameter changes to that of a different phase.
With the minimum windings ($n_1 = n_2 = 0$), we obtain the core structure 
of each conjugacy class by taking $f(r = 0) = 0$ as: 
(I)(II) $\Psi \propto \hat{S} ( 0, 0, 1, 0, 0 )^{T}$ and 
(III)(IV) $\Psi \propto \hat{S} ( 0, 0, 0, 1, 0 )^{T}$, {\it i.e.}, the core of (I) and (II) 
vortices have a finite spin-singlet pair amplitude ($A_{20} \neq 0$), 
and that of (III) and (IV) vortices have a finite magnetization ($S_{z} \neq 0$). 
The 1/3 and 2/3 vortices are a hallmark of the cyclic state with tetrahedral symmetry. 
The energetic stability of these fractional vortices in rotating potentials was 
examined in Ref. \cite{Huhtamaki09}

A salient feature of non-Abelian vortices emerges in their collision dynamics \cite{Kobayashi09}. 
When two Abelian vortices collide, there are three possibilities: 
(i) they reconnect themselves, (ii) they pass through each other, 
or (iii) they form a rung that bridges the two vortices, 
depending on the kinematic parameters and initial conditions. Usually, 
reconnection occurs due to the energetic constraint. 
However, when two non-Abelian vortices collide, only a rung can be formed; 
reconnection and passing through are topologically
forbidden because the corresponding generators do not commute with each other. 
In fact, the nonzero commutator of the two generators gives the generator 
of the rung vortex. Figure \ref{spin2phase}(c) illustrates a typical rung formation. 

In other studies, vortex molecules consisting of a bound pair of fractional vortices was 
discussed in Refs. \cite{Turner09,Huhtamaki09}.
The structure of vortex lattices in cyclic condensates was discussed in Ref. \cite{Barnett08}, 
where the transitions of the various lattice geometries were demonstrated 
as a function of applied magnetic field and temperature.

\section{Vortices in dipolar condensates}
A dipolar condensate refers to a condensate in which atoms interact 
via dipole-dipole interactions (DDIs), 
in addition to the usual s-wave contact interactions. For two particles 1 and 2 
with dipole moments along the unit vector ${\bf e}_{1}$ and ${\bf e}_{2}$, and 
relative position ${\bf r}$, the DDI has the form
\begin{equation}
V_{\rm dd}({\bf r}) =
\frac{c_{\rm dd}}{4\pi}\,\frac{{\bf e}_1 \cdot {\bf e}_2  - 3 ({\bf e}_1 \cdot \hat{\bf r}) 
({\bf e}_2 \cdot \hat{\bf r})}{r^3} .\label{ddp}
\end{equation}
The coupling constant $c_{dd}$ is given as $\mu_{0} \mu_{\rm d}^{2}$ for particles 
with a permanent magnetic dipole moment $\mu_{\rm d}$ ($\mu_{0}$ is the
magnetic permeability of vacuum) and $d^{2}/\epsilon_{0}$ for a permanent electric dipole moment $d$ 
($\epsilon_{0}$ is the permittivity of vacuum). Recently, BECs of chromium atoms 
have been created \cite{Griesmaier05}, in which the atoms exhibited a larger 
magnetic-dipole moment ($\mu_{\rm d} = 6 \mu_{\rm B}$) than typical alkali atoms 
($\mu_{\rm d} \simeq \mu_{\rm B}$). This enables the study of the effects 
of {\it anisotropic long-range} interactions in BECs. 
The rapid progress in studies of dipolar BECs can be seen in the review 
article of Ref. \cite{Lahayareview09}.

In particular, when the dipole moments are polarized under an external field along the $z$-axis, 
the interaction potential between two magnetic dipoles with ${\bf e}_{z}$ 
separated by ${\bf r}$ is given by $V_{\rm dd}({\bf r})
=(\mu_{0} \mu_{d}^{2}/4 \pi) ( 1 - 3 \cos^{2} \theta)/r^{3} $,
where $\hat{\bf e}_{z} \cdot \hat{\bf r} = \cos \theta$. Thus, the system is described by 
the scalar condensate wave function, and the DDIs contribute 
to the GP equation as a nonlocal mean-field potential 
\begin{equation}
i \hbar \frac{\partial \Psi}{ \partial t} = \left[ -\frac{\hbar^2 \nabla^2 }{ 2 m } + V_{\rm ex} 
+ g |\Psi|^{2} + \int d {\bf r}' V_{\rm dd}({\bf r}'-{\bf r}) |\Psi({\bf r}')|^{2} \right] \Psi. \label{DDIGP}
\end{equation}
Since the contact interaction $g=4\pi\hbar^2a/m$ can be tuned to zero 
by the Feshbach resonance technique, 
we can obtain novel quantum {\it ferrofluids} dominated by the dipole--dipole interaction \cite{Lahaye}. 

Before studying the vortex states, it is useful to examine the properties of non-rotating dipolar BECs. 
The principal effect of $V_{\rm dd}$ on the equilibrium shape of a trapped condensate 
is to cause distortion of its aspect ratio \cite{Santos00,Yi01}. Since the dipoles are aligned along the $z$-axis, 
the attractive interaction of dipoles becomes significant in a cigar-shaped trap, 
and the atomic cloud is elongated along the $z$-axis. 
Although this may appear counterintuitive, it can be understood by considering 
the energy density of the DDI. It has the anisotropic form of a saddle, with negative 
curvature along the direction of the dipoles \cite{Stuhler05}.
Because of attractive interactions, the condensate will eventually collapse, 
once the particle number exceeds some critical value $N_{c}$, similar to condensates 
with attractive contact interactions \cite{Donley}. 
For a pancake trap, the condensate is elongated similarly along the $z$-axis, but 
stable against the collapse because the DDI is mostly repulsive. 
The instability condition also depends on the strength of the contact interaction. 
These collapse dynamics were demonstrated by tuning the s-wave scattering length into the unstable regime, 
showing an anisotropic explosion characterized by $d$-wave symmetry \cite{Lahaye08}.

Another interesting feature of a dipolar BEC is that the Bogoliubov excitation 
spectrum exhibits a ``roton-maxon" behavior, caused by the attractive nature 
of the DDI. Santos {\it et al.}, found that, for a quasi-2D system, harmonically confined 
in the $z$-direction and free in the $x$ and $y$ directions, the excitation energy 
has a local maximum and minimum dependent upon the in-plane momentum $q$ \cite{Santos03}. 
Although this behavior is reminiscent of that found in the excitation spectrum 
of liquid helium II, its physical origin is somewhat different. 
This can be intuitively understood as follows. 
When the in-plane momenta $q$ are much smaller than the inverse size $L$ of the condensate 
in the $z$ direction, excitations have a 2D character. Because the dipoles are 
oriented along the $z$-axis, the interaction is effectively repulsive and 
the in-plane excitations become phonons. For $q \ll 1/L$, excitations begin 
to have a 3D character and the interparticle repulsion is reduced 
due to the attractive force in the $z$ direction. 
This decreases the excitation energy with an increase in $q$. 
The excitation energy reaches a minimum (roton) 
and then begins to increase with an further increase in $q$, 
eventually leading to a quadratic behavior $\propto q$. 

As the dipole interaction becomes stronger, 
the energy at the roton minimum decreases, and then reaches zero, 
representing an onset of instability. This ``roton instability" 
is probably towards the formation of a density wave. 
Ronen {\it et al.} numerically studied the equilibrium solutions for 
a pancake geometry, finding that the condensate always has unstable criterion 
of the dipolar interaction strength, even for the pancake-trap limit \cite{Ronen07}. 
They observed that, near the instability threshold in a certain range of the trap aspect ratio, 
the equilibrium density possesses ``biconcave" shape, where the density 
modulates radially and its local minimum 
appears at the center. This density modulation is associated with the roton instability. 
Note that this physical origin of the instability is different from that discussed 
in the dipolar collapse experiment; 
the roton instability exists even for a 2D pancake geometry. 
A more detailed discussion and references can be 
found in the review article \cite{Lahayareview09}.

\subsection{Vortices in spin-poralized dipolar BECs}
Here, we summarize some of the theoretical results on vortices in 
a spin-poralized dipolar BEC, in which the above features of the excitation spectrum 
have a strongly influence on vortex states. 

\subsubsection{Structure of a single-vortex}
The effect of dipolar interactions on the single vortex state was first studied by 
Yi and Pu by numerically solving Eq. (\ref{DDIGP}) \cite{Yi06}.
While there was no significant difference of the vortex structure 
for the repulsive contact interaction, 
the density profile exhibited a ``biconcave" structure around the vortex core 
when the condensate had attractive contact interactions ($a<0$) 
and was trapped in a highly prolate trap. 
More detailed numerical study was performed by Wilson {\it et al.} \cite{Wilson08}, who 
found that these biconcave density ripples around the vortex core emerge at milder 
trap aspect ratios and with purely dipolar interactions ($a=0$). 
By using perturbation theory, they related these density oscillations to 
the roton mode near the instability. 

\subsubsection{Vortex stability and dynamics}
The thermodynamic critical rotation frequency $\Omega_{c}$ for dipolar BECs 
in the TF limit was investigated by O'Dell and Eberlein \cite{ODell07}. 
The value of $\Omega_{c}$ decreased for a condensate in a pancake-shaped 
trap, while it increased in a cigar-shaped trap, compared to that of a conventional 
BEC. This is because, for the pancake trap, the DDI is repulsive on average, and 
results in the slight increase of both the axial size $R_{z}$ and the radial size $R_{\perp}$. 
Thus, $\Omega_{c}$ decreases since $\Omega_{c} \propto R_{\perp}^{-2}$. 
For the cigar-shaped trap, the DDI caused the slight increase in the axial size $R_{z}$ 
but decreased the radial size $R_{\perp}$, which increased $\Omega_{c}$. 
The TF approximation is invalid for pure dipolar BECs. The full 
numerical calculations for the single-vortex state were done by Abad {\it et al.} \cite{Abad09}, 
and their results were good agreement with Ref. \cite{ODell07}.

Another important issue is the critical rotation frequency for 
vortex nucleation in rotating dipolar BECs. Following the analysis 
for the conventional BEC \cite{Sinha}, Bijnen {\it et al.} studied the dynamical instability of a rotating 
dipolar BEC in the TF limit \cite{Bijnen07,Bijnen09}. The dynamically unstable region 
of the rotation frequency $\Omega$ versus trap aspect ratio $\omega_{z}/\omega_{\perp}$ 
nontrivially depends on the dipole interaction strength. 
Interestingly, the critical frequency $\Omega_{c}$ for a cigar-shaped trap 
can become larger than the onset of the dynamical instability of a rotating condensate. 
This is an intriguing regime, in which a rotating dipolar BEC is dynamically unstable 
but vortices will not enter. 

The BdG analysis for a single vortex state in dipolar BECs was done by Wilson {\it et al.} \cite{Wilson09}. 
As discussed in Sec. \ref{singlecompstability}, the single vortex state in a conventional BEC (without rotation)
is thermodynamically unstable, but dynamically stable. 
However, the vortex states in dipolar BECs can be dynamically unstable, depending on 
the dipolar strength and the aspect ratio of the trap potential.  
For a pancake trap, the instability is associated with the radial and angular 
roton modes, which induce a density wave and local collapse. 
For a cigar-shaped trap, the 3D character of the DDI becomes important. 
Thus, the vortex wave (Kelvin mode) is strongly affected by the DDI. 
Klawunn {\it et al.} showed that, when applying a periodic potential along the 
vortex line, which changes the effective mass along this direction, 
the dispersion of the Kelvin mode has a roton-like minimum \cite{Klawunn08}. 
As the dipolar interaction increases, this minimum can reach zero, 
forming an instability similar to the Donnelly-Glaberson instability 
(see Sec. \ref{Dynamicssinglevortex} for the discussion of vortex waves).  

\subsubsection{Vortex lattices}
Rapidly rotating dipolar BECs exhibit a rich variety of vortex phases characterized 
by different symmetries of the lattice structure \cite{Cooper05,Zhang05b,Komineas07}. 
Cooper {\it et al.} treated this problem within the LLL approximation \cite{Cooper05}. 
In particular, with increasing dipole interaction strength or with a high filling factor, 
the vortex lattice may undergo transitions between different symmetries: triangular, 
square, striped vortex crystal, and bubble state. In addition, for vortex lattices in double-well potentials, the competition between tunneling and interlayer DDI should lead 
to a quantum phase transition from a coincident phase to a staggered phase \cite{Zhang05b}. 

\subsection{Vortices in spinor dipolar condensates}
When the spin degree of freedom is taken into account for dipolar BECs, extremely 
rich physics appear. Let us consider the $F=1$ case. 
Typically, the energy associated with spin-exchange interactions 
is much smaller than that of contact (spin-preserving) interactions, i.e., $g_{n} \gg g_{s}$. 
Hence, the spin-exchange interaction may become comparable to the DDI. 
As a consequence, even for alkali spinor BECs (in particular $^{87}$Rb) the DDI plays 
a significant role \cite{Vengalattore08}. 

Under a uniform magnetic field ${\bf B}$, the second quantized Hamiltonian 
for spin-1 dipolar BECs is $\hat{H}=\hat{H}_{0}+\hat{H}_{\rm int}+\hat{H}_{\rm dd}$, 
where $\hat{H}_{0}$ and $\hat{H}_{\rm int}=\int d {\bf r}_{1} d {\bf r}_{2} 
V_{F=1} ( {\bf r}_{1} - {\bf r}_{2})$ is given by Eqs. (\ref{spin1singlehamilton}) 
and (\ref{spin1interaction}), respectively. 
The dipolar contribution becomes 
\begin{eqnarray}
\hat{H}_{\rm dd} = \frac{c_{\rm dd}}{8 \pi} \int \int \frac{ d{\bf r} 
d {\bf r}'}{|{\bf r}-{\bf r}'|^3} \sum_{\alpha, \alpha', \beta, \beta' = -1}^{1} 
\biggl[ \hat{\psi}_\alpha^{\dag}({\bf r}) \hat{\psi}_{\beta}^{\dag}({\bf r}')
{\bf F}_{\alpha \alpha'} \cdot {\bf F}_{\beta \beta'} 
\hat{\psi}_{\beta'}({\bf r}') \hat\psi_{\alpha'}({\bf r}) \nonumber \\ 
-3 \hat{\psi}_\alpha^\dag({\bf r}) \hat\psi_{\beta}^\dag({\bf r'}) ({\bf F}_{\alpha \alpha'} 
\cdot \hat{\bf r}_{12}) ({\bf F}_{\beta \beta'} \cdot \hat{\bf r}_{12}) 
\hat\psi_{\beta'}({\bf r}') \hat\psi_{\alpha'}({\bf r}) \biggr], \label{hspd}
\end{eqnarray}
where $\hat{\bf r}_{12}=({\bf r}-{\bf r}')/|{\bf r}-{\bf r}'|$ is a unit vector. 
The spin-exchange term and the dipolar term describe two types of
spin-dependent interactions. Contrary to the contact interaction, the DDI does not necessarily 
conserve the spin projection along the quantization axis due to the anisotropic character 
of the interaction. Note that the direction of the spin of the order parameter 
is not polarized by an external field. 
In such a situation, the system can develop a nontrivial spin texture due to the 
interplay and competition between these two terms.

\subsubsection{Einstein-de Haas effect}
If the atoms are initially prepared into a maximally polarized state, e.g., $m_{F} = - F$, 
contact interactions cannot induce any spinor dynamics due to the conservation of total magnetization $M$. 
DDIs, on the contrary, may induce a transfer into an $m_F + 1$ state. Here, the DDI conserves 
the total (spin $+$ orbital) angular momentum of the system. 
If the system preserves cylindrical symmetry around the quantization axis, 
this violation of the spin conservation is accompanied by a transfer of angular momentum from the 
spin to the orbital part. Due to this transfer, an initially spin-polarized dipolar 
condensate can dynamically generate vortices 
in a process resembling the Einstein-de Haas effect \cite{Kawaguchi06b,Santos06}. 

Unfortunately, the Einstein-de Haas effect is prevented in the presence of even a weak magnetic field, 
which favors the spin polarization. However, for a ferromagnetic spinor BEC such as $F = 1$ $^{87}$Rb BECs, 
a significant Einstein-de Haas effect may be expected under specific resonant 
conditions \cite{Gawryluk07}. The population transfer away from 
the initially prepared $m_F = 1$ state is typically very small, but can be significantly enhanced by applying 
a resonant magnetic field, such that the linear Zeeman energy of an atom in the $m_F = 1$ state is totally 
transferred into the kinetic energy of the rotating atom in $m_F = 0$. 

\subsubsection{Spontaneous circulation in the ground state}
The ground state structure of the dipolar spinor condensate was first studied by Yi {\it et al.} \cite{Yi04} 
under the single mode approximation (SMA), which assumes that all components share the same spatial 
wave function. For spinor BECs, the SMA is valid when the spin-dependent interaction is sufficiently weaker 
than the spin-independent interaction ($g_{n} \gg g_{s}$ for spin-1 case). 
However, in the presence of even weak DDIs, the dipole moments are spatially modulated due to the 
long-range anisotropic nature of the interactions, thus the application of the SMA is questionable. 
It is interesting that, in this system, the spatial spin variation, caused by the DDI, 
induces the supercurrent due to the spin-gauge symmetry, as seen in Eq. (\ref{spin1velocity}). 

Systematic numerical studies of the ground state are described in Refs. \cite{Yi06b,Kawaguchi06a}, 
revealing the spontaneous emergence of spin textures in the ground state. 
Yi and Pu constructed a phase diagram of the ground state of a spin-1 dipolar BEC with 
respect to the trap aspect ratio $\lambda$ and the dipolar interaction strength $c_{\rm dd}$, 
for both the ferromagnetic case and the antiferromagnetic case \cite{Yi06b}. 
For $c_{\rm dd}$ smaller than $g_{n}$, the SMA is valid. Otherwise, three 
distinct spin textures appear, which can be classified as: 
(a) for a pancake geometry ($\lambda >1$), the polar-core spin vortex with $(-1,0,1)$.
(b) for a cigar-shaped geometry ($\lambda < 1$), the coreless vortex with $(0,1,2)$, where the spins twist around the $z$-axis. 
(c) for a spherical geometry ($\lambda \sim 1$), the non-axisymmetric state.
Kawaguchi {\it et al.} studied ferromagnetic spin-1 BECs by varying $g_{s}$ and $c_{\rm dd}$, finding three
distinct phases: a polar-core vortex, a flower phase, and a chiral spin-vortex phase \cite{Kawaguchi06a}. 
Hence, dipolar spinor BECs can have spontaneous circulation in the ground state, 
even without rotation.

\section{Quantum turbulence}

As discussed in the previous chapter, the study of the turbulent state in quantum fluids and its relation to classical turbulence (CT) is an intriguing physical problem.
Although the study of quantum turbulence (QT) has a long history, only superfluid $^4$He and $^3$He systems have been used to realize QT until recently.
Recently, atomic Bose-Einstein condensates have become another candidate for QT research, since a turbulent state was realized in this system.

In this section, we briefly review some of theoretical studies and an important experimental work for QT in atomic BECs.

\subsection{Theoretical work on quantum turbulence in atomic BECs}
A major problem in the study of QT in atomic BEC is the difficulty of applying a velocity field to generate the turbulent state.
Thus, alternative methods are required to realize QT.

Berloff and Svistunov suggested the realization of QT in the dynamics of the formation of a BEC from a strongly degenerate nonequilibrium gas \cite{Svistunov2001,Berloff2002}.
Such dynamics can also be described by the GP equation in the framework of a classical field description.

Parker and Adams suggest the emergence and decay of turbulence in an atomic BEC under a simple rotation, starting from a vortex-free equilibrium BEC \cite{Parker}.
Starting from the vortex-free steady state with a constant potential, they performed a numerical simulation of the GP model with realistic experimental parameters for $^{87}$Rb BECs, and showed that the total dynamics can be divided into four distinct stages as follows: (i) Fragmentation - the quadrupolar oscillation of the condensate breaks down, ejecting energetic atoms to form an outer cloud, (ii) symmetry breaking - the twofold rotational symmetry of the system is broken, allowing the rotation to couple to additional modes, thereby injecting energy rapidly into the system, (iii) turbulence - a turbulent cloud containing vortices and high energy density fluctuations (sound field) is formed, and (iv) crystallization - the loss of energy at short length scales coupled with vortex-sound interactions allows the system to relax into an ordered lattice. In the third region of turbulence, the system produces a Kolmogorov energy spectrum \eqref{eq-Kolmogorov-law} demonstrating a classical analog of QT.

To realize steady QT, Kobayashi and Tsubota suggested combining rotations around two axes \cite{Kobayashi2007}.
Starting from a vortex-free initial state, they performed numerical simulations of the GP model with realistic parameters for $^{87}$Rb BEC, and obtained a steady turbulent state with no crystallization, but with highly-tangled quantized vortices.
The incompressible kinetic energy spectrum satisfies the Kolmogorov law in the inertial range.

White {\it et al.} numerically investigated QT in atomic BECs, and found nonclassical velocity statistics \cite{White2010}.
The probability density function (PDF) of velocity is another important statistical measure to study turbulence, and for the case of CT, the PDF obeys a Gaussian distribution \cite{Vincent1991,Noullez1997,Gotoh2002}.
They performed numerical simulations of the GP model \eqref{eq-gp} using realistic experimental parameters for a $^{23}$Na condensate.
The method to create turbulence was "phase-imprinting" in the condensate.
They calculated the PDF of the superfluid velocity, and found that the velocity statistics were non-Gaussian and had a power-law dependence.
Their result is consistent with the high-velocity tails found experimentally by Paoletti {\it et al.} in turbulent superfluid $^4$He \cite{Paoletti08}.

\subsection{Experimental work on quantum turbulence in atomic BECs}

Recently, a turbulent state was realized in atomic BECs by two methods.
Weiler {\it et al.} performed a rapid quench of an $^{87}$Rb gas through the BEC transition temperature \cite{Weiler08}, which is a similar situation to that examined by Berloff and Svistunov \cite{Berloff2002}.
Through the high density fluctuation regime (weak turbulence) in a short period, several vortices and anti-vortices were formed to make the system turbulent.
This experiment was interpreted in the previous section (subsection \ref{subsec-rapid-quench}).

\begin{figure}[htb]
\centering
\includegraphics[width=0.85\linewidth]{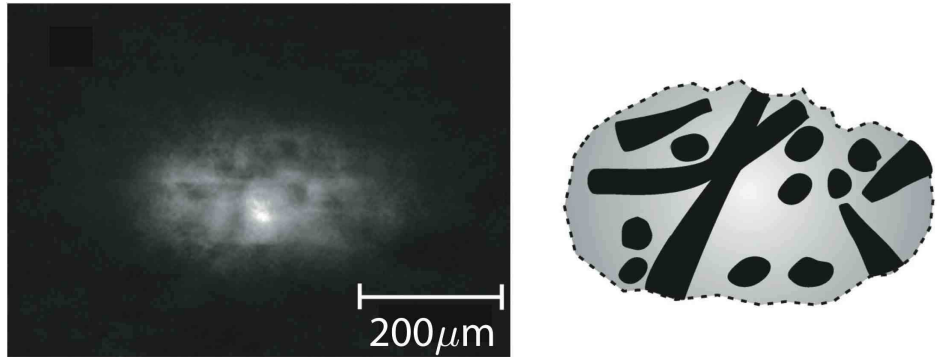}
\caption{\label{fig-Henn-2} Atomic optical density after 15 ms of free expansion (left) and a corresponding schematic diagram showing the inferred distribution of vortices obtained from the image (right) \cite{Henn2009b} [Henn, Seman, Roati, Magalh\~aes, and Bagnato: Phys. Rev. Lett. {\bf 103} (2009) 045301, reproduced with permission. Copyright 2009 by the American Physical Society].}
\end{figure}

The turbulent state created in the above method strongly depended on the initial uncontrollable thermal state.
Furthermore, the turbulent state is merely intermediate, and the final state that contains only a few vortices is not QT.
As a method with better control of the turbulence, Henn {\it et al.} introduced an external oscillatory perturbation to an $^{87}$Rb BEC \cite{Henn2009a,Henn2009b}.
This oscillating magnetic field was produced by a pair of anti-Helmholtz coils which were not perfectly aligned to the vertical axis of the cigar-shaped condensate.
Additionally, the components along the two equal directions that result in the radial symmetry of the trap were slightly different.
This oscillatory field induced a coherent mode excitation in a BEC.
For small amplitudes of the oscillating field and short excitation periods, dipolar modes, quadrupolar modes, and scissors modes of the BEC were observed, but no vortices appeared.
Increasing both parameters, vortices grew in number, eventually leading to the turbulent state.
In the turbulent regime, they observed a rapid increase in the number of vortices followed by proliferation of vortex lines in all directions, where many vortices with no preferred orientation formed a vortex tangle (Fig. \ref{fig-Henn-2}).
Another remarkable feature is that a completely different hydrodynamic regime was followed: the suppression of aspect ratio inversion during free expansion, despite the asymmetric expansion (from a cigar-shaped to pancake-shaped) of the usual quantum gas of bosons, or the isotropic expansion of a thermal cloud.
Although the theoretical understanding of this effect remains incomplete, it represents a remarkable new effect in atomic superfluids.
The emergence of QT with vortex reconnections, excitation of Kelvin waves, and the existence of cascade phenomena are expected to be directly observed in this well-controllable QT system in the near future.

\chapter{Summary and conclusions}
In this chapter, we have reviewed recent topics on quantized vortices in superfluid helium and atomic Bose-Einstein condensates.  Quantized vortices were discovered in superfluid $^4$He in the 1990s. However, they have recently grown in importance, for two reasons.  The first reason is that the research of QT entered a new era since the mid 1950s. A primary interest is to understand the relationship between CT and QT. The second reason is the realization of atomic BECs. Modern optical techniques have enabled the direct visualization of quantized vortices, and multi-component BECs have further enriched the world of quantized vortices.

The research of vortices in classical fluids has a very long history. When Leonardo da Vinci drew his famous sketches of water turbulence in the 16th century, he certainly recognized that turbulence consists of many eddies of different sizes.  It was in 1858 that Helmholtz proposed the idea of a vortex filament. However, it is not straightforward to understand the role of vortices in the classical hydrodynamics of turbulence. One reason is that vortices in classical fluids are unstable and not very well defined. Quantum condensed systems create more stable and well-defined vortices, namely quantized vortices.

This field is currently very active, attracting the interest of scientists from many different fields. We hope that this chapter will contribute to further development and progress in this fascinating field of physics.


\newpage
\thebibliography{999}

\bibitem{PLTP} Halperin, W. P. and Tsubota, M. eds. (2008).
{\it Progress in Low Temperature Physics} Vol. 16. Elsevier, Amsterdam.

\bibitem{TsubotaJPSJ} Tsubota, M. (2008). Quantum turbulence. {\it J. Phys. Soc. Jpn.}, {\bf 77}, 111006(1-12).

\bibitem{Kapitza} Kapitza, P. (1938). Viscosity of liquid helium below the $\lambda$ point. {\it Nature}, {\bf 141}, 74.

\bibitem{Allen} Allen, J. F. and Misener, A. D. (1938). Flow of liquid helium II. {\it Nature}, {\bf 141}, 75.

\bibitem{Landau} Landau, L. (1941). The theory of superfluidity of helium II. {\it J. Phys. U. S. S. R.}, {\bf 5}, 71-90.

\bibitem{Tisza} Tisza, L. (1938). Transport phenomena in helium II. {\it Nature}, {\bf 141}, 913.

\bibitem{GM} Gorter, C. J. and Mellink, J. H. (1949). On the irreversible processes in liquid helium II. {\it Physica}, {\bf 15}, 285-304.

\bibitem{London} London, F. (1938). On the Bose-Einstein condensation. {\it Phys. Rev.}, {\bf 54}, 947-954.

\bibitem{Onsager} Onsager, L. (1949). {\it Nuovo Cimento Suppl.}, {\bf 6}, 249-250.

\bibitem{Feynman} Feynman, R. P. (1955). Application of quantum mechanics to liquid helium. {\it Progress in Low Temperature Physics} Vol.1(Gorter, C. J. ed.). Amsterdam. North-Holland, 17-53.

\bibitem{HallVinen56a} Hall, H. E. and Vinen, W. F. (1956). The rotation of liquid helium II I. Experiments on the propagation of second sound in uniformly rotating helium II. {\it Proc. Roy. Soc. London}, {\bf A238}, 204-214.

\bibitem{HallVinen56b} Hall, H. E. and Vinen, W. F. (1956). The rotation of liquid helium II II. The theory of mutual friction in uniformly rotating helium II. {\it Proc. Roy. Soc. London}, {\bf A238}, 215-234.

 \bibitem{Vinen57a} Vinen, W.F. (1957). Mutual friction in a heat current in liquid helium II I. Experiments on steady heat currents,  {\it Proc. Roy. Soc. London}, {\bf A240}, 114-127.

\bibitem{Vinen57b} Vinen, W. F. (1957). Mutual friction in a heat current in liquid helium II. II. Experiments on transient effects. {\it  Proc. Roy. Soc. London}, {\bf A240}, 128-143.

\bibitem{Vinen57c} Vinen, W. F. (1957). Mutual friction in a heat current in liquid helium II III. Theory of mutual friction.  {\it Proc. Roy. Soc. London},  {\bf A242}, 493-515.

\bibitem{Vinen57d} Vinen, W. F. (1957). Mutual friction in a heat current in liquid helium II IV. Critical heat currents in wide channels.  {\it Proc. Roy. Soc. London},  {\bf A243}, 400-413.

\bibitem{Vinen61} Vinen, W. F. (1961). The detection of single quanta circulation in liquid helium II. {\it  Proc. Roy. Soc. London}, {\bf A260},  218-236.

\bibitem{Tough82} Tough, J. T. (1982). Superfluid turbulence. {\it Progress in  Low Temperature Physics} Vol. 8 (Gorter, C. J. ed.). Amsterdam. North-Holland, 133-220.

\bibitem{Schwarz85} Schwarz, K. W. (1985). Three-dimensional vortex dynamics in superfluid $^4$He: Line-line and line-boundary interactions.  {\it Phys. Rev. B}, {\bf 31}, 5782-5803.

\bibitem{Schwarz88} Schwarz, K. W. (1988). Three-dimensional vortex dynamics in superfluid $^4$He: Homogeneous superfluid turbulence.  {\it Phys. Rev. B}, {\bf 38}, 2398-2417.

\bibitem{Saffman} Saffman, P. G. (1992). {\it Vortex Dynamics}. Cambridge University Press, Cambridge. 

\bibitem{Tsubota00} Tsubota, M.,  Araki T.  and  Nemirovskii, S. K. (2000). Dynamics of vortex tangle without mutual friction in superfluid $^4$He. {\it Phys. Rev. B}, {\bf 62}, 11751-11762.

\bibitem{Adachi10} Adachi, H., Fujiyama, S. and Tsubota, M. (2010). Steady-state counterflow quantum turbulence: Simulation of vortex filaments using the full Biot-Savart law. {\it Phys. Rev. B}, \textbf{81}, 104511(1-7).

\bibitem{Boratav92} Boratav, O. N., Pelz, R. B. and Zabusky, N. J. (1992). Reconnection in orthogonally interacting vortex tubes: direct numerical simulations and quantifications. {\it Phys. Fluids A},  \textbf{4}, 581-605.

\bibitem{Koplik93} Koplik, J. and Levine, H. (1993). Vortex reconnection in superfluid helium. {\it Phys. Rev. Lett.},  \textbf{71}, 1375-1378.

\bibitem{Leadbeater01} Leadbeater, M., Winiecki, T., Samuels, D. C., Barenghi, C. F. and Adams, C. S. (2001). Sound emission due to superfluid vortex reconnection. {\it Phys. Rev. Lett.}, \textbf{86}, 1410-1413.

\bibitem{Ogawa02a} Ogawa, S., Tsubota, M. and Hattori, Y. (2002). Study of reconnection and acoustic emission of quantized vortices in superfluid by the numerical analysis of the Gross-Pitaevskii equation. {\it J. Phys. Soc. Jpn.}, \textbf{71}, 813-821.

\bibitem{Gross1961}
Gross, E. P. (1961). 
Structure of a quantized vortex in boson systems.
{\it Nuovo Cimento}, {\bf 20}, 454-457.

\bibitem{Pitaevskii1961}
Pitaevskii, L. P. (1961).
Vortex Lines in an Imperfect Bose Gas.
{\it Zh. Eksp. Teor. Fiz.}, {\bf 40}, 646-651 [{\it Sov. Phys. JETP}, {\bf 13}, 451-454].

\bibitem{PethickSmith} Pethick, C. J. and Smith, H. (2008). {\it Bose-Einstein condensation in dilute gases \/} (2nd edn.). Cambridge University Press, Cambridge. 

\bibitem{Vinen2007} Vinen, W. F. and Donnelly, R. J. (2007). Quantum Turbulence. {\it Physics Today}, \textbf{60}, 43-48.

\bibitem{Vinen2002} Vinen, W. F. and Niemela, J. J. (2002). Quantum Turbulence. {\it J. Low Temp. Phys.}, \textbf{128}, 167-231.

\bibitem{Svistunov1995} Svistunov, B. V. (1995). Superfluid turbulence in the low-temperature limit. {\it Phys. Rev. B}, \textbf{52}, 3647-3653.

\bibitem{Vinen2001} Vinen, V. F. (2001). Decay of superfluid turbulence at a very low temperature: The radiation of sound from a Kelvin wave on a quantized vortex. {\it Phys. Rev. B}, \textbf{64}, 134520(1-4).

\bibitem{Frisch1995} Frisch, U. (1995). Turbulence. {\it Turbulence}. Cambridge University Press, Cambridge.

\bibitem{Kolmogorov1941a} Kolmogorov, A. N. (1941). The local structure of turbulence in incompressible viscous fluid for very large Reynolds numbers. {\it Dokl. Akad. Nauk SSSR}, \textbf{30}, 299-303 (reprinted in 1991, {\it Proc. Roy. Soc. A}, \textbf{434}, 9-13).

\bibitem{Kolmogorov1941b} Kolmogorov, A. N. (1941). Dissipation of energy in the locally isotropic turbulence. {\it Dokl. Akad. Nauk SSSR}, \textbf{32}, 16-18 (reprinted in 1991, {\it Proc. Roy. Soc. A}, \textbf{434}, 15-17).

\bibitem{Richardson2007} Richardson, L. F. (2007). {\it Weather Prediction by Numerical Process} (2nd edn.) Cambridge University Press, Cambridge.

\bibitem{Kivotides2001a} Kivotides, D., Vassilicos, J. C., Samuels, D. C., and Barenghi, C. F. (2001). Kelvin Waves Cascade in Superfluid Turbulence. {\it Phys. Rev. Lett.}, \textbf{86}, 3080-3083.

\bibitem{Nore1997a} Nore, C., Abid, M., and Brachet, M. E. (1997). Kolmogorov Turbulence in Low-Temperature Superflows. {\it Phys. Rev. Lett.}, \textbf{78}, 3896-3899.

\bibitem{Nore1997b} Nore, C., Abid, M., and Brachet, M. E. (1997). Decaying Kolmogorov turbulence in a model of superflow. {\it Phys. Fluids}, \textbf{9}, 2644-2669.

\bibitem{Araki2002} Araki, T., Tsubota, M., and Nemirovskii, S. K. (2002). Energy Spectrum of Superfluid Turbulence with No Normal-Fluid Component. {\it Phys. Rev. Lett.}, \textbf{89}, 145301(1-4).

\bibitem{Kobayashi2005a} Kobayashi, M. and Tsubota, M. (2005). Kolmogorov Spectrum of Superfluid Turbulence: Numerical Analysis of the Gross-Pitaevskii Equation with a Small-Scale Dissipation. {\it Phys. Rev. Lett.}, \textbf{94}, 065302(1-4).

\bibitem{Kobayashi2005b} Kobayashi, M. and Tsubota, M. (2005). Kolmogorov Spectrum of Quantum Turbulence. {\it J. Phys. Soc. Jpn.}, \textbf{74}, 3248-3258.

\bibitem{Yepez2009} Yepez, J., Vahala, G., Vahala, L., and Soe, M. (2009). Superfluid Turbulence from Quantum Kelvin Wave to Classical Kolmogorov Cascade. {\it Phys. Rev. Lett.}, \textbf{103}, 084501(1-4).

\bibitem{Samuels1990} Samuels, D. C. and Donnelly, R. J. (1990). Sideband Instability and Recurrence of Kelvin Waves on Vortex Cores. {\it Phys. Rev. Lett.}, \textbf{64}, 1385-1388.

\bibitem{Mitani2003} Vinen, W. F., Tsubota, M., and Mitani, A. (2003). Kelvin-Wave Cascade on a Vortex in Superfluid $^4$He at a Very Low Temperature. {\it Phys. Rev. Lett.}, {\bf 91}, 135301(1-4).

\bibitem{Kozik2004} Kozik, E. and Svistunov, B. (2004). Kelvin-Wave Cascade and Decay of Superfluid Turbulence. {\it Phys. Rev. Lett.}, {\bf 92}, 035301(1-4).

\bibitem{Kozik2005} Kozik, E. and Svistunov, B. (2005). Scale-Separation Scheme for Simulating Superfluid Turbulence: Kelvin-Wave Cascade. {\it Phys. Rev. Lett.}, {\bf 94}, 025301(1-4).

\bibitem{Nazarenko2006} Nazarenko, S. (2006). Differential Approximation for Kelvin Wave Turbulence. {\it JETP Lett.}, {\bf 83}, 198-200.

\bibitem{Boffetta2009} Boffetta, G., Celani, A., Dezzani, D., Laurie, J., and Nazarenko, S. (2009). Modeling Kelvin Wave Cascades in Superfluid Helium. {\it J. Low Temp. Phys.}, {\bf 156}, 193-214.

\bibitem{L'vov2007} L'vov, V. S., Nazarenko, S. V., and Rudenko, O. (2007). Bottleneck crossover between classical and quantum superfluid turbulence. {\it Phys. Rev. B}, {\bf 76}, 024520(1-9).

\bibitem{Kozik2008} Kozik, E. and Svistunov, B. (2008). Kolmogorov and Kelvin-wave cascades of superfluid turbulence at $T=0$: What lies between. {\it Phys. Rev. B}, {\bf 77}, 060502(R)(1-4).




\bibitem{Kelvin1880} Thomson, W. (1880). Vibrations of a columnar vortex. {\it Phil. Mag.}, {\bf 10}, 155-168.

\bibitem{Hall1958} Hall, H. E. (1958). An Experimental and Theoretical Study of Torsional Oscillations in Uniformly Rotating Liquid Helium II. {\it Proc. Roy. Soc. London}, {\bf A245}, 546-561.

\bibitem{Hall1960} Hall, H. E. (1960). The rotation of liquid helium II. {\it Phil. Mag. Suppl.}, {\bf 9}, 89-146.

\bibitem{Zakharov1992} Zakharov, V. E., L'vov, V. S., and Fal'kovich, G. (1992). {\it Kolmogorov Spectra of Turbulence}, Springer, Berlin.

\bibitem{Connaughton2004} Connaughton, C. and Nazarenko, S. (2004). Warm Cascades and Anomalous Scaling in a Diffusion Model of Turbulence. {\it Phys. Rev. Lett.}, {\bf 92}, 044501(1-4).

\bibitem{Smith1993} Smith, M. R., Donnelly, R. J., Goldenfeld, N., and Vinen, W. F. (1993). Decay of Vorticity in Homogeneous Turbulence. {\it Phys. Rev. Lett.}, {\bf 71}, 2583-2586.

\bibitem{Maurer1998} Maurer, J. and Tabeling, P. (1998). Local investigation of superfluid turbulence. {\it Europhys. Lett.}, {\bf 43}, 29-34.

\bibitem{Stalp1999} Stalp, S. R., Skrbek, L., and Donnelly, R. J. (1999). Decay of Grid Turbulence in a Finite Channel. {\it Phys. Rev. Lett.}, {\bf 82}, 4831-4834.

\bibitem{Skrbek2000a} Skrbek, L., Niemela, J. J., and Donnelly, R. (2000). Four Regimes of Decaying Grid Turbulence in a Finite Channel. {\it Phys. Rev. Lett.}, {\bf 85}, 2973-2976.

\bibitem{Skrbek2000b} Skrbek, L. and Stalp, S. R. (2000). On the decay of homogeneous isotropic turbulence. {\it Phys. Fluids}, {\bf 12}, 1997-2019.

\bibitem{Stalp2002} Stalp, S. R., Niemela, J. J., Vinen, W. F., and Donnelly, R. J. (2002). Dissipation of grid turbulence in helium II. {\it Phys. Fluids}, {\bf 14}, 1377-1379.

\bibitem{Skrbek2003} Skrbek, L., Gordeev, A. V., and Soukup, F. (2003). Decay of counterflow He II turbulence in a finite channel: Possibility of missing links between classical and quantum turbulence. {\it Phys. Rev. E}, {\bf 67}, 047302(1-4).

\bibitem{Vinen2000} Vinen, W. F. (2000). Classical character of turbulence in a quantum liquid. {\it Phys. Rev. B}, {\bf 61}, 1410-1420.

\bibitem{Kivotides2007} Kivotides, D. (2007). Relaxation of superfluid vortex bundles via energy transfer to the normal fluid. {\it Phys. Rev. B}, {\bf 76}, 054503(1-12).

\bibitem{L'vov2006} L'vov, V. S., Nazarenko, S. V., and Skrbek, L. (2006). Energy Spectra of Developed Turbulence in Helium Superfluids. {\it J. Low Temp. Phys.}, {\bf 145}, 125-142.

\bibitem{Donnelly1991} Donnelly, R. J. (1991). {\it Quantized vortices in helium II}, Cambridge University Press, Cambridge.

\bibitem{Hinze1975} Hinze, J. O. (1975). {\it Turbulence} (2nd edn.) McGraw Hill, New York.

\bibitem{Walmsley2007} Walmsley, P. M., Golov, A. I., Hall, H. E., Levchenko, A. A., and Vinen, W. F. (2007). Dissipation of Quantum Turbulence in the Zero Temperature Limit. {\it Phys. Rev. Lett.}, {\bf 99}, 265302(1-4).

\bibitem{Walmsley2008} Walmsley, P. M. and Golov, A. I. (2008). Quantum and Quasiclassical Types of Superfluid Turbulence. {\it Phys. Rev. Lett.}, {\bf 100}, 245301(1-4).

\bibitem{Bradley06}  Bradley, D. I.,  Clubb, D. O.,  Fisher,  S. N., Guenault, A. M.,  Halay,  R. P.,  Matthews, C. J., Pickett, G. R., Tsepelin, V. and Zaki, K. (2006). Decay of Pure Quantum Turbulence in Superfluid $^3$He-B. {\it  Phys. Rev. Lett.}, \textbf{96}, 035301(1-4).

\bibitem{Eltsov2007} Eltsov, V. B., Golov, A. I., de Graaf, R., H\"anninen, R., Krusius, M., L'vov, V. S., and Solntsev, R. E. (2007). Quantum Turbulence in a Propagating Superfluid Vortex Front. {\it Phys. Rev. Lett.}, {\bf 99}, 265301(1-4).

\bibitem{Kobayashi2006} Kobayashi, M. and Tsubota, M. (2006). Decay of Quantized Vortices in Quantum Turbulence. {\it J. Low Temp. Phys.}, {\bf 145}, 209-218.

\bibitem{Tsubota2003} Tsubota, M., Araki, T., and Vinen, W. F. (2003). Diffusion of an inhomogeneous vortex tangle. {\it Physica B}, {\bf 329-333}, 224-225.

\bibitem{VinenPLTP} Skrbek, L. and Vinen, W. F. (2008). The use of vibrating structures in the study of quantum turbulence. {\it Progress in  Low Temperature Physics} Vol. 16 (Halperin, W.P. and Tsubota, M. edn.). Amsterdam. Elsevier, 195-246.

\bibitem{Jager95} J\"ager, J., Schuderer, B. and Schoepe, W. (1995). Turbulent and laminar drag of superfluid helium on an oscillating microsphere.  {\it Phys. Rev. Lett.}, \textbf{74}, 566-569.

\bibitem{Nichol04a}  Nichol, H. A., Skrbek, L., Hendry, P. C. and McClintock, P. V. E. (2004). Flow of He II due to an oscillating grid in the low-temperature limit. {\it Phys. Rev. Lett.},  \textbf{92}, 244501(1-4).

\bibitem{Nichol04b}  Nichol, H. A., Skrbek, L., Hendry, P. C. and McClintock, P. V. E. (2004). Experimental investigation of the macroscopic flow of He II due to an oscillating grid in the zero temperature limit. {\it Phys. Rev. E},  \textbf{70}, 056307(1-14).

\bibitem{Charalambous06} Charalambous, D.,  Hendry, P. C.,  McClintock, P. V. E.,  Skrbek, L. and Vinen, W. F. (2006). Experimental investigation of the dynamics of a vibrating grid in superfluid $^4$He over a range of temperatures and pressures,  {\it Phys. Rev. E},  \textbf{74}, 036307(1-10).

\bibitem{Bradley05a} Bradley, D. I., Clubb, D. O., Fisher, S. N., Gu\'enault, A. M., Haley, R. P., Matthews, C. J., Pickett, G. R. and Zaki, K. L. (2005). Turbulence generated by vibrating wire resonators in superfluid $^4$He at low temperatures. {\it J. Low Temp. Phys.}, \textbf{138}, 493-498.

\bibitem{Bradley05b} Bradley, D. I.,  Fisher,  S. N., Gu\'enault, A. M., Lowe, M. R.,  Pickett, G. R., Rahm, A. and  Whitehead, R. C. V.(2004). Emission of Discrete Vortex Rings by a Vibrating Grid In Superfluid 3He-B: A Precursor to Quantum Turbulence. {\it Phys. Rev. Lett.}, \textbf{95}, 035302(1-4).

\bibitem{Fisher01} Fisher,  S. N., Hale, A.J.,  Gu\'enault,  A. M. and  Pickett, G. R. (2001). Generation and Detection of Quantum Turbulence in Superfluid $^3$He-B.  {\it Phys. Rev. Lett.}, \textbf{86}, 244-247.

\bibitem{Bradley04} Bradley, D. I.,  Fisher,  S. N., Gu\'enault, A. M., Lowe, M. R.,  Pickett, G. R., Rahm, A. and  Whitehead, R. C. V.(2004). Quantum Turbulence in Superfluid 3He Illuminated by a Beam of Quasiparticle Excitations. {\it Phys. Rev. Lett.},  \textbf{93}, 235302(1-4).

\bibitem{Yano07} Yano, H., Hashimoto, N., Handa, A., Nakagawa, M., Obara, K., Ishikawa, O. and Hata, T. (2007). Motion of quantized vortices attached to a boundary in alternating currents of superfluid $^4$He. {\it Phys. Rev. B}, \textbf{75}, 012502(1-4).

\bibitem{Hashimoto07} Hashimoto, N.,  Goto, R., Yano, H.,  Obara, K., Ishikawa, O.  and Hata, T. (2007). Control of turbulence in boundary layers of superfluid $^4$He by filtering out remnant vortices.  {\it Phys. Rev. B},  \textbf{76}, 020504(R)(1-4).

\bibitem{Goto08} Goto, R., Fujiyama, S., Yano, H., Nago, Y., Hashimoto, N., Obara, K., Ishikawa, O., Tsubota, M. and Hata, T. (2008). Turbulence in boundary flow of superfluid $^4$He triggered by free vortex rings. {\it Phys. Rev. Lett.}, \textbf{100}, 045301(1-4).

\bibitem{Blazkova07b} M. Bla\v{z}kov\'{a}, M. \v{C}love\v{c}ko, V. B. Eltsov,  E.  Ga\v{z}o, R. de Graaf,  J. J. Hosio,  M. Krusius, D. Schmoranzer,  W. Schoepe, L. Skrbek,  P. Skyba,  R. E. Solntsev and W. F. Vinen: J. Low Temp. Phys. \textbf{150} (2008) 525-535.

\bibitem{Blazkova09}Bla\v{z}kov\'{a}, M., Schmoranzer, D., Skrbek, L. and Vinen, W. F. (2009). Generation of turbulence by vibrating forks and other structures in superfluid $^4$He. {\it Phys. Rev. B}, {\bf 79}, 054522(1-11).

\bibitem{Yano09} Yano. H., Ogawa, T., Mori, A., Miura, Y., Nago, Y., Obara, K., Ishikawa, O. and Hata, T. (2009). Transition to quantum turbulence generated by thin vibrating wires in superfluid $^4$He. {\it J. Low Temp. Phys.}, \textbf{156}, 132-144.  

\bibitem{Awschalom84} Awshalom, D. D. and Schwarz, K. W. (1984). Observation of a remnant vortex-line density in superfluid helium. {\it Phys. Rev. Lett.}, {\bf 52}, 49-52.

\bibitem{Ruutu97} Ruutu, V. M. H., Parts, \"U, Koivuniemi, J. H., Kopnin, N. B. and Krusius, M. (1997). Intrinsic and extrinsic mechanisms of vortex formation in superfluid $^3$He-B. {\it J. Low Temp. Phys.}, {\bf 106}, 93-164.

\bibitem{Hanninen07} H\"anninen, R.,  Tsubota, M. and Vinen, W. F. (2007). Generation of turbulence by oscillating structures in superfluid helium at very low temperatures. {\it Phys. Rev. B}, \textbf{75}, 064502(1-12).

\bibitem{Hanninen10} H\"anninen, R. and Schoepe, W. (2010). Universal onset of quantum turbulence in oscillating flows and crossover to steady flows. {\it J. Low Temp. Phys.}, \textbf{158}, 410-414.

\bibitem{Volovik03} Volovik, G. E. (2003). Classical and quantum regimes of the superfluid turbulence. {\it JETP Lett.}, \textbf{78}, 533-537.

\bibitem{Kopnin04} Kopnin, N. B. (2004). Vortex instability and the onset of superfluid turbulence. {\it Phys. Rev. Lett.}, \textbf{92}, 135301(1-4).

\bibitem{Schoepe04} Schoepe, W. (2004). Fluctuations and stability of superfluid turbulence at mK temperatures. {\it Phys. Rev. Lett.}, \textbf{92}, 095301(1-3).

\bibitem{Batchelor} Batchelor, G. K. (1967). {\it An introduction to fluid mechanics}. Cambridge University Press, Cambridge.

\bibitem{Hemmati09} Hemmati, A. M., Fuzier, S., Bosque, E. and Van Sciver, S. W. (2009). Drag measurement on an oscillating sphere in Helium II. {\it J. Low Temp. Phys.}, \textbf{156}, 71-83.

\bibitem{Fujiyama09} Fujiyama, S. and Tsubota, M. (2009). Drag force on an oscillating object in quantum turbulence. {\it Phys. Rev. B}, \textbf{79}, 094513(1-7).

\bibitem{Tsubota93} Tsubota, M. and Maekawa, S. (1993). Pinning and depinning of two quantized vortices in superfluid $^4$He. {\it Phys. Rev. B}, \textbf{47}, 12040-12050.

\bibitem{SciverPLTP} Van Sciver, S. W. and Barenghi, C. F. (2008). Visualization of quantum turbulence. {\it Progress in  Low Temperature Physics} Vol. 16 (Halperin, W.P. and Tsubota, M.eds.). Amsterdam. Elsevier, 247-303.

\bibitem{Zhang05a} Zhang, T. and  Van Sciver, S. W. (2005). Large-scale turbulent flow around a cylinder in counterflow superfluid $^4$He (He(II)),  {\it Nature Phys.}, \textbf{1}, 36-38.

\bibitem{Bewley06} Bewley, G. P.,  Lathrop, D. P. and Sreenivasan, K. R. (2006). Visualization of quantized vortices. {\it Nature}. \textbf{441}, 588.

\bibitem{Tsubota94} Tsubota, M. (1994). Capacity of a pinning site for trapping quantized vortices in superfluid $^4$He. {\it Phys. Rev.B},  \textbf{50}, 579-581.

\bibitem{Paoletti08} Paoletti, M. S., Fioroto, R. B.,  Sreenivasan, K. R. and Lathrop, D. P. (2008). Visualization of superfluid helium flow. {\it J. Phys. Soc. Jpn.}, \textbf{77}, 111007(1-7).

\bibitem{Poole05} Poole, D. R., Barenghi, C. F.,  Sergeev, Y. A. and Vinen, W. F. (2005). Motion of tracer particles in He II. {\it Phys. Rev. B}, \textbf{71}, 064514(1-16).

\bibitem{Sergeev06}Sergeev, Y. A., Barenghi, C. F. and Kivotides, D. (2006). Motion of micron-size particles in turbulent He II. {\it Phys. Rev. B},  \textbf{74}, 184506(1-5).

\bibitem{Kivotides07}Kivotides, D., Barenghi, C. F. and Sergeev, Y. A.  (2007). Collision of a tracer particle and a quantized vortex in superfluid helium: self-consistent calculation. {\it Phys. Rev. B},  \textbf{75}, 212502(1-4).

\bibitem{Kivotides08}Kivotides, D., Barenghi, C. F. and Sergeev, Y. A.  (2008). Interaction between particles and quantized vortices in superfluid helium. {\it Phys. Rev. B},  \textbf{77}, 014527(1-13).

\bibitem{Barenghi09} Barenghi, C. F. and Sergeev, Y. A.  (2009). Motion of vortex ring with tracer particles in superfluid helium. {\it Phys. Rev. B},  \textbf{80}, 024514(1-5).

\bibitem{Anderson95}
Anderson, M.H., Ensher, J.R., Matthews, M.R., Wieman, C.E., 
and Cornell, E.A. (1995).
Observation of Bose-Einstein Condensation in a Dilute Atomic Vapor. {\it Science}, {\bf 269}, 198-201.

\bibitem{Bradley}
Bradley, C.C., Sackett, C.A., Tollett, J.J., and Hulet, R.G. (1995). 
Evidence of Bose-Einstein Condensation 
in an Atomic Gas with Attractive Interactions. {\it Phys. Rev. Lett.}, {\bf 75}, 1687-1690.

\bibitem{Davis}
Davis, K.B., Mewes, M.-O., Andrews, M.R., van Druten, N.J., Durfee, D.S., Kurn, D.M., and Ketterle, W. (1995). 
Bose-Einstein Condensation in a Gas of Sodium Atoms. 
{\it Phys. Rev. Lett.}, {\bf 75}, 3969-3973.

\bibitem{Doneley}
Donnelly, R.J. (1991). 
{\it Quantized Vortices in Helium II}. Cambridge University Press, Cambridge.

\bibitem{PitaevskiiStringari}
Pitaevskii, L. and Stringari, S. (2003). 
{\it Bose-Einstein Condensation}. Oxford University Press, Oxford.

\bibitem{Matthews}
Matthews, M.R., Anderson, B.P., Haljan, P.C., Hall, D.S., Wieman, C.E., and Cornell, E.A. (1999). 
Vortices in a Bose-Einstein condensate. 
{\it Phys. Rev. Lett.}, {\bf 83}, 2498-2501.

\bibitem{Madison}
Madison, K.W., Chevy, F., Wohlleben, W., and Dalibard, J. (2000). 
Vortex formation in a stirred Bose-Einstein condensate. 
{\it Phys. Rev. Lett.}, {\bf 84}, 806-809.

\bibitem{Abo}
Abo-Shaeer, J.R., Raman, C., Vogels, J. M., and Ketterle, W. (2001). 
Observation of vortex lattices in Bose-Einstein condensates. 
{\it Science}, {\bf 292}, 476-479.

\bibitem{Madison01}
Madison, K. W., Chevy, F., Bretin, V., Dalibard, J. (2001). 
Stationary states of a rotating Bose-Einstein condensate: Routes to vortex nucleation.
{\it Phys. Rev. Lett.}, {\bf 86}, 4443-4446.

\bibitem{Hodby}
Hodby, E., Hechenblaikner, G., Hopkins, S. A., Marag\'{o}, O. M., and Foot, C. J. (2002). 
Vortex nucleation in Bose-Einstein condensates in an oblate, purely magnetic potential. 
{\it Phys. Rev. Lett.}, {\bf 88}, 010405(1-4).

\bibitem{Chevy}
Chevy, F., Madison, K.W., and Dalibard, J. (2000). 
Measurement of the angular momentum of a rotating Bose-Einstein condensate. 
{\it Phys. Rev. Lett.}, {\bf 85}, 2223-2227.

\bibitem{Haljan01}
Haljan, P. C., Anderson, B. P., Coddington, I., and Cornell, E. A. (2001). 
Use of surface-wave spectroscopy to characterize tilt modes of a vortex in a Bose-Einstein condensate. 
{\it Phys. Rev. Lett.}, {\bf 86}, 2922-2925.

\bibitem{Bretin03}
Bretin, V., Rosenbusch, P., Chevy, F., Shlyapnikov, G.V., and Dalibard, J. (2003). 
Quadrupole oscillation of a single-vortex Bose-Einstein condensate: Evidence for Kelvin modes. 
{\it Phys. Rev. Lett.}, {\bf 90}, 100403(1-4).

\bibitem{Hodby2}
Hodby, E., Hopkins, S.A., Hechenblaikner, G., Smith, N.L., and Foot, C.J. (2003). 
Experimental observation of a superfluid gyroscope in a dilute Bose-Einstein condensate. 
{\it Phys. Rev. Lett.}, {\bf 91}, 090403(1-4).

\bibitem{Anderson3}
Anderson, M.F., Ryu, C., Clad\'{e}, P., Natarajan, V., Vaziri, A., Helmerson, K., and Phillips, W.D. 
(2006). 
Quantized rotation of atoms from photons with orbital angular momentum. 
{\it Phys. Rev. Lett.}, {\bf 97}, 170406(1-4).

\bibitem{Haljan}
Haljan, P.C., Coddington, I., Engels, P., and Cornell, E.A. (2001). 
Driving Bose-Einstein-Condensate vorticity with a rotating normal cloud. 
{\it Phys. Rev. Lett.}, {\bf 87}, 210403(1-4).

\bibitem{Bretin04}
Bretin, V., Stock, S., Seurin, Y., and Dalibard, J. (2004). 
Fast rotation of a Bose-Einstein condensate. 
{\it Phys. Rev. Lett.}, {\bf 92}, 050403(1-4).

\bibitem{Cooperrev}
Cooper, N. R. (2008). 
Rapidly rotating atomic gases.
{\it Adv. Phys.}, {\bf 57}, 539-616.

\bibitem{Viefersrev}
Viefers, S. (2008). 
Quantum Hall physics in rotating Bose-Einstein condensates.
{\it J. Phys.: Condens. Matter}, {\bf 20}, 123202(1-14).

\bibitem{Bloch}
Bloch, I., Dalibard, J. and Zwerger W. (2008). 
Many-body physics with ultracold gases.
{\it Rev. Mod. Phys.}, {\bf 80}, 885-964.

\bibitem{Lin}
Lin, Y.-J., Compton, R. L., Perry, A. R., Phillips, W. D., Porto, J. V., and Spielman I. B. 
(2009). 
Bose-Einstein Condensate in a Uniform Light-Induced Vector Potential. 
{\it Phys. Rev. Lett.}, {\bf 102}, 130401(1-4). 

\bibitem{Lin2}
Lin, Y.-J., Compton, R. L., Jim\'{e}nez-Garc\'{i}a, K., Porto, J. V., and Spielman, I. B. (2009). 
Synthetic magnetic fields for ultracold neutral atoms. 
{\it Nature (London)}, {\bf 462}, 628-632.

\bibitem{Inouye2}
Inouye, S., Gupta, S., Rosenband, T., Chikkatur, A. P., G?rlitz, A., Gustavson, T. L., 
Leanhardt, A. E., Pritchard, D. E., and Ketterle W. (2001). 
Observation of vortex phase singularities in Bose-Einstein condensates. 
{\it Phys. Rev. Lett.} 87, 080402(1-4).

\bibitem{Chevy2}
Chevy, F., Madison, K. W., Bretin, V., and Dalibard, J. (2001). 
Interferometric detection of a single vortex in a dilute Bose-Einstein condensate. 
{\it Phys. Rev. A} 64, 031601(R)(1-4).

\bibitem{Bolda}
Bolda, E. L., and Walls, D. F. (1998). 
Detection of vorticity in Bose-Einstein condensed gases by matter-wave interference. 
{\it Phys. Rev. Lett.} 81, 5477-5480.

\bibitem{Blakie}
Blakie, P. B. and Ballagh, R. J. (2001). 
Spatially selective Bragg scattering: A signature for vortices in Bose-Einstein condensates. 
{\it Phys. Rev. Lett.} 86, 3930-3933. 

\bibitem{Muniz}
Muniz, S. R., Naik, D. S., and Raman, C. (2006). 
Bragg spectroscopy of vortex lattices in Bose-Einstein condensates. 
{\it Phys. Rev. A} 73, 041605(R)(1-4). 

\bibitem{Myatt}
Myatt, C. J., Burt, E. A., Ghrist, R. W., Cornell, E. A., and Wieman, C. E. (1997). 
Production of two overlapping Bose-Einstein condensates by sympathetic cooling. 
{\it Phys. Rev. Lett}., {\bf 78}, 586-589.

\bibitem{Hall1}
Hall, D. S., Matthews, M. R., Ensher, J. R., Wieman, C. E., and Cornell, E. A. (1998). 
Dynamics of component separation in a binary mixture of Bose-Einstein condensates. 
{\it Phys. Rev. Lett.}, {\bf 81}, 1539-1542. 

\bibitem{Stenger}
Stenger, J., Inouye, S., Stamper-Kurn, D. M., Miesner, H. -J., Chikkatur, A. P., and Ketterle, W. (1998). 
Spin domains in ground-state Bose-Einstein condensates.
{\it Nature (London)}, {\bf 396}, 345-348.

\bibitem{Barrett}
Barrett, M. D., Sauer, J. A., and Chapman, M. S. (2001). 
All-optical formation of an atomic Bose-Einstein condensate. 
{\it Phys. Rev. Lett.}, {\bf 87}, 010404(1-4).

\bibitem{Modugno2}
Modugno, G., Modugno, M., Riboli, F., Roati, G., and Inguscio, M. (2002). 
Two atomic species superfluid. 
{\it Phys. Rev. Lett.}, {\bf 89}, 190404(1-4). 

\bibitem{Schmaljohann}
Schmaljohann, H., Erhard, M., Kronj\"{a}ger, J., Kottke, M., Staa, S. van, Cacciapuoti, L., Arlt, J. J., 
Bongs, K., and Sengstock, K. (2004). 
Dynamics of F=2 spinor Bose-Einstein condensates. 
{\it Phys. Rev. Lett.}, {\bf 92}, 040402(1-4).

\bibitem{Chang}
Chang, M.-S., Hamley, C. D., Barrett, M.D., Sauer, J. A., Fortier, K. M., Zhang, W., You, L., 
and Chapman, M. S. (2004). 
Observation of spinor dynamics in optically trapped Rb-87 Bose-Einstein condensates. 
{\it Phys. Rev. Lett.}, {\bf 92}, 140403(1-4).

\bibitem{Kuwamoto}
Kuwamoto, T., Araki, K., Eno, T., and Hirano, T. (2004). 
Magnetic field dependence of the dynamics of Rb-87 spin-2 Bose-Einstein condensates. 
{\it Phys. Rev. A}, {\bf 69}, 063604(1-6).

\bibitem{Thalhammer}
Thalhammer, G., Barontini, G., De Sarlo, L., Catani, J., Minardi, F., 
and Inguscio, M. (2008). 
Double species Bose-Einstein condensate with tunable interspecies interactions. 
{\it Phys. Rev. Lett}, {\bf 100}, 210402(1-4).

\bibitem{Papp2}
Papp, S. B., Pino, J.M., and Wieman, C.E. (2008). 
Tunable miscibility in a dual-species Bose-Einstein condensate. 
{\it Phys. Rev. Lett}, {\bf 101}, 040402(1-4).

\bibitem{Vollhardt}
Vollhardt D., and Woelfle, P. (1990). 
{\it The Superfluid Phases of Helium 3}. Taylor and Francis, London.

\bibitem{supercond}
Bennemann, K. H. and Ketterson, J. B. eds. (2008).
{\it Superconductivity} Volume 1: Conventional and High Temperature Superconductors.
Springer-Verlag.

\bibitem{Girvin}
Girvin, S.M. (1998). {\it The Quantum Hall effect: Novel Excitations and Broken Symmtries, 
Proceedings of the Les Houches Summer School of Theoretical Physics}, 
Springer Verlag, Berlin, and Les Editions de Physique, Paris.

\bibitem{Volovik}
Volovik, G.E. (2003). {\it The Universe in a Helium Droplet}. Oxford University Press, Oxford.

\bibitem{Inouye}
Inouye, S., Andrews, M. R., Stenger, J., Miesner, H. -J., 
Stamper-Kurn, D. M., and Ketterle, W. (1998). 
Observation of Feshbach resonances in a Bose-Einstein condensate. 
{\it Nature (London)}, {\bf 392}, 151-154. 

\bibitem{Donley}
Donley, E. A., Claussen, R. R., Cornish, S. L., Roberts, J. L., Cornell, E. A., and Wieman, C. E. (2001). 
Dynamics of collapsing and exploding Bose-Einstein condensates. 
{\it Nature (London)}, {\bf 412}, 295-299.

\bibitem{Papp}
Papp, S. B., Pino, J. M., Wild, R. J., Ronen, S., Wieman, C. E., Jin, D. E., and Cornell, E. A. (2008).
Bragg spectroscopy of a strongly interacting Rb-85 Bose-Einstein condensate. 
{\it Phys. Rev. Lett.}, {\bf 101}, 135301(1-4).

\bibitem{Lahaye}
Lahaye, T., Koch, T., Frohlich, B., Fattori, M., Metz, J., Griesmaier, A., Giovanazzi, S. and Pfau, T. (2007). 
Strong dipolar effects in a quantum ferrofluid. 
{\it Nature (London)}, {\bf 448}, 672-675.

\bibitem{Donley2}
Donley, E.A., Claussen, N.R., Thompson, S.T. and Wieman, C.E. (2002). 
Atom-molecule coherence in a Bose-Einstein condensate. 
{\it Nature (London)}, {\bf 417}, 529-533.

\bibitem{Inguscio}
Inguscio, M., Ketterle, W., and Salomon, C. (2007), Eds,
{\it Ultra-cold Fermi Gases}, Proc. Int. School ``Enrico Fermi," 
Cource CLXIV (IOS Press, Amsterdam).

\bibitem{Chin}
Chin, C., Grimm, R., Julienne, P. and Tiesinga, E. (2010) 
Feshbach Resonances in Ultracold Gases.
Rev. Mod. Phys., to be published.

\bibitem{Fetterrev}
Fetter, A. L. and Svidzinsky, A. A. (2001). 
Vortices in a trapped dilute Bose-Einstein condensate.
{\it J. Phys. Condens. Matter}, {\bf 13}, R135-R194.

\bibitem{Kasamatsurev}
Kasamatsu, K. and Tsubota, M. (2008). 
Quantized vortices in atomic Bose-Einstein condensates. 
{\it Prog. Low Temp. Phys.}, {\bf 16}, (Halperin, W. P. and Tsubota, M. eds.) 
Amsterdam. Elsevier. 351-403. 

\bibitem{Fetterrev2}
Fetter, A. L. (2009). 
Rotating trapped Bose-Einstein condensates.
{\it Rev. Mod. Phys.}, {\bf 81}, 647-691.

\bibitem{Baym}
Baym, G., Pethick, C. J. (1995). 
Ground-state properties of magnetically trapped Bose-condensed rubidium gas. 
{\it Phys. Rev. Lett.}, {\bf 76}, 6-9.

\bibitem{Raman2}
Raman, C., Abo-Shaeer, J. R., Vogels, J. M., Xu, K., and Ketterle, W. (2001). 
Vortex nucleation in a stirred Bose-Einstein condensate. 
{\it Phys. Rev. Lett.}, {\bf 87}, 210402(1-4).

\bibitem{Coddington2}
Coddington, I., Haljan, P. C., Engels, P., Schweikhard, V., Tung, S., and Cornell, E. A. (2004). 
Experimental studies of equilibrium vortex properties in a Bose-condensed gas. 
{\it Phys. Rev. A}, {\bf 70}, 063607(1-11).

\bibitem{Engels}
Engels, P., Coddington, I., Haljan, P. C., and Cornell, E. A. (2002). 
Nonequilibrium effects of anisotropic compression applied to vortex lattices in Bose-Einstein condensates. 
{\it Phys. Rev. Lett.}, {\bf 89}, 100403(1-4).

\bibitem{Coddington}
Coddington, I., Engels, P., Schweikhard, V., and Cornell, E. A. (2003). 
Observation of Tkachenko oscillations in rapidly rotating Bose-Einstein condensates. 
{\it Phys. Rev. Lett.}, {\bf 91}, 100402(1-4).

\bibitem{Engels2}
Engels, P., Coddington, I., Haljan, P. C., Schweikhard, V., and Cornell, E. A. (2003). 
Observation of long-lived vortex aggregates in rapidly rotating Bose-Einstein condensates. 
{\it Phys. Rev. Lett.}, {\bf 90}, 170405(1-4).

\bibitem{Schweikhard}
Schweikhard, V., Coddington, I., Engels, P., Mogendorff, V. P., and Cornell, E. A. (2004). 
Rapidly rotating Bose-Einstein condensates in and near the lowest Landau level. 
{\it Phys. Rev. Lett.}, {\bf 90}, 170405(1-4).

\bibitem{Garcia-Ripoll01}
Garc\'{i}a-Ripoll, J.J., and P\'{e}rez-Garc\'{i}a, V.M. (2001). 
Vortex bending and tightly packed vortex lattices in Bose-Einstein condensates. 
{\it Phys. Rev. A}, {\bf 64}, 053611(1-7).

\bibitem{Modugno03}
Modugno, M., Pricoupenko, L., and Castin, Y. (2003) 
Bose-Einstein condensates with a bent vortex in rotating traps.
{\it Euro. Phys. J. D}, {\bf 22}, 235-257.

\bibitem{Aftalion03}
Aftalion, A., and Danaila, I. (2003). 
Three-dimensional vortex configurations in a rotating Bose-Einstein condensate.
{\it Phys. Rev. A}, {\bf 68}, 023603(1-6).

\bibitem{Rosenbusch02}
Rosenbusch, P., Bretin, V., and Dalibard, J. (2002). 
Dynamics of a single vortex line in a Bose-Einstein condensate. 
{\it Phys. Rev. Lett.}, {\bf 89}, 200403(1-4).

\bibitem{Dodd97}
Dodd, R. J., Burnett, K., Edwards, M., and Clark, C. W. (1997). 
Excitation spectroscopy of vortex states in dilute Bose-Einstein condensed gases. 
{\it Phys. Rev. A}, {\bf 56}, 587-590.

\bibitem{Rokhsar97}
Rokhsar, D. S. (1997). 
Vortex stability and persistent currents in trapped Bose gases. 
{\it Phys. Rev. Lett.}, {\bf 79}, 2164-2167.

\bibitem{Lundh}
Lundh, E., Pethick, C. J., and Smith, H. (1997). 
Zero-temperature properties of a trapped Bose-condensed gas: Beyond the Thomas-Fermi approximation. 
{\it Phys. Rev. A}, {\bf 55}, 2126-2131.

\bibitem{Svidzinsky}
Svidzinsky, A. A., and Fetter, A. L. (2000). 
Stability of a vortex in a trapped Bose-Einstein condensate. 
{\it Phys. Rev. Lett.}, {\bf 84}, 5919-5923. 

\bibitem{Feder}
Feder, D. L., Clark, C. W., and Schneider, B. I. (1999). 
Vortex stability of interacting Bose-Einstein condensates confined in anisotropic harmonic traps. 
{\it Phys. Rev. Lett.}, {\bf 82}, 4956-4959.

\bibitem{Isoshima}
Isoshima, T., and Machida, K. (1999). 
Instability of the nonvortex state toward a quantized vortex in a Bose-Einstein condensate under external rotation. 
{\it Phys. Rev. A}, {\bf 60}, 3313-3316.

\bibitem{Feder3}
Feder, D. L., Clark, C. W., and Schneider, B. I. (1999). 
Nucleation of vortex arrays in rotating anisotropic Bose-Einstein condensates. 
{\it Phys. Rev. A}, {\bf 61}, 011601(R)(1-4).

\bibitem{Dalfovo2}
Dalfovo, F., and Stringari, S. (2000). 
Shape deformations and angular-momentum transfer in trapped Bose-Einstein condensates. 
{\it  Phys. Rev. A}, {\bf 63}, 011601(R)(1-4).

\bibitem{Anglin}
Anglin, J. R. (2001). 
Local vortex generation and the surface mode spectrum of large Bose-Einstein condensates. 
{\it  Phys. Rev. Lett.}, {\bf 87}, 240401(1-4).

\bibitem{Williams2}
Williams, J.E., Zaremba, E., Jackson, B., Nikuni, T., and Griffin, A. (2002). 
Dynamical instability of a condensate induced by a rotating thermal gas.
{\it  Phys. Rev. Lett.}, {\bf 88}, 070401(1-4).

\bibitem{Simula2}
Simula, T. P., Virtanen, S. M. M., and Salomaa, M. M. (2002). 
Surface modes and vortex formation in dilute Bose-Einstein condensates at finite temperatures. 
{\it  Phys. Rev. A}, {\bf 66}, 035601(1-4).

\bibitem{Stringari}
Stringari, S. (1996). 
Collective excitations of a trapped bose-condensed gas. 
{\it Phys. Rev. Lett.}, {\bf 77}, 2360-2363.

\bibitem{Sinha}
Sinha, S., and Castin, Y. (2001). 
Dynamic instability of a rotating Bose-Einstein condensate. 
{\it Phys. Rev. Lett.}, {\bf 87}, 190402(1-4).

\bibitem{Reinisch}
Reinisch, G. (2007). 
Vortex-nucleating Zeeman resonance in axisymmetric rotating Bose-Einstein condensates. 
{\it  Phys. Rev. Lett}, {\bf 99}, 120402(1-4).

\bibitem{Tsubota}
Tsubota, M., Kasamatsu, K., and Ueda, M. (2002). 
Vortex lattice formation in a rotating Bose-Einstein condensate. 
{\it Phys. Rev. A}, {\bf 65}, 023603(1-4).

\bibitem{Kasamatsu}
Kasamatsu, K., Tsubota, M., and Ueda, M. (2003). 
Nonlinear dynamics of vortex lattice formation in a rotating Bose-Einstein condensate. 
{\it Phys. Rev. A}, {\bf 67}, 033610(1-14).

\bibitem{Penckwitt}
Penckwitt, A. A., Ballagh, R. J., and Gardiner, C. W. (2002). 
Nucleation, growth, and stabilization of Bose-Einstein condensate vortex lattices. 
{\it Phys. Rev. Lett.}, {\bf 89}, 260402(1-4).

\bibitem{Lobo}
Lobo, C., Sinatra, A., and Castin, Y. (2004). 
Vortex lattice formation in Bose-Einstein condensates. 
{\it Phys. Rev. Lett.}, {\bf 92}, 020403(1-4).

\bibitem{Kasamatsu2}
Kasamatsu, K., Machida, M., Sasa, N., and Tsubota, M. (2005). 
Three-dimensional dynamics of vortex-lattice formation in Bose-Einstein condensates. 
{\it Phys. Rev. A}, {\bf 71}, 063616(1-5).

\bibitem{Parker}
Parker, N. G., and Adams, C. S. (2005) 
Emergence and decay of turbulence in stirred atomic bose-einstein condensates. 
{\it Phys. Rev. Lett.}, {\bf 95}, 145301(1-4).

\bibitem{Wright}
Wright, T. M., Ballagh, R. J., Bradley, A. S., Blakie, P. B., and Gardiner, C. W. (2008). 
Dynamical thermalization and vortex formation in stirred two-dimensional Bose-Einstein condensates. 
{\it Phys. Rev. A}, {\bf 78}, 063601(1-22).

\bibitem{Jackson}
Jackson, B., McCann, J. F., and Adams, C. S. (1999). 
Vortex line and ring dynamics in trapped Bose-Einstein condensates. 
{\it Phys. Rev. A}, {\bf 61}, 013604(1-7)

\bibitem{McGee}
McGee, S. A., and Holland, M. J. (2001). 
Rotational dynamics of vortices in confined Bose-Einstein condensates. 
{\it Phys. Rev. A}, {\bf 63}, 043608(1-6).

\bibitem{Fedichev}
Fedichev, P. O., and Shlyapnikov, G. V. (1999). 
Dissipative dynamics of a vortex state in a trapped Bose-condensed gas. 
{\it Phys. Rev. A}, {\bf 60}, R1779-R1782.

\bibitem{Anderson00}
Anderson, B. P., Haljan, P. C., Wieman, C. E., and Cornell, E. A. (2000). 
Vortex precession in Bose-Einstein condensates: Observations with filled and empty cores. 
{\it Phys. Rev. Lett.}, {\bf 85}, 2857(1-4).

\bibitem{Feder01}
Feder, D. L., Svidzinsky, A. A. Fetter, A. L., and Clark, C. W. (2001). 
Anomalous modes drive vortex dynamics in confined Bose-Einstein condensates. 
{\it Phys. Rev. Lett.}, {\bf 86}, 564(1-4).

\bibitem{Simula08}
Simula, T. P., Mizushima, T., and Machida, K. (2008). 
Kelvin waves of quantized vortex lines in trapped Bose-Einstein condensates. 
{\it Phys. Rev. Lett.}, {\bf 101}, 020402(1-4).

\bibitem{Mizushima03}
Mizushima, T., Ichioka, M., and Machida, K. (2003). 
Beliaev damping and Kelvin mode spectroscopy of a Bose-Einstein condensate in the presence of a vortex line. 
{\it Phys. Rev. Lett.}, {\bf 90}, 180401(1-4).

\bibitem{Tsubota03}
Tsubota, M., Araki, T., and Barenghi, C. F. (2003)
Rotating superfluid turbulence. 
{\it Phys. Rev. Lett.}, {\bf 90}, 205301(1-4).

\bibitem{Takeuchi09}
Takeuchi, H., Kasamatsu, K., and Tsubota, M. (2009). 
Spontaneous radiation and amplification of Kelvin waves on quantized vortices in Bose-Einstein condensates. 
{\it Phys. Rev. A}, {\bf 79}, 033619(1-5).

\bibitem{Leanhardt}
Leanhardt, A. E., G\"{o}rlitz, A., Chikkatur, A. P., Kielpinski, D., Shin, Y., Pritchard, D. E., and Ketterle, W. 
(2002). 
Imprinting vortices in a Bose-Einstein condensate using topological phases. 
{\it Phys. Rev. Lett.}, {\bf 89}, 190403(1-4).

\bibitem{Shin}
Shin, Y., Saba, M., Vengalattore, M., Pasquini, T. A., Sanner, C., Leanhardt, A. E., Prentiss, M.,  
Pritchard, D. E., and Ketterle, W. (2004). 
Dynamical instability of a doubly quantized vortex in a Bose-Einstein condensate. 
{\it Phys. Rev. Lett.}, {\bf 93}, 160406(1-4).

\bibitem{Pu99}
Pu, H., Law, C. K., Eberly, J. H., and Bigelow, N. P. (1999). 
Coherent disintegration and stability of vortices in trapped Bose condensates. 
{\it Phys. Rev. A}, {\bf 59}, 1533-1537.

\bibitem{Mottonen03}
M\"{o}tt\"{o}nen, M., Mizushima, T., Isoshima, T., Salomaa, M. M., and Machida, K. (2003). 
Splitting of a doubly quantized vortex through intertwining in Bose-Einstein condensates. 
{\it Phys. Rev. A}, {\bf 68}, 023611(1-4).

\bibitem{Huhtamaki06}
Huhtam\"{a}ki, J. A. M., M\"{o}tt\"{o}nen, M., Isoshima, T., Pietil\"{a}, V., 
and Virtanen, S. M. M. (2006). 
Splitting times of doubly quantized vortices in dilute Bose-Einstein condensates. 
{\it Phys. Rev. Lett.}, {\bf 97}, 110406(1-4).

\bibitem{Mateo06}
Mu\~{n}oz Mateo, A., and Delgado, V. (2006). 
Dynamical evolution of a doubly quantized vortex imprinted in a Bose-Einstein condensate. 
{\it Phys. Rev. Lett.}, {\bf 97}, 180409(1-4).

\bibitem{Isoshima07}
Isoshima, T., Okano, M., Yasuda, H., Kasa, K., Huhtam\"{a}ki, J. A. M., Kumakura, M., 
Takahashi, Y. (2007). 
Spontaneous splitting of a quadruply charged vortex. 
{\it Phys. Rev. Lett.}, {\bf 99}, 200403(1-4).

\bibitem{Kawaguchi04}
Kawaguchi, Y., and Ohmi, T. (2004). 
Splitting instability of a multiply charged vortex in a Bose-Einstein condensate. 
{\it Phys. Rev. A}, {\bf 70}, 043610(1-7).

\bibitem{Yarmchuk79}
Yarmchuk, E. J., Gordon, M. J. V., and Packard, R. E. (1979). 
Observation of stationary vortex arrays in rotating superfluid helium.
{\it Phys. Rev. Lett.}, {\bf 43}, 214-217. 

\bibitem{Abrikosov}
Abrikosov, A. A. (1957). 
On the magnetic properties of superconductors of the second group. 
{\it Sov. Phys. JETP}, {\bf 5}, 1174-1182. 

\bibitem{Ho}
Ho, T. L. (2001). 
Bose-Einstein condensates with large number of vortices. 
{\it Phys. Rev. Lett.}, {\bf 87}, 060403(1-4).

\bibitem{Watanabe}
Watanabe, G., Baym, G., and Pethick, C. J. (2004). 
Landau levels and the Thomas-Fermi structure of rapidly rotating Bose-Einstein condensates. 
{\it Phys. Rev. Lett.}, {\bf 93}, 190401(1-4).

\bibitem{Cooper1}
Cooper, N. R., Komineas, S., and Read, N. (2004). 
Vortex lattices in the lowest Landau level for confined Bose-Einstein condensates. 
{\it Phys. Rev. A}, {\bf 70}, 033604(1-5).

\bibitem{Aftalion4}
Aftalion, A., Blanc, X., and Dalibard, J. (2005). 
Vortex patterns in a fast rotating Bose-Einstein condensate. 
{\it Phys. Rev. A}, {\bf 71}, 023611(1-11).

\bibitem{Sonin2}
Sonin, E. B. (2005). 
Ground state and Tkachenko modes of a rapidly rotating Bose-Einstein condensate in the lowest-Landau-level state. 
{\it Phys. Rev. A}, {\bf 72}, 021606(R)(1-4).

\bibitem{Aftalion5}
Aftalion, A., Blanc, X., and Nier, F. (2006). 
Vortex distribution in the lowest Landau level. 
{\it Phys. Rev. A}, {\bf 73}, 011601(R)(1-4).

\bibitem{Cozzini06}
Cozzini, M., Stringari, S., and Tozzo, C. (2006). 
Vortex lattices in Bose-Einstein condensates: From the Thomas-Fermi regime to the lowest-Landau-level regime. 
{\it Phys. Rev. A}, {\bf 73}, 023615(1-4).

\bibitem{Butts}
Butts, D. A. and Rokhsar, D. S. (1999). 
Predicted signatures of rotating Bose-Einstein condensates. 
{\it Nature (London)}, {\bf 397}, 327-329.

\bibitem{Kavoulakis}
Kavoulakis, G. M., Mottelson, B., and Pethick, C. J. (2000). 
Weakly interacting Bose-Einstein condensates under rotation. 
{\it Phys. Rev. A}, {\bf 62}, 063605(1-10).

\bibitem{Vorov}
Vorov, O. K., Isacker, P. V., Hussein, M. S., and Bartschat, K. (2005). 
Nucleation and growth of vortices in a rotating Bose-Einstein condensate. 
{\it Phys. Rev. Lett.}, {\bf 95}, 023406(1-4).

\bibitem{Cozzini03}{2003}
Cozzini, M., and Stringari, S. (2003). 
Macroscopic dynamics of a Bose-Einstein condensate containing a vortex lattice. 
{\it Phys. Rev. A}, {\bf 67}, 041602(R)(1-4).

\bibitem{Fischer}
Fischer, U. R., and Baym, G. (2003). 
Vortex states of rapidly rotating dilute Bose-Einstein condensates. 
{\it Phys. Rev. Lett.}, {\bf 90}, 140402(1-4). 

\bibitem{Baym3}
Baym, G., and Pethick, C. J. (2004). 
Vortex core structure and global properties of rapidly rotating Bose-Einstein condensates. 
{\it Phys. Rev. A}, {\bf 69}, 043619(1-9).

\bibitem{Watanabe06}
Watanabe, G., Gifford, S. A., Baym, G., and Pethick, C. J. (2006). 
Structure of vortices in rotating Bose-Einstein condensates. 
{\it Phys. Rev. A}, {\bf 74}, 063621(1-9).

\bibitem{Tkachenko2}
Tkachenko, V. K. (1966). 
Stability of vortex lattices.
{\it Sov. Phys. JETP}, {\bf 23}, 1049-1056. 

\bibitem{Baym03}
Baym, G. (2003). 
Tkachenko modes of vortex lattices in rapidly rotating Bose-Einstein condensates. 
{\it Phys. Rev. Lett.}, {\bf 91}, 110402(1-4).

\bibitem{Cozzini04}
Cozzini, M., Pitaevskii, L. P., and Stringari, S. (2004). 
Tkachenko oscillations and the compressibility of a rotating Bose-Einstein condensate. 
{\it Phys. Rev. Lett.}, {\bf 92}, 220401(1-4).

\bibitem{Sonin05}
Sonin, E. B. (2005). 
Continuum theory of Tkachenko modes in rotating Bose-Einstein condensate. 
{\it Phys. Rev. A}, {\bf 71}, 011603(R)(1-4).

\bibitem{Mizushima04}
Mizushima, T., Kawaguchi, Y., Machida, K., Ohmi, T., Isoshima, T., and Salomaa, M. M. (2004). 
Collective oscillations of vortex lattices in rotating Bose-Einstein condensates. 
{\it Phys. Rev. Lett.}, {\bf 92}, 060407(1-4).

\bibitem{Baksmaty04}
Baksmaty, L. O., Woo, S. J., Choi, S., and Bigelow, N. P. (2004). 
Tkachenko waves in rapidly rotating Bose-Einstein condensates. 
{\it Phys. Rev. Lett.}, {\bf 92}, 160405(1-4).

\bibitem{Fetter}
Fetter, A. L. (2001). 
Rotating vortex lattice in a Bose-Einstein condensate trapped in combined quadratic and quartic radial potentials. 
{\it Phys. Rev. A}, {\bf 64}, 063608(1-6).

\bibitem{Lundh2}
Lundh, E. (2002). 
Multiply quantized vortices in trapped Bose-Einstein condensates. 
{\it Phys. Rev. A}, {\bf 65}, 043604(1-6).

\bibitem{Kasamatsu1}
Kasamatsu, K., Tsubota, M., and Ueda, M. (2002). 
Giant hole and circular superflow in a fast rotating Bose-Einstein condensate. 
{\it Phys. Rev. A}, {\bf 66}, 053606(1-4).

\bibitem{Jackson04}
Jackson, A. D., Kavoulakis, G. M., and Lundh, E. (2004). 
Phase diagram of a rotating Bose-Einstein condensate with anharmonic confinement. 
{\it Phys. Rev. A}, {\bf 69}, 053619(1-7).

\bibitem{Fetter05}
Fetter, A. L., Jackson, B., and Stringari, S. (2005). 
Rapid rotation of a Bose-Einstein condensate in a harmonic plus quartic trap. 
{\it Phys. Rev. A}, {\bf 71}, 013605(1-9).

\bibitem{Danaila05}
Danaila, I. (2005). 
Three-dimensional vortex structure of a fast rotating Bose-Einstein condensate with harmonic-plus-quartic confinement. 
{\it Phys. Rev. A}, {\bf 72}, 013605(1-6).

\bibitem{Fu06}
Fu, H., and Zaremba, E. (2006). 
Transition to the giant vortex state in a harmonic-plus-quartic trap. 
{\it Phys. Rev. A}, {\bf 73}, 013614(1-14).

\bibitem{Blanc08}
Blanc, X., and Rougerie, N. (2008). 
Lowest-Landau-level vortex structure of a Bose-Einstein condensate rotating in a harmonic plus quartic trap. 
{\it Phys. Rev. A}, {\bf 77}, 053615(1-8).

\bibitem{Tung06}
Tung, S., Schweikhard, V., and Cornell, E. A. (2006). 
Observation of vortex pinning in Bose-Einstein condensates. 
{\it Phys. Rev. Lett.}, {\bf 97}, 240402(1-4).

\bibitem{Reijnders04}
Reijnders J. W., and Duine, R. A. (2004). 
Pinning of vortices in a Bose-Einstein condensate by an optical lattice. 
{\it Phys. Rev. Lett.}, {\bf 93}, 060401(1-4).

\bibitem{Pu05}
Pu, H., Baksmaty, L. O., Yi, S., and Bigelow, N. P. (2005). 
Structural phase transitions of vortex matter in an optical lattice. 
{\it Phys. Rev. Lett.}, {\bf 94}, 190401(1-4).

\bibitem{09Kasamatsu}
Kasamatsu, K. (2009). 
Uniformly frustrated bosonic Josephson-junction arrays. 
{\it Phys. Rev. A}, {\bf 79}, 021604(R)(1-4).

\bibitem{10Williams}
Williams, R. A., Al-Assam, S., and Foot, C. J. (2010). 
Observation of vortex nucleation in a rotating two-dimensional lattice of Bose-Einstein condensates. 
{\it Phys. Rev. Lett.}, {\bf 104}, 050404(1-4).

\bibitem{Berezinskii}
Berezinskii, V. L. (1972). 
Destruction of long-range order in one-dimensional and two-dimensional systems 
possessing a continuous symmetry group. II. quantum systems
{\it Sov. Phys. JETP}, {\bf 34}, 610-616.

\bibitem{Kosterlitz}
Kosterlitz, J. M., and Thouless, D. J. (1973). 
Ordering, metastability and phase transitions in two-dimensional systems.
{\it J. Phys. C}, {\bf 6}, 1181?1203.

\bibitem{Stock05}
Stock, S., Hadzibabic, Z., Battelier, B., Cheneau, M., and Dalibard, J. (2005). 
Observation of phase defects in quasi-two-dimensional Bose-Einstein condensates. 
{\it Phys. Rev. Lett.}, {\bf 95}, 190403(1-4).

\bibitem{Hadzibabic06}
Hadzibabic, Z., Kr\"{u}ger, P., Cheneau, M., Battelier, B., and Dalibard, J. (2006). 
Berezinskii-Kosterlitz-Thouless crossover in a trapped atomic gas. 
{\it Nature (London)}, {\bf 441}, 1118-1121.

\bibitem{Polkovnikov}
Polkovnikov, A., Altman, E., and Demler, E. (2006). 
Interference between independent fluctuating condensates. 
{\it Proc. Natl. Acad. Sci. USA}, {\bf 103}, 6125-6129.

\bibitem{Simula3}
Simula, T. P., and Blakie, P. B. (2006). 
Thermal activation of vortex-antivortex pairs in quasi-two-dimensional Bose-Einstein condensates. 
{\it Phys. Rev. Lett.}, {\bf 96}, 020404(1-4).

\bibitem{Kibble76}
Kibble, T. W. B. (1976). 
Topology of cosmic domains and strings. 
{\it J. Phys. A}, {\bf 9}, 1387-1398.

\bibitem{Zurek85}
Zurek, W. H. (1985). 
Cosmological experiments in superfluid helium?
{\it Nature (London)}, {\bf 317}, 505 - 508.

\bibitem{Bauerle96}
B\"{a}uerle, C., Bunkov, Y. M. Fisher, S. N., Godfrin, H., and Pickett, G. R. (1996). 
Laboratory simulation of cosmic string formation in the early universe 
using superfluid $^{3}$He.
{\it Nature (London)}, {\bf 382}, 332-334.

\bibitem{Ruutu96}
Ruutu, V. M. H., Eltsov, V. B., Gill, A. J., Kibble, T. W. B., Krusius, M., Makhlin, Y. G., Pla\c{c}ais, B., 
Volovik, G. E., Xu, W. (1996). 
Vortex formation in neutron-irradiated superfluid $^3$He as an analogue of cosmological defect formation.
{\it Nature (London)}, {\bf 382}, 334-336.

\bibitem{Weiler08}
Weiler, C. N., Neely, T. W., Scherer, D. R., Bradley, A. S., Davis, M. J., and Anderson, B. P. (2008). 
Spontaneous vortices in the formation of Bose-Einstein condensates. 
{\it Nature (London)}, {\bf 455}, 948-951.

\bibitem{Gardiner02}
Gardiner, C. W., Anglin, J. R., and Fudge, T. I. A. (2002). 
The stochastic Gross-Pitaevskii equation. 
{\it J. Phys. B}, {\bf 35}, 1555-1582.

\bibitem{Gardiner03}
Gardiner, C. W. and Davis, M. J. (2003). 
The stochastic Gross-Pitaevskii equation: II. 
{\it J. Phys. B}, {\bf 36}, 4731-4753.

\bibitem{Bradley08}
Bradley, A. S., Gardiner, C. W. Davis, M. J. (2008). 
Bose-Einstein condensation from a rotating thermal cloud: Vortex nucleation and lattice formation. 
{\it Phys. Rev. A}, {\bf 77}, 033616(1-14).

\bibitem{Scherer}
Scherer, D. R., Weiler, C. N., Neely, T. W., and Anderson, B. P. (2007). 
Vortex formation by merging of multiple trapped Bose-Einstein condensates. 
{\it Phys. Rev. Lett.}, {\bf 98}, 110402(1-4).

\bibitem{Kasamatsu02}
Kasamatsu, K., and Tsubota, M. (2002). 
Vortex generation in cyclically coupled superfluids and the Kibble-Zurek mechanism. 
{\it J. Low Temp. Phys.}, {\bf 126}, 315-320.

\bibitem{Feder00}
Feder, D. L., Pindzola, M. S., Collins, L. A., Schneider, B. I., and Clark, C. W. (2000). 
Dark-soliton states of Bose-Einstein condensates in anisotropic traps. 
{\it Phys. Rev. A}, {\bf 62}, 053606(1-11). 

\bibitem{Carretero08}
Carretero-Gonz\'{a}lez, R., Whitaker, N., Kevrekidis, P. G., and Frantzeskakis, D. J. (2008). 
Vortex structures formed by the interference of sliced condensates. 
{\it Phys. Rev. A}, {\bf 77}, 023605(1-8). 

\bibitem{Carretero}
Carretero-Gonz\'{a}lez, R., Anderson, B. P., Kevrekidis, P. G., Frantzeskakis, D. J., and Weiler, C.N. (2008).
Dynamics of vortex formation in merging Bose-Einstein condensate fragments. 
{\it Phys. Rev. A}, {\bf 77}, 033625(1-11). 

\bibitem{Ruben}
Ruben, G., Paganin, D. M., and Morgan, M. J. (2008). 
Vortex-lattice formation and melting in a nonrotating Bose-Einstein condensate. 
{\it Phys. Rev. A}, {\bf 78}, 013631(1-9).

\bibitem{Leanhardt2}
Leanhardt, A.E., Shin, Y., Kielpinski, D., Pritchard, D. E., and Ketterle, W. (2003). 
Coreless vortex formation in a spinor Bose-Einstein condensate. 
{\it Phys. Rev. Lett.}, {\bf 90}, 140403(1-4).

\bibitem{Schweikhard2}
Schweikhard, V., Coddington, I., Engels, P., Tung, S., and Cornell, E. A. (2004). 
Vortex-lattice dynamics in rotating spinor Bose-Einstein condensates. 
{\it Phys. Rev. Lett.}, {\bf 93}, 210403(1-4).

\bibitem{Sadler}
Sadler, L. E., Higbie, J. M., Leslie, S. R., Vengalattore, M., and Stamper-Kurn, D. M. (2006). 
Spontaneous symmetry breaking in a quenched ferromagnetic spinor Bose-Einstein condensate. 
{\it Nature (London)}, {\bf 443}, 312-315.

\bibitem{Leslie09}
Leslie, L. S., Hansen, A., Wright, K. C., Deutsch, B. M., and Bigelow, N. P. (2009). 
Creation and Detection of Skyrmions in a Bose-Einstein Condensate. 
{\it Phys. Rev. Lett.}, {\bf 103}, 250401(1-4).

\bibitem{Maddaloni00}
Maddaloni, P., Modugno, M., Fort, C., Minardi, F., and Inguscio, M. (2000). 
Collective oscillations of two colliding Bose-Einstein condensates. 
{\it Phys. Rev. Lett.}, {\bf 85}, 2413-2417.

\bibitem{Mertes07}
Mertes, K. M., Merrill, J. W., Carretero-Gonzalez, R., Frantzeskakis, 
D. J., Kevrekidis, P. G., and Hall, D. S. (2007). 
Nonequilibrium dynamics and superfluid ring excitations in binary Bose-Einstein condensates. 
{\it Phys. Rev. Lett.}, {\bf 99}, 190402(1-4).

\bibitem{Miesner99}
Miesner, H.-J., Stamper-Kurn, D. M., Stenger, J., Inouye, S., Chikkatur, A. P., and Ketterle W. (1999). 
Observation of metastable states in spinor Bose-Einstein condensates. 
{\it Phys. Rev. Lett.}, {\bf 82}, 2228-2231.

\bibitem{Ho2}
Ho, T. -L., and Shenoy, V. B. (1996). 
Binary mixtures of bose condensates of alkali atoms. 
{\it Phys. Rev. Lett.}, {\bf 77}, 3276-3279.

\bibitem{Tim98}
Timmermans, E. (1998). 
Phase separation of Bose-Einstein condensates. 
{\it Phys. Rev. Lett.}, {\bf 81}, 5718-5721. 

\bibitem{Ao98}
Ao, P., and Chui, S. T. (1998). 
Binary Bose-Einstein condensate mixtures in weakly and strongly segregated phases. 
{\it Phys. Rev. A}, {\bf 58}, 4836-4840.

\bibitem{Williams99}
Williams, J. E.  and Holland, M. J. (1999). 
Preparing topological states of a Bose-Einstein condensate. 
{\it Nature (London)}, {\bf 401}, 568-572. 

\bibitem{Ripoll00}
Garc\'{i}a-Ripoll J. J., and P\'{e}rez-Garc\'{i}a, V. M. (2000). 
Stable and unstable vortices in multicomponent Bose-Einstein condensates. 
{\it Phys. Rev. Lett.}, {\bf 84}, 4264-4267.

\bibitem{Skrybin00}
Skryabin, D.V. (2000). 
Instabilities of vortices in a binary mixture of trapped Bose-Einstein condensates: Role of collective excitations with positive and negative energies. 
{\it Phys. Rev. A}, {\bf 63}, 013602(1-10). 

\bibitem{Leonhardt00}
Leonhardt, U. and Volovik, G. E. (2000). 
How to create an Alice string (half-quantum vortex) in a vector Bose-Einstein condensate. 
{\it JETP Letters}, {\bf 72}, 46-48. 

\bibitem{Mueller04}
Mueller, E. J. (2004). 
Spin textures in slowly rotating Bose-Einstein condensates. 
{\it Phys. Rev. A}, {\bf 69}, 033606(1-14). 

\bibitem{Kasamatsu05}
Kasamatsu, K., Tsubota, M., Ueda, M., 
Spin textures in rotating two-component Bose-Einstein condensates
(2005) {\it Phys. Rev. A}, {\bf 71}, 043611(1-14). 

\bibitem{Anderson77}
Anderson, P. W., and Toulouse, G. (1977). 
Phase slippage without vortex cores: vortex textures in superfluid $^{3}$He. 
{\it Phys. Rev. Lett.}, {\bf 38}, 508-511. 

\bibitem{Mermin76}
Mermin, N. D., and Ho, T. L. (1976). 
Circulation and angular momentum in the a phase of superfluid Helium-3. 
{\it Phys. Rev. Lett.}, {\bf 36}, 594?597.

\bibitem{Rejanbook}
Rajaraman, R. (1989). {\it Soliton and Instantons}, North-Holland, Amsterdam.

\bibitem{Matthews99}
Matthews, M. R., Anderson, B. P., Haljan, P. C., Hall, D. S., Holland, M. J., Williams, J. E., 
Wieman, C. E., and Cornell, E. A. (1999). 
Watching a superfluid untwist itself: Recurrence of Rabi oscillations in a Bose-Einstein condensate. 
{\it Phys. Rev. Lett.}, {\bf 83}, 3358-3361.

\bibitem{Kasamatsu04}
Kasamatsu, K., Tsubota, M., and Ueda, M. (2004). 
Vortex molecules in coherently coupled two-component Bose-Einstein condensates. 
{\it Phys. Rev. Lett.}, {\bf 93}, 250406(1-4).

\bibitem{Son02}
Son, D. T., and Stephanov, M. A. (2002). 
Domain walls of relative phase in two-component Bose-Einstein condensates. 
{\it Phys. Rev. A}, {\bf 65}, 063621(1-10).

\bibitem{Mueller02}
Mueller, E. J., and Ho, T.-L. (2002). 
Two-component Bose-Einstein condensates with a large number of vortices. 
{\it Phys. Rev. Lett.}, {\bf 88}, 180403(1-4). 

\bibitem{Keceli06}
Ke\c{c}eli, M., and Oktel, M. \"{O}. (2006). 
Tkachenko modes and structural phase transitions of the vortex lattice of a two-component Bose-Einstein condensate. 
{\it Phys. Rev. A}, {\bf 73}, 023611(1-17).

\bibitem{Kasamatsu03}
Kasamatsu, K., Tsubota, M., and Ueda, M. (2003). 
Vortex phase diagram in rotating two-component Bose-Einstein condensates. 
{\it Phys. Rev. Lett.}, {\bf 91}, 150406(1-4). 

\bibitem{Woo07}
Woo, S. J., Choi, S., Baksmaty, L. O., and Bigelow, N. P, (2007). 
Dynamics of vortex matter in rotating two-species Bose-Einstein condensates. 
{\it Phys. Rev. A}, {\bf 75}, 031604(R)(1-4). 

\bibitem{Kasamatsu09}
Kasamatsu, K., and Tsubota, M. (2009). 
Vortex sheet in rotating two-component Bose-Einstein condensates. 
{\it Phys. Rev. A}, {\bf 79}, 023606(1-7). 

\bibitem{Ueda10}
Ueda, M., and Kawaguchi, Y. (2010). 
Spinor Bose-Einstein condensates. 
{\it arXiv:1001.2072}. 

\bibitem{Ohmi98}
Ohmi, T. and Machida, K. (1998). 
Bose-Einstein condensation with internal degrees of freedom in alkali atom gases. 
{\it J. Phys. Soc. Jpn.}, {\bf 67}, 1822-1825. 

\bibitem{Ho98}
Ho, T. L. (1998). 
Spinor Bose condensates in optical traps. 
{\it Phys. Rev. Lett.}, {\bf 81}, 742-745. 

\bibitem{Ho96}
Ho, T. L., and Shenoy, V. B. (1996). 
Local spin-gauge symmetry of the Bose-Einstein condensates in atomic gases. 
{\it Phys. Rev. Lett.}, {\bf 77}, 2595-2599. 

\bibitem{Zhou01}
Zhou, F. (2001). 
Spin correlation and discrete symmetry in spinor Bose-Einstein condensates. 
{\it Phys. Rev. Lett.}, {\bf 87}, 080401(1-4). 

\bibitem{Murata07}
Murata, K., Saito, H., and Ueda, M. (2007). 
Broken-axisymmetry phase of a spin-1 ferromagnetic Bose-Einstein condensate. 
{\it Phys. Rev. A}, {\bf 75}, 013607(1-10). 

\bibitem{Nakahara00}
Nakahara, M., Isoshima, T., Machida, K., Ogawa, S., Ohmi, T. (2000). 
A simple method to create a vortex in Bose-Einstein condensate of alkali atoms. 
{\it Physica B}, {\bf 284}, 17-18. 

\bibitem{Isoshima00}
Isoshima, T., Nakahara, M., Ohmi, T., Machida, K. (2000). 
Creation of a persistent current and vortex in a Bose-Einstein condensate of alkali-metal atoms. 
{\it Phys. Rev. A}, {\bf 61}, 063610(1-10). 

\bibitem{Nakahara08}
Pietil\"{a}, V., M\"{o}tt\"{o}nen, M., Nakahara, M. (2008). 
Topological vortex creation in spinor Bose-Einstein condensates. 
{\it Electromagnetic, Magnetostatic, and Exchange-Interaction 
Vortices in Confined Magnetic Structures} (Kamenetskii, E. O. eds.). Kerala. 
Transworld Research Network, 297-329.

\bibitem{Ogawa02b}
Ogawa, S. -I., M\"{o}tt\"{o}nen, M., Nakahara, M., Ohmi, T., and Shimada, H. (2002). 
Method to create a vortex in a Bose-Einstein condensate. 
{\it Phys. Rev. A}, {\bf 66}, 013617(1-7). 

\bibitem{Kumakura06}
Kumakura, M., Hirotani, T., Okano, M., Takahashi, Y., Yabuzaki, T. (2006). 
Topological formation of a multiply charged vortex in the Rb Bose-Einstein condensate: 
Effectiveness of the gravity compensation. 
{\it Phys. Rev. A}, {\bf 73}, 063605(1-7). 

\bibitem{Mottonen07}
M\"{o}tt\"{o}nen, M., Pietil\"{a}, V., Virtanen, S. M. M. (2007). 
Vortex pump for dilute Bose-Einstein condensates. 
{\it Phys. Rev. Lett.}, {\bf 99}, 250406(1-4). 

\bibitem{Xu08}
Xu, Z. F., Zhang, P., Raman, C., and You, L. (2008). 
Continuous vortex pumping into a spinor condensate with magnetic fields. 
{\it Phys. Rev. A}, {\bf 78}, 043606(1-8). 

\bibitem{Isoshima01}
Isoshima, T., Machida, K., and Ohmi, T. (2001). 
Quantum vortex in a spinor Bose-Einstein condensate.  
{\it J. Phys. Soc. Jpn.}, {\bf 70}, 1604-1610. 

\bibitem{Isoshima02}
Isoshima, T., and Machida, K. (2002). 
Axisymmetric vortices in spinor Bose-Einstein condensates under rotation.  
{\it Phys. Rev. A}, {\bf 66}, 023602(1-8). 

\bibitem{Yip99}
Yip, S. K. (1999). 
Internal vortex structure of a trapped spinor Bose-Einstein condensate.
{\it Phys. Rev. Lett.}, {\bf 83}, 4677-4681. 

\bibitem{Mizushima02a}
Mizushima, T., Machida, K., Kita, T. (2002). 
Mermin-Ho vortex in ferromagnetic spinor Bose-Einstein condensates.
{\it Phys. Rev. Lett.}, {\bf 89}, 030401(1-4). 

\bibitem{Mizushima02b}
Mizushima, T., Machida, K., and Kita, T. (2002). 
Axisymmetric versus nonaxisymmetric vortices in spinor Bose-Einstein condensates.  
{\it Phys. Rev. A}, {\bf 66}, 053610(1-8). 

\bibitem{Martikainen02}
Martikainen, J. P., Collin, A., and Suominen, K. A. (2002). 
Coreless vortex ground state of the rotating spinor condensate.
{\it Phys. Rev. A}, {\bf 66}, 053604(1-6). 

\bibitem{Bulgakov03}
Bulgakov, E. N. and Sadreev, A. F. (2003). 
Vortex phase diagram of F=1 spinor Bose-Einstein condensates.
{\it Phys. Rev. Lett.}, {\bf 90}, 200401(1-4). 

\bibitem{Mizushima04b}
Mizushima, T., Kobayashi, N., and Machida, K. (2004).
Coreless and singular vortex lattices in rotating spinor Bose-Einstein condensates. 
{\it Phys. Rev. A}, {\bf 70}, 043613(1-10).

\bibitem{Saito06}
Saito, H., Kawaguchi, Y., and Ueda, M. (2006). 
Breaking of chiral symmetry and spontaneous rotation in a spinor Bose-Einstein condensate.
{\it Phys. Rev. Lett.}, {\bf 96}, 065302(1-4). 

\bibitem{Ji08}
Ji, A.-C., Liu, W. M., Song, J. L., and Zhou, F. (2008). 
Dynamical creation of fractionalized vortices and vortex lattices. 
{\it Phys. Rev. Lett.}, {\bf 101}, 010402(1-4). 

\bibitem{Chiba08}
Chiba, H. and Saito, H. (2008). 
Spin-vortex nucleation in a Bose-Einstein condensate by a spin-dependent rotating trap. 
{\it Phys. Rev. A}, {\bf 78}, 043602(1-5). 

\bibitem{Hoshi08}
Hoshi, S. and Saito, H. (2008). 
Magnetization of a half-quantum vortex in a spinor Bose-Einstein condensate. 
{\it Phys. Rev. A}, {\bf 78}, 053618(1-6). 

\bibitem{Pietila08}
Pietil\"{a}, V., M\"{o}tt\"{o}nen, M., Virtanen, S. M. M. (2007). 
Stability of coreless vortices in ferromagnetic spinor Bose-Einstein condensates. 
{\it Phys. Rev. A}, {\bf 76}, 023610(1-7). 

\bibitem{Takahashi09}
Takahashi, M., Pietil\"{a}, V., M\"{o}tt\"{o}nen, M., Mizushima, T., and Machida, K. (2009). 
Vortex-splitting and phase-separating instabilities of coreless vortices 
in F=1 spinor Bose-Einstein condensates. 
{\it Phys. Rev. A}, {\bf 79}, 023618(1-10).

\bibitem{Kita02}
Kita, T., Mizushima, T., and Machida, K. (2002). 
Spinor Bose-Einstein condensates with many vortices. 
{\it Phys. Rev. A}, {\bf 66}, 061601(R)(1-4).

\bibitem{Reijnders04b}
Reijnders, J. W., van Lankvelt, F. J. M., Schoutens, K., Read, N. (2002). 
Rotating spin-1 bosons in the lowest Landau level. 
{\it Phys. Rev. A}, {\bf 69}, 023612(1-23).

\bibitem{Saito07a}
Saito, H., Kawaguchi, Y., Ueda, M. (2007). 
Topological defect formation in a quenched ferromagnetic Bose-Einstein condensates. 
{\it Phys. Rev. A}, {\bf 75}, 013621(1-10). 

\bibitem{Lamacraft07}
Lamacraft, A. (2007). 
Quantum quenches in a spinor condensate. 
{\it Phys. Rev. Lett.}, {\bf 98}, 160404(1-4).

\bibitem{Uhlmann07}
Uhlmann, M., Schutzhold, R., Fischer, U. R. (2007).
Vortex quantum creation and winding number scaling in a quenched spinor Bose gas.  
{\it Phys. Rev. Lett.}, {\bf 99}, 120407(1-4).

\bibitem{Damski07}
Damski, B. and Zurek, W. H. (2007).
Dynamics of a quantum phase transition in a ferromagnetic Bose-Einstein condensate.  
{\it Phys. Rev. Lett.}, {\bf 99}, 130402(1-4).

\bibitem{Saito07b}
Saito, H., Kawaguchi, Y., Ueda, M. (2007). 
Kibble-Zurek mechanism in a quenched ferromagnetic Bose-Einstein condensate. 
{\it Phys. Rev. A}, {\bf 76}, 043613(1-10).

\bibitem{Ciobanu00}
Ciobanu, C. V., Yip, S.-K., and Ho, T. L. (2000)
Phase diagram of F=2 spinor Bose-Einstein condensates.
{\it Phys. Rev. A}, {\bf 61}, 033607(1-5).

\bibitem{Turner2007} 
Turner, A. M., Barnett, R., Demler, E., and Vishwanath, A. (2007). 
Nematic Order by Disorder in Spin-2 Bose-Einstein Condensates. 
{\it Phys. Rev. Lett.}, {\bf 98}, 190404(1-4).

\bibitem{Song2007} 
Song, J. L., Semenoff, G. W., and Zhou, F. (2007). 
Uniaxial and Biaxial Spin Nematic Phases Induced by Quantum Fluctuations. 
{\it Phys. Rev. Lett.}, {\bf 98}, 160408(1-4).

\bibitem{Ueda02}
Ueda, M. and Koashi, M. (2002)
Theory of spin-2 Bose-Einstein condensates: Spin correlations, magnetic response, and excitation spectra.
{\it Phys. Rev. A}, {\bf 65}, 063602(1-22).

\bibitem{Tojo2009} 
Tojo, S., Hayashi, T., Tanabe, T., Hirano, T., Kawaguchi, Y., Saito, H., and Ueda, M. (2009). 
Spin-dependent inelastic collisions in spin-2 Bose-Einstein condensates. 
{\it Phys. Rev. A}, {\bf 80}, 042704(1-7).

\bibitem{Semenoff07}
Semenoff, G. W., and Zhou, F. (2007).
Discrete symmetries and 1/3-quantum vortices in condensates of F=2 cold atoms. 
{\it Phys. Rev. Lett.}, {\bf 98}, 100401(1-4).

\bibitem{Barnett08}
Barnett, R., Mukerjee, S., and Moore, J. E. (2008)
Vortex lattice transition in cyclic spinor condensates.
{\it Phys. Rev. Lett.}, {\bf 100}, 240405(1-4).

\bibitem{Turner09}
Turner, A. M. and Demler, E. (2009)
Vortex molecules in spinor condensates.
{\it Phys. Rev. B}, {\bf 79}, 214522(1-24).

\bibitem{Kobayashi09}
Kobayashi, M., Kawaguchi, Y., Nitta, M., and Ueda, M. (2009)
Collision dynamics and rung formation of non-Abelian vortices.
{\it Phys. Rev. Lett.}, {\bf 103}, 115301(1-4).

\bibitem{Huhtamaki09}
Huhtam\"{a}ki, J. A. M., Simula, T. P., Kobayashi, M., and Machida, K. (2009)
Stable fractional vortices in the cyclic states of Bose-Einstein condensates.
{\it Phys. Rev. A}, {\bf 80}, 051601(R)(1-4).

\bibitem{Griesmaier05}
Griesmaier, A., Werner, J., Hensler, S., Stuhler, J., and Pfau, T. (2005). 
Bose-Einstein condensation of chromium. 
{\it Phys. Rev. Lett.}, {\bf 61}, 160401(1-4). 

\bibitem{Santos00}
Santos, L., Shlyapnikov, G. V., Zoller, P., and Lewenstein M. (2000). 
Bose-Einstein condensation in trapped dipolar gases. 
{\it Phys. Rev. Lett.}, {\bf 85}, 1791-1794.

\bibitem{Yi01}
Yi, S. and You. L. (2001). 
Trapped condensates of atoms with dipole interactions. 
{\it Phys. Rev. A}, {\bf 63}, 053607(1-14).

\bibitem{Stuhler05}
Stuhler, J., Griesmaier, A., Koch, T., Fattori, M., Pfau, T., Giovanazzi, S., 
Pedri, P., and Santos, L. (2005). 
Observation of dipole-dipole interaction in a degenerate quantum gas. 
{\it Phys. Rev. Lett.}, {\bf 95}, 150406(1-4).

\bibitem{Lahayareview09}
Lahaye, T., Menotti, C., Santos, L., Lewenstein, M., and Pfau, T. (2009). 
The physics of dipolar bosonic quantum gases. 
{\it Rep. Prog. Phys.}, {\bf 72}, 126401(1-41). 

\bibitem{Lahaye08}
Lahaye, T., Metz, J., Fr\"{o}hlich, B., Koch, T., Meister, M., Griesmaier, A., Pfau, T., 
Saito, H., Kawaguchi, Y., and Ueda M. (2008). 
d-wave collapse and explosion of a dipolar Bose-Einstein condensate. 
{\it Phys. Rev. Lett.}, {\bf 101}, 080401(1-4). 

\bibitem{Santos03}
Santos, L., Shlyapnikov, G. V., and Lewenstein M. (2003). 
Roton-maxon spectrum and stability of trapped dipolar Bose-Einstein condensates.
{\it Phys. Rev. Lett.}, {\bf 90}, 250403(1-4).

\bibitem{Ronen07}
Ronen, S., Bortolotti, D. C. E., and Bohn, J. L. (2007). 
Radial and angular rotons in trapped dipolar gases. 
{\it Phys. Rev. Lett.}, {\bf 98}, 030406(1-4).

\bibitem{Yi06}
Yi, S., and Pu, H. (2006). 
Vortex structures in dipolar condensates. 
{\it Phys. Rev. A}, {\bf 73}, 061602(R)(1-4).

\bibitem{Wilson08}
Wilson, R. M., Ronen, S., Bohn, J. L., and Pu, H. (2008). 
Manifestations of the roton mode in dipolar Bose-Einstein condensates. 
{\it Phys. Rev. Lett.}, {\bf 100}, 245302(1-4).

\bibitem{ODell07}
O'Dell, D. H. J., and Eberlein, C. (2007). 
Vortex in a trapped Bose-Einstein condensate with dipole-dipole interactions. 
{\it Phys. Rev. A}, {\bf 75}, 013604(1-11).

\bibitem{Abad09}
Abad, M., Guilleumas, M., Mayol, R., Pi, M., and Jezek, D. M. (2009). 
Vortices in Bose-Einstein condensates with dominant dipolar interactions. 
{\it Phys. Rev. A}, {\bf 79}, 063622(1-9). 

\bibitem{Bijnen07}
van Bijnen, R. M. W., O'Dell, D. H. J., Parker, N. G., Martin, A. M. (2007). 
Dynamical instability of a rotating dipolar Bose-Einstein condensate. 
{\it Phys. Rev. Lett.}, {\bf 98}, 150401(1-4). 

\bibitem{Bijnen09}
van Bijnen, R. M. W., Dow, A. J., O'Dell, D. H. J., Parker, N. G., Martin, A. M. (2009). 
Exact solutions and stability of rotating dipolar Bose-Einstein condensates in the Thomas-Fermi limit. 
{\it Phys. Rev. A}, {\bf 80}, 033617(1-11).

\bibitem{Wilson09}
Wilson, R. M., Ronen, S., and Bohn, J. L. (2009). 
Stability and excitations of a dipolar Bose-Einstein condensate with a vortex. 
{\it Phys. Rev. A}, {\bf 79}, 013621(1-7).

\bibitem{Klawunn08}
Klawunn, M., Nath, R., Pedri, P., and Santos, L. (2008). 
Transverse instability of straight vortex lines in dipolar Bose-Einstein condensates. 
{\it Phys. Rev. Lett.}, {\bf 100}, 240403(1-4).

\bibitem{Cooper05}
Cooper, N. R., Rezayi, E. H., and Simon, S. H. (2005). 
Vortex lattices in rotating atomic bose gases with dipolar interactions. 
{\it Phys. Rev. Lett.}, {\bf 95}, 200402(1-4).

\bibitem{Zhang05b}
Zhang, J., and Zhai, H. (2005). 
Vortex lattices in planar Bose-Einstein condensates with dipolar interactions.
{\it Phys. Rev. Lett.}, {\bf 95}, 200403(1-4).

\bibitem{Komineas07}
Komineas, S., and Cooper, N. R. (2007). 
Vortex lattices in Bose-Einstein condensates with dipolar interactions beyond the weak-interaction limit. 
{\it Phys. Rev. A}, {\bf 75}, 023623(1-5).

\bibitem{Vengalattore08}
Vengalattore, M., Leslie, S. R., Guzman, J., Stamper-Kurn, D. M. (2008). 
Spontaneously modulated spin textures in a dipolar spinor Bose-Einstein condensate.
{\it Phys. Rev. Lett.}, {\bf 100}, 170403(1-4). 

\bibitem{Kawaguchi06b}
Kawaguchi, Y., Saito, H., Ueda, M. (2006). 
Einstein-de Haas effect in dipolar Bose-Einstein condensates.
{\it Phys. Rev. Lett.}, {\bf 96}, 080405(1-4). 

\bibitem{Santos06}
Santos, L. and Pfau, T. (2006). 
Spin-3 chromium Bose-Einstein condensates.
{\it Phys. Rev. Lett.}, {\bf 96}, 190404(1-4). 

\bibitem{Gawryluk07}
Gawryluk, K., Brewczyk, M., Bongs, K., Gajda, M. (2007). 
Resonant Einstein-de Haas effect in a rubidium condensate.
{\it Phys. Rev. Lett.}, {\bf 99}, 130401(1-4). 

\bibitem{Yi04}
Yi, S., You, L., and Pu, H. (2004). 
Quantum phases of dipolar spinor condensates. 
{\it Phys. Rev. Lett.}, {\bf 93}, 040403(1-4). 
 
\bibitem{Yi06b}
Yi, S. and Pu, H. (2006). 
Spontaneous spin textures in dipolar spinor condensates.
{\it Phys. Rev. Lett.}, {\bf 97}, 130404(1-4). 

 \bibitem{Kawaguchi06a}
Kawaguchi, Y., Saito, H., Ueda, M. (2006). 
Spontaneous circulation in ground-state spinor dipolar Bose-Einstein condensates.
{\it Phys. Rev. Lett.}, {\bf 97}, 130404(1-4). 

\bibitem{Yarmchuk1982} Yarmchuk, E. J. and Packard, R. E. (1982). Photographic Studies of Quantized Vortex Lines. {\it J. Low Temp. Phys.}, {\bf 46}, 479-515.

\bibitem{Svistunov2001} Svistunov, B. (2001). in {\it Quantized Vortex Dynamics and Superfluid Turbulence} Vol. 571 (Barenghi, C. F., Donnelly, R. J., and Vinen, W. F. ed.). Springer-Verlag, Berlin, 327-333.

\bibitem{Berloff2002} Berloff, N. G. and Svistunov, B. (2002). Scenario of strongly nonequilibrated Bose-Einstein condensation. {\it Phys. Rev. A}, {\bf 66}, 013603(1-7).

\bibitem{Kobayashi2007} Kobayashi, M. and Tsubota, M. (2007). Quantum turbulence in a trapped Bose-Einstein condensate. {\it Phys. Rev. A}, {\bf 76}, 045603(1-4).

\bibitem{White2010} White, A. C., Barenghi, C. F., Proukakis, N. P., Youd, A. J., and Wacks, D. H. (2010). Nonclassical Velocity Statistics in a Turbulent Atomic Bose-Einstein Condensate. {\it Phys. Rev. Lett.}, {\bf 104}, 075301(1-4).

\bibitem{Vincent1991} Vincent, A. and Meneguzzi, M. (1991). The spatial structure and statistical properties of homogeneous turbulence. {\it J. Fluid Mec.}, {\bf 225}, 1-20.

\bibitem{Noullez1997} Noullez, A., Wallace, G., Lempert, W., Miles, R. B., and Frisch, U. (1997). Transverse velocity increments in turbulent flow using the RELIEF technique. {\it J. Fluid Mec.}, {\bf 339}, 287-307.

\bibitem{Gotoh2002} Gotoh, T., Fukayama, D., and Nakano, T. (2002). Velocity field statistics in homogeneous steady turbulence obtained using a high-resolution direct numerical simulation. {\it Phys. Fluids}, {\bf 14}, 1065-1081.

\bibitem{Min1996} Min, I. A., Mezi\'c, I., and Leonard, A. (1996). L\'evy stable distributions for velocity and velocity difference in systems of vortex elements. {\it Phys. Fluids}, {\bf 8}, 1169-1180.

\bibitem{Henn2009a} Henn, E. A. L., Seman, J. A., Ramos, E. R. F., Caracanhas, M., Castilho, P., Ol\'impio, E. P., Roati, G., Magalh\~aes, D. V., Magalh\~aes, K. M. F., and Bagnato, V. S. (2009). Observation of vortex formation in an oscillating trapped Bose-Einstein condensate. {\it Phys. Rev. A}, {\bf 79}, 043618(1-5).

\bibitem{Henn2009b} Henn, E. A. L., Seman, J. A., Roati, G., Magalh\~aes, K. M. F., and Bagnato, V. S. (2009). Emergence of Turbulence in an Oscillating Bose-Einstein Condensate. {\it Phys. Rev. Lett.}, {\bf 103}, 045301(1-4).


\endthebibliography
\end{document}